\documentclass[namedreferences]{SSRv}

\newcommand{\bibs}{\bibitem[\protect\citeauthoryear{}{}]{} }
\newcommand{\ang}{\AA}
\newcommand{\gapprox}{\lower.4ex\hbox{$\;\buildrel >\over{\scriptstyle\sim}\;$}}
\newcommand{\lapprox}{\lower.4ex\hbox{$\;\buildrel <\over{\scriptstyle\sim}\;$}}
\def\arcsec{\hbox{$^{\prime\prime}$}}
\def\arcmin{\hbox{$^{\prime}$}}
\def\etal{{\sl et al.}\ }
\def\copyright{{\ooalign{\hfil\raise.07ex\hbox{c}\hfil\crcr\mathhexbox20D}}}
\def\refe#1   {\par\noindent\hangindent0.5cm {#1}} 
\def\ca#1#2{\par\noindent{#1}{\hfill {#2}}}
\def\cb#1#2{\par{#1}{\hfill {#2}}}
\def\cc#1#2{\par{\small \hskip1cm#1}{\hfill {#2}}}
\def\aaohc{{\sl Ann. Ast. Obs. Harvard College}\ }
\def\aap {{\sl A\&A}\ } 
\def\aapr{{\sl A\&AR}\ } 
\def\acta{{\sl Acta Astron.}\ }
\def\acu {{\sl Astron. Circ. USSR no.}\ }
\def\advaa{{\sl Adv.A\&A}\ }
\def\aj  {{\sl Astron. J.}\ }
\def\ams {{\sl Ann. Math. Stud.}\ }
\def\al  {{\sl Astron. Lett.}\ }
\def\an  {{\sl AN}\ }
\def\apj {{\sl ApJ}\ } 
\def\anyas{{\sl An.. NY Acad. Sci.}\ }
\def\apjs{{\sl ApJS}\ } 
\def\app {{\sl Astroparticle Physics}\ }
\def\apss{{\sl Ap\&SS}\ }
\def\araa{{\sl ARA\&A}\ }
\def\arep{{\sl Astron. Rep.}\ }
\def\as{  {\sl American Scientist}\ }
\def\asr {{\sl Adv. Space Res.}\ }
\def\astl{{\sl Astron. Lett.}\ }
\def\astr{{\sl Astron. Rep.}\ }
\def\aya {{\sl AYA}\ }
\def\azh {{\sl A. Zh.}\ }
\def\basi{{\sl BASI}\ }
\def\beb {{BMC Evol. Bio.}\ }
\def\caosp{{\sl Contr. Astron. Obs. Skalnate Pleso}\ }
\def\cas {{\sl Communic. Asteroseismology}\ }
\def\chjaa{{\sl Chin. J. A\&A}\ }
\def\cr  {{\sl Comptes Rendues}\ }
\def\epsl{{\sl Earth Pl. Sci. Lett.}\ }
\def\gb  {{\sl Genome Bio.}\ }
\def\geo {{\sl Geology}\ }
\def\grl {{\sl GRL}\ }
\def\hpa {{\sl Helv. Phys. Acta}\ }
\def\ica {{\sl Icarus}\ }
\def\jaa {{\sl J. Astrophys. Astron.}\ }
\def\jahh{{\sl J. Astron. Hist. \& Heritage}\ }
\def\jacs{{\sl J. Am. Chem. Soc.}\ }
\def\jcp {{\sl J. Chem. Phys.}\ }
\def\jcs {{\sl J. Chem. Soc.}\ }
\def\jetpl{{\sl JETP Lett.}\ }
\def\jfe {{\sl J. Fusion Energy}\ }
\def\jgr {{\sl J. Geophys. Research}\ }
\def\jrasc{{\sl JRASC}\ }
\def\kfn {{\sl Kinematika i Fizika Nebesnykh Tel}\ }
\def\lrsp{{\sl Living Reviews in Solar Physics}\ }
\def\mnras{{\sl MNRAS}\ } 
\def\msai{{\sl Mem. Soc. Astr. It.}\ }
\def\njp {{\sl New J. Phys.}\ }
\def\ns  {{\sl New Scientist}\ }
\def\oleb{{\sl Origins of Life \& Evol.~Biospheres}\ }
\def\pa  {{\sl Pop. Astron.}\ }
\def\nat {{\sl Nature}\ } 
\def\pasj{{\sl Publ. Astron. Soc. Japan}\ }
\def\pasp{{\sl Publ. Astron. Soc. Pacific}\ }
\def\pdao{{\sl Publ. Dom. Astrophys. Obs.}\ }
\def\pepi{{\sl Phys. of Earth \& Plan. Interiors}\ }
\def\phm {{\sl Phil. Mag.}\ }
\def\phys{{\sl Physica}\ }
\def\pire{{\sl Proc. IRE}\ }
\def\ppnp{{\sl Proc. Part. Nucl. Physics}\ }
\def\pr{{\sl Physical Review}\ }
\def\prl{{\sl Physical Review Letters}\ }
\def\prd{{\sl Physical Review D}\ }
\def\prs{{\sl Proc. Roy. Soc.}\ }
\def\rmaa{{\sl Rev. Mex. A\&A}\ }
\def\rmsc{{\sl Rev. Mex. Ser. Conf.}\ }
\def\rmp{{\sl RMP}\ }
\def\rp{{\sl RP}\ }
\def\sa {{\sl Soviet Anstronomy AJ}\ }
\def\saj{{\sl Serbian Astronomical J.}\ }
\def\sam{{\sl Scientific American}\ }
\def\sci{{\sl Science}\ }
\def\sca{{\sl Spectrochimica Acta}\ }
\def\ssr{{\sl Space Science Reviews}\ }
\def\ssyr{{\sl Solar System Research}\ }
\def\sp {{\sl Solar Phys.}\ } 
\def\zap{{\sl Zs. f. Ap.}\ }

\begin{document}
\begin{article}
\begin{opening}
\rm

\title{ASTROPHYSICS IN 2006}

\author{Virginia Trimble$^1$,
	Markus J.~Aschwanden$^2$,
	and Carl J.~Hansen$^3$}

\runningauthor{TRIMBLE ET AL.}
\runningtitle{ASTROPHYSICS IN 2006}

\institute{$^1$	Department of Physics and Astronomy, University of California, 
		Irvine, CA 92697-4575,
     		Las Cumbres Observatory, Santa Barbara, CA:
		\email{vtrimble@uci.edu}\\
	   $^2$ Lockheed Martin Advanced Technology Center, 
		Solar and Astrophysics Laboratory,
  		Organization ADBS, Building 252,
  		3251 Hanover Street, Palo Alto, CA 94304:
		\email{aschwand@lmsal.com}\\
	   $^3$	JILA, Department of Astrophysical and Planetary Sciences, 
		University of Colorado, Boulder CO 80309:
		\email{chansen@jila.colorado.edu}}

\date{Received ... : accepted ...}

\begin{abstract}
The fastest pulsar and the slowest nova; the oldest galaxies and the youngest
stars; the weirdest life forms and the commonest dwarfs; the highest energy
particles and the lowest energy photons. These were some of the extremes of
Astrophysics 2006.  We attempt also to bring you updates on things of
which there is currently only one (habitable planets, the Sun, and the
universe) and others of which there are always many, like meteors and
molecules, black holes and binaries.
\end{abstract}


\end{opening}


\sl
\section*{CONTENTS}

\ca{1. Introduction}{6}
\cb{1.1 Up}{6}
\cb{1.2 Down}{9}
\cb{1.3 Around}{10}
\ca{2. Solar Physics}{12}
\cb{2.1 The solar interior}{12}
\cc{2.1.1 From neutrinos to neutralinos}{12}
\cc{2.1.2 Global helioseismology}{12}
\cc{2.1.3 Local helioseismology}{12}
\cc{2.1.4 Tachocline structure}{13}
\cc{2.1.5 Dynamo models}{14}
\cb{2.2 Photosphere}{15}
\cc{2.2.1 Solar radius and rotation}{15}
\cc{2.2.2 Distribution of magnetic fields}{15}
\cc{2.2.3 Magnetic flux emergence rate}{15}
\cc{2.2.4 Photospheric motion of magnetic fields}{16}
\cc{2.2.5 Faculae production}{16}
\cc{2.2.6 The photospheric boundary of magnetic fields}{17}
\cc{2.2.7 Flare prediction from photospheric fields}{17}
\cc{2.2.8 Sunspots}{17}
\cb{2.3 Chromosphere and transition region}{18}
\cc{2.3.1 Chromospheric abundances}{18}
\cc{2.3.2 Heating of the chromosphere}{18}
\cc{2.3.3 Chromospheric oscillations}{19}
\cc{2.3.4 Microflaring in the transition region}{20}
\cb{2.4 Corona}{21}
\cc{2.4.1 Coronal atomic physics}{21}
\cc{2.4.2 The coronal magnetic field}{21}
\cc{2.4.3 Magnetic helicity}{22}
\cc{2.4.4 Elementary and composite coronal loops}{23}
\cc{2.4.5 MHD Modeling of coronal loops}{23}
\cc{2.4.6 Coronal heating}{24}
\cc{2.4.7 Coronal oscillations and waves}{25}
\cc{2.4.8 Coronal holes}{26}
\cc{2.4.9 Coronal streamers}{26}
\cb{2.5 Filaments and prominences}{27}
\cc{2.5.1 Quiescent filaments}{27}
\cc{2.5.2 Eruptive filaments}{28}
\cb{2.6 Flares}{28}
\cc{2.6.1 Preflare magnetic field configuration}{28}
\cc{2.6.2 Magnetic field change during flares}{29}
\cc{2.6.3 Magnetic reconnection}{29}
\cc{2.6.4 Particle accleration}{30}
\cc{2.6.5 RHESSI observations}{31}
\cc{2.6.6 Oscillations and waves in flares}{32}
\cc{2.6.7 Self-organized criticality in solar flares}{33}
\cb{2.7 Coronal mass ejections (CMEs)}{33}
\cc{2.7.1 CME initiation and magnetic field configuration}{33}
\cc{2.7.2 Propagation of CMEs}{34}
\cc{2.7.3 CME-initiated waves}{35}
\cc{2.7.4 Geo-effectiveness of CMEs}{35}
\cc{2.7.5 Solar energetic particles}{36}
\cb{2.8 Heliosphere}{37}
\cc{2.8.1 Solar wind}{37}
\cc{2.8.2 Interplanetary radio bursts}{38}
\cc{2.8.3 Termination shock}{38}
\cb{2.9 Solar-Terrestrial relations}{38}
\cc{2.9.1 Global warming, ozone, and cloud-free skys}{39}
\cc{2.9.2 Geomagnetic storms}{39}
\cc{2.9.3 Solar activity cycle}{39}
\ca{3. Galaxies near and far}{40}
\cb{3.1 Home sweet home: The Milky Way and Local Group}{40}
\cb{3.2 The Local Group grows}{41}
\cb{3.3 MW $\neq$ M31}{42}
\cb{3.4 Minor members}{43}
\cb{3.5 The Milky Way and other spirals}{44}
\cc{3.5.1 Arms}{44}
\cc{3.5.2 Bars}{45}
\cc{3.5.3 Disks}{45}
\cc{3.5.4 Stellar populations}{46}
\cc{3.5.5 Bulges and halos}{46}
\cb{3.6 Interactions}{47}
\cb{3.7 The first galaxies and how they grow}{48}
\cb{3.8 Broad brushes}{49}
\cb{3.9 Some special galaxies}{52}
\cb{3.10 Residual problems}{54}
\cb{3.11 Clusters of galaxies}{55}
\cb{3.12 Between the galaxies}{57}
\ca{4. All turtles that on Earth do dwell}{58}
\cb{4.1 Inside the Earth}{58}
\cb{4.2 Atmospheres and climates}{60}
\cb{4.3 Other species}{61}
\cb{4.4 And the wisdom to know the difference}{62}
\cb{4.5 Let us now praise famous persons}{66}
\cb{4.6 Oh, dear. I'm really rather glad I didn't say that!}{68}
\cc{4.6.1 The minor awards}{69}
\cc{4.6.2 The minor malfunctions}{70}
\cc{4.6.3 With a little help from our friends}{72}
\ca{5. Interstellar matter, star formation, young stellar objects, and chemical evolution}{74}
\cb{5.1 The interstellar medium}{74}
\cb{5.2 Star formations}{77}
\cc{5.2.1 First things and modes}{78}
\cc{5.2.2 Changes with redshift}{78}
\cc{5.2.3 Initial conditions and their causes}{79}
\cc{5.2.4 The initial mass function}{80}
\cc{5.2.5 Stars of large and small masses}{81}
\cb{5.3 Young stellar objects}{81}
\cb{5.4 Chemical evolutions}{82}
\cc{5.4.1 The largest scales}{83}
\cc{5.4.2 Galaxies and stellar populations}{83}
\cc{5.4.3 Individual processes and nuclides}{85}
\ca{6. Stars}{87}
\cb{6.1 Location, location, location}{87}
\cb{6.2 Ordinary things about ordinary stars}{88}
\cb{6.3 Stellar structure and evolution calculations}{89}	
\cb{6.4 The chemically peculiar stars}{91}
\cb{6.5 Brown dwarfs}{92}
\cb{6.6 Real time stellar evolution}{93}
\cb{6.7 Motions in space}{95}
\cb{6.8 Stellar rotation and activity}{96}
\cb{6.9 Pulsating stars}{98}
\cb{6.10 Sad tails of the deaths of stars}{100}
\cb{6.11 White dwarfs}{101}
\cb{6.12 Single black holes}{103}
\ca{7. Astrobiology}{104}
\cb{7.1 Reviews, Books, and Awards}{104}
\cb{7.2 SETI Lives}{105}
\cc{7.2.1 The ATA}{106}
\cc{7.2.2 BOINC}{107}
\cc{7.2.3 The Drake equation}{108}
\cb{7.3 Updates, Mendings, and Miscellaneous}{109}
\cc{7.3.1 Water on Mars?}{109}
\cc{7.3.2 Archeal musings}{109}
\cc{7.3.3 Let's have a little fun}{110}
\ca{8. More than $1$ but less than $10^6$}{111}
\cb{8.1 Binary stars}{111}
\cc{8.1.1 Binaries in stellar populations}{111}
\cc{8.1.2 Binary statistics and triples}{112}
\cc{8.1.3 Binary evolution}{114}
\cc{8.1.4 Cataclysmic variables}{115}
\cc{8.1.5 Binary black holes}{117}
\cb{8.2 Star clusters}{118}
\cc{8.2.1 Open clusters, moving groups, and star streams}{118}
\cc{8.2.2 KREDOS}{119}
\cc{8.2.3 Globular Clusters}{120}
\cb{8.3 Other sorts of star clusters}{123}
\ca{9. Silent, upon an peak in DARIEN: Exo and endo planets}{124}
\cb{9.1 Exoplanets}{124}
\cc{9.1.1 Extrema}{124}
\cc{9.1.2 Search methods}{125}
\cc{9.1.3 Statistical properties, hosts vs. planets}{126}
\cc{9.1.4 Formation mechanisms}{127}
\cb{9.2 Endoplanets and the rest of the solar system}{128}
\cc{9.2.1 Dust and meteors}{128}
\cc{9.2.2 Comets and asteroids}{129}
\cc{9.2.3 Moons}{132}
\cc{9.2.4 Major planets, general confusion, and colonel Deshafy}{134}
\ca{10. Biggish bangs}{136}
\cb{10.1 Supernovae, their remnants, and gamma ray bursts}{136}
\cc{10.1.1 Supernovae}{136}
\cc{10.1.2 Some other things about supernovae}{138}
\cc{10.1.3 Supernovae remnants}{139}
\cc{10.1.4 Gamma ray bursters}{140}
\cb{10.2 The pulsars of the nations}{141}
\cc{10.2.1 More pulsars singles and singularities}{142}
\cc{10.2.2 Binary neutron stars}{145}
\cb{10.3 Black blogs: QSOs, AGNs, and all}{147}
\cc{10.3.1 More certain answers}{147}
\cc{10.3.2 Less certain answers}{149}
\cc{10.3.3 An absorbing topic}{151}
\ca{11. OAO}{152}
\cb{11.1 Countdown}{152}
\cb{11.2 First}{159}
\cb{11.3 Extrema}{161}
\cb{11.4 Astronomical extrema}{162}
\cb{11.5 Familiar physics, expected effects, and wonderful widgets}{165}
\cc{11.5.1 Physical principles}{165}
\cc{11.5.2 Processes}{167}
\cc{11.5.3 Catalogs}{168}
\cc{11.5.4 Sites and observatories}{169}
\cc{11.5.5 Widgets}{169}
\ca{12. Cosmology}{172}
\cb{12.1 A child's garden of cosmological models}{172}
\cb{12.2 The conventional universe}{176}
\cb{12.3 Distance indicators and gravitational lensing}{179}
\cb{12.4 Dark matter}{180}
\cb{12.5 Dark energy}{182}
\cb{12.6 The backgrounds}{182}
\cb{12.7 Cosmic rays}{186}
\cb{12.8 Very large scale structure and streaming}{187}
\ca{13. Misteaks were made}{190}
\cb{13.1 We done it}{190}
\cb{13.2 They done it}{191}
\ca{Acknowledgements}{195}
\ca{References}{196-239}

\clearpage

\rm
\section{INTRODUCTION}

{\sl Astrophysics in 2006} modifies a long tradition by moving to a new journal,
which you hold in your (real or virtual) hands.  The fifteen previous articles in
the series are referenced occasionally as {\sl Ap91} to {\sl Ap05}
below and appeared in volumes 104-118 of
{\sl Publications of the Astronomical Society of the Pacific}.  The ground rules are
fairly simple: we read a lot, decide what we think is important, and tell you about
it. Used in compiling Sections 3-6 and 8-13 were the issues that arrived as paper
between 1 October 2005 and 20 September 2006 of 
{\sl Nature}, {\sl Physical Review Letters},
{\sl Science}, {\sl Astrophysical Journal} (plus {\sl Letters} and {\sl Supplements Series}), 
{\sl Monthly Notices of the Royal Astronomical Society}, 
{\sl Astronomy and Astrophysics}, {\sl Astronomical Journal}, 
{\sl Acta Astronomica}, {\sl Revista Mexicana Astronomia y Astrofisica}, 
{\sl Astrophysics and Space Science}, {\sl Astronomy Reports}, 
{\sl Astronomy Letters}, {\sl Astronomische Nachrichten}, 
{\sl Publications of the Astronomical Society of Japan}, 
{\sl Publications of the Astronomical Society of the Pacific}, 
{\sl Journal of Astrophysics and Astronomy}, 
{\sl Bulletin of the Astronomical Society of India}, 
{\sl Contributions of the Astronomical Observatory Skalnate Pleso}, and {\sl IAU Circulars}.
Read less systematically (sometimes because only sporadically available) and cited
irregularly were {\sl Observatory}, {\sl Astrofizica}, 
{\sl New Astronomy} (plus {\sl Reviews}), 
{\sl Journal of the American Association of Variable Star Observers}, 
{\sl Astronomy and Geophysics}, {\sl Mercury}, {\sl New Scientist}, 
{\sl Science News}, {\sl American Scientist}, {\sl Scientometrics}, 
{\sl Monthly Notes of the Astronomical Society of South Africa}, and 
{\sl Journal of the Royal Astronomical Society of Canada}. 
Additional journals provided material for Sections 2 and 7, and
are mentioned there.

Many of the discussions start with (but move in random directions from) some
particularly exciting item, which is described as deserving a green circle (the identifier
in the senior author's notebook) or gold star or red flag.  It is not required that
the reader agree with us (though experience indicates that the authors cited quite
often do).

\subsection{Up}

Some of these are launches, some first lights or extended operations, and others
miscellaneous good news.

\begin{itemize}
\item[o] {ESA's Venus Express was launched from Baikonur on 9 November 2005 and entered orbit
on 11 April, though apparently with a spectrometer mirror jammed ({\sl Science} {\bf 312}, 827).}

\item[o] {SALT (South African Large Telescope) was officially opened in November, 2005
({\sl Nature} {\bf 438}, 18; also reporting the curious factoid that Richard Woolley was the
only South African born director of SAAO to date; and yes, they missed the chance to
appoint a very good one this time around). First light comes later.}

\item[o] {Auger caught its first ultrahigh energy cosmic rays in November 
({\sl Nature} {\bf 438}, 270).
The facility is safely back in Argentina from an unknown site in Chile, to which one
of your authors accidentally transported it in an earlier review.}

\item[o] {Japan began putting CO$_2$ samplers on commercial flights in November 
({\sl Nature} {\bf 438}, 266)
with the intention of providing closely spaced four-dimensional data.}

\item[o] {The secular (as opposed to {\sl ApXX}) year ended with the launch of the first satellite of
the European GPS equivalent on 28 December.}

\item[o] {Stardust returned safely on 15 January ({\sl Nature} {\bf 459}, 255), 
though what it was carrying
was really comet dust, from Wild 2, and the connection with Woolley, just mentioned,
is in name only.}

\item[o] {New Horizons took off for Pluto on 19 January.}

\item[o] {A successful launch of ASTRO-F, the second Japanese infrared satellite on 21 February
led to its being renamed AKARI. First light was on 13 April, and an all-sky survey
is underway.}

\item[o] {FUSE was back up and working in February 2006.}

\item[o] {Cerro Pachon was chosen as the site for the Large Synoptic Survey Telescope in
March ({\sl Nature} {\bf 441}, 397).  LSST will not be lonesome, since Gemini South and SARS
are already there. We confess to having rooted for San Pedro Martir.}

\item[o] {April was an uncruel month, with initial operations of CARMA (the union of radio
millimeter facilities made out of OVRO and BIMA, {\sl Nature} {\bf 441}, 141), the initiation of
optical SETI at Oak Ridge Observatory (Harvard), and the first 10 dishes of the Allen
(SETI)  Telescope Array in place at Hat Creek.  Rebuilding of Mount Stromlo Observatory
is in progress ({\sl Science} {\bf 312}, 684), with the Great Melbourne Telescope to become  Sky
Mapper and the reconstructed near infrared integral field spectrograph already dispatched 
to Gemini.}

\item[o] {The operation of SoHO was extended to 2009 in a May decision ({\sl Nature} {\bf 441}, 562).  It
sometimes discovers sun-grazing comets as well as So's and Ho's.}

\item[o] {Construction began at about the same time on Antares, a French neutrino detector
under the Mediterranean ({\sl Science} {\bf 312}, 1305).  International collaborations are hard at
work trying to figure out how to distinguish French neutrinos from those of other
nationalities.  We suspect it has something to do with the sauce.}

\item[o] {PAMELA took off from Baikonur on 15 June ({\sl Nature} {\bf 441}, 920).  Its job is to look for
positrons and anti-protons in cosmic rays from any source.}

\item[o] {MAGIC (which detects TeV photons by Muggles\footnote{Anyone who hasn't heard that
muggles are non-magicians in the series of novels centered on Harry Potter probably
isn't going to like this review very much anyway.} methods) has seen a source 
known also to
HESS and INTEGRAL ({\sl ApJ} {\bf 637}, L41).  As it is the largest single-dish atmospheric Cerenkov
detector and the authors ranged from Albert to Zapatero, this item could have appeared
in several other sections.}

\item[o] {The Japan Aerospace Exploration Agency (JAXA) launched the sun-observing
satellite Solar-B on September 23 and renamed it to Hinode, the Japanese word for
sunrise. Its sharp vision with 0.2 arc-second resolution makes it like a ``Hubble
for the Sun'' (NASA press release).}

\item[o] {Awards given in September by the {\sl Astronomical Society of the Pacific} included
outreach and amateur achievement to colleagues from Iran and the Czech Republic.}

\item[o] {AREX, the Atacama Pathfinder Experiment at Chajnantor produced a package of 20
scientific letters ({\sl A\&A} {\bf 454}, L13 and following). We would have called the facility a
trial balloon for ALMA, but even our linguistically elastic minds rebelled at that
bit of cognitive dissonance.}

\item[o] {The World/International Year of Astronomy in 2009 will also see Darwin's 200th
birthday (12 February, just like A. Lincoln) and the 150th anniversary of publication
of {\sl Origin of the Species}. The competition will surely be good for science education
in general, though perhaps bad for astronomy specifically.  Students have reminded us
that, in addition  to  the first telescope observations, 1609 also witnessed the
publication of Kepler's {\sl New Astronomy} (which made him something like the 14th
Copernican in world history).}

\item[o] {NASA announced a selection of Discovery Program concepts for future study.
They are an asteroid sample return, Venus chemistry and dynamics orbiter, mapping
of gravitational fields on the moon, and possible continuing use of DeepImpact to fly
to a second comet or visit Temple 1 and look for changes since it Deeply Impacted.}
\end{itemize}

\subsection{Down} 

We have not succeeded in putting these in chronological order, and indeed were
in some cases undecided whether a particular entity belonged here or in 1.3.

\begin{itemize}
\item[o] {The last Western Union telegram was sent on 27 January 2006 ({\sl Los Angeles Times},
8 February, p. B13).  The first came on 24 May 1844 from Samuel F.B. Morse. Nokia will
keep its SMS (Short Message Service) in Morse.  What this means for astronomy is that
if NAS Domestic Secretary Abbott were attempting to stage the Curtis-Shapley debate
this year, he would have to figure out some other way to invite the participants.}

\item[o] {China has stopped selling lunar real estate for \$37 per acre ({\sl Nature}
{\bf 438}, 269).
We missed our chance to buy some but have hopes that stones from the Great Wall might
make their appearance on the open market, like the ones from London Bridge some years
ago (yes; it serves as a candle snuffer).}

\item[o] {Marshall Field disappeared into Macy's this year and will be mourned by all who
remember {\sl The Imperturbability of Elevator Operators} by S. Candlestickmaker.
Overcivility may never happen again ({\sl Hecht's Filene's, Woodward and Lathrop, Robinsons},
and {\sl The May Company}, where Mary Livingston worked, are other recent losses).}

\item[o] {Italy is said to be eliminating 20,000 tenure track jobs for young researchers 
({\sl Science} {\bf 310}, 761) 
which was described as somehow an improvement. We didn't know they had 
that many.}

\item[o] {A language dies every ten days with its last fluent speaker ({\sl Nature}
{\bf 438}, 148)}.

\item[o] {We caught at least two papers being retracted out from under their first authors,
who either disagreed or were not consulted ({\sl Nature} {\bf 437}, 940; {\sl Science}
{\bf 310}, 49;
{\sl Science} {\bf 310}, 425).  One of the first authors concerned plans to sue.}

\item[o] {Folklore describes trying to catch up with an ill-founded rumor as like trying to
chase down all the feathers from a pillow torn open in a high wind, but we have not
actually seen any feather chasers contradicting the gossip column version ({\sl Science}
{\bf 313}, 1032; {\sl Nature} {\bf 442}, 859) version of the departures of Wesley Huntress, Eugene
Levy, and Charles Kennel from the NASA Science Advisory Comittee. Their error
seems to have been in advising that science continue to be done.}

\item[o] {In contrast, while George Coyne has indeed retired as director of the Vatican
Observatory and been replaced by a younger Jesuit astronomer and AAS member, Jose
Gabriel Funes ({\sl Nature} {\bf 442}, 970; {\sl Science} {\bf 313}, 1031), 
the gossip columns attributed
far too much discredit to various parties in what was merely a long-intended,
orderly succession of leadership. We have it from the horse's mouth, or anyhow the
horse's email.}

\item[o] {SMART-1, the first European moon orbiter, plunged into the moon on 2 September, as
planned.  Genesis came down on 8 September.}

\item[o] {Hayabusa (``Falcon''), which was supposed to rendez-vous with asteroid Itokawa, is the
most difficult to classify, and was the subject of at least half a dozen news items
through the year. A series of eight short papers ({\sl Science} {\bf 312},  1326) testifies that
it got there and imaged the asteroid as a sort of rubble pile (R.A. Lyttleton would
have been pleased), but it gave its life in the process, and, ``drained of its
lifeblood'' ({\sl Science} {\bf 311}, 1859), it will not return with the samples that were the main
purpose of the mission ({\sl Nature} {\bf 439}, 137).  The same news item also recaps the demises
of some other recent Japanese missions, Nozumi to Mars (December 2005), Midori-II
(an Earth observatory), and the second try at ASTRO-E, called Suzuku.}

\item[o] {{\sl Sky and Telescope} has been employee-owned since its 1941 founding by the Federers.
But both numbers of subscriptions and ad pages had declined 20\% or more from 2000 to
2006, And, by unanimous vote of shareholders, it was sold on 10 February to {\sl New Track
Media LLC} for a price not disclosed to the public or to non-share-holding employees.
Immediately after the vote to sell, about one-quarter of the 50-plus employees were laid
off. The long-occupied buildings at 49 Bay Street were sold to {\sl AAVSO} a couple of 
months after the end of the index year.}
\end{itemize}

\subsection{Around} 

The {\sl American Physical Society} will keep its name ({\sl Science} {\bf 310}, 1612), 
but ``you're
not going to see it anymore,'' according to the President, but only assorted logos
incorporating {\sl APS}, except on publications intended for its own members.  On the happy
side, an out-of-period editorial in {\sl Physical Review Letters} makes clear that they will
still accept papers on paper, equaling in this respect the overcivility of {\sl Nature} 
({\sl Nature} {\bf 437}, 952).

The University of Sussex will keep its chemistry department ({\sl Nature} {\bf 441}, 397),
though the new arrangement (the  British word ``scheme'' is liable to be misunderstood
by speakers of American) for determining the amount of government support each university
will receive ({\sl Nature} {\bf 441}, 917) may soon undermine it again.

Hard to know whether an obituary is a good thing or not.  It means, generally, that
you are dead, though typically no deader than you were before (despite some evidence
that an American scientist isn't really dead until {\sl The New York Times} says
so).  But we are sorry that {\sl Physics Today} will be publishing many fewer (whether physicists
or astronomers) in the future ({\sl Physics Today} 58, No.{\bf 10}, p. 10). 
There will be an on-line
archive instead, in which we hope we will not see you for a long, long time.

ESA, not (yet) anything like as fickle as NASA, has delayed the launch of COROT
({\sl Nature} {\bf 442}, 970) to change rockets, and postponed some
other missions, but also extended XMM and INTEGRAL to 2010.

As for the green spiral of Archimedes that contains our notes on NASA, the clearest
statement from the director has been that the Moon and Mars have priority over science
funding ({\sl Nature} {\bf 441}, 134). Within that science funding, it seems that JWST and HST will
eat the lion's share, though JWST's launch may retreat to 2018 ({\sl Science} {\bf 310}, 1594),
and the short-fall in the Shuttle budget was initially \$5G for fiscal 2006 ({\sl Science}
{\bf 310},
957).  Within those priorities, the Mars Sample Return has nevertheless be delayed
indefinitely ({\sl Science} {\bf 311}, 1540), with other delays for Terrestrial Planet Finder, Space
Interferometry Mission, and all ...  And, though science is, in some sense, getting
a different \$5G per year ({\sl Nature} {\bf 440}, 127), four missions that were ready to start
construction have been sent back to the scrum line to recompete for one new start
({\sl Science} {\bf 311}, 1540).  They are LISA (the interferometric detector for long wavelength
gravitational radiation), JDEM (the Joint Dark Energy Mission, for which there are, in
turn, three contenders), Constellation-X, and BlackHole Finder.  The only sort of
individual action that seems likely to affect these circumstances is political voting,
writing to the appropriate people in government, perhaps seeking to become a person in
government, for the total cost of JWST will be something like one million dollars per
full member of the AAS, and very few of us could pay our share.


\section{SOLAR PHYSICS}

\subsection{The Solar Interior}

\subsubsection{From Neutrinos to Neutralinos}

After the solar neutrino problem has been solved thanks to
the three-flavor detection with the Sudbury Neutrino Observatory
in 2001, while Raymond Davis Jr. garnered the Noble Price in Physics
2002 for it, and the uncrowned John Bahcall died in 2005, it became
a bit more quiet around the solar neutrinos. 
Sophisticated power-spectrum analysis of solar neutrino fluxes exhibits 
quasi-periodic modulations at a frequency of 9.43 yr$^{-1}$ (38.7 days) 
(Sturrock et al.~2005; Sturrock and Scargle 2006), a result that was 
attributed to r-mode oscillations. 

A new hotly-debated problem is the elemental abundance in the solar 
interior, raising the question whether the solar interior is metal-poor 
(e.g., in Mg, Si, Fe) or metal-rich (e.g., in Ne; Bahcall et al.~2005) 
relative to its surface, which can be constrained 
by solar neutrino flux measurements that include the effects
of neutrino flavor mixing (Gonzalez 2006a). [By the way, the solar metallicity 
dropped also almost by a factor of two (Z=0.0122) in new 3D hydrodynamic 
models of the photosphere (Asplund et al.~2006a).]

As a side note, muonic neutrinos can also be used to detect neutralino
annihilation, as found in the galactic center, where neutralinos 
(the lightest stable super-symmetric particle) are considered as a good 
candidate for cold dark matter, but no excess of muonic neutrinos has been
found in the sun (Ackermann et al.~2006). 

\subsubsection{Global Helioseismology}

The almost 10-year time series of GONG and SoHO/MDI data have proven to be
sensitive enough to probe subtle solar dynamo features, such as the flow
patterns of the solar convection zone (Howe et al.~2006). SoHO/MDI analysis 
shows also that the ``helioseismic radius'' (related to the subsurface 
stratification) seems to vary strongest just below the solar surface, around 
0.995 $R_{\odot}$, and the radius of deeper layers between 0.975 and 0.99
$R_{\odot}$ seem to vary in phase with the 11-year cycle (Lefebvre 
and Kosovichev 2005). 

Gravity waves (g-modes) are still not detected yet, although low-frequency
propagating gravity waves and high-frequency standing modes are predicted
from theoretical simulations (Rogers and Glatzmaier 2005).

\subsubsection{Local Helioseismology}

Local helioseismology is performed now with an increasing number of techniques
(Gizon and Birch 2005):
Fourier-Hankel decomposition, ring-diagram analysis (Hernandez et al.~2006), 
time-distance helioseismology (Jensen et al.~2006), 
helioseismic holography (Zhao and Kosovichev 2006), 
and direct modeling. 
Applications of local helioseismology include large-scale flows 
(Ha\-nasoge et al.~2006), meridional circulation (Hernandez et al.~2006;
Zaatri et al.~2006), 
the solar-cycle dependence of the internal structure (Juckett 2006; 
Serebryanskiy and Chou 2005; Verner et al.~2006),
perturbations associated with regions of magnetic activity, 
and solar supergranulation (Green and Kosovichev 2006).

A major improvement in the local helioseismology ``travel-time method'' is 
the use of the Born approximation in the 3D inversion of the sound speed 
below a sunspot, which takes finite-wavelength effects into account and 
thus is more accurate (Couvidat et al.~2006; Gizon et al. 2006) and more 
appropriate (Duvall et al.~2006) than the previously used ray-path approximation.
Helioseismic holography methods start to resolve subphotospheric layers
with a vertical thickness as thin as $\approx 1$ Mm (Braun and Birch 2006).   
The helioseismic propagation of sunquakes has been simulated with a
compressible fluid model and is in agreement with observations within 10-20\%
(Podesta 2005).

\subsubsection{Tachocline Structure}

How is the tachocline, the thin transition layer between the convection
zone and radiation zone at $\approx 0.7 R_{\odot}$ formed? MHD simulations
show that meridional flows with speeds of about 10 m s$^{-1}$ penetrating
to a maximum depth of 1000 km below the convection zone are able to
generate an almost horizontal magnetic field in the tachocline region so
that the internal field is almost totally confined to the radiative zone
(Kitchatinov and R\"udiger 2006). Simulations reveal that strong axisymmetric
toroidal magnetic fields (about 3000 G) are realized within the lower
stable layer (Browning et al.~2006), while peak toroidal fields up to 
100,000 G have been simulated in the overshoot tachocline 
(Dikpati et al.~2006). The tachocline is extremely stable.
3D simulations show that any part of the tachocline that is radiative
is found to be hydrodynamically stable against small perturbations 
(Arlt et al.~2005). However, the global MHD shallow-water instability
of differential rotation and toroidal field bands in the tachocline is
still thought to provide a mechanism for the formation and evolution of
active longitudes seen in synoptic maps (Dikpati and Gilman 2005).
Active longitudes are now also explained by ``stroboscopic effects'',
where a rotating active region is illuminated by an activity wave
propagating from mid-laditudes to the equator, mimicking this way 
a differential rotation (Berdyugina et al.~2006). 

Simulating the rise of cylindrical fluxtubes through the convective zone
predicts a ring of reverse current helicity on the periphery of active
regions, which seems to agree with observations (Chatterjee et al.~2006).
The concept of fluxtubes, however, can break down when magnetic field
lines wander chaotically over a large volume (Cattaneo et al.~2006)
or undergo asymmetric vortex shedding (Cheung et al.~2006). 
Disconnection of emerging fluxtubes from their parent magnetic structures
in the tachocline is thought to take place in a depth of 2000-6000 Mm
below the photosphere (Sch\"ussler and Rempel 2005).
Tiny entropy variations in the order of $\approx 10^{-5}$ in the tachocline
are thought to influence the differential rotation, by causing latitudinal
temperature variations of about 10 K (Miesch et al.~2006). 

\subsubsection{Dynamo Models}

Dynamo models are driven by the motion of conducting fluids, which can
amplify energy at small scales when magnetic fields are stretched. One
such mechanism is the turbulent cascade, which also transfers magnetic
helicity from large to small scales (Alexakis et al.~2006), but the
turbulent transport is anisotropic (Kim 2005a). The current helicity of
observed active regions seems to follow a hemispheric rule, but violations
of this rule seem to increase with the anchor depth of active regions
(Zhang et al.~2006c). Statistics of 17,200 vector magnetograms revealed
that strong fields ($B>1000$ G) show that both alpha and current helicity
present a sign opposite to that of weak fields ($B<500$ G, which follow
the hemispheric rule), suggesting that the solar dynamo produces opposite
helicity signs in the (weak) mean field and in the (strong field) fluctuations
(Zhang 2006). Opposite polarities were also found for strong and weak 
field regions of coronal holes (Zhang et al.~2006a).

Flow fields in the solar interior have quite different responses to the
stochastic fluctuations of the Reynolds stress, and thus consequently the
time scale for the replenishment of the differential rotation ($\approx 10$
years) is nearly 4 orders of magnitude longer than the time scale for the
replenishment of the meridional flow (Rempel 2005). Interestingly, the
meridional flow is found to be directed toward the poles in the longer
sunspot cycles and largely toward the equator in the shorter cycles
(Javaraiah and Ulrich 2006). Now we think that meridional flow is central
to the cycle reversal and to the dynamo itself (Sheeley 2005).

Theoretical simulations of a Babcock-Leighton-type flux-transport dynamo
model manage to correctly predict the relative peaks of 8 solar cycles
(cycle \# 16-23), which also demonstrate that the high-latitudinal fields
have a memory of at least 3 cycles (Dikpati and Gilman 2006). A key
parameter is the magnetic Prandtl number, which, if small, can reproduce
``grand minima'' or prolonged phases of significantly reduced magnetic
activity (Bushby 2006).

\clearpage

\subsection{Photosphere}

\subsubsection{Solar Radius and Solar Rotation}

Previous measurements of the solar radius have claimed temporal variations,
but recent studies confirm the constancy of the solar radius, independent
of the solar cycle or other factors (Lefebvre et al.~2006) and blame
previous inconsistencies of astrolabe measurements on ``data analysis biased
by theoretical preconcpetions, by empirical results which without scientific
arguments are considered as canonical references and by over-interpretation
of casual agreements between visual and CCD astrolabe results'' (Noel 2005).
Alternatively, the solar radius can also be determined from helioseismology,
particularly from f-modes, which have been measured with a precision of 
$10^{-5}$,
but this method was found not to be accurate enough to detect changes in the
solar radius (Sofia et al.~2005). 

Synoptic maps using Carrington Coordinates have been in use for at least
hundred years, but we find only now a paper that corrects the non-synchronicity
of the spherical wrapping for differential rotation (Ulrich and Boyden 2006).
On the other side, there are still papers published that correct the 
differential rotation rate evaluated by Galileo Galilei in 1612 
(Casas et al.~2006). Over the last century (1874-1981) a secular
deceleration (with 3$\sigma$ significance) of the solar rotation rate has 
been detected (Brajsa et al.~2006). 

\subsubsection{Distribution of Magnetic Field Strengths}

A statistical study quantified the distribution of magnetic field strengths
from 0 to 1800 Gauss in the quiet sun. The magnetic energy density was found
to amount to a significant fraction of the kinetic energy of the granular
motions at the base of the photosphere ($\gapprox 15\%$), and the quiet-sun
photosphere was found to have far more unsigned magnetic flux and magnetic
energy than the active regions and the network together (Dominguez Cerdena
et al.~2006). The mean magnetic field strength of the quiet sun is estimated
to be 20 G, but high-resolution simulations of the 
Zeeman signal render it as a lower limit (Sanchez-Almeida 2006). 
On the other side, collisional depolarization in the MgH lines of the
``second solar spectrum'' yields an upper limit of 10 G for the hidden magnetic
field in the quiet sun photosphere, assuming the simplest case of microturbulence
that fills the entire upflowing photospheric volume (Asensio-Ramos and 
Trujillo-Bueno 2005). 

\subsubsection{Magnetic Flux Emergence Rate}

Why do coronal holes form? Emerging magnetic flux from the solar interior
occurs in form of magnetic bipoles and thus contains closed fields,
but there exists currently no theoretical model that explains the transition
from closed fields generated in low-latitude active regions to open fields
in coronal holes during the poleward diffusion of the solar cycle.
While the diffusive
transport of open fields was predicted to depend on the rate of emergence of
new magnetic flux in a recent model (Fisk 2005), analysis of SoHO/MDI and EIT
data indeed revealed that the emergence rate in coronal holes is about a
factor of two lower than in adjacent quiet-sun regions (Abramenko et al.~2006).
The poleward meridional flow can also be enhanced by magnetic reconnection
processes (Cohen et al.~2006).

\subsubsection{Photospheric Motion of Magnetic Fields}

The random walk of the footpoints of magnetic field lines in the photosphere,
caused by the subphotospheric magneto-convection was always considered as a
key ingredient of coronal heating. However, although active regions display 
orders of magnitude higher heating rates, the random motion tracked from
so-called ``moving magnetic features'' was found to be slower in active
regions than in the quiet-sun network (with a mean of 1.11 km s$^{-1}$), 
and thus the heating rate does not scale proportionally to the random motion  
velocity (M\"ostl et al.~2006). Standard local correlation tracking methods
were found to be inconsistent with the magnetic induction equation
(Schuck 2005, 2006). However, combining horizontal and line-of-sight velocity
measurements, the 3D convection pattern of emerging flux regions could be
reconstructed and a depth of 600 km and a turn-over time of two hours was
found (Kozu et al.~2006), while other authors arrived at a similar granular
cell model from inversion of high-resolution slit spectrogram time series
(Puschmann et al.~2005). 

Regarding magneto-convection on larger scales, 
MHD simulations of small-scale magnetoconvection show that fluxtubes
collapse to kG field strengths that form intermittent
intergranular lanes and collect over the boundaries of the underlying
mesogranular scale cells (Stein and Nordlund 2006; De Wijn et al.~2005).
The advected magnetic flux in the mesogranular boundaries are very long-lived,
they have an average lifetime of about 9 hours (De Wijn et al.~2005).
The spatial clustering of granules, which occur on the size of a supergranulation
cell, is now also explored with ``hexagonal normalized information entropy''
(Berilli et al.~2005). 
Measurements of supergranulation rotation rates have been criticized to be 
too high due to projection effects (Hathaway et al.~2006).

\subsubsection{Faculae Production}

High-resolution observations shed some new light on the physical mechanism
of faculae production. The emergence of faculae from the
sub-photospheric turbulent boundary layer, which show a
characteristic hairpin substructure, are thought to be produced by a fluid
dynamic instability, called ``vortex/shear layer interaction'', 
causing stretching of the vortices and spiral-sheating that explains also
other observed features like ``hot walls'' and ``dark lanes'' (Falco 2006).
Other high-resolution observations (0.1\arcsec) revealed the rapid evolution
of faculae, including merging, splitting, rapid motion, apparent fluting, 
and possible swaying (De Pontieu et al.~2006). 

\subsubsection{The Photospheric Boundary of Magnetic Fields}

The coronal magnetic field is generally calculated from extrapolations
of the field measured at the photospheric boundary. In the photosphere,
strong magnetic flux is concentrated at the boundaries of supergranulation
cells, also called network lanes in line emission. However, the vertical
expansion of the network boundary with height is still a challenge for
modeling and depends sensitively on the surrounding mixed-polarity 
background fields, deviating significantly from the funnel expansion model
(Aiouaz and Rast 2006). 

\subsubsection{Flare Prediction from Photospheric Fields ?}

Since magnetic energy seems to be the ultimate energy source for solar
flares and the photospheric magnetic field can be measured easily, 
a method for predicting flares based on photospheric magnetogram data
was always considered as highly desirable. Traditionally,
classifications of sunspot morphology helped, where a $\delta$-spot configuration
has the highest probability for flaring. Recent studies track the evolution
of twisting and shearing in fast emerging magnetic flux regions that lead
up to a $\delta$-spot and calculate the related helicity injection. It is
found that the strong shear motion produces more magnetic helicity than
the twisting motion (Liu and Zhang 2006). The sign of the injected
helicity in active regions is found to be predominantly uniform
(Pariat et al.~2006). Rapid changes of the magnetic gradient along
a flaring neutral line were found during flares, thought to be indicators
of magnetic reconnection events (Wang 2006a).

\subsubsection{Sunspots}

The details on the street maps of solar sunspots are getting better and sharper. 
Diameters of 170 km are inferred for umbral dots (Sobotka and Hanslmeier 2005),
100-300 km for magnetic fluxtubes in sunspot penumbrae (Borrero et al.~2006), 
300-500 km for dark penumbral filaments (Rimmele and Marino 2006), 
and sizes of 600-1000 km for moving magnetic features around sunspots, 
with average lifetimes of about 1 hr (Hagenaar and Shine 2005).
Vertical magnetic field gradients of 0.5-1.5 G km$^{-1}$ and current densities 
of $\pm40$ mA m$^{-2}$ have been measured in sunspots (Balthasar 2006),
while a magnetic scale height of 6900 km was measured in coronal heights,
where the field drops from 1750 G to 960 G in a height range from 8000 km
to 12,000 km (Brosius and White 2006). However, the 3D magnetic topology
of a sunspot is not as simple as a potential field of a vertical coil;
Flux ropes of opposite helicity were found to coexist in the same spot
and to form complex topologies with opposite-sign torsions (Socas-Navarro 2005),
not to talk about the interlocking-comb structure in the outer penumbra
(Thomas et al.~2006).

The understanding of the dynamics in sunspots requires a lot of velocity 
measurements. Doppler shifts in penumbrae were found to increase with depth 
in the photosphere up to 1.5 km s$^{-1}$ (Bellot Rubio et al.~2005), and the
velocity stratification was found to be consistent with the model of
penumbral fluxtubes channeling the Evershed flow (Bello Rubio et al.~2006).
Upflows with 2 km s$^{-1}$ were observed above the sunspot umbra and
downflows with 4 km s$^{-1}$ above magnetic pores (Zuccarello et al.~2005).
Doppler shifts in a bright plume above a sunspot revealed downflows of
up to 52 km s$^{-1}$ (Brosius and Landi 2005), and outflowing (coronal) jets with
speeds of up to 300 km s$^{-1}$ were spotted near a sunspot (Lin et al.~2006).

We are used to hearing about typical temperatures of 4500-6000 K in sunspots,
but much lower values have been inferred in absorbing layers of molecular lines 
in sunspot umbrae, as low as 1240 K (Bagare et al.~2006). 

Spectropolarimetric studies of the oscillation power in sunspots suggest that
the 3-min period is a direct signal of upward propagating slow magnetoacoustic
shock waves, rather than beats or overtones of the global 5-min oscillations
(Centeno et al.~2006; Marsh and Walsh 2006; McEwan and De Moortel 2006). 
The 3-min oscillations are not confined to the umbra or
plume, but can be seen in many bright locations outside the umbra, but are
often too weak to be seen in an unfiltered signal (Lin et al.~2005).
Running penumbral waves and umbral flashes are now considered as closely
related oscillatory phenomena in sunspots (Tziotziou et al.~2006). 

\subsection{Chromosphere and Transition Region}

\subsubsection{Chromospheric Abundances}

Revisions of elemental abundances in the chromosphere depend crucially on
theoretical 3D hydrodynamic models of convection, such as the solar carbon
monoxide (CO) and OH molecules in the midphotosphere (Ayres et al.~2006),
or the solar metallicity (Asplund et al.~2006a). 

The broad emission line in the solar spectrum near 117 nm detected by SUMER 
has been identified as broad autoionization transitions of neutral sulfur 
(Avrett et al.~2006).

\subsubsection{Heating of the Chromosphere}

How is the chromosphere heated? By dissipation of strong acoustic shocks,
magnetic reconnection in nanoflares, or by dissipation of convection-driven 
Pederson currents? 
Well, high-frequency waves have been detected with TRACE
in the frequency range of 5-50 mHz and the integrated acoustic energy flux was 
estimated to be 255 W m$^{-2}$, which is about a factor 10 too low to heat the
chromosphere (Fossum and Carlsson 2006). 2D MHD simulations, however, reveal
that upward propagating acoustic waves can produce significant shock heating 
by creation of vortical motions at the edges of fluxtubes (Hasan et al.~2005),
something that escaped previous 1D MHD simulations (Ulmschneider et al.~2005).
Also the inclined magnetic field lines at the boundaries of supergranules
have been identified as ``portals'' of low-frequency ($<5$ mHz) magnetoacoustic
waves that can heat the chromosphere with a four times higher energy flux than
that carried by the high-frequency ($>5$ mHz) acoustic waves 
(Jefferies et al.~2006).	

Alternatively, a scenario of upward-propagating fast-mode
MHD waves (of mHz frequencies) that trigger the Farley-Buneman plasma
instability was proposed, which produces anomalous resistivity and wave
energy dissipation that can heat the chromosphere as well as absorb the
p-modes in magnetic regions (Fontenla 2005). The coupling between the
high plasma-$\beta$ in the chromosphere with the low plasma-$\beta$ in
the transition region is thought to be most effiicient to generate currents
driven by resistive stresses (Ryutova 2006). 
Also simulated was how
large-amplitude shear Alfv\'en waves excite quasi-electrostatic waves
driven by the modified two-stream instability that can heat the chromosphere
(Sakai and Saito 2006). If two current loops with counterhelicity collide
in the upper chromosphere they can launch jets upward into the corona
(Sakai et al.~2006), perhaps as observed in the form of a surge with an upward
speed of 50 km s$^{-1}$ (Tziotziou et al.~2005) or 240 km s$^{-1}$
(Van Noort et al.~2006).

\subsubsection{Chromospheric Oscillations}

Intermittent chromospheric and transition region oscillations with periodicities
in the 4-7 min range as well as coronal oscillations in the 2-5 min range
were detected in He I, O V, Mg IX, with time delays of $27\pm5$ s for He I
with respect to O V, which was interpreted in terms of compressive waves
that propagate downward from the transition region to the chromospheric
network, possibly generated by coronal nanoflares (G\"om\"ory et al.~2006).
Vice versa, oscillations in the transition region were detected in the 
Ne VIII line above a supergranular cell of the quiet sun chromospheric network, 
possibly generated by chromospheric nanoflares (Gontikakis et al.~2005).
Besides these detections of chromospheric oscillations in UV, radio observations
with BIMA in mm wavelengths (85 GHz) show also consistent periods of 5 min
in the network and 3 min in the internetwork (White et al.~2006).

A different type of oscillation, interpreted in terms of fast-mode MHD
oscillations in the kink mode, has also been detected in spicules, 
with periods of 35-70 s, wavelengths of 3500 km, and heights of 3800-8700 km,
suspected to be excited by granular motions (Kukhianidze et al.~2006). 

\subsubsection{Microflaring in the Transition Region}

Heating and cooling of
the chromosphere and lower transition region seems to occur so rapidly,
that one observer concludes, based on the physical disparity of hotter and 
cooler structures, that the time scale of heating and cooling are shorter
than 20 s (Doschek 2006). Also theoretical hydrodynamic simulations show
that time scales of 20 s, heated in intervals of 100 s, and spatial heating
scales in the order of 1 Mm produce the closest agreement with observed
structures of the quiet Solar EUV transition region (Spadaro et al.~2006).  
Repetitive occurrence of explosive (speak ``reconnection'') events at 
a coronal hole boundary were detected with an initial period of 3 min
that increased to 5 min, probably triggered by transverse kink oscillations
of a neighbored (closed field) fluxtube (Doyle et al.~2006a). 
A high-velocity downflow with a speed of 75 km s$^{-1}$ was detected in O V,
supposedly originating from a reconnection event in the transition region
(Doyle et al.~2006). 
The radiative and kinetic energies of so-called ``Ellerman bombs'' were
estimated to be $10^{26}-5\times 10^{27}$ erg, which puts them into the
category of ``submicroflares'' (Fang et al.~2006).
Spectro-polarimetric observations identify ``Ellerman bombs'' as strong
downflows in a hot layer between the upper photosphere and the lower chromosphere
(Socas-Navarro et al.~2006a).
Microflares with energies of $10^{26}-10^{27}$ erg have been detected
in Fe XIX with SUMER, at temperatures of $T>6$ MK (Wang et al.~2006e).
Microflares are also detected simultaneously in hard X-rays and radio,
with a thermal spectrum in the energy range of $\approx 3-10$ keV 
and a nonthermal spectrum above (Kundu et al.~2006).
To simplify the cluttered nomenclature of microflaring phenomena, 
it is gratifying to see that a unification of ``blinker events" and 
``macrospicules'' seems to be in sight (Madjarska et al.~2006).

Mircoflaring with dissipated energies in the range of $10^{26}-10^{28}$ erg
has also been inferred from strongly peaked differential emission measure
distributions ($T\approx 10^7$ K, $n_e \lapprox 10^{13}$ cm$^{-3}$) of
stars (Cargill and Klimchuk 2006). On the other hand, a detailed analysis
of the first vacuum ultraviolet (VUV) emission line profile originating
from the chromosphere through the transition region all the way up to
the corona, obtained with SUMER from the sun and compared with stellar
spectra, led to the conclusion that the broadened line wings are a
consequence of structures in the magnetic chromospheric network,
rather than microflaring (Peter 2006). As a consequence, $\alpha$Cen
is found to have a considerably higher amount of magnetic flux
concentrated in the chromospheric magnetic network than the sun
(Peter 2006). 

\subsection{Corona}

\subsubsection{Coronal Atomic Physics}

Resonance scattering (e.g., Sahal-Brechot and Raouafi 2006) 
has often been invoked to explain the disagreement
between the observed and predicted Fe XVII line ratios, but recent
laboratory measurements seem to rule out this interpretation, which
makes the corona more transparent and explains the fuzziness of 
coronal Fe XV images (at 284 \ang ) simply by a filling factor of
crowded loops (Brickhouse and Schmelz 2006; Keenan et al.~2006).
Our understanding of the
atomic physics of coronal emission lines is also reassured by the
excellent agreement between the Fe XIV green line index and a model index
based on EUV differential emission measure (DEM) maps (Cook et al.~2005),
except that the discrepancies for the Fe XVIII and Fe XIX still exist in
the CHIANTI database (Landi and Phillips 2006). And the
neon abundance of stellar X-ray spectra and solar interior models is 
still in trouble with photospheric abundances (Young 2005).

Of course, although we are not sure whether the coronal plasma has Maxwellian 
temperature distributions, there is always the possibility of non-thermal
kappa distributions (Dzifcakova 2006). Some models even claim that
the prominent temperature anisotropy of the velocity distributions of O VI
lines established by SoHO/UVCS could be ambiguous (Raouafi and Solanki 2006).
Strangely, a normal coronal loop with a typical 10\% He abundance at 
the well-mixed chromospheric footpoint can theoretically accumulate 
in the coronal part 10 times more helium than the hydrogen density, 
if you give it enough time, say 1-3 days (Killie et al.~2005), which seems 
to rule out the existence of such long-lived (steady-state) ``quiescent'' 
loops. Of course, the ubiquitous flows would prevent any such long-lived steady state.
``$\Pi\alpha\nu\tau\alpha\ \rho\epsilon\iota$'' (everything flows), 
as the old Greeks say! 

\subsubsection{The Coronal Magnetic Field}

Although potential field extrapolations are easy, cheap, and fast to
``guestimate'' the coronal magnetic field, they often do not match the
observed coronal structures seen in EUV and soft X-rays, in particular
at interesting locations such as sites of current sheets, filaments, 
flares, and CMEs. This challenge has been taken up by a number of modelers
to come up with better magnetic field computation methods, preferentially
nonlinear force-free (NLFF) extrapolations. New MHD simulations demonstrate that 
the currents associated with emerging fluxtubes are exclusively field-aligned 
and thus the coronal field is force-free (Leake and Arber 2006).
New efforts quantify the 
performance of force-free methods using known solutions (Barnes et al.~2006),
comparing optimization, magnetofrictional, Gradi-Rubin based 
(Amari et al.~2006), and Green's function-based methods (Schrijver et al. 2006), 
the Wheatland-Sturrock-Roumeliotis method (Inhester and Wiegelmann 2006), 
the direct boundary integral formulation method (Yan and Li 2006),
the minimum dissipation rate method (Hu and Dasgupta 2006), 
the resolution of the $180^\circ$ degree ambiguity (Metcalf et al.~2006),
application of the virial theorem (Wheatland and Metcalf 2006),
using stereoscopic (Wiegelmann and Inhester 2006) or tomographic
constraints (Kramar et al.~2006), 
preprocessing of vector magnetograph data at the bottom boundary 
(Wiegelmann et al.~2006), but good solutions require also the knowledge 
of the lateral and top boundaries of the computation box 
(Wiegelmann et al.~2006a). Actually, also MHD effects such as the 
temperature distribution at the coronal base (i.e., at the lower boundary
of the computation box) affect the extrapolated magnetic field (Hayashi et 
al.~2006). Also the magnetic field of individual
coronal loops still poses a most intriguing enigma, because the
observed loop structures do not fan out with height as much as the 
theoretically calculated linear force-free extrapolation predicts 
(Lopez Fuentes et al.~2006).

\subsubsection{Magnetic Helicity}

Magnetic helicity became an important parameter to quantify 
the complexity of
the sheared, twisted, or writhed magnetic field compared with its
lowest energy state (potential field). Also it plays a keyrole in
MHD because magnetic helicity is almost perfectly conserved on a 
timescale less than the global diffusion time, so it is an ideal
hydromagnetic invariant. However, direct measurements of the 
magnetic helicity in the photosphere, corona, and in interplanetary 
space became feasible only recently, either from the electric
current structures observed in photospheric fields, or from the
geometry/morphology/topology of filaments, prominences, and fluxropes.  
There are at least three definitions of the magnetic helicity: (1) the
self helicity of the closed field, (2) the mutual helicity between the
closed field and reference field, and (3) the vacuum helicity (self-helicity
of the reference field) (Regnier et al.~2006). 
For closed field lines, the mutual helicity is their relative winding around  
each other, also known as ``Gauss linkage number'' (Demoulin et al.~2006). 
Interestingly, the complex topology of an active region (i.e., the 
mutual helicity between the closed and potential fields) can be more 
important for the eruption than the twist (i.e., the self-helicity of
the closed field) (Regnier and Canfield, 2006). Also expelled plasmoids were 
found to have external linkages to the flux system (Archontis et al.~2006).
A theoretical difficulty in studying solar helicity is the free gauge 
of the magnetic vector potential (Low 2006), although a gauge-independent
formulation has been proposed for the solar dynamo (Subramanian and 
Brandenburg 2006). 

\subsubsection{Elementary and Composite Coronal Loops}

Why is it so hard to study the physics of coronal loops?
The first problem is an observational one, namely to isolate a clean
uncontaminated loop strand from the background soup of thousand other
loops, and the second problem is a theoretical one, that we have no
consensus about the heating function, nor on its spatial or temporal
properties. In order to deal with the first (observational) problem, the 
best bet nowadays is to use at least three different temperature filters 
(since two-filter ratios are not unique anyway; Weber et al.~2005). 
A TRACE triple-filter analysis yielded 
three criteria to discriminate between elementary and composite
loops, namely: (1) iso-thermality within $dT\lapprox 0.2$ MK,
(2) small loop widths of $w\lapprox 2$ Mm, and (3) a faint contrast
of $\lapprox 30\%$ relative to the background flux (Aschwanden 2005).
The first two conditions might even give us a clue that the heating
occurs in the well-mixed transition region, because this would explain
the isothermality and cross-section of coronal fluxtubes as a result of
magnetic connectivity to the photospheric (convective) granulation cells 
(Aschwanden and Nightingale 2005). 
Coronal loops may have even very sharp edges no thicker
than a few meters, if we want to explain the depolarization of metric 
radio bursts that occur by reflection off boundary layers no thicker
than about a wavelength (Melrose 2006). 
Images from the SoHO/CDS instrument
have a temperature diagnostic over a larger temperature range than TRACE, 
but the insufficient spatial resolution of CDS never resolves a 
multi-thermal bundle of elementary loop strands, as revealed  by TRACE 
(e.g., see Gontikakis et al.~2006a), 
and thus always yields a broad differential emission measure (DEM)
or temperature distribution for these composite loops (see, e.g., see 
Brosius 2006; Schmelz and Martens 2006; Reale and Ciaravella 2006). 

\subsubsection{MHD Modeling of Coronal Loops}

Spectroscopic studies of coronal loops have detected
a short-lived (redshifted) siphon flow in a cold loop with a speed of 
120 km s$^{-1}$ (Doyle et al.~2006b), siphon flows in a cool loop
seen in Ne VIII (Gontikakis et al.~2006a), downflows in both EUV and H$\alpha$
(De Groof et al.~2005), blueshifts of $\gapprox 30$ km s$^{-1}$ in a 
transequatorial loop (Brosius 2006), 
and a statistical nonthermal
velocity of $\approx 17$ km s$^{-1}$ in heights of $\approx 150$ Mm
(Singh et al.~2006). Forward modeling of the corona with a full MHD
code, driven by random footpoint motion and braiding of the magnetic field,
actually demonstrates that intermittent heating causes rapid
variabilities of the Doppler shifts, but comparatively slow changes
in the evolving emissivity, which calls for faster spectroscopic
observations (Peter et al.~2006). Nanoflare models predict high-speed
evaporative upflows, which may be manifested as nonthermal broadening
in spectral line profiles (Patsourakos and Klimchuk 2006). 

Modeling the heating function of a loop is a bit more tricky, because
the spatial and temporal dependence of the heating function is subject
to theoretical models, and thus even a close reproduction of the 
observed temperature and density evolution of a loop does not allow
for a unique identification of the heating function. Nevertheless,
new models have been tried using intermittent but spatially uniformly 
distributed nanoflares caused by MHD turbulence (Reale et al.~2005), or
filamentary loop systems with spontaneous reconnection at tangential
discontinuities (Petrie 2006).
Another hydrostatic simulation of loops with a volumetric heating 
function of $\epsilon_H \propto \langle B \rangle /L$, where $\langle B \rangle$ is the 
magnetic field strength $B$ averaged over the loop length $L$,
consistent with the field-braiding reconnection model of Parker (1983),
was found to yield the closest agreement with the observed brightness
of coronal loops in soft X-ray temperatures (Warren and Winebarger 2006),
but the authors surmise that this static model ignores the intermittent
heating seen in EUV and that the retrieval of the heating function
is not unique either. 
Even studies on the detailed time evolution and relative delays in
different temperature filters brought up troubling results, such as
observed lifetimes that are much longer than expected from theoretical
cooling times, which could not be modeled with either steady or
intermittent heating (Ugarte-Urra et al.~2006), i.e., cooling times
in the order of hours (e.g., Reale and Ciaravella 2006). 
A time-dependent reconstruction of hydrodynamic loop simulations was
even attempted with Kalman filtering of a tomographically sampled
corona (Frazin et al.~2005).

\subsubsection{Coronal Heating}

There seems a consensus has been formed that the convection-driven random
footpoint motion in the photosphere provides the main energy source for
heating of coronal loops: One study calculates the index of the magnetic
power spectrum and the magnetic energy dissipation rate of turbulent 
convection and finds a clear correlation with the coronal soft X-ray flux
(Abramenko et al.~2006a). Stokes polarimetry reveals that the footpoints
of hot and cool loops are located near the periphery of small magnetic
concentrations, such as pores and azimuth centers, having a field strength
of 1000-1800 G and a spatial size of $2\arcsec - 5\arcsec$, i.e. 1500-3500 km (Nagata et
al.~2006). 

Several studies scrutinized the popular nanoflare heating model
of Parker. One study found that random footpoint \`a la Parker alone does 
not induce current-sheet formation, without orchestrated motions involving
topological features in the background such as nulls or separators
(Craig and Sneyd 2005). A reviewer of coronal heating models proposes
a modification of Parker's scenario, where a secondary instability sets in
after saturation of the primary (tearing-mode) instability, which 
becomes nonlinear and transitions to turbulence and produces intense
heating pulses with durations of $\approx 100$ s and energies of 
$10^{24}-10^{25}$ erg (Klimchuk 2006). 

There are still the two main camps of solar heating theories: the
DC (direct current) camp, and the AC (alternate current) or wave heating camp.
Although the DC camps seems to gain momentum with Parker's original or
modified approach, the wave heating camp argues that high-frequency
($\gapprox 0.3$ Hz) fast modes could heat the corona (Kumar et al.~2006;
Ofman 2005), 
that there is evidence of magnetoacoustic waves in off-limb polar
regions from time series analysis of SoHO/CDS data (O'Shea et al.~2006),
a non-negligible fraction of Alfv\'en wave energy is dissipated inside
the corona if both reflection and transmission is properly included
(Malara et al.~2005),
or that the observed correlation of the heating rate with the total unsigned
magnetic flux in a decaying active region could be better explained with
wave heating rather than nanoflaring (Milligan et al.~2005),
since the total power to accelerate the solar wind scales linearly with
the magnetic flux anyway (Schwadron et al.~2006).
We think that
the DC vs. AC debate can be reconciled by proper territory division, if we
donate active regions to the DC models, and coronal holes (together with
the fast solar wind) to the AC models,
which leaves us with the quiet corona as unclaimed territory, but there are
many large-scale connections to active regions, even in form of
transequatorial loops, which ensure that the quiet corona is mostly heated
from the exhausts of active regions. 

Besides the tug-a-war between the DC and AC camps, there is no shortage
of fresh ideas for new coronal heating theories, such as ``reverse
conduction up temperature gradients'' (Ashbourn and Woods 2006),
Rechester-Rosenbluth diffusion with cross-field transport (Galloway et al.~2006), 
a nonmodal cascade self-heating wave-heating process (Shergelashvili 
et al.~2006), Alfv\'enic wave heating near nullpoints and separatrices
where waves accumulate (McLaughlin and Hood 2006), 
or the application of a conventional laboratory MHD generator 
to the solar corona (Tsiklauri 2005, 2006, 2006a). The jury is still out
on these new cases. 

\subsubsection{Coronal Oscillations and Waves}

New studies on oscillations and waves in the solar corona explore
mostly second-order effects and the physics of wave damping: 
the effects of curvature on radially polarized fast modes, which bring
out besides the well known MHD (kink and sausage) modes new families of
vertical, swaying (longitudinal), and rocking modes (Diaz et al.~2006),
the effects of intermittent and random heating on oscillatory patterns
of moving plasma blobs (Mendoza-Briceno and Erdelyi 2006),
eigen-modes of magnetic coronal arcades (Mikhalyaev 2006), 
strong and weak damping of slow MHD modes in hot loops (Pandey and
Dwivedi 2006), the evolution of damped transverse oscillations into
torsional oscillations (Terradas et al.~2006), 
the propagation of MHD waves in magnetically twisted fluxtubes
(Erdelyi and Carter 2006), 
the damping in vertically polarized fast MHD kink mode oscillations
(Verwichte et al.~2006b), the effects of
thermal conduction on acoustic waves in coronal loops (Bogdan 2006),
or viscous damping in the unbounded corona (Kumar and Kumar 2006).

Numerical simulations studied the resonant damping of fast MHD kink
modes, finding a coupling between quasi-modes and resistive Alfv\'en
waves in inhomogeneous loops (Arregui et al.~2005), elliptical loops 
which have different damping rates and some MHD oscillations 
modes that behave differently (Diaz 2006), or the modes excited
by impulsively generated MHD waves (Ogrodowczyk and Murawski 2006;
Selwa et al.~2006). Other studies investigated the 
excitation of trapped and leaky modes (Terradas et al.~2005),
and the damping by wave leakage and tunneling 
(Brady et al.~2006; Cally 2006; Verwichte et al.~2006, 2006a). 
Arguments against wave 
leakage to be unphysical because of initial-value calculations 
(Ruderman and Roberts 2006) were countered to be incorrect (Cally 2006).

Upward propagating waves along loops anchored near sunspots were detected
with 5-min as well as 3-min periods, having propagation speeds ($\approx
100$ km s$^{-1}$) corresponding to (longitudinal) acoustic waves, and
carrying insufficient energy ($\approx 300$ erg cm$^{-2}$ cm$^{-1}$) to
heat the corona, thought to be driven by leakage of the global 5-min 
p-modes (McEwan and DeMoortel 2006). The same result was reported four years
earlier, but a new detail was the fact that adjacent loop strands with 
diameters of a few 1000 km oscillate independently for short time periods. 

\subsubsection{Coronal Holes}

Magnetic funnels in coronal holes are believed to be the primary heating
source of the fast solar wind, by viscous and resistive dissipation of
Alfv\'en waves (Dwivedi and Srivastava 2006; Ofman 2005). 
Outflow velocities in solar plumes have been measured out to 2.4 solar radii
with SoHO/UVCS, and the mass flow was found to decrease with height
(Gabriel et al.~2005).
MHD modeling of coronal funnels with different forms of the heating 
function suggests that they can be discriminated from different 
variations of the blue-shift across the funnel position (Aiouaz et al. 2005). 

The boundaries of coronal holes are places with most likely reconnection
(or disconnection) between open and closed field lines, and thus they
are thought to be the sources of (gradual) solar energetic particles
(SEP) (Shen et al.~2006) and $^3$He-rich events (Wang et al.~2006i;
Pick et al.~2006). 

\subsubsection{Coronal Streamers}

Streamers are associated with the folds in the heliospheric current 
sheet, but additional features are needed to describe its 3D structure,
such as additional folds of the neutral line with secondary plasma
sheets (Saez et al.~2005).
The morphology of coronal streamers were generally described in terms
of a system of cusp-shaped arches with a single radially-oriented
ray (or ``stalk'') on top, but now there is evidence for double
rays, which raises some interesting questions about potential
instabilities (Eselevich and Eselevich 2006). A magnetic island 
may form that travels outward, driven by the diamagnetic force due
to the radial magnetic field gradient, spit out like a melon seed
(Rappazzo et al.~2005). The density is found to vary by one order of
magnitude within the streamer structure (Thernisien and Howard 2006).
Flow speeds at the boundaries of streamers
are unmeasurable at 2.5 solar radii and ramp up to 100 km s$^{-1}$ 
at 5 solar radii (Suess and Nerney 2006). 

\subsection{Filaments and Prominences}

\subsubsection{Quiescent Filaments}

We continue to study the structure and stability of filaments in the
quiescent stage (before some of them become eruptive and self-destructive), 
like doctors taking the pulse of critically ill patients. We are still
questioning how filaments can stably be supported by magnetic fields with dips, 
and whether the dips can  be generated by sufficiently strong twisting of
magnetic fluxtubes (Anzer and Heinzel 2006; Heinzel and Anzer 2006), 
or by a multi-step process
of photospheric shear, reconnection, and formation of quasi bald patch
separatrix surface (Aulanier et al.~2006; Fan and Gibson 2006).
First observations of bald patches in a filament channel and at a barb
endpoint were actually reported from the THEMIS instrument 
(Lopez Ariste et al.~2006).
Although called ``quiescent'', filaments can be subject to substantial
``hick-ups'' during their mass loading. 
Mass transport at much higher velocity ($\gapprox 30$ km
s$^{-1}$) then ever has been detected, with long periods ($\ge 80$ min), 
and large distances ($\gapprox 40,000$ km) (Jing et al. 2006).
Even slower oscillations with periods of 5-6 hrs were detected,
probably corresponding to the slow MHD kink mode (Pouget et al.~2006), 
and repeated motions in a loop near a prominence were also observed
(Kucera and Landi 2006).
Numerical MHD simulations actually show that condensations in nearly
horizontal fluxtubes are most likely to produce transient high-speed
motions, and as a consequence, elongated threads, supporting low-twist
models and thermal nonequilibrium as the origin of prominence
condensation (Karpen et al.~2006, 2006a). Temperatures of 8300 K and
7600 K were measured with SUMER in prominences (Parenti et al.~2005),
as well as a first prominence atlas with 550 line profiles was also
presented from SUMER (Parenti et al.~2005a). 

\subsubsection{Eruptive Filaments}

The identification of the driving force that disrupts filaments and CMEs,
a prerequisite for space weather prediction, is a
million-dollar question. In the model of Chen and Krall it is the
toroidal Lorentz hoop force that drives the filament eruption in the
magnetic structure underlying a CME, which was successfully tested
from the altitude dependence of the acceleration in 13 eruptive
prominences (Chen et al.~2006c). Other equilibrium loss mechanisms
include supercritically twisted flux ropes undergoing kink instability
(Gibson and Fan 2006, 2006a),
also known as torus instability in spheromaks (Kliem and T\"or\"ok 2006),
which can be driven by the helicity injection of rotating sunspots
(Tian and Alexander 2006; De Moortel and Galsgaard 2006),
or octopole background configurations with a magnetic energy larger
than the fully open field energy (Ding and Hu 2006, Ding et al.~2006), 
or the magnetic Rayleigh-Taylor instability at the top of an emerging
fluxtube (Isobe et al.~2006), which does need to involve a coronal nullpoint
(Li et al.~2006b). 

In one case an oscillatory motion has been observed in a polar crown
filament before it erupted, suggesting that the eruption was triggered
by fast magnetic reconnection rather than by slow photospheric shearing
(Isobe and Tripathi 2006). Another study finds a high correlation between
the flux rope acceleration and the thermal energy release rate, even  
when the reconnection rate is fast (Reeves 2006). 
The triggering of a filament eruption and CME seems to involve a much
larger environment than previously thought, for instance there appears
a transequatorial filament in the famous Bastille Day event
that appears to play a key role
(Wang et al.~2006k). After the filament eruption, dimming regions
are often observed witch mark the evacuated feet of the disrupted flux rope
(Jiang et al.~2006a). 

\subsection{Flares}

The fact that we organize the following part of this review into two
sections entitled ``flares'' and ``CMEs'' does not mean that we consider
these phenomena as two different animals, it merely helps us to organize
the information at the lower (coronal base) and upper part (extended 
corona and heliosphere) of the same complex phenomenon that should
rather be called ``corona-magnetic instability'' or ``coronal catastrophe''. 

\subsubsection{Preflare Magnetic Field Configuration}

The flare productivity is correlated with measures of the magnetic
complexity and nonpotentiality, such as the maximum horizontal
magnetic field gradient (Wang et al.~2006l), the length of the 
neutral line, the number of singular points (Cui et al.~2006), 
the total magnetic energy dissipated over the active region area
(Jing et al.~2006a),
the ``effective distance'', which describes the degree of the 
isolation or mutual penetration of the two magnetic polarities 
of an active region (Guo et al.~2006),
or the maximum unsigned zonal and meridional vorticity components
of active regions (Mason et al.~2006). 
Persistent strong horizontal and vertical shear flows
along the neutral line were observed during 5 hours before an X10 flare
(Deng et al.~2006).

Quasi-separatrix layers (QSL), broadly defined as volumes in which field 
lines locally display strong gradients of connectivity, are considered as
preferential locations of current sheet formations, with the strongest
currents developing at the thinnest QSLs, in the hyperbolic fluxtube
separator (Aulanier et al.~2005). Preflare magnetic configurations are
often so complex that the principal trigger cannot be identified: In an
X17 flare two pre-events were observed, one related to a localized flux
emergence, and another involving a large-scale quadrupolar reconnection
(Schmieder et al.~2006). In another event such
complex multi-loop interactions were observed which could be consistent with 
either emerging, colliding, or magnetic breakout (Sui et al.~2006).
During one flare it was concluded that in the initial phase was clear
evidence for tether-cutting reconnection, while the later evolution
followed the standard flare model or magnetic break-out model
(Yurchyshyn et al.~2006). Evidence for the magnetic breakout process
was also found from remote brightenings that required this kind of
large-scale magnetic coupling (Liu et al.~2006f). 

\subsubsection{Magnetic Field Change During Flares}

The shear angle of conjugate H$\alpha$ and hard X-ray footpoints
was found to decrease during a flare, indicating the magnetic
relaxation of the sheared magnetic field (Ji et al.~2006a).
A statistical study revealed 42 sites (during 15 X-class flares) 
where the magnetic field changed by a median value of 90 G within 10 min
during the main flare phase (Sudol and Harvey 2005). 
An interesting diagnostic of the magnetic field in flare loops can 
be provided by applying Alfv\'en soliton models to radio observations of
parallel-drifting frequency bands, which yielded in one case
magnetic field strengths of 130-270 G (Wang and Zhong 2006). 

\subsubsection{Magnetic Reconnection}

Should we not call this ``magnetic newconnection'', since the magnetic
topology changes to a new configuration that is rarely a re-arrangement
back to a previously existing combination. Moreover, since entropy is
increasing in all dissipative processes, a new magnetic configuration
is not a reversible process, so we can not re-connect to an old
state with higher energy. Every theoretician probably agrees that
a magnetic instability represents a switch from an unstable high energy
state into a low energy state, because electromagnetic energy is dissipated. 
In the framework of ``general magnetic
reconnection'' (GMR) theory there exists at least one field line that
has an integrated parallel electric field that is non-zero and the
reconnection rate is given by the maximum value of the potential
of the parallel electric field along this special magnetic field line,
independent on the exact magnetic topology (Hesse et al.~2005). 

Investigating specific magnetic topologies that lead to magnetic
reconnection we find numerical MHD simulations of 3D reconnection
driven by rotational footpoint motions, which build up strong currents
concentrated along separatrix surfaces (De Moortel and Galsgaard 2006),
3D MHD simulations of an emerging bipolar region, where the 
initially rising (horizontal) current sheet evolves from a tangential
discontinuity to a rotational discontinuity with no null surface and
produces high-speed and high-temperature jets (Archontis et al.~2005),
dynamic 3D reconnection in a separator geometry with two null points
(Pontin and Craig 2006), or Hall effects on dynamic magnetic reconnection
in an X-type neutral point (Senanayake and Craig 2006) and in a
Sweet-Parker current sheet (Cassak et al.~2006). 

From the observational side we hear about spectroscopic detections of 
coronal bidirectional reconnection inflows ($\pm 3$ km s$^{-1}$) 
above the top of a flare loop (Hara et al.~2006), about measurements
of the non-dimensional reconnection rate $v_{in}/v_A\approx
0.02-0.07$ in three flares, which is found to be consistent with
fast Petschek reconnection (Isobe et al.~2005; Noglik et al.~2005), 
about a statistical Yohkoh
study of reconnection inflow velocities ($v_{in}\approx 10$ km s$^{-1}$)
and Alfv\'enic outflow velocities ($v_A\approx 10^3-10^4$ km s$^{-1}$),
yielding reconnection rates of $v_{in}/v_A \approx 10^{-3}-10^{-2}$
(Nagashima and Yokoyama 2006), and about a statistical SoHO/EIT study of
reconnection inflow velocities ($v_{in}\approx 3-40$ km s$^{-1}$),
where also a correlation of the CME speed with the inflow velocity
was found (Narukage and Shibata 2006). There are also reports about
simultaneous hard X-rays, gamma-rays, and radio bursts, interpreted
in terms of acceleration in reconnection outflow shocks (Mann et al.~2006).

\subsubsection{Particle Acceleration}

Once the 3D geometry of magnetic reconnection regions is known,
one can play endless games by populating them with particles and
watching how they get accelerated in the electromagnetic fields. 
Simulations of this kind revealed how particles are accelerated in 
3D magnetic nullpoints (Dalla and Browning 2006), 
in 2D X-points with a single X-line (Hannah and Fletcher 2006),
in force-free fields with anomalous resistivity (Arzner and Vlahos 2006), 
in strong turbulence regions (Arzner et al.~2006), in particular 
for $^3$He and $^4$He ions (Liu et al.~2006e),
in spatially intermittent turbulence regions (Decamp and Malara 2006),
in stochastic current sheets of stressed fields (Turkmani et al.~2006), 
in 3D Harris reconnecting current sheets where particles follow chaotic,
regular, and mirror-type regular orbits (Gontikakis et al.~2006),
or in fan 3D reconnection geometries (Litvinenko 2006). 
The highest energies of accelerated, gamma-ray producing particles
reported this year went up to $>200$ MeV, produced by neutral 
pion decay (Kuznetsov et al.~2006).

\subsubsection{RHESSI Observations}

One enigma raised by recent RHESSI observations is the initial
altitude decrease of flare X-ray looptop sources before changing
to the commonly observed upward expansion of the postflare loop
system. Detailed modeling of the X3.9 flare on 2003 Nov 3 suggests
that the observed properties can be reproduced with thermal
bremsstrahlung originating in a collapsing magnetic trap in the
cusp of a standard 2D reconnection model (Veronig et al.~2006;
Karlicky and Barta 2006; Giuliani et al.~2005). 
In one flare also the thermal flare loop
seen with TRACE in 195 \ang\ was observed to shrink along and to
oscillate subsequently (Li and Gan 2006), in another a simultaneous
loop shrinkage and expansion occurred (Khan et al.~2006).  
Sometimes the thermal hard X-ray looptop sources 
bifurcate distinctly when moving towards the two conjugate footpoints 
(Sui et al.~2006a). Other times, coronal hard X-ray blobs
bubble out above the flare loops, suggesting a vertical current sheet
that undergoes tearing-mode instability and produces magnetic islands
this way (Sui et al.~2005).

Another enigma is the lack of detected hard X-ray flare ribbons 
which are expected to trace out the UV ribbons, since both emissions
are excited by nonthermal electrons of similar energy in the chromosphere,
and both emissions vary in a synchronized way as confirmed with the VUSS-L 
and SONG instruments onboard CORONAS-F (Nusinov et al.~2006).
One interpretation was the disparity of spatial scales of
interacting flux systems during separator reconnection, where
the compact loops are more conducive to produce hard X-rays, while
the footpoints of the more extended loops mark the UV double ribbons
(Alexander and Coyner 2006). No problem was found for white-light
continuum emission, which closely follows the hard X-ray footpoints
(Chen and Ding 2006). White-light flares have also for the first time
been observed in near-infrared (Xu et al.~2006a).

A novel discovery made by RHESSI is the photospheric albedo of ``reflected''
hard X-ray emission (e.g., Kontar et al.~2006), although the inversion
of hard X-ray spectra is fundamentally method-dependent 
(e.g., Prato et al.~2006; Brown et al.~2006e).  
The effect of the photons backscattered in the
photosphere on the determination of the electron flux spectrum has been 
clearly demonstrated in a flare with an unusually flat ($\gamma=1.8$)
hard X-ray spectrum (Kasparova et al.~2005). 

Another first is a cospatial and cotemporal hard X-ray and EUV observation
of a chromospheric evaporation event using RHESSI and SoHO/CDS, which
showed high upflow velocities ($\approx 230$ km s$^{-1}$) in Fe XIX at
high temperatures and much lower downflow velocities ($\approx 40$ km
s$^{-1}$) in the cooler He I and O V lines (Milligan et al.~2006).
In contrast, a completely different behavior was observed in an event 
with much weaker electron precipitation, where thermal conduction is thought
to be the driving mechanism, resulting into significantly gentler upflow 
velocities (Milligan et al.~2006a).

A novel spectral feature that could not be observed before RHESSI are 
the 6.7 keV and 8 keV iron Fe XXV lines, which serve as a convenient
diagnostic for flare plasma temperatures and Fe/H abundances. In a
statistical study of 27 flares, the Fe abundance was found to be
consistent with coronal (Feldman) abundances (i.e., 4 times the photospheric
value) in 17 events (Phillips et al.~2006). 

Filling factors of $\approx 20\%-50\%$ are inferred in the cooling phase
of flares (Teriaca et al.~2006). The flare loops observed in soft X-rays
are likely to consist of many threads that are heated and
are cooling independently, so that the observed light curve is just
an envelope with a misleading decay time. Some long-lived hot coronal 
structures have been found during the postflare phase with CORONA-F/SPIRIT
that cannot be explained with a simple plasma cooling process 
(Grechnev et al.~2006), and nonuniform brightness features along the
loops have even blamed on the failure of Spitzer heat conduction
(Phillips et al.~2005).  Multithread hydrodynamic models were able to 
place a typical heating time scale of $\approx 200$ s for individual 
threads (Warren 2006). 

Energies deposited by hard X-ray producing electrons were estimated
to be of order 10\%-80\% of the magnetic energy release rate
(Lee et al.~2006b). The thermal flare energy estimated from soft X-rays
was found to correlate with the total duration, peak flux, and radiated
energy or radio emission (Benz et al.~2006). 

The first polarimetry of solar flare emission at 0.2-1.0 MeV gamma-ray
energies found only marginal polarization in the order of $\approx
10\%-20\%$ (Boggs et al.~2006).
After the RHESSI discovery of displaced hard X-ray and gamma-ray
(2.2 MeV neutron-capture line) sources a year ago, more cases with
similar displacements were reported, indicating some bifurcation in
acceleration/propagation of electrons and ions (Hurford et al.~2006).

Positron annihilation in solar flares can occur in a wide variety of
environments, from fully ionized, partially ionized, to neutral
(Murphy et al.~2005).

\subsubsection{Oscillations and Waves in Flares}

Oscillating loops are commonly seen after the impulsive phase of a flare
in EUV as well as in soft X-rays (Mariska 2006), 
but rarely in hard X-ray emission,
probably because the nonthermal hard X-rays originate at the footpoints
of flare loops, where the amplitude of transverse as well as longitudinal
MHD mode oscillations is almost
zero. However, damped oscillations with periods of 2-4 min and damping 
times of several tens of minutes were observed with RHESSI in 25 keV
hard X-ray emission, thought to be excited by super-Alfv\'enic beams
in the vicinity of the reconnection region (Ofman and Sui 2006). 
Oscillation signatures are also observed in radio wavelengths, which
seem to have a preference for 5-min periods, and thus could be excited
by global p-mode oscillations (Kislyakov et al.~2006). 
Since MHD oscillations modulate gyrosynchrotron emission they may be
detectable even on stars (Nakariakov and Melnikov 2006; Nakariakov et al.~2006).

\subsubsection{Self-Organized Criticality in Solar Flares}

What solar flares have in common with earthquakes is the same powerlaw-like
frequency distribution of sizes, interoccurrence times, and the same
temporal clustering, following the Omori law as found for seismic
aftershock sequences (De Arcangelis et al.~2006), independent of
intensity thresholds (Baiesi et al.~2006; Paczuski et al.~2005).
A flare time profile represents a cluster of many elementary ``avalanches'',
similar to the sequence of earthquake aftershocks,
with time scales down to 2-3 s found in H$\alpha$ (Qiu and Wang 2006),
and down to 4-60 ms in radio (Magdalenic et al.~2006).
Flares in H$\alpha$ also show an excess of coincidences for time intervals
of $<10$ min, indicating ``sympathetic flaring'' coupled by coronal
disturbances propagating from one active region to another (Wheatland 2006).

\subsection{Coronal Mass Ejections (CMEs)}

\subsubsection{CME Initiation and Magnetic Field Configuration}

Everybody agrees that some catastrophe is triggered in an unstable
coronal magnetic field configuration that leads up to the eruption of
a flux rope or CME, but there
exists a variety of scenarios and theoretical models for the 
trigger or driver mechanism, such as shearing of the bipolar background field 
(Chen et al.~2006d; Jacobs et al.~2006), 
breakout-like and thether-cutting reconnections (Zhang et al.~2006b),
the toroidal Lorentz hoop force which predicts a
maximum acceleration when the height of the erupting flux rope matches
the footpoint separation (Chen et al.~2006e), 
the total free (non-potential) energy which is found to be 
a stronger determinant than the overall twist or helicity alone
(Falconer et al.~2006), the kink-mode instability (Inoue and Kusano 2006;
Gibson and Fan 2006, 2006a; Kliem and T\"or\"ok 2006; Birn et al.~2006), 
magnetic flux emergence in ``plasma sheets'', i.e., at the boundary between 
two open field areas with the same polarity (Liu and Hayashi 2006), or 
photospheric flux cancellation at the polarity inversion lines between
two bipolar magnetic regions, leading to ``interchange reconnection'' 
(Mackay and van Ballegooijen 2006a, 2006b; Welsch 2006). 
The magnetic morphology of the ``magnetic breakout model'' was 
observed in a flare, but the acceleration of
the CME front already 1 hr before the apparent field opening disagrees
with the CME initiation mechanism predicted by the breakout model 
(Bong et al.~2006). 
A conjecture was raised that there is an upper bound of magnetic helicity
that can be accumulated in a force-free field, before a CME erupts
(Zhang et al.~2006a). Although we think that the magnetically unstable
region of a flare or CME is confined to a single active region, a much larger
global connectivity was found for the Bastille-Day 2000 event, where a 
transequatorial filament seemed to play a key role (Wang et al.~2006m).
In the overall, it is doubtful whether the pre-CME magnetic field
configuration reveals all necessary or even sufficient criteria to predict
a CME (Alexander 2006). Nevertheless, quiescent cavities are thought to 
represent
the calm before the storm (Gibson et al.~2006), and the total reconnection
flux was found to be highly correlated with the CME speed (Qiu and Yurchyshyn 2005).

\subsubsection{Propagation of CMEs}

CMEs or prominences can now be automatically detected in white light
(Qu et al.~2006), as well as in EUV, once
they propagate above the solar limb (Foullon and Verwichte 2006). 
CME speeds are measured between 260 and 2600 km s$^{-1}$ (Manoharan 2006;
Pohjolainen and Lehtinen 2006),
and most of the acceleration within 20 solar radii is accomplished by 
the background solar wind (Nakagawa et al.~2006).
CMEs propagate not always away from the sun, sometimes parts of them
move back, detected either as dark or as bright downflows (Tripathi et al.~2006).
Further out to distances of 5 AU, CMEs are decelerated all the way,
but not as strongly as predicted by the ``snow plough'' model or the
``aerodynamic drag'' model, so it is suspected that a low-latitude
coronal hole could provide some additional driving force (Tappin 2006).

Prominences are thought to be the best tracer of the helical flux ropes
that drive a CME (Filippov et al.~2006). A combination
of SoHO and Ulysses data allowed to track the formation of a current sheet
in the aftermath of a CME down from the corona all the way out to interplanetary
distances (Bemporad et al.~2006), out to 5.4 AU (Wang et al.~2005a) 
and 20 AU (Liu et al.~2006g).
Tracking earth-bound CMEs were thought to
provide also a forecast for interplanetary shocks (Howard et al.~2006), but
this conclusion was vehemently countered (Gopalswamy et al.~2006) and
defended (Howard 2006).

CMEs can overtake each other, an interaction that was also referred to as 
``CME cannibalism''. Numerical MHD simulations of such violent processes, 
even launched 10 hrs apart, show that the two merged shocks form a stronger
and faster one with a discontinuity between the old and new downstream region
(Lugaz et al.~2005). The brightness of a radio type II burst was found to be
five times greater when one shock overtakes another one (Sakai et al.~2006a). 

Masses of CMEs were compared with the masses of quiescent prominences,
eruptive prominences, and surges, all in the range of $\approx 10^{14}-10^{15}$ g
when estimated from EUV absorption features (Gilbert et al.~2006a), but much more
massive with $\approx 10^{17}$ g when estimated from SMEI white-light tomography
(Jackson et al.~2006). 

\subsubsection{CME-Initiated Waves}

Global waves, which  propagate over the entire surface of the sun
(EIT waves) after the launch of a CME, were found to have a well-defined
period (Ballai et al.~2005). A slightly different phenomenon was observed
on 1998 June 13 with TRACE (we are afraid that someone is going to call it
``TRACE waves''), which exhibited nonuniform propagation, unlike the
circular fronts seen in ``EIT waves'', but the wave front displayed a
nice Gaussian profile, suggesting a single propagating compression
front (Wills-Davey 2006). And after we finished reading the paper we
were very relieved to learn that all the perceived dissimilarities between
``EIT waves'' and ``TRACE waves'' can be explained in terms of the
different observing cadences used by both instruments. Theoretical
MHD simulations elucidated the difference between Moreton waves and
EIT waves, which showed that a pair of slow and fast MHD waves propagate
both in upward and downward direction after initiation of a pressure pulse
(Wu et al.~2005). We don't know whether the slow and fast MHD wave interact
in an EIT wave, but a coupling has been considered for the solar wind
(Zaqarashvili and Roberts 2006). Interestingly, coronal holes were found
to stop Moreton waves and EIT waves (Veronig et al.~2006). Or do the
waves just run into a vacuum? One thing that has been cleared up is that
EIT waves are only generated during CMEs, but not by pressure pulses from
flares (Chen 2006). On the other side, 80\% of radio type II bursts have
been found to be associated with CME origins (Cho et al.~2005; Shanmugaraju
et al.~2006), driven
by rising soft X-ray loops as observed in some cases (Dauphin et al.~2006;
Lehtinen et al.~2005),
although no correlation between the speeds of soft X-ray ejecta and CMEs
was found (Kim et al.~2005b). In some cases, Moreton waves, type II radio bursts,
and the CME were found to start simultaneously (Liu et al.~2006f). 

\subsubsection{Geo-Effectiveness of CMEs}

The total probability of geo-effectiveness for frontside halo CMEs
was evaluated to be 40\% (Kim et al.~2005a), some missed or grazed the Earth
even from eastern near disk center locations (Wang et al.~2006k), 
and full-halo CMEs from
the same active region do not necessarily have the same degree
of geo-effectiveness (Liu et al.~2006i). Halo CMEs, which are mostly
Earth-directed, also tend to have higher CME speeds (Lara et al.~2006). 
The geo-effectiveness, i.e., the prediction of geomagnetic storms,
seems to be strongly dependent on the southward orientation of the
magnetic field in the CME source region, and the predictions based on the
coronal flux rope model seem to be much better than those based
on force-free field models (Kang et al.~2006). The arrival time of
solar storms at Earth can be predicted with an accuracy as good as $\approx$
3\% (Xie et al.~2006). 

Actually 86\% of the
magnetic clouds have a leading field that is consistent with the CME
source region, and thus the field orientation is well conserved through
the heliosphere (Kang et al.~2006), although the magnetic orientation 
was found to rotate slowly by $\approx 50^\circ$ during its propagation
over 1 AU in one case (Krall et al.~2006). Particular CMEs with large
motional electric field and large dynamic pressure were found to
effectively increase the ring current ions and radiation belt electrons
in the Earth's magnetosphere (Miyoshi and Kataoka 2005).

\subsubsection{Solar Energetic Particles}

The arrival time of solar energetic particles can be even faster
than the travel time along the Parker spiral, because some random
paths have been found to be real ``fastlanes'' (Pei et al.~2006).

Even the CME-related shocks are thought to be powerful accelerators
of electrons and ions (Sokolov et al.~2006), 
and a detailed timing study proved in one case that the
detected near-relativistic protons originated at the flare site (Simnett 2006).
Energetic proton beams with energies over 20 GeV are believed to be
detected in some SEP events (Wang and Wang 2006), as evidenced by
measurements from a tracking muon telescope (Nonaka et al.~2006). 
The spectrum up to 4 GeV
was fitted by a shock acceleration spectrum in the Bastille-Day 2000 event
(Bombardieri et al.~2006).

Solar proton events are generally associated with fast CMEs, but an exception
was found for a slower CME ($v\approx 800$ km s$^{-1}$; Cliver 2006).
The mass of CMEs seems to be the most dominant characteristic for production
of SEP events (Kahler and Vourlidas 2005).
SEP events can also be produced in fast solar wind regions (i.e., in coronal
holes) and there is no requirement for those associated CMEs to be
significantly faster (Shen et al.~2006).

The ratio of $^3$He/$^4$He has been found to be enhanced by a factor of
up to 150 over the solar wind value, but the enhancement was attributed to
the seed population rather than to the CME-driven shock acceleration
(Desai et al.~2006). 

You do not necessarily need to build particle detectors to count energetic
protons from flares and CMEs; you can simply count the ``snow storms'' at the
edges of the EIT CCD camera (Didkovsky et al.~2006).

\subsection{Heliosphere}

\subsubsection{Solar Wind}

The SoHO/UVCS observations have revealed surprisingly large temperatures,
outflow speeds, and anisotropic velocity distributions for minor ions, which
revolutionized our understanding of the acceleration and heating of the
solar wind (Kohl et al.~2006). They can essentially only be explained in terms
of ion-cyclotron resonance interactions (Marsch 2006; Li and Habbal 2005), 
which likely dissipate energy fluctuations of MHD turbulent cascades 
(Markovskii et al.~2006). 
Turbulence might be created by supradiffusion via cross-field displacements
(Ragot 2006a,b; Zimbardo et al.~2006).
Interplanetary turbulence can be efficiently probed with two-point
measurements from multiple spacecraft such as Wind, ACE, and Cluster
(Matthaeus et al.~2005) or from the characteristic Kolmogorov frequency spectrum
(Podesta et al. 2006).
The interaction of Alfv\'en waves with solar wind
particles might be further complicated by MHD wave refraction
(Mullan and Smith 2006) and wave reflections (Suzuki and Inutsuka 2006). 
Alternatively, ultra-fine filamentary structures could also account for the
density fluctuations and interplanetary scintillations detected with
radio propagation measurements in the solar wind (Woo 2006).

Fast solar wind streams ($>500$ km s$^{-1}$) seem to be dominated by
fluctuations quasi-parallel to the local magnetic field, while slow streams,
which appear to be more fully evolved turbulence, are dominated by
quasi-perpendicu\-lar fluctuations (Dasso et al.~2005).

Intermittent anisotropic contributions were detected for the first time
at all scales in the frequency range of $1-10^{-6}$ Hz in the fast solar wind
(Bigazzi et al.~2006), probably produced by Alfv\'en wave heating of heavy
ions (Hellinger et al.~2005). Long-period ($\approx 170$ min) intensity
fluctuations have also been observed with SoHO/SUMER in the solar wind
above coronal holes (Popescu et al.~2005).

The slow solar wind (originating from streamers) is energized out to at least
2.7 solar radii (Antonucci et al.~2006), but the mass flow rate in plumes
decreases with height due to some mass transfer to interplume regions
(Gabriel et al.~2005). The power available to accelerate the solar wind
is found to scale linearly with the magnetic flux (Schwadron et al.~2006). 

The location and size of coronal holes (Robbins et al.~2006) as well as
the energetics of Alfv\'en waves (Suzuki 2006) bear 
information to predict the solar wind speed, up to 8.5 days in advance
of their arrival at Earth.

Solar wind speeds can be estimated from the initial speed and acceleration
of CMEs, because a dragging force is acting on the speed difference between
the CMEs and ambient plasma (Nakagawa et al.~2006). 

The solar wind does not flow steadily, there is evidence for
reconnection events from detected Petschek-type exhausts (Gosling et al.~2006)
and from cross-field displacement of magnetic field lines (Ragot 2006a,b). 
Besides reconnection outflows, 
there are also relative flow speed differences between alpha particles
and protons created, which can excite both magnetosonic and oblique Alfv\'en modes
(Lu et al.~2006b).
Magnetic reconnection, although thought to be bursty in the magnetosphere
and in solar flares, can also occur as a large-scale process, as an X-line
extending over 350 Earth radii, as testified by the detection with three 
spacecraft (Phan et al.~2006). 

The velocity distributions of electrons associated with solar
impulsive electron events are found to be nearly isotropic, probably
isotropized by nonresonant pitch-angle scattering of Alfv\'en waves 
(Lu et al.~2006a; Ragot 2006). 

\subsubsection{Interplanetary Radio Bursts}

Realistic simulations of interplanetary type III radio bursts include now
inhomogeneities in the solar corona and interplanetary space, microscale
quasilinear and nonlinear processes, intermediate-scale driven ambient
density fluctuations, the large scale evolution of electron beams, bursty 
Langmuir and ion sound waves, bidirectional propagation, fundamental and 
harmonic emission, where the latter is found to dominate (Li et al.~2006a). 

Interplanetary type II bursts were found to have a universal characteristic
in the sense that their bandwidth-to-frequency ratio ($\Delta \nu/\nu \approx
0.3$) is approximately constant out to 30 solar radii, even their
bandwidth decreases (Aguilar-Rodriguez et al.~2005).

An unusually highly circularly polarized solar radio type IV burst was
observed with the Wind spacecraft, and was 100\% circularly
polarized and observed for six days, while a similar event was
detected only some 26 years ago (Reiner et al.~2006).

\subsubsection{Termination Shock}

Voyager 1 has crossed the termination shock of the solar wind at a
distance of 94 AU from the sun, where low-energy ions abruptly increase
at the shock according to the Rankine-Hugoniot pressure relationship
(Fisk et al.~2006), anisotropic particle beams are generated 
(Gloeckler and Fisk 2006), and modulations of anomalous protons 
(Langner et al.~2006). The magnetic field in the heliosheath was 
measured by Voyager 1 to be as low as 0.05 nT (Burlaga et al.~2006).

\subsection{Solar-Terrestrial Relations}

There is now an end-to-end Space Weather Modeling Framework
(SWMF) that includes everything from the solar corona, eruptive
filaments, inner heliosphere, solar energetic particles, global
magnetosphere, inner magnetosphere, radiation belt, ionosphere
electrondynamics to the upper atmosphere in a high-performance
coupled model (Toth et al.~2005).

\subsubsection{Global Warming, Ozone, and Cloud-Free Skys}

In case somebody worries that the currently hyped global warming
has anything to do with our sun, there is a reassuring {\sl Nature} paper
out there that tells us that all luminosity changes of the sun are
accounted for by sunspots and faculae and that they did not 
contribute to the accelerated global warming over the
past 30 years at all (Foukal et al.~2006). 

However, solar irradiance variations around the Ly$\alpha$ emission
line near 121.6 nm range were found to vary up to 50-100\% during a 
solar cycle, which affects the stratospheric chemistry and controls 
production and destruction of ozone (Krivova et al.~2006).

If you heard that the number of days with cloud-free skys increased in China
over the last 5 decades, you still might not decide to move there,
because air pollution and pan evaporation increased, which produced 
a fog-like haze that
reflected and absorbed radiation from the sun and resulted in less solar
radiation reaching the surface, despite concurrent increasing trends
in cloud-free sky over China (Qian et al.~2006). 

\subsubsection{Geomagnetic Storms}

Only one out of the 18 very strong geomagnetic storms during the solar
cycle 23 is believed to be produced by a high-speed stream that compressed
an average-speed interplanetary coronal mass ejection (ICME) and intensified
its internal southward magnetic field (Dal Lago et al.~2006).

Different magnitudes of solar flares 
were found to influence the VLF signal amplitude in the 
Earth-ionosphere waveguide in such specific ways, that their
GOES (soft X-ray flux) class (C, M, X) can be classified from the
response of the ionosphere
(Grubor et al.~2005). The ionospheric effects of solar flares were also
derived from ground-based global positioning system (GPS) receivers
(Liu et al.~2006h). Responses in the Earth's thermosphere are most
dramatic in enhanced infrared emission from nitric oxide (NO), while
emission from carbon dioxide and atomic oxygen remain almost constant
during a solar storm (Mlynczak et al.~2005).

\subsubsection{Solar Activity Cycle}

Forecasting of the solar cycle becomes more and more of an urge
after we became so dependent on (or vulnerable from) cell phones, 
hospital beepers, power brown-outs, GPS, polar airplane routes, etc.

Forecasts for the next solar cycle were regarded from ``fair'' with a
sunspot number of $80\pm30$ (Schatten 2005) to ``high'' with
a sunspot number of $160\pm25$ (Hathaway and Wilson 2006), and
actually the majority (27 out of 30) voted for a ``strong'' cycle 24 
in a poll called by NOAA. 
The sun will tell us the verdict of cycle 24 in 4 years! 
The current cycle 23 was considered to be very low in flare activity
compared with cycle 22 (Atac and Ozguc 2006), although the rate of large
flares was relatively high in the declining phase (Bai 2006).
Some pessimistic long-term prospects even expect another
Maunder minimum around the year 2040$\pm$10 (Abdussamatov 2005).

In modeling the sunspot cycle it was found that the Gnevyshev and Ohl
rule (discovered in 1948), which allows to predicting an odd cycle from the
preceding even cycle, is often violated, but it was found that the period
of violations is correlated with the sun's retrograd orbital motion about
the centre of mass of the Solar system (Javaraiah 2006).
Also the solar wind speed at 1 AU was found to vary with the solar cycle
(Watari et al.~2005).
Of geophysical importance may also be that the tilt of the heliospheric
current sheet is strongly correlated with the sunspot number
(Pishkalo et al.~2006), i.e., the ``ballerina skirt'' is flat
during the solar minimum and highly warped during the solar maximum,
as you expect for slower and faster pirouettes.

If you think that the solar activity we have seen since Galileo's first sunspot
drawings is typical for our sun, you are mistaken. Paleomagnetic time series
analyzed over the last 7000 years find that the magnetic dipole moment
was significantly lower in the past and that the sun spends only 2-3\% of
the time in a state of high activity as seen in the modern era
(Usoskin et al.~2006). What exciting times we live in!

\section{GALAXIES NEAR AND FAR} 

Many of the hind inputters (acknowledged at the end of Section 13) awarded the
electronic equivalent of a  gold star to recent discoveries of additional satellite
galaxies, tidal tails, and star streams around the Milky Way and M31, so that is where
we are going to start, with a slight modification of a small green circle rather than
a gold star as the symbol of wowness (both because that was the symbol used this year
in the greenest author's notebook and because gold pens seemed to have disappeared
with the Lindy\footnote{
As her father generously said, when he worked for PaperMate and she bought a bunch
of Lindys because they came in more interesting colors, ``Lindy makes a good pen.''
Sadly, like the Kaiser-Frazer, Johnsons Baby Cream, and the Los Angeles area commercial FM
classical radio station, they are no more.  The demise of Marshall Field appears
elsewhere.}

\subsection{Home sweet home: The Milky Way and Local Group} 

Both were jam-packed with exciting papers in 2006.  This may have something to do
with the author whose slogan is PLAN AHEa$_d$ having accidentally tried to cram all 84
Milky Way papers into the upper left corner of notebook page 53 (spiral galaxies), and
all 54 Local Group papers into the lower left corner of p. 65 (clusters of galaxies,
part I, leaving page 66 completely blank). The papers of most obvious timeliness deal
with how the Galaxy and the galaxies were assembled and include comparisons between
the Milky Way and Andromeda (which became more alike in some ways and more different
in others during the year).

But the coveted green circles go (a) to Kalirai et al. (2006) for,  it would seem,
having finally found the true, R$^{-2}$ halo of M31 and separating it from the R$^{1/4}$ bulge,
which required looking outside 30 kpc and made  M31 rather more Milky in its population
structure, except that the bulge is much bigger [with implications for the mass of
the M31 black hole, whose X-ray luminosity is at most 10$^{36}$ ergs s$^{-1}$ 
(Garcia et al.~2005), but whose mass is $1.1-2.3\times 10^8$ M$_{\odot}$ 
(Bender et al.~2005)] and (b) to
Loeb et al.~(2005) for a clever indirect determination of the true transverse speed of
M31 relative to us from the requirement that M33 not have been torn apart by a close
passage.  The answer is 100$\pm$20 km s$^{-1}$, comparable with the radial velocity 
of approach,
confirming the temptation in lectures to say, 
given that the two quantities must be rather similar,
``oh, let's just assume they are equal,'' when trying to calculate when they will meet
each other. Loeb et al.'s time scale is $5-10$ Gyr for the MW and M31 dark halos to
start mingling.

\subsection{The Local Group Grows} 

Hubble's initial Local Group already contained at least six galaxies (MW, M31,
M33, L\&SMC, NGC 6822) before Shapley (1938) started adding dwarf spheroidals.  How many
are there really?  Oh, a couple dozen used to be a safe answer, but one must now say
at least three dozen.  Two new ones were added in index year 2006, dwarf spheroidal
companions to the Milky Way in the directions of (after a brief pause to look up the
nominative) Canes Venatici (Zucker et al.~2006) and of Bootes (Belokurov et al.~2006)
and we will take as given appropriate puns about new Bootes and in the boat.

Most controversial remains the discovery a couple of years ago of something on past
the stars of Canis Major.  It was first announced as a disintegrating satellite, and a
couple of papers continued to maintain that view (Dinescu and Belmont 2005; Bragaglia et
al.~2006) or even a picture of a satellite still gravitationally bound together (Martinez
Del Gado et al.~2005).  Other authorities held with equal firmness that we are
seeing a projection of the warped outer Galactic disk (Momany et al.~2006).
Lopez-Corredoira (2006) opines that, on the basis of existing data, either is possible;
and Bellazzini et al.~(2006) conclude that both are present $-$ stars in the southern
galactic warp on top of a dwarf galaxy.  Consultation with an auxiliary index indicates
that this is the first time in {\sl Ap06} that we will be voting with the ``both please'' camp.

How many LG members is this?  Even L.G. Maven this year was reduced to ``many.''
And the inventories are surely not complete (McConnachie and Irwin 2006), in the sense that
most of the 16 Andromeda satellites now known are on our side (geographically, not
militarily, and presumably because the ones on the other side are fainter and more
likely to suffer some extinctions\footnote{
``Suffer'' is surely an exageration, but ``to be extincted'' sounds even worse both as
as a condition and as English grammar.}.                                     
And 20\% or so of the Milky Way supply, especially
at low galactic latitude, remain to be found.

Just how many equatorial companions the Milky Way has bears on one of our favorite
effects, Holmberg I, his conclusions that satellites tend to avoid parent major axes,
at least in the plane of the sky. The current majority view (e.g., Yang et al.~2006; Shaw
et al.~2006) is that companions actually prefer the major axis  (again in two dimensions).
Yang et al. conclude that Holmberg simply had too few data points (218 satellites of 58
hosts) and bad luck.  Such three dimensional information as we have is interestingly
complementary.  Both the Milky Way and M31 have most of their satellites lying in a
single plane or great circle (Koch and Grebel 2006; Libeskind et al.~2005 on M31 and MW,
respectively), for which possible explanations include (a) the break-up of a single
progenitor,  (b) a prolate dark matter halo, or (c) accretion along one or a few 
large-scale filaments, and cases for each are made in papers this year.

A comparable number of papers dealt with alignments of galaxies on somewhat larger
scales, of which we mention only Hu et al.~(2006) on the local supercluster, for the
pleasure of noting that they cite not only Eric Holmberg and Gerard de Vaucouleurs (who
knew what they were seeing even if not all their contemporaries agreed with them) but
even William Herschel (who did not).

So much for Holmberg Effect I.  What is II?  It is the tendency for galaxies in
close pairs to resemble each other more than randomly chosen pairs.  It, too, lost ground
during the year, Franco-Balderas et al.~(2005) saying that there is no such concordance
in their sample.

And if you would please be kind enough to ask whether our Galaxy and the adjacent
big one exemplify Holmberg II, you will provide just the right introduction to the next
subsection.

\subsection{MW $\neq$ M31} 

We begin by claiming that there are three possible relationships:  same, different,
and scalable.  The green circle items at the beginning of Section 3 are both examples of
``scalable'' - M31 has a much bigger bulge than our Galaxy, but with the same density
profile and the same ratio to central black hole mass.  Darnley et al.~(2006) have
assumed a sort of scalability of nova rates to get $34^{+15}_{-12}$    
for the Milky Way from $65\pm16$
for Andromeda (per year in each case, from the POINT-AGAPE survey), so it would be wrong
to claim this as an example.  The five Galactic novae reported in index year (IAUCs
8607, 8671, 8673, 8700, 8697, plus one in the LMC IAUC 8135) set anyhow a firm lower
limit.

Properties and traits indexed under ``like'' include (a) the X-ray source populations
(Trudolyubov et al.~2005), (b) the existence of star streams and dying/dead satellites
owing to the central galaxy being a great big bully (Geehan et al.~2005; Font et al.~2006;
Piatek et al.~2006, commenting on Lynden-Bell and Lynden-Bell 1995, and Munoz et al.~2005)
and don't forget the entity called the Large Magellanic Stream (Connors et al.~2006), and
(c) the masses implied by at least part of the (flat) rotation curves according to
Carignan et al.~(2006), for which you will need your log tables, as they report $3.4\times
10^{11}$ M$_{\odot}$
for M31 out to 35 kpc and $4.9 \times 10^{11}$ for the MW out to 50 kpc; but 50/35 =
4.9/3.5 very nearly.  They refrain from extrapolating, but Battaglia et al.~(2005) opt
for a total Milky Way mass of $6-30 \times 10^{11}$ M$_{\odot}$ out to 120 kpc 
on the basis of velocities
of 240 halo objects and a range of models. Dehnen et al.~(2006) roughly concur.  Of these
three items,  (a) and (b) seem plausible, and (c) rather more surprising, given the
differences in disk, bulge, and halo populations (Athanassoula and Beaton 2006).

The harder you look, however, the more differences you find, for instance (a)
Andromeda is down to only about 2\% gas, of which less than 10\% is H$_2$ (Nieten et al.~2006;
Worthey et al.~2005, the latter noting the M31 is also more chemically mature, more nearly
fit by a closed box model of chemical evolution, and more nearly all disk stars out even
to 40 kpc), (b) Andromedal star formation is largely  confined to an offset ring at about
10 kpc with only rather spotty, scruffy arms taking off from the end of the bar (Gordon et
al.~2006), while young Galactic stars are a good deal more widely scattered; that you
haven't noticed this before by comparing face-on images of the two galaxies arises only
because there are very few face-on images of either to be had, and (c) at least five more
papers pointing out differences in the star formation histories of spheroids/bulges, thick
disks, thin disks, and halos of the two (Sarajedinin and Jablonka 2005; Brown et al.~2006;
Allende Prieto et al.~2006; Fulbright et al.~2006; Schuster et al.~2006; Olsen et al.~2006;
Font 2006).  Very crudely, while both galaxies show evidence  for both monolithic
central collapse and later accretion of small galaxies and gas, and for star formation over
a range of chemical compositions, M31 is slanted toward more monolithic and more early
star formation as well as toward having its disk, H I, and relatively metal rich stars
extend further out (Ibata et al.~2005).

\subsection{Minor Members} 

Concerning M33 we note only that it has a bunch of stellar populations with halo
composition ``not unlike'' M31  (McConnachie et al.~2006, the first detection of individual
stars in this halo) and a disk population that extends somewhat older than ours
(Sarajedini et al.~2006), but then even the poor old dwarf spheroidal in Sagittarius
has at least two populations of red giants with different metallicity (Monaco et al.~2005).

The LMC turned up in 11 papers, of which the most striking are of the form ``not much.''
First, the OGLE data base finds that only 0.7\% of the B dwarf stars are binaries with
periods less than 10 days (Mazeh et al.~2006), less than a tenth of the Galactic
allowance (and comparable with the deficiency of close binaries in 47 Tuc, 
Albrow et al.~2001). This is striking only to those of us who are old enough to remember when
metal-poor populations weren't supposed to have any binaries, until they were declared
normal a couple of decades ago. Second, though there are RR Lyrae stars where the
halo ought to be, Subramanian (2006) says they were made in the disk and puffed up,
leaving us still with no evidence for a truly stellar halo.

The SMC had a green circle paper, whose implications are perhaps also negative.
Harris and Zaritsky (2006) report that it is essentially a low luminosity spheroid with
appearance distorted by a modest amount of recent star formation.  Notice that star
formation being described as ``modest'' is not quite so silly as some of the other 
astronomical entities to which the word is applied, since it does indeed sometimes hide
behind dust (Takeuchi et al.~2006).

\subsection{The Milky Way and Other Spirals} 

Topics on which multiple papers were indexed included the nature and origin of the
stellar populations in halos, bulges, thick and thin disks (perhaps not the same for all
spirals); properties of spiral arms (including spacing of molecular clouds along them);
disk warps and their causes; bars; and halo shapes.  We will touch on all of these, not
perhaps equally or quite in that order.

{\sl 3.5.1. Arms.} 
We start with the nearly obvious (and rapidly depart therefrom)
almost all Galactic giant molecular clouds are near spiral arms (Stark and Lee 2006),
and in two galaxies examined by Elmegreen et al.~(2006), not only are they in the arms or
rings, but the gas clumps are nearly of a standard size and regularly spaced.  Perhaps a
gravitational instability opine the authors.

Some arms have spurs (Dobbs and Bonnell 2006) presumably so that they can rotate
faster, or anyhow faster than the gas out of the plane, so say Barnabe et al.~(2006) and
Fratelli and Binney (2006).  Just how fast is that?  In the range $200-220$ km s$^{-1}$ say
Kiera et al.~(2005, looking at planetary nebulae), or $228\pm21$ km s$^{-1}$ 
says Branham (2006,
from OB stars excluding Gould's Belt) near the solar circle. There are also gas speeds,
and Hernandez et al.~(2005) explain how you can use them, plus some other morphological
data, to extract pattern speeds.  Unfortunately, they assume a distance from the Galactic
center to the solar circle of 8.5 kpc (an IAU standard of the past), rather than the
geometrically-determined 7.5 kpc (Eisenhauer et al.~2005), which agrees with at least
some numbers based on distributions of stars  (Nishiyama et al.~2006), though not all
(Groenewegen and Bommart 2005). We suspect that, given the geometrical value, $R_0$
(like the speed of light) should be declared known and future papers be regarded as
efforts to find out whether the star populations are complete and understood.

Why are there arms anyhow?  Well, once upon a time there were three wizards named
Lin, Shu, and Soliton (but see Jalali and Hunter 2005 for an update); and a signature of
their mechanism is deviations from purely circular motion (Popova and Loktin 2005;
Fresneau et al.~2005).

Where are the Galactic arms?  Both interior to and exterior to the solar circle
(Sofue 2006 and Levine et al.~2006 respectively for the details).  Or, ``where the
magnetic field is counterclockwise'' according to Han et al.~(2006, though other years
have seen other answers to that question).  Or ``where the magnetic fields are strong''
at least for Grand Design spirals (Shetty and Ostriker 2006, indexed originally under
``Woltjer lives'',  til  we discovered the authors hadn't cited him\footnote{
And we are reasonably certain neither author was yet born when that paper was
published about 1958, but the Rotund Dipole, whom you will meet properly later, would
know.}.

\smallskip
{\sl 3.5.2 Bars.} 
Of the year's where/why/what are the bars papers, you get only three
answers: (1) the expected ``where the halo is losing'' its fight to stabilize (Curir et
al.~2006), (2) the unexpected ``where there is extra dark matter'' (Colin et al.~2006),
and (3) the incomprehensible  where there is a ``bundle composed of all the invariant
manifolds for all possible energies, as well as all the orbits driven by them''  
(Romero-Gomez et al.~2006).  And we think that Lyapunov (1949) lives in here somewhere
as well.  Both are calculations rather than observations.

\smallskip
{\sl 3.5.3 Disks.} 
Where/why are the outer parts of many spiral disks warped or even
scallopped\footnote{These are American-A scahllops, like the ones on edges of skirts.  
Only when you eat them are they the British-A scawllops.}?
Pertubations by companion galaxies  (Weinberg and Blitz 2006), by material
falling in  (Shen and Sellwood 2006), by self-gravitation (Saha and Jog 2006a), by magnetic
fields (Levine et al. 2006a), perhaps by any of the above (Saha and Jog 2006b), or
perhaps by none of the above (Sanchez-Salcedo 2006).

Really, of course, there are many disks, (a) a dust disk (with scale height =
125 pc, among the thinnest of the galactic ones), which shares the warps of CO, H I,
and all  (Marshall et al.~2006), (b) a dark matter disk (Bienayme et al.~2006) though not 
much of one, and we oscillate back and forth through it at $P = 42 \pm 2$ Myr, not 35 Myr,
(c) a maximal disk, meaning one with as large a mass to light ratio at the center, in the
baryonic component, as data will sustain (Kassin et al.~2006), (d) a non-maximal disk, 
meaning one where the dark matter halo is significant even at the center, which is to
say that they are not all the same  (Berentzen and Shlosman 2006), (e) a thin  disk,
meaning that of gas and stars recently formed from the gas (Ryden 2006), and  (f) a thick
disk and we must take a deep breath and start a new sentence, gasping the last of this
paragraph with the note that the scale height for nearby open clusters is only 53 pc
(Yoshi 2006).

Thick disks rotate more slowly than the thin component according, unopposed, to
Vallentari et al.~(2006), who deduce quick heating of a precursor distribution of stars.
More substructure and discontinuities in thick and thin disk kinematics appear in
Alcobe and Cubarsi (2005) who used Hipparcos data and came up with an analog of the
Titius-Bode law (sending us to our thesauri looking for a suitably innocuous-looking
Greek-derived epithet).   Rectangular parallelepipeds,  flummery being of Welsh origin,
like the Keen Amateur Dentist.

\smallskip
{\sl 3.5.4 Stellar Populations} 
Confirming what most of us thought we had known all
along, the largest star sample examined for both kinematics and composition found the
stars of the thick disk conveniently intermediate between those of the halo and the
thin disk  (Allende Prieto et al.~2006), not to mention differences from the corresponding
populations in M31 and an absence of either radial or vertical composition gradients
in the Galactic thin disk. Not to be outdone, Schuster et al.~(2006), using only 1223
stars, managed to distinguish two thick disk populations and several in the halo, once
again using criteria of age, metallicity, and kinematics.  Even the least skeptical
author ended that notebook entry with, ``really?''

Even for the simplest set of two disk populations, there would seem to be room for
multiple origins; and indeed there are two classes of theories $-$ one in which the thick
disk stars were once part of a thin disk and got heated or puffed up (Wyse et al.~2006),
and one in which the two were always separate. Thick = accreted stars and thin = accreted
gas (Yachim and Dalcanton 2006) is one possibility. It is, however, worth reminding
ourselves that current thick disk stars are chemically different from those in our current
satellite galaxies  (Pritzl et al.~2005). And nor can the current globular clusters or halo
field stars have come from populations like the current satellites (where the four-fold
repetition of ``current'' is deliberate).  Indeed you are invited to anticipate 
Font et al.~(2006a) in saying that some field stars come from satellites accreted long ago, 
which had only early star formation, deficient in iron-production by Type Ia supernovae, and
so large values of [$\alpha$/Fe].

\smallskip
{\sl 3.5.5 Bulges and Halos,} 
a word or two before you go. First, these are defined so that
stars that pass near the solar system belong to the halo population, the bulge being further
in (though it would be different if we lived in Andromeda\footnote{
For starters, you would call your companion galaxy Centaurus, and it would be about
M 131, because there is more fuzzy stuff in that direction. Rectangular parallelepipeds.},
where the bulge extends
very much further out). And, of course, you, or we, or somebody must keep track of the
difference between the halo of stars and the halo of dark matter, the latter being both
considerably more extensive (Narayan et al.~2006) and considerably more massive (Pizgano
et al.~2005), and, it would seem more nearly spherical (Phelps et al.~2005 on the stellar
halo with $c/a = 0.16$ vs. Belokurov et al.~2006a on the dark matter halo, nearly spherical).
But we reserve the Queen Anne is Dead award for the rediscovery of the existence of halo
stars in the RR Lyrae population (Maintz and de Boer 2005); unless you prefer the recognition
that satellites still recognizable as such arrived later on average than those that have
already been torn apart (Moore et al.~2006) or the reasseveration that the dark halo of
the Milky Way is not made mostly of MACHOs (Calchi Novati et al.~2005).

Or, if you would just as soon not have to deal with bulges at all, you could move to
one of the 15\% of a large sample of edge-on SDSS spirals that don't have them (Kautsch et
al.~2006).

\subsection{Interactions} 

Think of this, if you wish, as the future of the Local Group. It is interestingly easy
to arrange triplets of papers into syllogisms:  Major premise: most ellipticals form from
mergers (Rothberg and Joseph 2006).  Minor premise: bulges of disk galaxies and
ellipticals are a single family distinguished only by mass (Davies 2006).  Conclusion:
most bulges form from mergers (Moorthy and Holtzman 2006). Like many syllogisms, this
approaches perilously close to a tautology, since the ones that form from disk stars
are called pseudobulges (Galaz et al.~2006; Fisher 2006, who then rather spoils the
whole thing by declaring that real bulges come from single star bursts rather than from
mergers of existing units, and ditto for ellipticals).

In general one expects mergers to make galaxies more spheroidal than the input
pieces (Boylan-Kochin et al.~2006, though of course MWiN\footnote{
{\bf M}ore {\bf W}ork {\bf i}s {\bf N}eeded. If you sometimes say it yourself, 
feel free to borrow our abbreviation.}. But it also apparently
happens that disks survive interactions (Lee et al.~2006) and disks can be rebuilt
after interactions (Puech et al.~2006). The latter becomes less surprising if you
note the requirement that there still be a good deal of gas and net angular momentum.
Indeed with enough of both, disks can even be the natural produce of mergers
(Robertson et al.~2006).

Another standard interaction product is the polar ring galaxy, and in the case of
NGC 922 we actually see the companion that dropped through (Wong et al.~2006). Other
cases, including the classic Cartwheel, have been claimed as mere disk instabilities
(Griv 2005).  And a third category consists of galaxies with rings formed by late, slow
gas in-drizzle (Iodice et al.~2006) either from another galaxy or along a filament of
the large scale structure (Mareo 2006; Reshetnikov et al.~2006). About half of them
are really polar disks anyway (Theis et al.~2006).

When we were children, virtually all polar ring galaxies were S0's, but this year
there is an H I polar ring considerably smaller than the distribution of carbon stars
nearly perpendicular to it (Battinelli et al.~2006); and it used to be a Dwarf
Irregular member of the Local Group.

If accreted gas with the wrong net angular momentum makes its way down to the
center of a galaxy without turning entirely into stars, it becomes a counter-rotating
core, whose presence in about 10\% of edge-on galaxies implies that a good many others
must have had late-addition gas more or less in the right direction (Bureau and Chung
2006).  The small galaxies that have thick disks including OB stars sound like a
possible product of gas arriving in slightly the wrong direction (Seth et al.~2005).

Gentle readers who gently read about the re-assembling of the Antikythera
device (Wright 2005) may been pleasantly surprised that they didn't seem to have
any parts left over.  We have been less lucky with the Milky Way and feel that we must
cram in, whether they fit or not, (a) the fact that it is made of stars, and has been 
known to be so at the intelligent lay-painter level since at least early 1609 (Kemp 2006),
(b) the probability that its magnetic field has components both from a large scale
dynamo and from local events like supernovae and pulsars  (Goldreich and Sridhar 2006) 
and is highly intermittant toward the center (Boldyrev and Yusef-Zadeh 
2006)\footnote{
And no  we are still not used to the idea that ``intermittant'' can mean ``spotty in space''
as well as ``spotty in time''.  Well, our intelligence may well be intermittant in both.},
and (c) the odd asymmetry that makes newly discovered stars, previously known open
clusters, and 2MASS sources commoner at galactic latitude and galactic coordinates
(b and z) below the plane than above (Mercer et al.~2005).  The solar system itself is
above the local galactic plane by 23 pc (Joshi 2005), leaving us with the desire to
holler ``selection effect,'' but Mercer et al. offer neither this explanation nor any other.

\subsection{The First Galaxies and How They Grew} 

The Early Universe must have loved dwarf galaxies, because she made so many of 
them\footnote{
Other versions of this line concern other sorts of entities, with the formation process
variously attributed to The Creator (who apparently had a remarkable fondness for
beetles), G.d (and the Common Man, according to Will Rogers), and condensed matter
physicists.}, 
especially at the beginning. This is not what astronomers expected long ago
(meaning 15 or so years) and so is not what they saw.  In the days when ``top down''
formation seemed the best bet, we looked for ``primordial galaxies,'' meaning $10^{12}$ or
so solar masses of baryons undergoing a first, enormous burst of star formation (of which
giant elliptical galaxies and the largest spiral spheroids would be the descendents) and
radiating absolutely oodles of Ly $\alpha$ photons; and we were collectively puzzled
at not finding them ({\sl Ap95}, Sect. 8).  Instead, as we all now know, halos start small
(though the first to form are now in big, dense structures) and grow by mergers and
accretion (both of additional halos and of gas), forming stars as and when the gas gets
dense enough and cool enough.

We do not see these $z \gapprox 10$ entities of 
$10^8$ M$_{\odot}$ with $10^7$ M$_{\odot}$ of stars, though Read
et al.~(2006) point out that they must be the ``basic baryonic building blocks'' of which
the Milky Way should have got $100 - 10$ still present as dwarf companions and 90 broken
up into the stellar halo.

A slightly later phase is observable according to Panagia et al.~(2005), who point
out that known $z = 6.5$ galaxies could already have ionized the medium around them at
$z = 15\pm 5$.  The building blocks duly build, and by $z = 2 - 3.5$ there are galaxies as
bright as the most impressive ones around now, but they exhibit more irregular shapes $-$
tadpoles, clump-clusters, chains and such $-$ which today are found only among dwarf galaxies
(Elmegreen et al.~2005; Elmegreen et al.~2005a; Wadadekar et al.~2006; Straughn et 
al.~2006).  The logical chain of progression from here is via processes (mostly in the realm
of the theorist) to types of galaxies  (frequently observed), landing in a small pile of
unsolved problems that have been around for a decade or so.  In the present climate, your
process had better be one that can occur in a $\Lambda$ CDM universe, or publication may be
difficult.

\subsection{Broad Brushes} 

Grand scenarios are attempts to start with something like those fundamental building
blocks of halos and gas and follow them through mergers, star formation, central black
hole growth, and gastrophysical feedback, to end up with populations, correlations,
clustering and all that resemble present reality.  We caught three during the year, Hopkins
et al.~(2006), Bowers et al.~(2006), and Mori and Umemura (2006). We belong to the ``well,
it will probably be all right on the night'' school of thought (and didn't worry much
about dress rehearsals or studying for exams either), though one must acknowledge
recognizable discrepancies, like clusters of galaxies that are slower rotating, rounder, and
more centrally condensed than real ones  (Nagashima et al.~2005; Avila-Rees et al.~2005).
But keep in mind that these and some other such broad-brush paintings have very few pixels
(mass points) per galaxy.

The concept that dominated the 2005 papers on formation and evolution of galaxies
(or anyhow our discussion, {\sl Ap05}, Sect. 4) was undoubtely ``downsizing'' $-$ that is, at any
given moment, the oldest, most metal rich stars reside in large structures that have pretty
much finished star formation, which continues in smaller objects.  Once the idea is
lodged in a reviewer's mind, it becomes an irresistable indexing term, with 28 entries
this year and another dozen papers under closely-related concepts like ``passive evolution'',
$-$ which means that a galaxy mostly just fades away after its last, big, star-triggering
merger $-$ and merger histories.  You will get a subset, including a few papers that
illustrate downsizing of non-galaxies, and maybe a dissident or two.

It has perhaps already become necessary to explain why the concept and the data
summarized by it initially seemed so surprising.  If little things came first and have
been merging more or less at random as they found each other, and the merger process is
ongoing, then surely the largest things around should be the ones being assembled this
week (whether a $z = 3$ week or a $z = 0$ week). This is not what we see (Nelan et al.~2005
on the fundamental plane vs. redshift).  Instead, the biggest, oldest, reddest star
populations seem to have formed earliest, no matter what redshift you look at  (Conselice
2006; De Lucia et al.~2006).

Things that you might also reasonably describe as downsizing occur for (a) chemical
evolution (Maier et al.~2006),  (b) the mass of the entities typically being added to
bigger ones (Brough et al.~2006a),  (c) where in a cluster of galaxies most of the star
formation is located (Collobert et al.~2005), (d) where and when star formation is quenched
(Cooper et al.~2006a),  (e) properties of X-ray emitting groups of galaxies  (Mulchaey et
al.~2006),  (f) both formation (Hopkins et al.~2006a) and turn-off of AGNs and their black
holes (Hasinger et al.~2005;  Scannapieco et al.~2005),  (g) the onset of passive evolution
(Franceschini et al.~2006),  (h) the assembly of galaxies  (Cimatti et al.~2006), and (i)
L/M of luminous blue galaxies  (Lee et al.~2006a).

One of the implications is that the dense central parts of the biggest galaxies, which
is where observations are likely to be concentrated\footnote{
Why look where there isn't any light? A point of view that arguably prevented a
pre-discovery of the CMB by a couple of students who were told to go measure the radio
sky background temperature and who looked where it was brightest instead of where it was
faintest.} were put together so early that the
product can look like that of a monolithic collapse (Lintott et al.~2006).

And yes, there is a circled green word for 2006 as well.  It is ``dry mergers'', not
quite a new idea, because the last couple of {\sl ApXX}'s have mentioned that stars can be older
than the year of final assembly of the galaxies in which they now live, but a newish way
of saying it.  ``Dry'' in this context means, with little or no diffuse gas around  (and the
only meaning of the word of which we entirely approve). Not having picked up on the phrase
until part way into the index year, we cannot promise to have found every paper in which
the process appears.

Kauffmann et al.~(2006) point out that star formation shuts down when the stellar
surface mass density reaches about $3 \times 10^8$ M$_{\odot}$ kpc$^{-3}$, 
so that the galaxy is thereafter
bulge dominated, and subsequent mergers will not include a burst of star formation (or
we suppose, of AGN feeding), say Papovich et al.~(2006).  But mergers do continue to occur
assembling larger galaxies  (Bell et al.~2006).  The products will include giant elliptical
galaxies with boxy isophotes  (Naab et al.~2006), vs.  smaller disky ones from spiral
mergers, and the most luminous field giant ellipticals now around (van Dokkum 2005).
If you decide to rootle through the literature looking for another problem to which dry
mergers might be the answer, we might decide to point you toward Bregman et al.~(2006a), who
have found some ordinary-looking NGC ellipticals for which the infrared spectral
signatures from AGB stars and the optical absorption lines from bluer stars give ages that
differ by 2 Gyr. Well, perhaps they were put together from pieces whose dominant star
formation  epochs occurred at different times, and the AGB stars really are older.  At
least a single burst of star formation lasting at most a Gyr seems unlikely.

What about the products and problems?  The galaxy word of the year for 2004 was
bimodality, reflecting a discovery made in giant SDSS samples that the distributions of
color and, correlated, of luminosity and mass, seemed to be double peaked, with a rapid
transition near halo M = $3 \times 10^{11}$ M$_{\odot}$, L = $5 \times 10^9$ L$_{\odot}$,
velocity dispersion = 88 km s$^{-1}$,
and star mass = $1-2 \times 10^{10}$ M$_{\odot}$. This must still be more or less true, 
because additional
supporting data surfaced and (the real test!) a number of theorists came forward to explain
the bimodality.

On the data side, Mateus et al.~(2006) found that the double-peaked distribution of
colors carries over to ages and total mass in stars  (and the correlations are an aspect of
downsizing), while Forbes (2005) reported that the two sorts of galaxies also have different
populations of globular clusters, with, in particular, only the bluer of two cluster
populations present in the low mass, blue galaxies.

The five explanatory papers we caught all invoke some sort of gas-based feedback and
are listed here in approximate chronological order of publication: Keres et al.~(2005,
gas arriving cold along filaments for small mass galaxies,  quasi-spherically and shocked
to high temperature for big ones), Menci et al.~(2005, a hierarchical model of galaxy
formation), Cooray and Cen (2005, effect of re-ionization on Jeans mass), Driver et al.~(2006,
cold vs. hot gas inflow), and Shankar et al.~(2006, transition in feedback from stars to
AGN for large masses).

A couple of apparent dissents are really reminders that, in theory and in practice,
there is a continuum of mass in stars and associated properties  (Gallazzi et al.~2006),
including the bulge to disk ratio (Allen et al.~2006a), rather than two truly different
sorts of galaxies.

Another rather sudden change of appearance within a continnum (of halo masses, near
$10^{13}$ M$_{\odot}$) 
is the transition from single galaxies to groups (Humphrey et al.~2006).  The
fossil groups  (where you have to look hard to see that anything except the biggest one is
still there) also belong to this mass range (Cypriano et al.~2006, on the second confirmed
X-ray case).

Incidentally,  if you would like to resume the search for primordial galaxies in the
pre-1995 sense,  Spitzer Space Telescope  (Borys et al.~2006) and SCUBA (Huang et al.~2005)
have found some mighty bright galaxies from when the world was less than half its present
age, but most of the expected Ly $\alpha$ doesn't get out. There is a similar IRAM sample
(Tacconi et al.~2006).

\subsection{Some Special Galaxies} 

Pick a favorite galaxy.  Any galaxy.  Aw, go on.  Well, no, you cannot have an isolated
galaxy, because there aren't any (Brosche et al.~2006). What seem to be isolated 
star-forming dwarfs are, they say, merely the brightest members of small groups.

An XBONG? Nope, you'll have to ask for it by circumlocution - extra nuclear dust
in plane hides emission lines  (Rigby et al.~2006).  Endiphel is at least pronouncible!

You want a red spiral?  OK, but you will have to go back to $z = 1-2$, where Roche et
al.~(2006) found some, along with spheroidals and mergers, among EROs  (extremely red
objects), an observation duly confirmed by theory (Fritze-v. Alvensleben and Bicker 2006).

An E+A?  Yagi et al.~(2006, and an earlier paper during the year which the most
elliptical author forgot to record) point out that these result from galaxy-wide starbursts.

An cD?  The one on offer (at $z = 0.19$) has its X-ray surface brightness profile tracing
the optical (Wang et al.~2005), which,  if the visible light is stars and the X-rays hot
gas is perhaps a little odd.

The closest galaxy with more than $10^{11}$ M$_{\odot}$ of H I remaining?  It belongs to 
Donley
et al.~(2006), and you will need to borrow their radio telescope to study it, because it
is in the Zone of Avoidance.

An ultracompact dwarf?  Some doubt is cast on their existence as a separate class in
Sect. 8, but Mieske et al.~(2006) is a nice ``both please'' conclusion, advocating
merged young massive star clusters in the Fornax cluster and tidal stripping in Virgo.
The Virgo ones were reported by Jones et al.~2006a), who advocate tidal thrashing in all
cases (by NGC 1399 in Fornax and mostly by M87 in Virgo).  Defining size is less than
100 pc and $M_B =-10.7$ to $-12.9$.

An early-type galaxy? NGC 5128  (Cen A) is the nearest really big one, so no surprise
that it was the first to have its core helium burning (horizontal branch, red clump) stars
resolved (Rejkuba et al.~2005).

A gE with little or no dark matter?  Sorry, no.  This is another false alarm that went
away.  NGC 3379 has its fair share (Pellegrini and Ciotti 2006; Pierce et al.~2006).
Curiously,  these two papers describe themselves as refuting two different false alarms:
Romanowski et al.~(2004) for Pierce et al., and Cappellari et al.~(2006) for Pellegrini and
Ciotti.  We cite the supposedly refuted because the refuting authors do not seem (from
their acknowledgements sections) to have consulted with them. But we will admit in everyone's
defense that you sometimes have to look very far out in giant Ellipticals to get to the
part that is dark matter dominated (Fukazawa et al.~2006).

E and S0 galaxies with cool or cold interstellar material?  Oh, all right, but you
can have much (Kaneda et al.~2005 on PAHs; Morganti et al.~2006 on H I, up to $10^{6-9}$
$M_{\odot}$
in 8 of 12 galaxies).  The gas is often misaligned and could come either from a dIrr merger
or from smooth cold accretion.
What makes S0 galaxies in general? Well, the gas must be sent somewhere $-$ inward to
make a final starburst, outward to make intergalactic  medium, or perhaps both  (Cortes
et al.~2006).  And it must be done quite slowly, taking more than 3 Gyr (Cortes et al.),
or quite quickly in less than 1 Gyr (Moran et al.~2006).  And the removal must be ongoing,
because some S0s today have less gas than would have accumulated from star losses  (Sage
and Welch 2006). Well, they weren't all the same S0's.

A truly young dwarf, experiencing its first burst of star formation?  What never?
No never  (Aloisi et al.~2005).  What never?  Well, hardly ever (Pustilnik et al.~2006,
on DDO 68, with no detectable stars older than 900 Myr.  It is in a void and almost as
metal poor as I Zw 18.)

A former gassy dIrr that becomes a gasless dSph except for star loss  (Marleau et al.~2006;
Bouchard et al.~2005)?  Well, sort of, in that we occasionally see the gas leaving
(Tarchi et al.~2005).  But you can never get rid of the disk completely,  so ultra-compact
dwarfs cannot be made this way  (Mastropietro  et al.~2005).  And indeed 5-10\% of 476
Virgo dE's examined have clear disk, bar, or S arm structures (Lister et al.~2006). The
authors call these dEdi  (perhaps a nickname for a friend who dislikes the given name
Daedalus and is slightly capital-challenged).  This got roughly a green half circle.
That present dwarf ellipticals were, at moderate redshift, compact blue galaxies  (Crawford
et al.~2006) is called the Butcher-Oemler effect.  Mayer et al.~(2006) ascribe gas
departure to tidal erosion and ram pressure in clusters, but do not mention 
supernova-driven winds.

Andromeda IX ?  Why would anyone want this companion of smallest known surface
brightness?  Perhaps for its M/L which is 93, though with large error bars  (Chapman et 
al.~2005).

As for the traditional, pure dwarf spheroidals of the Milky Way, you are getting
Fornax\footnote{Not to be confused with the cluster of the same name; and you should see what
the author still in active teaching harness did to the homewark records of Chr. Wang
and Chn. Wang in Physics 20A.}, take it or leave it.   It has a nice, comfortable middle
range M/L = 15  (Walker et al.~2006), and, perhaps a bit strangely,  forms a bridge to our
residual problems section, because,  say Goerdt et al. (2006) its central profile must
have a core  (rather than cusp) shape, or else its globular cluster would have sunk to
the center in a few Gyr.  To provide such a core, the authors require significant amounts
of warm dark matter, with particle mass less than 0.5 keV.

\subsection{Residual Problems} 

Bouncing off Fornax, which is either part of the problem or part of the solutions to
the core-cusp dilemma (and not very elastic anyhow), we hit the U Ma dwarf spheroidal,
with M/L of at least 300 (Kleyna et al.~2005).  It is the most DM dominated galaxy to
date, suggesting that there could be many additional companion/satellite galaxies around
the Milky Way with M/L = infinity or thereabouts, one of several possible solutions to
the missing satellite problem.

As has been the case in several previous years, a majority of the 2006 papers did
not bemoan the irreconcilable differences of core/cusp and missing satellites but proposed
solutions (15 c/c;  5 missing satellites) that they regarded as promisng. Not, of course,
all the same solutions.  Rationing here operates on the press-releases principle: ``on the
other hand, supporters of $2+2=5$ said...'', so we point out a solution or two and one
reservation each.

Reionization will certainly reduce the number of small halos that can acquire enough
gas to make stars, and all is OK (Wyithe and Loeb 2006; Dobler and Keeton 2006, who note that
details of strong lensing reveal substructure for four radio sources at more or less the
expected level.  Problems remain ($2 + 2 = 4 {1\over 2}$?) an the other hand according to 
Razoumov and Norman (2006).

The view that SuperWIMPs as the dark matter can take care of everything (Cembranos et
al.~2005) sits outside the current range of astronomical thought, and indeed those
simulations have not been done with as fine mass resolution etc. as the standard 
$\Lambda$ CDM ones (which reached points of $7 \times 10^6$ M$_{\odot}$, 
Diemand et al.~2005).

Attempting to evade the core/cusp discrepancy, hard working observers have actually
found some structures cusped at least in surface brightness  (Pellegrini 2005 on E/S0
galaxies; Patel et al.~2006 on brightest  member galaxies in X-ray clusters; Voigt and
Fabian 2006).  But  (another $4 {1 \over 2}$, we suppose) there is still serious 
disagreement for very blue LSB galaxies  (Zackrisson et al.~2006).

And your confuson pair for the day has two papers suggesting that there is more
substructure in some observed objects than predicted in some calculations (Chen et 
al.~2006b on locations of satellite galaxies; Amara et al.~2006 on substructure in 
gravitational
lenses).  This is the opposite of the ``excess small scale structure'' problem.

\subsection{Clusters of Galaxies} 

Since we began this discussion by opining that the Early Universe must have loved
dwarfs  (because she made so many of them), it seems appropriate to dive in here with a
second cluster found  (like Virgo) whose members are about 80\% dwarfs  
(NGC 5846, Mahdavi et
al.~2005).  It sounds like the same is true of Coma (Adami et al.~2006), though for the
brighter sources in the field, the single largest category among the 60,000 catalogees
is foreground stars  (dropping to only 10\% or less at $I = 20$).  The unluckiest cluster of
the year has to be poor, old ($z = 0.29$) RX J0646.4+4204  (Kotov et al.~2006), which has the
misfortune to be behind M31.

More fortunate clusters can be observed in many more ways to tackle other sorts of
questions.  This year we have selected two residual problems  (M/L ratios and cooling flows)
and one item of selective inattention (intergalactic matter) for examination.

Within living memory, one hesitated to admit that the several cluster mass indicators
(radial velocity dispersions, X-ray properties, and weak lensing maps) didn't always
agree, lest some Foe of the Dark sneer,  ``Well,  if you guys can't agree about how much there
is, maybe you are all wrong, and there isn't any.''  That battle has been more or less won
(owing, as usual,  to the aging of the anti-camp, relative to their general condition
at the time of the 1961 Santa Barbara conference, Neyman, Page and Scott 1961). It is,
therefore, time now, we think,  to regard these discrepancies as an opportunity to
learn about the full range of heating mechanisms for X-ray gas  (of which more in a few
lines), deviations from Virial equilibrium in clusters as they form from smaller entities,
and invasive fore and background fuzzies in lensing studies  (Clowe et al. 2006a).

Data and analyses to start the thought processes appear, for instance, in:
Dietrich et al.~(2005, lensing mass less than the other two), Jee et al.~(2006, X-ray
mass less than lensing), Khosroshahi et al.~(2006, dynamical M = twice the X-ray),
Rasia et al.~(2006, X-ray underestimates from neglected physics), Baughan et al.~(2006
two  mergers in progress in J0152.7-1357 at $z = 0.89$), and Chandran and Dennis (2006, on
cosmic ray heating and convection).

Considerations of how all the X-ray gas might be heated lead directly into ``residual
problem two,'' cooling flows, the ancient and honorable idea that,  given the temperatures and
densities at the centers of many clusters, the gas should radiatively cool itself and flow
inward (owing to loss of pressure support) in much less than a Hubble time. We suspect that
this isn't happening both because there is not much evidence for cool gas in cluster cores
or for stars that might have formed from it recently, and because of the improbability of
our living in the Last Days of Fabian et al.~(2006)\footnote{
We have tried hard to persuade ourselves that Fabian is the metrical equivalent of
Pompeii in the way that Catullus's mistresses were the metrically equivalent Clodia                  
and Lesbia, and there will be a small prize for the first reader who correctly remembers
which was the real and which the poetic one.  The Faustian Acquaintance is not eligible
for this prize for historical reasons: he probably knew them both.}
for so many clusters.

At present, this appears to be an ``all of the above'' territory.  That is, in some
cases there are large uncompensated flows,  in some rather modest flows that can feed
a cool gas or young star supply, and in some significant heating sources that keep the
gas in rough equilibrium.  Bohringer et al.~(2005) discuss some that are dumping
1000 M$_\odot$ yr$^{-1}$ or more into the cluster center.  Bregman et al.~(2006b) 
have done some
re-estimating that makes a real  cooling rate more like 30 M$_{\odot}$ yr$^{-1}$, 
and Hicks and Mushotzhy
(2005) and Salome et al.~(2006) discuss star formation (indicated by UV luminosity)
using the gas as it arrives.

And, looking again at the apparently large, real cooling rate clusters, the culler
of the literature also finds a number of processes that can probably add up to a
comparably large heating rate.  Some authors indicate that their mechanism is the entire
answer, others that it is not, but if you are not one of them, you are allowed to take into
account  (a) occasional mergers  (Chatzikos et al.~2006),  (b) infall of individual galaxies
(Jetha et al.~2006),  (c) heating by cosmic rays  (Chandran 2005),  (d) dynamical
friction (Kim et al.~2005),  (e) conduction from outer hot gas (Pope et al.~2005), and
(e) several different ways. AGNs, usually at cluster centers, might pay back the debt they
owe for gas accreted from the cooling flow by tending to stop it (Fabian et al.~2006;
Forman et al.~2006; Buote et al.~2005; McNamara et al.~2006; Kraft et al.~2006), and at
least another half dozen papers.  Not all are optimists, with Brighenti and Mathews (2006)
of the opinion that cooling flows remain a problem.

Self-evidently, hot X-ray emitting gas is one of the substances to be found between
cluster members, and Ferrara et al.~(2005) is one of many reference year papers reminding
us all that the gas has, on average, something like 1/3 solar metallicity, much of the
pollution (like most of the star formation in big elliptical galaxies of the cluster sort)
having happened a long time ago.

Small clusters whose virial temperature is not enormous can also have a good deal of
cold gas wandering around on its own (Bouchard et al.~2006 on the Local Group; Trinchieri
et al.~2005 on Stephan's quintet, where virtually all the gas, cold, warm, and hot, is
outside the galaxies).

\subsection{Between the Galaxies} 

What else is there?  Zwicky (1933) suggested small, faint galaxies and diffuse gas,
both of which indeed there are,  though neither is much of the dark matter he posited.
And your most forgetful author had to be reminded yesterday (it was a Thursday) that Fritz
himself had actually gone out looking for visible light from between the galaxies and
thought he had seen it  (Zwicky 1951), with special thanks to science writer Richard Panek
for the reminder and to Da Rocah and Mendes de Oliveira (2005) for saving us from having to
go back into the past further than last December to find the reference.  They looked in
three Hickson compact groups and found that the intragroup halos contained 46, 11, and 0\%
of the total light.   Other index year reports included Mihos et al.~(2005), who found
that such light is largely structured and is due to tidal streamers, tails, bridges, and
such, and Sun et al.~(2005), who describe stars between M81 and M82 as most probably
dragged out of the galactic disks but possibly formed in situ after neutral hydrogen gas
had been dragged out.

There were also a number of observations of intracluster light made between 1951 and
2006 which have now been confirmed by theory.  Various sorts of mergers ought to dump
anything from 10 to 50\% of the stars present at onset into the space between the galaxies
(Sommer-Larson 2006a; Stanghellini et al.~2006).

To be seen far from the galaxy where you formed, you need both to live a long time
(100 kpc at 500 km s$^{-1}$ = $2 \times 10^8$ yrs) 
and to be sturdy enough to hold together during
the ejection process.  Planetary nebulae are a classic intergalactic population, and you
might suppose that they would fail both tests, consisting as they do of tenuous gas that
dissipates in $10^4$  years or thereabouts, but of course it was the progenitor stars that
escaped, while the nebulae formed very recently (Vallaver and Stanghellini 2005 on Virgo).
There ought also to be intracluster supernova remnants  (with the same sort of history).
Maoz et al.~(2005) predict 10-150 in Virgo, depending on how much circumstellar gas is
around to be shocked.  Some may already have been seen optically as intergalactic H II
regions, and a radio search should be possible.

The planetary nebulae and supernova remnants imply in due course intergalactic white
dwarfs and neutron stars.  What about black holes?  Well, some should be expected say
Libeskind et al.~(2006), but only $2-3$\% of the cluster's total BH mass allotment, and
the authors don't quite say how a search might be carried out. Also predicted but not yet
seen are ``tramp novae'' (Shara 2006), of which LSST should find MANY and globular clusters
(Bassino et al.~2006a, a prediction for the Fornax cluster).  Both pass the ``I hope you
live a long time'' test, but we are not quite so sure about their ability to hold together
if violently accelerated.

Non-thermal radio emisson can extend beyond the galaxies responsible for accelerating
of electrons and (we suppose) generation of magnetic fields (De Bruyn and Brentjens 2005).
The decision has yet to be made between a top-down scenario, in which a truly cosmological
seed field gets stirred around to greater strength and a bottom up one, where fields
produced in early stars, GRBs, and active galaxies get blown out and, again, stirred around.
The proceedings of a conference on the subject (Beck 2006) begin with Rees (2006a) voting
for the first fields coming after the first non-linear structures (bottom up).  We know,
in any case, that supercluster scales must be achieved (Xu et al.~2006, a rotation measure
detection toward the Hercules and Perseus-Pisces clusters).  The field is partially
coherent on 500 kpc scales and weaker than 0.3 $\mu$G if the electron supply is the one they
expect from WHIM and radio galaxy leakage. A couple of the processes suggested to occur
on the cluster scale  (Medvedev et al.~2006; Subramanian et al.~2006a; Ensslin and Vogt 2006)
are, at any rate, not obviously inconsistent with the observations, which is as nice as we
get about magnetic fields on a day when the Rounded Dipole\footnote{
The Rounded Dipole is a new-comer to this series (who can stand for several of the
colleagues cited in Sect. 4), and is not to be confused with the Keen Amateur Dentist,
who is very slender, The Faustian Acquaintance, whose picture appears in 
{\sl Nature} {\bf 442}, 238, or The Medical Musician, who is rotund and bipolar.}
has sent us one of his preprints.

\section{ALL TURTLES THAT ON EARTH DO DWELL} 

May 23rd was World Turtle Day, but, sadly, the only actual news item was the death
of Harriet the Tortoise (Anonymous 2005a) at the age of about 175; and it would seem that
she never met Darwin (Nicholls 2006), because he never visited Santa Cruz Island, and she did
very little travelling in those days,  though she died in an Australian zoo.  Seekers after
extended mnemonics for the sequence of spectral types may wish to make use of long-lived
tortoises,  for instance in the form, Ordovician Barnacles And Fossil Gastropods Killed
My Long-lived Tortoise Yesterday.

\subsection{Inside the Eartn} 

Since we intend to end up back on the surface with astronomers and other species, it
probably makes sense to start with the earth as a whole and its interior.  The pear
shape was advertized around 1960 as a significant discovery from the space program, but
Danson (2006) opines that this was known earlier.  Various wobbles of the earth's rotation
axis relative to the ``fixed stars'' are also not a recent discovery, but Chandler would
surely have been surprised by the four separate components  (with periods close to a day,
close to a year, and several years) reported by Vondrak (2005).  The details can be
predicted only a year or two in advance (Anulenko et al.~2006).

The degree of inhomogeneity of the bits and pieces from which the earth formed has
been debated over the years, but Wood et al.~(2006) are of the opinion that it all started
with planetesimals that already had iron cores.  The solid core is still growing at the
expense of the fluid iron around it, which should slow terrestrial rotation, though
not enough to compete with the longer days due to tidal drag of the moon, etc.  (Denis
et al.~2006).  The amount is 2-7 $\mu$s/century.

Basic core formation came before the giant impact that made the moon, and a modern
calculation (Wood and Halliday 2005) confirms that Kelvin was about right in his estimate
($3 \times 10^7$ yr) of the time it would have taken the earth to cool from an initially high
temperature,  if that were the way it had been done. The terrestrial magnetic field, blamed
on the core, has been dropping about 5\% per century since 1840 (after 2.5 centuries of
constancy), as a result of patches of reverse flux in the southern hemisphere, but it is
not about to flip  (Constable and Korte 2006).

Mantle convection was a single-zone phenomenon this year (Albarede 2005), but with
a phase change in the perovskite in the bottom few hundred kilometers  (Hutko et al.~2006
and about three similar papers).  Mantle plumes were fighting for their lives  (or anyhow
existence) as the index year closed  (Hirano et al.~2006; Dziewoksi 2006).  Contributions
in support were to have been accepted at a site off the coast of Hawaii, but it was
damaged  by the same earthquake that put a good many of our larger telescopes on the
orthopedic sick list in October.

Apparently mantle convection started dragging crustal plates around as soon as crust
existed  (4.5 Gyr ago according to a conference report, Anonymous 2006k), or perhaps even a
bit sooner,  since the oldest crust claimed may not reach quite that far back 
(Tsuyoshi et al.~2006; Harrison et al.~2005).

Craters have been forming when stuff hit ever since (but also eroding much faster than
on the moon, Mercury, Mars, or other barer surfaces).  We caught papers claiming periodicities
of 35,  58, and 77 Myr or 24,  35, and 42 Myr (but have lost the references), at most one of
which can be our oscillation period back and forth through the galactic plane.  Svensmark
(2006)  says 31, 142, and 200 Myr and 3 Gyr on the basis of encoded cosmic ray flux records.
Presumably earthquakes are also as old as the crust and plate tectonics, but somehow it
is always the recent ones that get most attention (Bilham 2006, on escalating death rates
$-$ a statement about demographics, not geophysics), or even the future ones 
(Fialko et al.~2006, 
an examination of the southern part of the San Andreas fault, where the accrued
slip deficit is at least $9-10$ meters, comparable with the maximum documented anywhere). This
paper received a little black circle to remind us to renew our earthquake insurance, but
not tell the agent why.  Dietrich and Perion (2006) are of the opinion that the topographies
of earth and Mars (and now, we suppose, Titan) are different enough that you can infer the
existence of life on the first.

\subsection{Atmospheres and Climates} 

Then the air arrives, and the most disputed issue we caught in the fiscal year was just
how early there was enough O$_2$ to care about and whether its abundance was fairly steady
once it formed.  Omoto et al.~(2006) favor early O$_2$, peaking at 2.76 and 2.93 Gyr ago, and
Knauth (2006) other possible interpretations for data before 2.4 Gyr BP. The present
terrestrial atmosphere (observed with moonshine, which was a highlight several years ago)
shows reasonable evidence for biological input  (Turnbull et al.~2006).

Not all current changes are anthropogenic.  The Afar Rift Valley in east Africa is
currently opening (Wright et al.~2006) with no help or hindrance from East Africans past
or present.  And the unusual turbidity of the atmosphere after 1815 came from
the Tambora volcano eruption that year  (Stothers 2005).  Incidentally, the turbidity was
measured by the faintness of the moon in eclipse  (invisible on 10 January 1816 in the 
extreme case), a method suggested by Danjon (1920).

We indexed 37 papers from the reference year dealing with climate change, 
possible effects on time scales
of decades to centuries, probabilities of a sequence of unfortunate events arising from
anthropogenic global warming, and so forth, about the same number as in each of several
previous years, when a dozen or so got cited.  But it is at least as urgent to keep
in mind that  many short-term disasters arise from (and could be mitigated by) local human
activities, including (a) the rapid withdrawal of ground water (Dixon et al.~2006 on the
sinking of New Orleans, but the same applies to Venice and to now-often-flooded coastal
zones of the Phillipines and other countries that are even less well prepared than the US is),
(b) ill-considered land use  (Jackson et al.~2005 on the effects of tree plantations on
stream flow, soil salinity and soil acidity), and (c) loss of biodiversity (Biesmeijer
et al.~2006, on birds and bees,  for pollination, not sex education).

In the very long term, of course, annual cycles (aka weather), longer ones from the
Earth's orbit (Milankovitch), and effects of living creatures will all give way to the
inexorable brightening of the sun (Huybers and Curry 2006).  The only survivors among the 
eukaryotes will be the fungi and worms that can still reproduce at $50^\circ$ C 
(Girguis and Lee 2006).

\subsection{Other Species} 

We green-circled several inhuman, or at least non-human, papers, including the macaques
who are better at recognizing faces than at least one of your authors (Tsao et al.~2006;
Kanwisher 2006), which is not a particularly high standard, and the confused elderly
octopus who tried to attack a small submarine  (Cosgrove 2006).  According to Duchaine
(2006), something like 2\% of the population has difficulty recognizing faces  (most without
realizing that other folks are better at it),  though the test was done with photographs,
which the one of your authors most likely to attack a small submarine thinks are much
easier than live humans, perhaps because they rarely say ``You don't remember me, 
do you?''.

Among other confused species we noted  (1) chimps and humans, who may have continued
to interbreed after the initial separation  (Patterson et al.~2006),  (2) birds that need
to recalibrate their magnetic compasses but are unable to get up in time to do so with
polarization of sunlight at dawn (Mulheim et al.~2006), (3) the crocodiles and 
scallops\footnote{
This sort is pronounced scawllops, unlike the scahllops around the edges of spiral galaxy
disks, in Sect.~3.5.3.}
classified as fish by the Australian government (Anonymous 2006l),  (4) a flagellate
in the process of acquiring a new green plasmid as a symbiont  (Okamoto and Inouye 2005), (5)
the tunicates, sea squirts, cephalochordates,  lancelets, hemichordates,  sea acorns, and
all currently in competititon to be our closest relatives  (Delsuc et al.~2006), that is,
if you classify yourself  primarily by your backbone,  (6) the fishapod (titaalik roseae,
whom we briefly mistook for the author and close collaborator of Scidmouse), the first
terrestrial creature to be able to adopt a limp-wristed pose, because the first to have
wrists  (Daeschler et al.~2006),  (7) stramovaris, which was trying hard to become a
ctenophor (Conway Morris 2006), presumably to demonstrate that he could pronounce
it,  (8) castorocavda lutrasilimilis  (Ji et al.~2006), whose name makes clear he was undecided
about whether to become a beaver (because someone had to have a broad, flat tail, even in
the Jurassic) or a river otter,  (9) stromatolites, whose biogenic nature has been doubted
(though never by us) and has been reaffirmed  (Allwood et al.~2006), and  (10) malaria and
TB, which are losing out to AIDS and lower respiratory infections in disability-adjusted
years of potential life lost  (Yager et al.~2006).

On the advancing edge of humanity, we note  (l) Pithecanthropus  erectus, whose name
(vs. Homo erectus) still lives, though the species does not  ({\sl Nature} {\bf 438}, 1099),  (a)
evidence that homo habilis, with a BMI of 27.8  (Dennell 2006) would probably not have
passed the bar stool test, while (3) Neanderthal, who overlapped in time and place with
H. sapiens sapiens  (Gravina et al.~2005; Mellars 2006) probably would have.  We express
no opinion this year on whether he attempted to take advantage of this, nor on the status
of H. floriensis, who was too small to get up on a bar stool anyhow, and would probably
have been carded\footnote{
As were we just a few weeks ago in Chicago (ORD) airport.  Luckily the passport was
still in hand from the struggle through security (why do you think we needed that drink?).
Not that any reader is likely to need to know this, but the local supermarkets do not
accept passports as evidence of age and require a drivers license.  Oh.  The bar stool
test. Surely we've told before. Human means that if suitably dressed, given lots 
of money, and perched in a barstool in Las Vegas, the personage will attract potential
mates. The original version was less gender-neutral.  And probably funnier.}.

As we approach the neolithic,  the first plant to lay down its life for a noodle was
millet  (Anonymous 2006m). The first domesticated plant may have been a fig (Kislev et 
al.~2006). And if they had been anything like the figs Grandmother Farmer made into jam,
that would have been the end of the agricultural revolution right there.  Instead,  it
seems that other earlier Farmers went on to domesticate six key species still in use
(emmer and einkorn wheat, barley, lentils, chick peas, and flax), plus some others tried
and abandoned more or less permanently (chenopod. marsh elder) or temporarily (oats,
rye, squash, sunflower; Weiss et al.~2006).  The spread of agriculture could have been
accomplished either by the transport of just plant genetic material or of human genetic
material as well, invasive males being one possibility ({\sl Science} {\bf 310}, 964).

Still more recently, over-enthusiastic taxonomists have contributed to confusion
about distinctness of plant species (Riesenberg et al.~2006), and pharmaceutical companies
have discovered that it is not a good investment to develop antibiotics, because new ones
are used sparingly (Nathan 2006). Of course, if the old ones had been used sparingly,
we would not be in such desperate need of new ones! Meanwhile, the FDA regulates 
25 cents out of every consumer dollar spent in the USA (Kennedy 2005).

\subsection{And the Wisdom to Know the Difference}

In which we examine an assortment of human activities  (mostly scientific),  some of
which came out better than others.  Given that the average scientist's work week is about
80 hours (Anonymous 2006n), you would think we would do better.  Forty papers were
nevertheless,indexed under ``oops'' and related concepts like ``consider the alternative''
and ``the faint praise award.''  On balance we think we are in favor of
\begin{itemize}
\item[o] {The {\sl American Physical Society} sticking to a planned meeting in New Orleans 
(Anding 2005)
and retaining its present name, vs. {\sl American Physics Society}, though only under
pressure of corporate law (Cohen 2005).}
\item[o] {Open access literature for {\sl APS} to be achieved by something very much like the 
reimposition
of page chages, (Anonymous 2006o), though at only \$975-1300 per article we think they
may lose their physical (or physics?) shirts.}
\item[o] {A range of ways of evaluating Journal impact factors that allow each of {\sl Reviews of
Modern Physics}, {\sl Applied Physics Letters}, and {\sl Physical Review Letters} to come first in
one of them (Bohlen 2006).}
\item[o] {The website of the Royal Society (London) offering ``Fellows Bedroom Service''  
(according to the 18 February issue of New Scientist).}
\item[o] {A metric for picking top universities that selects 17 American ones,  
Cambridge, Oxford, and the Swiss Federal Institute of Technology (ETH)\footnote{Through
whose doors have passed, generally in both directions, great scientists from Aschwanden
to Zwicky. Oh, and that fellow Einstein. - V.T.} (Hirst 2006).}
\item[o] {Some thought being given to how the next US astronomical decade report ({\sl Astronomy 
and Astrophysics in the 2010's} presumably) might be carried out more fruitfully than the
last one  (Anonymous 2006p).  A few details there are off  (e.g., it was the Bahcall report
for the 1990s that directly involved the largest number of astronomers).  And we are a
little uncomfortable with the ``them that has gets'' Matthew tendencies implied.  Luckily
the appearance of the NSF ``senior review'' is out of period  (though some citation data
were available to  writers, Trimble 2006a).}
\item[o] {The conclusion that excessive secrecy really does impede progress  (Agar 2006). The
report concerns early UK and US developments in computing (Bletchly Park vs. ENIAC).}
\item[o] {Two astronomers already on the faculty (titular at least - is this like the titular
bishop of Titipu?) of UC Merced  ({\sl ApJ} {\bf 647}, 1040, author list).}
\item[o] {The recognition that excessive artificial light is bad for many creatures beside 
astronomers (Rich and Longcore 2006), but sad that the few countries that have darkened in the last
decade or so are in bad shape in other ways  (Ukraine, Moldava for example; Elvidge 2006).}
\item[o] {The Biblioteca Alexandrina  ({\sl Nature} {\bf 439}, 913),  in which everybody seems to be looking at
books rather than computer terminals.}
\item[o] {The six questions about the future of chemistry ({\sl Nature} {\bf 442}, 500) which are very 
different
from the sorts posed by David Mermin and Vitaly Ginzburg for physics and astrophysics
a few years ago, but ``if indeed chemistry still exists'' could be a real issue,  given
the nibbling at the edges by molecular biology, materials science, and so forth.}
\item[o] {An increasing trend toward asking joint authors of a paper to note who did what 
({\sl Nature}
{\bf 440}, 392).  We were scolded for trying to do this last year,  though not by {\sl Nature}.}
\item[o] {Discovery of a bandwagon effect in success of new music (Salganik et al.~2006) 
through
a controlled experiment in downloads. In the astronomical case, one might track the
change in numbers of preprints on a particular topic for the six months or so after
some high profile person produces his first.  The sort of new music in the experiment
might not be recognized by the two of your three authors who bought from Deadalus
last year the complete works of Mozart and Bach.}

\item[o] {Early recognition of the need for both foxes and hedgehogs in science: ``It is probably
best for mankind that the research of some investigators should be conceived within a
narrow compass while others pass more rapidly throuh a more extensive sphere of
research.'' ({\sl Nature}, {\bf 438}, 291, a contemporary judgement of Thomas A. Young, of the
wave theory of light, partial decipherment of Egyptian hieroglyphs, and much else).}
\end{itemize}

Neither approval nor disapproval seems appropriate for the following facts of
life:

\begin{itemize}
\item[o] {Astronomers probably don't often get rich from patents, but it may be no use 
even trying,
since the US patent Office in May 2000 was $10^6$ applications behind (Lucas 2006).}
\item[o] {In China, 7.4\% of the population carries the surname Li, 7.2\% Wang, and 6.8\% Zhang
({\sl Nature} {\bf 439}, 125), and given that the number of graduate students from China (and India)
in the US is now back up more or less to normal (Blumenthal 2006), you may want to rethink
how you alphabetize your class lists.  We have, however, been privately advised by a
Nanjing colleague who carries the name that Wang in English conceals two very different
Chinese characters, one of which is a good deal less common as a surname than the other.}
\item[o] {Hubbert's peak is nearly upon us, especially for non-OPEC countries (Campbell 2005).
But if Germany fueled the Luftwaffe on liquids from coal during WW II, presumably we
could too, at a cost of about \$ 50 per barrel, says the quotee.}
\item[o] {Of {\sl Nature}'s most downloaded papers, 89\% are in biological sciences 
({\sl Nature} {\bf 440},
23 March, p. xvii).}
\end{itemize}

And, third, here is a set of facts and opinions that prompted reactions like ``Ugh!''
and ``Ugh-faugh'',  the comparative.  The superlative is ``Ugh-faust.''

\begin{itemize}
\item[o] {``No one would require someone from Massachusetts to collaborate with someone 
from South
Dakota''  (Winnacher 2006, who is the new secretary-general of the European Research
Council and was addressing the requirement for collaboration among European countries).}
\item[o] {The Russian Academy of Sciences will force institute directors to retire at 55
(Ananchenko 2005).  The mean age of the Academicians is 69.  We foresee a new
exodus.}
\item[o] {A scheme for ranking UK universities  (Anonymous 2006q) will indeed be good for our
friends, but seems excessively Matthew-ish. In particular, making government funding
partially proportional to that received from industry and charities will benefit Oxford
and Cambridge.  An editor ({\sl Nature} {\bf 440}, 581) thinks it reasonable.  Perhaps.  Just out
of period, we ourselves were quoted as having said something was ``reasonable,'' when
the actual remark was ``as good as one could reasonably have expected.''  (and we have
tried hard to learn the lesson taught by The Keen Amateur Dentist that, if one doesn't
expect much, one is less likely to be disappointed).}
\item[o] {The beyond-decimation of Bell Telephone Labs scientific staff from 3200 in 1998
to 1000 in 2005  (Kim 2006).  We have just used the 2007 AAS directory that arrived
yesterday to check that neither the institution nor anyone employed in Holmdel is
listed.  That the Bell name is no longer even over the door of Alcatel-Lucent labs is
post-period, along with a further reduction in research staff.}
\item[o] {The first e-only paper in mainstream astronomical literaure  (Anders et al.~2006) has
only the abstract on paper.  We think it is about methods of photometry.}
\item[o] {The exponentially rising quantity of space debris  (Liou and Johnson 2006).  They
especially advise you not to try to live at 900 or 1450 km above the earth's surface.}
\item[o] {We recorded something like a double handful of instances of age and gender bias in
various branches of science, and, in the interests of minimizing the incidence of new
enemies, quote mostly the historical ones. ``Mrs. Clerke ... sometimes let her sympathies
limit her range of vision in the field of stellar research''  ({\sl Nature} {\bf 440}, 616, quoting
a 19 March 1906 review of one of her books.  A {\sl Nature} {\bf 437}, 963 quote from a 12
October 1905 book review attributes the unfortunate views to a Prof. Shaler of Harvard
but at least Shaler was impugning environmental rather than genetic influences in saying
that women were unlikely to become successful scientists.  And as long ago as 15
October 1955, J.B.S. Haldane was objecting to forced retirement of scientists at some
arbitrary age ({\sl Nature} {\bf 437}, 963).}
\item[o] {NASA images of Titan and Mars are ``fed through a very particular stylistic filter''
(Ball
2005) intended, we suspect, to make the landscapes look exotic, but not too exotic.}
\item[o] {And, if it is true that human productivity peaks where the average temperature is
$10^\circ$ C 
(Proc. NAS 103,  10.1073, quoted in {\sl Science} {\bf 311}, 1347), then it is surely because they
are trying to earn enough to move to some more clement clime!}
\end{itemize}

\subsection{Let Us Now Praise Famous Persons} 

Of the 49 individuals whose names appeared attached as other than authors to indexed
papers, some had good things happen to them, some not so good.  Here are a lucky 13,
with the G/NSG labels removed on the advice of Bernard Shaw, who said,  ``Do not do unto
others as you would have them do unto you.  Their tastes may be different.''

\begin{itemize}
\item[o] {The female vice-president of the International Union of Pure and Applied Chemistry, who
was to have become the first female President in January 2008 has resigned over a
funding fuss at her home university ({\sl Science} {\bf 313}, 31).  She is Kazako Matsumoto. And let it
be said that the IAU has only just acquired its first female President  (Catherine
Cesarsky), and IUPAP is not there yet. Indeed at the most recent hand-over of offices,
the outgoing IUPAP president looked at the new membership for Commission 19  (Astrophysics)
and said, ``too many women.''  It was half, and though suffering from a slight digestive
disturbance, he appeared to be sincere (and for our purposes, nameless).}
\item[o] {Whether you get on a postage stamp seems to be determined by the Fickle Finger of Fate.
In India, Bose,  Bhabha, and Chandra yes, Raman no.  In Slovakia Dionyz Ilkovic yes
({\sl Nature} {\bf 438}, 1089).}
\item[o]{Giovanni Schiaparelli's middle name was Virginio (Mazzuato 2006), which, like ``it
seems that you are the son of the Mikado'' (W.S. Gilbert) cannot be said just to have
happened.}
\item[o] {Farinelli (the most famous operatic castrato) is to be dug up for examination 
({\sl Nature}
{\bf 442}, 230), and Copernicus (for whom E. P. Hubble's cat was named) has had his face 
reconstructed  in most unflattering fashion  ({\sl Nature} {\bf 438}, 1067).  Neither will mind.}
\item[o] {Darwin is going to be asked to share his 200th birthday  (12 February 1809) and 150th
anniversary of publication of the {\sl Origin of the Species} with International Year of
Astronomy. He won't mind either, but astronomy is liable to lose out to the biologists
(cf. percentage of top downloads from {\sl Nature} above!) in festivities, or so says the
keen amateur successor to the first dentist  (Coppa et al.~2006).}
\item[o] {Danish astrophysicist  Anja Andersen has won quite a nice prize ({\sl Science} {\bf 310}, 1765),
though not, we admit, quite so nice as the \$1.4 million Templeton Price received this
year by John Barrow ({\sl Nature} {\bf 440}, 396).}
\item[o] {Unflattering things were said about Heber Doust Curtis (Setti 2006) for not accepting
gravitational deflection of starlight in 1919-23, nor had he really by 1942. But don't
forget that the IAU abolished its Commission on Relativity in 1925, because there
wasn't really anything else for it to do.}
\item[o] {Peter Debye had his name removed from an institute and a prize  ({\sl Science} {\bf 311}, 1236, and
several later editorials and letters). In a brief,  serious moment we remark that that
sort of de-naming can be done by a small committee, but only the physics community
acting together could take his length away from him.}
\item[o] {Thomas Robinson of the University of Miami, Florida ({\sl Science} {\bf 312}, 1871) reports that
he is not after all descended from Ghengis Khan.  But he had already admitted that he
hadn't done much pilaging lately anyhow.}
\item[o] {Macbeth's name has been given to an effect wherein a threat to one's moral purity
induces
the need to cleanse the body  (Zhong and Liljenqvist 2006).  Yes, they have done experiments,
and they mean, we think, Lady M.}
\item[o] {Robin Corbet has been micro-eponymized in a ``milli-Corbet diagram'' ({\sl ApJ} {\bf 638}, 966,
footnote).  It displays correlations of the properties of millisecond pulsars  (vs.  the
AGNs in a kilo-milli-Corbet diagram).}
\item[o] {Harry Collins managed to pass as a worker in general relativity 
({\sl Nature} {\bf 442}, 8), based
on answers to questions about gravitational radiation compared to those of real
relativists. We suspect you won't entirely concur with either set of answers, but this
was our green circle paper on the topic.}
\item[o] {{\sl Physics Today} finally resolved its differences with terminated staff member Jeff Schmidt
(Brodsky 2006), who was awarded back wages and benefits, rehired, and then immediately
resigned.}
\item[o] {Tutankhamen drank white wine as well as red (Lamuela-Raventos 2006).}
\item[o] {David Baltimore was hailed as ``arguably the most eminent voice in all of American
science'' ({\sl Nature} {\bf 439}, 891).  They gave no indication of his wine preferences, but his
successor as President of Caltech, Jean-Lou Chameau is a French-born civil engineer
({\sl Nature} {\bf 441}, 562).}
\end{itemize}

\subsection{Oh, Dear.  I'm Really Rather Glad I Didn't Say That!} 

Acronyms come first, then some neologisms, some previously-unannounced awards
and an assortment of mispeaks, not to be confused with the misteaks of Sect. 13.
Pronouncible, or nearly so, acronyms numbered about 20, ranging from Alpaca and Angstrom
to SEXCLAS and SLUGS.  Not all were decoded by their users, and we were most puzzled
by the computer code DJEHUTY (Dearborn et al.~2005), because it is a relatively poorly
known variant of the name of Thut, Tut, or Thoth.  Some coinages were quite ambitious:
EPIC = European Photometric Imaging Camera ({\sl ApJ} {\bf 637}, 699), ELVIS = Extragalactic Lensing
VLBI Imaging Survey ({\sl ApJ} {\bf 648}, 73), MATISSE = Matrix Inversion for Spectral Synthesis
({\sl MNRAS} {\bf 370}, 141),  and ANGSTROM = ANdromeda Galaxy STellar RObotics Microlensing program
({\sl MNRAS} {\bf 365}, 1099).  Others sounded modest, at least by comparison.  SLUGS = SCUBA Local
Universe Galaxy Survey  (MMRAS 364, 1253), RATS = RApid Temporal Survey  ({\sl MNRAS} {\bf 371},
975), one of whose discoveries, RAT J 0455+1308 is an EC 14026 star, and FRED = Fast
Rise, Exponential Decay ({\sl ApJ} {\bf 643}, 276).

Aristotle is supposed to have said that the first step toward knowledge is to call
things by their right names (no, we don't know what he called Pluto, or Plato). And if the right
name wasn't previously to be found in dictionaries, now is your chance to pick one.

Zeptog measurements  ({\sl Nano Letters} {\bf 6}, 583)

Hectospec  (Fabricant et al.~2005)

Isopedic  ({\sl MNRAS} {\bf 364}, 475), which we still think should mean ``having the same feet''.

Separatrix (source not recorded, but is it perhaps a polite synonym for what used to be
called a female co-respondent?)

Astrocladistic and descent with modification (it was applied to types of cataclysmic
variables)

Semigeostrophic and Stratorotational instabilities ({\sl MNRAS} {\bf 365}, 85)

Cloudshine ({\sl ApJ} {\bf 636}, L105)

Pixie dust  ({\sl ApJ} {\bf 637},  774)

Dicotron ({\sl A\&A} {\bf 445},  779)

Ter-annual and Quadri-annual ({\sl A\&A} {\bf 446}, 346).  Hint: think which of semi-annual
and bienniel is which.

Kriging ({\sl MNRAS} {\bf 369}, 84, a footnote)

Bad quaterions  ({\sl ApJ} {\bf 637}, 886, Table 2)

Higher criticism ({\sl MNRAS} {\bf 369}, 598).  We had always associated this with studies of
Sherlock Holmes, but it is actually an effective statistic to detect non-Gaussianity.
Suprising and Unexpected high incidence of ``suprising'' and ``unexpected'' in the scientific
literature  (Jasienski 2006).

\subsubsection{The Minor Awards}

In addition to our major LDDFJR and Berlinski Awards (Sect.~11) here are a few lesser
ones, coming, appropriately,  from the author who has just been told by her colleagues on
the executive committee of the {\sl APS Forum on History of Physics} that they expect their
certificates of appreciation this year to be signed by Isaac Newton.

\begin{itemize}
\item[o] {The resume normal speed award to Zagury (2006) for showing that the background visible
light in the Milky Way is just forward scattering of starlight.}

\item[o] {The faint praise award to {\sl AJ} {\bf 131}, 2550, achnowledgements section which thanks ``the
anonymous referee for comments and suggestions that have contributed to the quality of
this paper,''  and perhaps to EB ({\sl MNRAS} {\bf 368}, 1716) who thanks the Israeli Army ``for
hospitality during the last month of this project.''  And we note that {\sl Nature} receives
12\% of its submissions from Israel but only 1\% of its referees'  reports  ({\sl Nature} 27
April, p. xv).}
\item[o] {The under the wrong lamppost award  (no, not the one where Lili Marlene is waiting, but
the one under which the drunkard is looking for the wallet he dropped elsewhere) was a
tie: Wellhouse et al.~(2005), whose goal was to find pre-magnetic CVs, but who started
with a catalog of magnetic white dwarfs and looked for 2MASS etc counterparts, which
presupposes the WD is the brighter of the two stars and its spectrum not too messed up
to allow detection of magnetic field effects; and Petit et al.~(2006) who set out to
look for irregular moons of Uranus and Neptune but discovered 66 new outer solar system
objects.  They actually found three of the desired moons as well, vs. no pre-magnetic CVs
from the first project.}
\item[o] {The short memory award to an editorial ({\sl Nature} {\bf 437}, 790) proclaiming that ``experience has
shown that launching sibling research journals strengthens {\sl Nature}.''  Does anybody but
us remember {\sl Nature - Physical Sciences}?}
\item[o] {The red spot on Jupiter award to Hopkins et al.~(2006), whose models of galaxy formation
and evolution explain everything else except ...}
\item[o] {The consider the alternatives award, previously given to closed-box models of chemical
evolution of the universe, has three contenders in 2006: (1) Pulsar emission coming from
somewhere between the surface and the light cylinder ({\sl A\&A} {\bf 445}, 779, Introduction),  (2)
``Neither LTE nor non-LTE models fit''  ({\sl A\&A} {\bf 455}, 318; we are betting on some different sort
of non-ness), and (3) A high ratio of radio to X-ray luminosity, which can be attributed
either to efficient radio production or to inefficient X-ray production (Hardcastle et
al.~2006).}
\item[o] {The didn't your mother tell you to say please and thank you award to half a dozen
examples
each of folks or papers who should have been cited but were not and of folks who should
have been consulted and thanked but were not.  Suggestions on how to call attention to
these without adding to the offenders' citation counts would be welcome!}
\item[o]{And the two game wardens, seven hunters, and a cow award  goes to Kulkarni and Rau 
(2006), 
Rao et al.~(2006), and Rykof et al.~(2005) for their assorted searches and surveys
addressed to optical transients, orphan afterglows, and so forth which have yielded an
assortment of cataclysmic variables and a dense foreground fog of flare stars.}
\end{itemize}

\subsubsection{The Minor Malfunctions} 

Some of these are actual acts of omission or commission, most others inept phrasing
or descriptions, often the result of a shortage of envelop backs.  But we begin with
honest appreciation of honest authors ({\sl ApJ} {\bf 642}, 842, abstract) who explain that ``due to
human error,  intensive monitoring did not begin until 43 minutes after peak magnification.''
The unmonitored was gravitational microlens event OGLE 2003-BLG-343, which was caught
at a magnification of 1200, but the peak was 8000, and could have yielded tighter limits
on small planets or, of course, conceivably a discovery.

Also on the observational front, an apparent absorption line at 8 keV in the spectrum
of 4C +74.26 is actually an emission line of Cu (K$\alpha$) produced by the circuit board 
in the
background spectrum that was subtracted (Ballantyne 2005).  We are reminded every time we
read something like this of the seemingly never resolved issue of X-ray features from gamma
ray bursts that once indicated $10^{12}$ G magnetic fields on the surfaces of nearby neutron
stars as the sources.  And the infrared counterparts of SCUBA galaxies seen at 8 $\mu$ with
SST are a reasonable mix of active nuclei and star formers (Ashby et al.~2006), but 11
of 17 are not the optical counterparts reported earlier.

Computations are, of course, not exempt.  Roche lobe overflow in close binaries can
develop instabilities due entirely to time steps being too big (Buning and Ritter 2006).
Sometimes even arithmetic malfunctions. ``The line (of Japanese emperors) has been continuous
for 100 generations or since about 500 AD'' (Nakahori 2006), which, at 15 years per generation
implies a good deal of precocity no longer found in that royal family.  Conceivably a good
many successions by younger brothers are included as ``generations.'' A few errors stand on
the edge between arithmatic and anhistoricity. Two Egyptian examples: (1) Tut's drinking
of white wine was dated to 2700 BCE (Lamuela-Raventos 2006) rather than something like 1700
BCE.  Medinet Habu is also 18th Dynasty, so when McClain (2006) described it as having
``survived 21 centuries in the desert'' he was perhaps tampering with the climate.

The only numbers worth a pocket calculator are, a plasma colleague once said, sums
of money,  for instance  (Pielke 2006), ``shuttle launches cost \$G each, vs. \$400 million
NASA estimated.''  G for Grand once meant \$1000. If only it were so.  According to
{\sl Science} ({\bf 312}, 174) the ``goverment's main science agency (is) USGS'', whose budget is
undoubtly
many \$G, but we would bet on NIH for many more \$G, and a vote based on accomplishments would
be invidious.  A pocket calculator may be needed to sort out the following: ``{\sl Nature}
{\bf 443}, 246  incorrecly stated that the device could accelerate electrons to 0.15\% of their
initial speed.  That number actually refers to the change of kinetic energy of the electron
bundle'' ({\sl Nature} {\bf 443}, 383). Pretty feeble either way, we think.

Of all parts of carrying out research,  from having an idea,  to scaring up money and
colleagues, to the bitter end of going uncited, the least pleasant part is surely the
proof reading.  This perhaps accounts for (a) a category of galaxy described in the
abstract as red and in the summary as blue  (Blanton et al.~2005), (b)  ``the tau meson or
K$^+$ as it is now known'' ({\sl Nature} {\bf 440}, 162, in an obituary of Richard Dalitz),  (b) a
fabulous image of M20 with SST and an inadequate caption (Rho et al.~2006); it is a
class zero object with multiple protostars, but it looks as if west is up, and what are
those blue-white dots?? and (c) the description by Lai et al. (2006) of ``inhomogeneous H I
reionization'' and ``incomplete H II reionization'', reminding us that, many years ago,
stellar spectroscopist R. P. Kraft remarked that hydrogen was fairly easy to ionize once
but almost impossible to ionize more than once.  He II reionization was presumably
intended.

Our closing examples are more or less historical (or, rather, not historical)
\begin{itemize}
\item[o] {``Long before `Smoke Gets in Your Eyes' astronomers had to contend with interstellar
dust''
(Speck 2005). Well, Trumpler found the dust in 1930, and the song dates from 1933
(and yes we know both verses).}
\item[o]  {``The identification of stellar mass black holes with X-ray telescopes in the 1960s''
(Dyson 2005, in a book review), but 1972 is the earliest one could possibly claim,
and there were doubters for another decade or so.}
\item[o] {Mauve was 150 years old in 2006 and discovered by William Perkins at age 18, but the
picture  ({\sl Nature} {\bf 400}, 429) shows a very bright purple.  Perhaps the word has faded.
Early aniline dyes were VERY vivid.}
\item[o] {Alfred Russel Wallace gets accused of enough things of which he was conceivably guilty
(psychic research, mislocating the solar system at the center of the Galaxy) without
attacking ``his interpretation of the canal-like structures appearing on contemporary
photographs of Mars''  (Smith 2006).  But the bone of contention was that the features
were reported only by observers looking through telescopes during rare, brief instants
of excellent seeing and did not appear in photographs. Only recently have imaging
devices become fast enough to record what Pickering,  Schiaparelli, Lowell, and others
thought they saw.}
\item[o] {``The only supernova explosion, 1987a, so far observed through modern telescopes''
({\sl Nature} {\bf 441}, 32).  One is tempted to say, like the wife in a classic {\sl New Yorker} cartoon
``they got your weight wrong too.''  At that 1987A got off easier than the supposed
SNR OA 184, which is a mere H II region,  ionized by an 07.5V star (Foster et al.~2006).
And 1987a was Comet Levy.}
\item[o] {A few other reclassifications: the microquasar candidate IRXS J162848.1-41524 is really
a K3 IV + K7-M V binary (Torres et al.~2005).  OGLE sc5-2859 is not the 3rd longest
microlensing event, but a classical nova (Afonso et al.~2006).}
\end{itemize}

\subsection{With a Little Help from Our Friends} 

Each year, anything from 10 to 50 folks respond to requests for suggestions of
highlights of the past year.  The request is intepreted in interestingly different ways,
ranging from ``My gut  feeling is that, for the first time in my career,  the most exciting
things may be in the fields of solar system studies and the discovery of planets rather
than in my own realm of the extragalactic'' to a list of 17 papers on arguments for the
respondent's own ideas, of which he is the senior or sole author.
The ``he'' is deliberate.
None of the responses of that sort were from women,  though this may be a selection effect
$-$ 10\% of 10\% of 50 is 0.5  (and you must supply your own hermaphrodiate joke).  This year
the author who collects most of these responses thought it only fair to be sure that at
least one paper or topic proposed by each respondent get mentioned.  A good many of the
suggestions already had green circles, red rectangles, or other notebook entries; some
were out of period  (perhaps in {\sl Ap07}) or otherwise not in the data base.  But here is a
stab at the rest.

\begin{itemize}
\item[o] {Diffuse interstellar bands attributed to magnesium tetrabenzoporphyrin, pyradine, and
related molecules  (Johnson 2006).}
\item[o] {Repulsive interactions between neutrons as the primary source of solar and stellar 
energy rather than hydrogen fusion (Manuel 2006).}
\item[o] {Correlation between stellar mass and correct value of the mixing length parameter
(Yildiz et al.~2006). It is in the direction of $\alpha$ increasing with mass and is required
by accurate data on binaries in the Hyades.}
\item[o] {A demonstration that the neutron source is $^{22}$Ne ($\alpha$ ,n)$^{25}Mg$ 
in at least some 4-8 M$_{\odot}$
stars from the existence of Rb-rich AGB stars  (Garcia-Hernandez et al.~2006).  The
secret to finding the right stars was to pick ones associated with heavily obscured
OH maser sources with large expansion velocities.}
\item[o] {An extraordinary richness of modes in roAp stars (whose very existence was a highlight
within  the time frame of {\sl ApXX}), whose explication requires both layering of the ions
in whose spectra the various frequencies show up and strong interaction with the local
magnetic fields  (Handler et al.~2006; Kurtz et al.~2006; Kochukhov 2006).}
\item[o] {A ``blue tilt'' in the color-magnitude relation for blue globular clusters, suggesting
that self-enrichment is a widespread phenomenon in them, not just confined to Omega
Cen and a few clones (Brodie and Strader 2006).}
\item[o] {A seemingly final resolution of the old problem of what emits the ridge of X-ray photons
coming from the galactic plane.  Not, it turns out, diffuse hot gas or the well known
species of NS or BHXRBs, but mostly cataclysmic variables and active stars like
RS CVn binaries  (Revnivtsev et al.~2006).  Both must be considerably more abundant
(perhaps a factor 10) there than here to do it all.}
\item[o] {Gazillions of things from Spitzer Space Telescope, most of which are to be found with
the subjects to which they pertain (and not always credited to SST; we wear a different
hat to track productivity of telescopes), but Humphreys et al.~(2006) report that the
Hubble-Sandage variable A in M33, which erupted in 1950, has developed a 
pseudo-photosphere of ejecta, so that the bolometric luminosity has held up, but in the
infrared, rather than visible.}
\item[o] {A small, weird variable star class (meaning the stars are weird, the class is small
and the authors are neither, though one could perhaps make a case for other combinations)
comprising V838 Mon, M31 RV, V4332 Sgr, perhaps Nova CK Vul (1670, on which spectral
information is sparse), and Nova V1148 Sgr  (1943).  A critical point is that they do not
all arise from the same sort of stellar population.  V 838 Mon belongs to a very young
cluster, while M31 Red Variable is part of the old bulge population (Bond and Siegel 2006).
The result has been a pleasing diversity of scenarios,  including stellar mergers  (Tylenda
and Soker 2006), some sort  of mass-losing red giant  (De Guchi et al.~2005, since V 838 Mon
now has an SiO maser), novae or last helium shell flash models held over from 2005, and
planetophagia (Retter et al.~2006), with three meals on record for V838.  Bond and
Siegel cautiously concur.}
\end{itemize}

\section{INTERSTELLAR MATTER, STAR FORMATION, YOUNG STELLAR OBJECTS, AND CHEMICAL EVOLUTION}

This section could also perhaps be called ``from diffuse baryons to the r process,''
and in coupling at least the first two items, we follow the lead of Wu et al (2005), who
have declared that the ``basic unit of star formation (from the Milky Way to high z) is
the dense core'' which they trace with HCN.  Interstellar matter, medium, or ISM will here
mean gas and dust, because the cosmic rays and magnetic fields appear in Sect. 3 and 12.

\subsection{The Interstellar Medium}

The dusty green circle of the year is the conclusion that grains are more like
solid solutions of organics in silicates than layers (Freund and Freund 2006).  This includes
grains condensed from gas phases, and the paper has some very good chemical appendices
for those of us whose last exposure to solid solutions was an 11th grade project called
``The Christmas Colloid.''

A traditional '`dust'' issue is the extent to which it is the same everywhere.  Not
entirely, is the answer.  It always reduces the gas abundances of heavy elements (called
depletion), but the patterns are different in the Milky Way, LMC, and SMC (Sofia et al.~2006;
Cox et al.~2006).  Notoriously the 2175 \ang\ feature is strong in some places (Noll and
Pierini 2005 on reddened high redshift sources) and non-existent in others (Chen et al.~2006a
on GRB host dust).  As for the nature of the 2175 absorber, we caught a vote for
the plasmon band of 7-ring aromatic C$_{24}$H$_x$ (Duley 2006), one for fullerenes (Iglesias-Groth
2006), and one for biological materials (Wickramasinghe et al.~2005).  Remarkably,
the feature has been known and worried about for more than 40 years (Stecher 1965).

Lots of dust (we won't vote on all, most, much) forms in the expanding atmospheres
of cool asymptotic giant branch stars, and you get different stuff depending on whether
C/O in the atmosphere is greater than or less than one (Steinfadt et al.~2005 on
UY Cen). Grain shapes are complex and arguably more fractal than merely porous (Min et
al.~2006), though they compactify with prolonged ion (GCR) bombardment (Palumbo 2006, a
laboratory result).  And our favorite lab dust of the year is lizardite (Hofmeister and
Bowey 2006).  This is not an example of biological dust, but rather layering of a
serpentine mineral and sheet silicate.  The authors call it Mg$_{2.95}$F$_{0.05}$Si$_2$O$_5$ 
(OH)$_4$ but do not expect it to come when called.

As for the gas phases (and starting with cool), we seem to have entered a new golden
era for discovery of interstellar molecules.  We will not attempt a full list but focus
on a few favorites, some of which expose the latent thought that, eventually, one has to
understand the structure of a molecule to be able to pronounce its name with the accent on
the right syllable.  From simple to complex:

H$_3^+$ is more important than you might expect (Geballe and Oka 2005). It was discovered by
J.J. Thomson (1911).

H$_2$O$_2$ definitely exists in the lab, in the atmospheres of Europa and some other moons, and
perhaps in the ISM (Zheng et al. 2006a), a laboratory paper which we indexed under
``bleaching of the stars.''

CF$^+$ is new (Neufeld et al.~2006), as is C$_3$O (Tenenbaum et al.~2006). And we tiptoe past
the acetone holding our noses (Friedel et al.~2005) to reach

CH$_3$C$_5$N, methyldyanodiacetylene, confirmed with the Greenbank Telescope (Snyder et al.~2006),
which counts as a good thing, since such molecules were among the drivers for the design.

CH$_3$CONH$_2$, acetamine (Hollis et al.~2006), the largest interstellar molecule with a peptide
bond (which is not the same as saying you should eat it; indeed we suspect that rather
few of these ISM denizens are green in any Gore-ish sense, however many circles we may
award them).

CH$_3$OC$_2$H$_5$, ethyl methyl ether (Fuchs et al.~2005), and some aspects of propanol 
(Maeda et
al.~2006), which is what comes after methanol and ethanol.  What comes after that is
presumably a deuce of a hangover.  One might reasonably suppose that CH$_3$CH$_2$CH$_2$CH$_2$OH
would be called butanol, since CH$_3$CH$_2$CH$_2$OH is propanol, but don't buy it on our say-so.

Interstellar gas has structure on many different scales.  The most puzzling statement
we read in this territory (Garrod et al.~2006) is that molecular clouds consist of transient
cores, which form, grow  to high density, and decay back into the background.  We had really,
truly been counting on them to form stars, so that we could go on to the next subsection.

Semi-regular spacing of giant molecular clouds (and hence of young star formation
regions) along spiral arms, which puzzled us back in Sect. 3, has been explained by a
magnetic Jeans instability that can build GMCs up to $3 \times 10^7$ M$_\odot$, spaced apart by 
about 10 times the Jeans length (Kim and Ostriker 2006).

Undoubtedly the two most confusing ISM phases we encountered during the year were the
cold neutral (Stanimirovic and Heiles 2005) at 80 K and the warm diffuse molecular (Henkel et
al.~2005) at 55 K.  Notice that you should wear your woolies in either case.

Gas in spiral halos occurs in a wider range of temperature and density phases than you
would expect (including H$_2$, Gillmon and Shull 2006), because star formation (and death) in
the disk feeds in (Tullman et al.~2006; Shelton 2006).  Even in the disk, nominally unstable
phases are well represented.  Hennebelle and Passot (2006) say that Alfv\'en waves keep up the
supply, as well as making cold gas very fragmented.

Our immediate surroundings include some WHiM (warm-hot interstellar, as opposed to
warm-hot intergalactic medium) according to Van Dyke Dixon et al.~(2006).  And at the
hottest end, we find gas of $9 \times 10^7$ K (8 keV) in the core of the Milky Way.  It is real
and energetically puzzling, say Belmont et al.~(2005), and required the capabilities of a
fully-operational Astro-E2 for its elucidation.  Well, someday perhaps there will be an
Astro-E3.

Most of the galactic gas is, of course, neutral H I, and if what you care about is
converting it to stars, then the most important issues are keeping up the supply and
avoiding pressure sources that might impede collapse or contraction of clouds.  The good news
on the first front is that we are not actually losing gas $-$ it is a fountain, not a wind
(Keeney et al.~2006).  As for possible fresh supplies, for many of the past 15 years we
have quoted various experts on whether the  high velocity clouds are (as William A.
Fowler said many years ago) S0Bs diluting the heavy elements that stars burn out their
hearts to make.  About 10 papers this year, including a ``no'' from Miville-Deschenes et 
al.~(2005) whose observations with SST and the GBT revealed a cloud with dust (hence not fresh
unpolluted gas), and a ``yes'' from Sommer-Larsen (2006), and sometimes the arrival triggers
star formation (Casusa et al.~2006) allowing us at last to move on toward contracting
clouds.  There we will discover that the strongest support comes from magnetic fields of
10-20 $\mu$G in small H I clouds (Sarma et al.~2005). This is a good thing, because we know
what to do about it\footnote{
Compare the chap who deliberately took his cold out into sleety  weather so it would
turn to pneumonia, which can be cured. Antibiotics or ambipolar diffusion, as the case
may be.}.

\subsection{Star Formation} 

Our two index pages for star formation list more than 120 pages, but having got here
by way of magnetic fields and one possible triggering mechanism, those are perhaps the
places to start.

The star formation process absolutely must dissipate dynamically important magnetic
fields to get even as far as a protoplanetary disk, or the average new star would have
B $\approx 10^7$ G vs. the $10^{3-4}$ G seen in T Tauris (Galli et al.~2005). Happily, we can
sometimes see that gravity is winning (Girart al.~2006) from the relative orientation of
disks and field lines.  But, say Thompson et al.~(2006), star bursts (Arp 220 in their
case) are different in having thermal pressure much larger than magnetic.

Is star formation triggered by something that causes relatively rapid compression of
some volume of gas?  Quite often, apparently yes.  On the largest scales come galaxy
mergers and interactions (Menbel et al.~2005 on NGC 4038/39; Chiosi et al.~2005 on 
SMC/LMC; Alonso et al.~2006).  Next are spira1 arms (Bonnell et al.~2006a; Smith et 
al.~2005a) both assembling GMCs and urging them to collapse.  The passing shock stirs the
ISM, and next come cloud-cloud collision, on which Looney et al.~(2006) blame the
formation of the cluster around BD +40$^\circ$ 4124. And finally we get radiation-driven
implosions (Urquhart et al.~2006) and other processes that act on the scale of single
clouds and star clusters, so that star formation propagates across a region (Wilking et
al.~2005; Li and Smith 2005), enabling the process to be quite extended in time as well as
in space (Moriarty-Schieven et al.~2006), and perhaps producing clusters for which
different age indicators will give different answers (Sect. 8).  But sometimes, say
Whitmore et al.~(2005), it's just the clouds' time and nothing shocking is needed\footnote{
We, if not you, are reminded of the scene between Julie Jordan and Billie Bigelow
in Carousel.  BB: Look at the blossoms. (which are falling around them). JJ. The wind 
brings em down.  BB:  Ain't no wind tonight.  JJ: Just their time, I guess.  And
yes, she falls shortly thereafter.}

Having lots of gas around is a necessary condition for star formation, but apparently
not a sufficient one (Begum et al.~2006 on GMRT H I maps of relatively small galaxies).
And before going on to detailed considerations, many of which involve theoretical input,
we pause at our first green circle star formation paper, Marcel et al.~(2006) on
fragmentation of molecular clouds to cores.  They point out that the minimum core mass you
find and the peak of the log normal N(M) go down as the spatial resolution of your
calculation becomes finer.  We think this is a bit like needing a very fine grid in weather
forecasting in order to find the most extreme wind speeds, drenching rains, and so on.

Some standard questions to which there was at least one answer this year include:
When and how does star formation begin?  How does the global rate depend on redshift, and
what are the global processes?  What are the proper initial conditions in a molecular cloud
with which to start calculating (it being not unlikely that you will get out more or less
what you put in)? What makes the IMF?  Do high and low mass stars form in more or less the
same way?  Is star formation in the universe (or in specific galaxies) really just about
over?  This is what we,  in our Copernican days, used to call the ``last gasp problem.''
We will take these more or less in order.

\subsubsection{First Things and Modes}

The earliest vote we caught was for star formation beginning by z $\approx$ 20 (Chary et 
al.~2005). What did it consist of?  No small stars at all, at least up to 10$^{-4}$ of solar
metallicity (Tumlinson 2006). Or, if you prefer, stars down to as small as 1 M$_\odot$ with
cooling from HD (Shchekinov and Vasiliev 2006) and/or dust by Z = $10^{-6}$ solar (Schneider et
al.~2006), and/or outward transport of angular momentum by magnetic fields (Silk and Langer
2006).

In a global sense, there are perhaps two rather different modes of star formation, the
one now common in disks and small spheroids, where negative feedback comes from supernovae
and the products are the disk stars we see about us (Silk 2005), and an earlier one
belonging to massive spheroids, where feedback comes from AGN jets and can be positive if
the jet concentrates the gas supply.  Predictions include a flatter IMF for the latter,
spheroid star mass proportional to (black hole)$^{2/3}$, the biggest black holes first, and
super-Eddington outflow as the cause of strong radio sources.

\subsubsection{Changes with Redshift}

Assessing the redshift dependence of star formation rates requires deciding which
of MANY indicators you propose to rely on.  And, before going on to anything with numbers
(the customary units are M$_{\odot}$ yr$^{-1}$ Mpc$^{-3}$), we remind the one reader in seven 
for whom this is calculator-stays-in-the-pocket-day\footnote{
A respected colleague who eventually succumbed to Alzheimers disease remarked during
the process that the first warning he had received was a fairly sudden decline in facility
at mental arithmatic, so the once-a-week check is not entirely silly.  You get a choice
of three possible silly associated remarks:  (1) How will the younger generation be able
to tell?  (2) We may have told you this in some previaus {\sl Ap XX} but cannot remember, 
or (3) we hope the Internal Revenue Service will understand that the AMT form was filled 
out on CSITP day.} that, for H = 70 km s$^{-1}$ Mpc$^{-1}$ and $\Omega_b = 0.04$,
if all the baryons were in stars, then the baryon density is $4 \times 10^{-31}$ 
g cm$^{-3}$ 
or $6 \times 10^9$ M$_\odot$ Mpc$^{-3}$, and in a universe that began forming stars about 
13 Gyr ago, the
average rate cannot be larger than 0.45 M$_{\odot}$ yr$^{-1}$ Mpc$^{-3}$, 
which should be adjusted downward
to allow for the 3/4 or more of baryons that are not currently in stars.  At a slightly
less quantitative level, it is also worth remembering that there are many more years per
redshift bin at small redshift.

Indicators that someone invoked during the year include mid and far infrared, H I,
R band surface brightness, H$\alpha$, H$\beta$, PAH emission, 6 cm radio emission, [O II],
[C II], X-rays, Ly$\alpha$, UV continuum, high mass X-ray binaries, supernovae, and OB stars.
Obviously some can apply only for nearby, resolved galaxies, others only when you actually
understand the physics of, e.g., radio emission and can eliminate active nuclear
contributions.  The salient points we caught were:
\begin{itemize}
\item[o]{discordance even for the SMC, where the stars are more or less laid out before us,
but you can get $0.05-0.4$ M$_{\odot}$ yr$^{-1}$ (Shtykovikiy and Gilvanof 2005),}
\item[o]{some gain in confidence if you compare mid IR and 6 cm maps (Vogler et al.~2005 
on M83),}
\item[o]{factors of two uncertainty after correcting H$\alpha$ for the extinction implied 
by other fluxes (Moustakas et al.~2006),}
\item[o]{a vote for H$\alpha$ + 24 $\mu$m to seek out unobscured plus obscured star formation 
(Perez-Gonzales et al.~2006, who find about half and half\footnote{
Half and half is a good guess when dividing things about which very little is known, but
there is now considerable mathematical theory for cutting cakes of which some portions are
more desirable than others (Mirsky 2007).  And no, we never figured out who got the
piece of the SN1006 anniversary cake at Prague that had the supernova on it.  The most
frosted author helped cut and serve and so got a corner piece with lots of frosting.}.
Other authors find more
obscuration in one context or another (Lamers et al.~2006a) and others less 
(Rodriguez-Fernandez et al.~2006).}
\end{itemize}

Please keep all this salt (and frosting) in mind as we tell you that the local,
current rate is 0.018 M$_{\odot}$ yr$^{-1}$ Mpc$^{-3}$ (Iglesias-Paramo et al.~2006); 
that it rises as about
$(1+z)^3$ to $z=1$ (Doherty et al.~2006); is crudely flat at $5-15$ times the local level over
$z=1-4$ (Colbert et al.~2006; Thompson et al.~2006; Wadadekar et al.~2006; 
Wang et al.~2006e); and was smaller again at $z=4-6$ (Thompson et al.~2006).

Do we then live in a dying universe?  Yes, more or less.  Indeed a few small galaxies
have enough gas to last another Hubble time at their present star formation rates
(Alonso-Garcia et al.~2006; Fukugita and Peebles 2006), but, as we said before, citing a
different authority, gas is not enough (Warren et al.~2006), though it certainly helps
(Komogi et al.~2005).

\subsubsection{Initial Conditions and Their Causes}

Perhaps the most important thing to be said is that there is no such thing as a
homogeneous molecular cloud upon which you can then operate (with shocks, turbulence,
magnetic fields, or models thereof).  They have lots of density and velocity structure from
the get-go (Heitsch et al.~2005).  And yes, it is fairly easy to arrange the initial
conditions of your calculation so that they both resemble observed clouds (to the very
limited extent we can measure that substructure) and produce stellar populations like ones
seen (Stanke et al. 2006; Krumholz et al.~2005; Sanchez et al.~2006; Rathbone et al.~2006).

We think, however, we also caught four papers saying nearly the opposite:
Ballesteros-Paredes et al.~(2006, high Mach number makes lots of little bits), 
Vig et al.~(2006, 
the distribution of core masses is N(M) $\propto M^{-5}$ for 14-22 M$_{\odot}$, 
which is certainly
not a standard IMF), Clark and Bonnell (2006, the N(M) for clumps is neither due to gravity
nor similar to the standard IMF),  Reid and Wilson (2006, clump masses range from 0.2 to
120 M$_{\odot}$, which sounds promising, but peak at 4 M$_\odot$ which is much less so).

With the votes so far running at 4:4 for and against some basic understanding of
star formation, we went back to our notes to hunt for a tie-breaker, and found a study of
the Orion B star formation in which the clump size distribution peaks at 
$10^{-3}$ M$_{\odot}$
(Johnstone et al.~2006, meaning that some further agglutinative process will be needed)
and a calculation in which more power on large scales yields more small stars (Goodwin
et al.~2006).  This would seem to be one of each, so short of going into extra innings
we must simply tackle the problem from the other end, the Initial Mass Function itself.

\subsubsection{The Initial Mass Function}

Two groups of observers focussed on the low mass end.  Levine et al.~(2005b) tell us
that not all young clusters are the same, and Oasa et al.~(2006) find that, in one specific
region, there is no evidence for sharp changes in slope of the IMF either at the mass
where deuterium fusion becomes the only game in town (0.085 M$_\odot$ or thereabouts) or at
the mass where not even deuterium fusion is important (0.03 M$_\odot$ perhaps). 
And two groups
of theorists tell us that the median mass will be the thermal Jeans mass for ambient
conditions (Bate 2005) and that it is important to get the gas cooling physics correct 
(Bonnell et al.~2006).

Well, as Martin Schwarzschild is supposed to have said, in connection with student
complaints that a particular distinguished professor had both an inaudible voice and
unreadable writing, ``at least there is no contradiction.''  In case you should want one
we leave the last word in this subsection to Mouschovias et al.~(2006), who  put forward
31 arguments to show that ambipolar diffusion is more important than turbulence in the
context of star formation.  They state that the project had no external support, but
{\sl The Astrophysical Journal} is thanked for waiving page charges.

\subsubsection{Stars of Large and Small Mass}

Interest in this issue arises largely from the assumption that the formation of
stars of less than, say, 5 M$_\odot$ is well understood, plus the desire to know whether bigger
ones do it the same way, despite the risk of being brighter than their Eddington limits
during later accretion phases.  The basic ``same'' mechanism is a core onto which
accretion continues from a disk, while angular momentum and magnetic field are
being removed by collimated jets top and bottom.

Of many papers on either side, we bring you two ``sames'' (Fuller et al.~2005
on observed infall for massive cores at rates of $2-10 \times 10^{-4}$ M$_{\odot}$ yr$^{-1}$,
and Wolff et al.~2006 on continuity of angular momentum per unit mass over the full range of Orion
stars), two ``differents'' (Sollins and Ho 2005; Alvarez and Hoare 2005, on the rarity of
collimated jets belonging to high-mass cores), and a triumphant ``both please'' from
Peretto et al.~(2006) reporting the merger of two or more class 0 sources at the center
of a collapsing protocluster, centered on a massive turbulent core.  We suppose this
probably also bridges to three related results:  (1) the IMF of big stars joins smoothly
to that of small clusters near 100 M$_\odot$ (Dopita et al.~2006),  (2) sometimes it takes a
whole big cloud to make a whole big cluster (Keto et al.~2005), and (3) only large clusters
contain massive stars (Massi et al.~2006).  This is meant to be slightly more informative
than the thought that you cannot make a 100 M$_{\odot}$ star from a 50 M$_\odot$ cloud, 
because the
authors summed the observed N(M)'s of six clusters that, if they were one, would probably
have had a 22 M$_\odot$ star but actually reach only to 10 M$_\odot$.

Logically at this point, we should advance either to star clusters or young stellar
objects.  In fact YSOs come next, and clusters live in Sect. 8.

\subsection{Young Stellar Objects}

The earliest class, 0 (``zero''), emerging from the interstellar material, and the
processes of star formation last only $2-6 \times 10^5$ yr (Froebrich et al.~2006), 
which is something like the free-fall time at the density, $10^{-9}$ g cm$^{-3}$, 
when collapse sets in.
Class 0's by definition are supposed to derive all their energy from gravitational
collapse/contraction, and, say Terebey et al.~(2006), the age of a protostar is defined
as the time since the onset of cloud collapse ($10^5$ yr for TMC-1). Almost before you know
it, the objects  have both infall and outflow (Chandler et al.~2005 on IRAS 16293-2442),
and pretty soon all your favorite sorts of YS0s are there.  We picked just one paper on
each of our favorite sorts this year:

\begin{itemize}
\item[o]{FUOrs, a ``both please'' paper, in the sense that Kinner et al.~(2006) say that 
the X-rays come mostly from magnetic activity, but also from accretion.}
\item[o]{The very first Herbig-Haro object, HH1, for which spectroscopy shows that at least 
1/3 of the metals are still in dust (Nisini et al.~2005).}
\item[o]{T Tauri stars, a difficult choice among a double handful of papers, but the coveted
colored squiggle goes to the three different ways they power their outflows (Ferreira et 
al.~2006), self-collimated disk winds or jets, pressure-driving, and blobs ejected from the
magnetosphere.}
\item[o]{The Herbig Ae/Be stars, more massive relatives of the T Tauris, grade into ordinary
Be stars say Pogodin et al.~(2006) concerning HD 52267 which is a binary consisting of
a 20 M$_\odot$ BOe main sequence star and a 5 M$_\odot$ pre-MS companion.}
\end{itemize}

Most YSOs have residual accretion disks, which last of order $10^7$ yr (Jayawandhana
et al.~2006); have accretion rates proportional to core mass (Natta et al.~2006; Dullemond
et al.~2006; Gregory et al.~2006; careful bean-counters will notice that this is one from
each of the giant Journals, so it must be true); and are rare by the time you reach the
weak-lined T Tauri stars (Padgett et al.~2006).  This last is not actually a tautology,
because the emission lines come from a gas disk, and the authors were stalking infrared
excesses from dust disks.  An implication is that very cautious beans can escape being
counted (in estimates of star formation rate, for instance), just as the late worm does
not get eaten.

Accretion and outflow naturally spin the stars up and down (not necessarily
respectively), and having plowed through the correlations in {\sl Ap 02} (Sect. 3.2) we will
say in 2006 only that the correlations remain complex (Fallscheer and Herbst 2006, on
effects of disk locking) and that kiloGauss fields are common (Yang et al.~2005 on
TW Hya at $2.61\pm 0.23$ kG), and many uncited papers.

Several groups presented evolutionary tracks for pre-MS stars, but the weirdness award
(a pinkish-green circle) goes to Marquest et al.~(2005) for a very strange little extra loop
for $1.8-1.9$ M$_\odot$ stars between $\log T = 3.92$ and 3.94 and $\log L/L_\odot = 1.05$
to 1.15.  The
cause is convective overshooting; the loop crosses the standard evolutionary track; and
we have difficulty imagining data precise enough to trace it out in an observer's HR
diagram.

\subsection{Chemical Evolution}

With arithmatic inevitability, by the time you read this, it will be 2007, the 50$^{th}$
anniversary of Burbidge, Burbidge, Fowler, and Hoyle (1957) or B$^2$FH, generally thought of
as the beginning of our  modern understanding of nucleosynthesis and chemical evolution
in the universe, though it was, from a different point of view, a synthesis of a great
deal of work from the preceding decade. They focussed on attributing the full range of
elements and isotopes to well-defined processes (except that called X, which turned out to
be a combination of the early universe and cosmic ray spallation) occurring in stars and
supernovae.  To a very considerable extent, the processes they identified are still seen
as dominating production of the stuff we live with.  Major developments since concern
the assembly of stellar populations and whole galaxies from individual stars and reactions
so as to end up with the right compositions, colors, gas fractions, numbers of past
mergers, and much else all at the same, present, time.

\subsubsection{The Largest Scales}

First presumably comes the whole universe, and we start by trying to understand what
it means that the theoretical atom counters lead us to expect more than the observational
atom counters can find of both baryons and metals (Lehner et al.~2006; Vikhlinin 2006a;
Bouche et al.~2005).  A unified solution would say that both are lurking, well mixed,
in low density phases that are difficult to probe with either emission or absorption
features. Now, since baryons are all left from $t \approx 0$, while metals have been produced
continuously in stars, which are generally found in galaxies, this doesn't really sound
terribly promising.  Metals should cling to galaxies, should they not?  Well do they?
Gradients there certainly are, with the inner parts of large galaxies more metal rich than
the outskirts (Vorobyov 2006, which happens to concern the Milky Way).  The case is already
not so clear for X-ray clusters (Sanders and Fabian 2006).

As we step outside the densest regions and explore using assorted kinds of QSO
absorption clouds, you have to dig out from under the effects of depletion (Rodriguez et 
al.~2006a) and downsizing (Zwaan et al.~2005) to find the 2006 answer, ``some of them do and
some of them don't''.  Metals cling to galaxies more firmly than the baryons do for
Scannadico et al.~(2006) and Simcoe et al.~(2006), but less so for Pieri et al.~(2006) and
Vladilo and Peroux (2005). Kapferer et al.~(2006) endorse situational ethics, because
mergers can erase previous clinging behavior.  Just for a moment, one thinks that an
answer might come from the relative fractions of gas turned into stars and blown out by
star forming galaxies, but both the papers we caught this year that attempt such an
estimate (blow-out wins in both) start with models of chemical evolution, otherwise known
as affirming the consequent (Erg et al.~2006; Lanfranchi et al.~2006).

\subsubsection{Galaxies and Stellar Populations}

On the more modest scale of one galaxy at a time, modelers for decades stumbled
against the so-called  G dwarf problem.  The simplest possible model is a large (well,
galaxy-size) box of gas that eventually turns itself into stars, which make heavy elements,
explode, and enrich in a self-contained fashion, and you drop in (or print out) from
time to time and look at where things have got to for both stars and gas.  Such a model
for the disk of the Milky Way has always predicted more stars with less than 10\% of solar
metallicity than we find among stars (e.g. G dwarfs) that live as long as the age of the
disk.  With gas blowing out and falling into galaxies all over the place, not to mention
mergers, perhaps this should never have been a surprise.

Ought we, alternatively, then to be surprised that many stellar populations are
fairly well fit by closed box models?  Not really, so long as metals and other gas are
lost or merged or gained together.  The closed boxes of the year included (a) gas-rich dwarf
galaxies (Van Zee et al.~2006),  (b) spirals near the center of the Virgo cluster (Dors
and Copeti 2006), noted by Shields (1991); and our excuse-making minds immediately suggest
``well, any gas available to fall in there must already be fairly metal rich,'' (c) M31
(Worthey et al.~2005), and (d) perhaps even the bulge of our own Milky Way (Fulbright et
al.~2006).

The next scale down includes parts of galaxies and stellar populations.  We all firmly
think that cosmic gas started out with no heavy elements (except a scrap of lithium).
Stars formed from that are said to belong to Population III and must have made the first
metals.  No such stars have ever been seen, and calculations of the nucleosynthetic
products are bedeviled by  major uncertainties about the amount and importance of (a)
rotation (Chiappini et al.~2006),  (b) mass loss (Smith and Owocki 2006), and (c) 
non-LTE\footnote{
This is one abbreviation that cannot be expanded to less than a paragraph.  LTE is
local thermodynamic equilibrium $-$ meaning that all temperatures can be taken to be the
same at a sing1e location in a star.  Non-LTE means that they cannot, with radiation
temperature typically higher than kinetic temperature in regions of low gas density.
But it is the equilibrium not the locality that is being denied, so non-local thermodynamic
equilibrium means the opposite of what is intended.  Feel free to contemplate other
examples, for instance how one might expand the attitude of being anti-NASA.)}
(Collet et al.~2006).  The range of masses making up Pop III is also poorly determined
and addressed elsewhere  (Johnson and Bromm 2006 for a quick summary).

As a result, there remain disagreements about (1) how much pristine gas will remain
to form ``late Pop III'' stars by $z = 3$  (quite a lot says Keel, 2006; very little say
Jimenez and Haimon 2006 looking at FUSE data, and the difficulty in trusting the observations
fully is that we necessarily look where stars have formed) and (2) whether there is
evidence in the composition of extreme population II stars for the products of Pop III
(pair instability) supernovae (Meynet et al.~2006 no; Frebel et al.~2006 maybe).
Because the most metal poor stars are rare, faint, and difficult to analyze (because
weak-lined), we do not yet have a sample with enough stars to say that their average
represents the sum of a bunch of inputting events $-$ each individual star having been
fed by only one or few supernovae (Aoki et al.~2005).

Coming closer to the present, things get much better.  Metallicity really does
increase as time goes on, in more or less the expected way (Rocha-Pinta et al.~2006;
Lamareille et al.~2006) and with the number of stars that have given their lives to
provide heavy elements (Lee et al.~2006).

\subsubsection{Individual Processes and Nuclides}

At the still more detailed level of particular elements, nuclides, and processes,
there is a near anti-correlation between abundances and numbers of papers, but we will
attempt to buck the trend and start with common stuff. It has become exceedingly difficult
to measure the primordial helium abundance, both because all the entities that had it
are gone and because the measurers  cannot be unaware of what the theorists expect
them to find.  Two
values  reported during the year touch each other and the predictions within their
error bars, $Y_p = 0.244 \pm 0.004$ (Holovatyi and Melekh 2005; Fukugita and Kawasaki 2006).  
How the helium content of the universe has changed since depends very heavily on which bit of
the universe you ask, and it is a topic in which we take a modest interest, having devoted
a prehistoric paper or two to trying to make the predictions of output from massive
population II stars match (some one set) of the observations.

This year, $\Delta Y/ \Delta Z$ included (a) about 20 between generations of stars in the 
globular
cluster $\omega$ Cen (Maeder and Meynet 2006),  (b) $8.8 \pm 4.6$ (Holovatyi and Melekh (2005),
and
(c) $1.1 \pm 1.4$ from metal-poor, gas-rich galaxies (Fukugita and Kawasaki 2006),
a result which
has the curious property of taking in a decrease of Y with time. Our value was 3,
inconsistent with all of the above.

Nothing exciting seems to have happened with regard to the common things, carbon,
oxygen, and the iron peak (Froehlich et al.~2006), while the ratio of alpha elements to
iron peak may well depend on stellar rotation (Tsujimoto 2006).

You are not, perhaps, used to thinking of $^3$He as an abundant nuclide (especially if
you have tried to buy some lately), though it is, compared to just about everything except
other forms of hydrogen and helium.  Two lines of evidence concur that it is made in
intermediate mass stars and ejected thereby.  Balser et al.~(2006) have found a second
planetary nebulae with elevated  $^3$He/$^4$He$=2 \times 10^{-3}$; and Busemann et al.~2006 
report that
the local interstellar abundance is larger than the BBN prediction.  The presence of
flourine in planetary nebulae (Zhang and Liu 2005) shows that it, too, gets out of AGB stars,
where it is produced in some sense at the expense of Na and Al (Smith et al.~2005).

And, before we go on to heavier elements, yet another look at lithium.  Traditionally
some comes from the early universe, and the amount is that seen as a plateau of Li/B
vs. effective temperature in population II stars.  This may still be much of the truth, but we
caught enough quibbles and confusions of sufficient complexity that they have been upgraded
from (a), (b), (c) to separate sentences.  There is some discord between calculations and
observations concerning the extent to which Li is depleted in pre-MS evolution (Montalban
and D'Antona 2006, on Pop I, of course, but similar problems may well have afflicted Pop II).

Downward diffusion occurs during main sequence evolution, leaving one rather uncertain
about what the plateau value really is or what it means (Korn et al.~2006; Charbonel
2006).  The idea can be traced back to Aller and Chapman (1960).  Aoki et al.~(2006) ask
whether stars with very small Fe/H, large C/Fe, and detectable Li belong to the plateau.
And Yoshida et al.~(2006) say that both Li and Be will be produced in supernovae, at
a rate we suspect cannot be very large, since it depends on the strength of neutrino
oscillation.  To regain your confidence in the conventional view, please visit Charbonel
and Primai (2006).  Our notes on this say ``beaten to death,'' but both the authors and the
lithium will surely be back next year.

What about $^6$Li, which is supposed to be a product of cosmic ray spallation of
CNO?  Some made in situ by active stars say Christian et al.~(2005) on the K dwarf GJ 117.
And some must have been made before the most metal poor Pop II stars, since there is also
a $^6$Li plateau (Asplund et al.~2006).  The authors suggest pregalactic shocks or 
decaying/annihilating dark matter particles as the cause. Beryllium remains plateauless, 
though it does not just track iron (Boesgaard and Novicki 2006).

Nitrogen is supposed to be mostly secondary (made from the C \& O products of helium
fusion during CNO cycle hydrogen fusion).  Each year, however, we record a bit of 
observational evidence for primary nitrogen (Nava et al.~2006).  And the source this year is
massive stars in star bursts (Izotov et al.~2006a).

The cute little green circle of nucleosynthesis went to Diehl et al.~(2006) for showing
that $^{26}$Al must be largely made in massive stars, because it co-rotates with the 
thin disk
of the Milky Way  (based on high resolucion spectra from INTEGRAL). It comes, say
Limongi and Chieffi (2006), to a considerable extent from the explosive burning of C and Ne
in stars of 11-100 M$_\odot$ (rather than from AGBs for instance) as was predicted by Arnett 
(1977).

Of the heavy element processes, r (rapid capture of neutrons on Fe seeds) appears
first in the history of the universe because it can all be done in one massive star
(Ivans et al.~2006, an observation of HD 22170; Nishimura et al.~2006, a calculation for
a 13 M$_\odot$ star).

The s process comes later, because the star doing it must have some iron (etc.) when
it forms.  The most metal poor star for which s process abundances patterns have been
seen is CS 30322 023 at [Fe/H] = -3.4 (Masseron et al.~2006).  It also has the smallest
surface gravity, $\log g = -0.3$, for any well established Pop II star.

A few metal-poor stars are relatively rich in both s (slow) and r (rapid) capture
nuclides.  Jonsell et al.~(2006) suggest the poor things may have been zapped by both a
nearby pre-formation supernova and mass transfer from a now defunct AGN companion.

Finally comes the poor old p-process (meaning proton capture, or neutron removal,
or perhaps even transformation of n's to p's where many weak interactions sum up to fairly
strong).  One is used to thinking of it as secondary (acting on r seeds) or even tertiary
(acting on s seeds), but the part at A$<110-120$ apparently comes from a neutrino-driven
wind and can be primary (Wanajo et al.~2006), while the heavier stuff comes from secondary
processes in the O and Ne shells when a shock wave from a supernova core collapse
passes through (Pruet et al.~2006).  Goriely et al.~(2005) would like to make some
additional p (no, we didn't say that; it's just your twisted mind reading it) from helium
accretion on white dwarfs below the Chandrasekhar limit.  We wonder a bit about the
difficulties of getting it out (and we didn't say that either).

Hanging right off the end, we have the first detection of indium outside the solar
system.  This comes from a sing1e line of In I at 4511 \ang , the same as in the sun, though
for 42 stars (Gonzalez 2006).  The pattern of In/Fe vs. Fe/H says that it is mostly an
r product, as expected.  Why should you care?  Well,  once you get outside the solar
system, you might still want to be able to produce InSb (pronounced ``inz-bee'') detectors
for infrared radiation.  Have we checked that extra-solar system Sb is known to exist?
No, but you could look it up.

\section{STARS} 

Most definitions of astronomy mention something about stars, including, we guess,
the offering this morning from a student in Physics 20A, who described astronomy and
astrology as two different ways of studying the stars. And no, his request for additional
points on a paper that mentioned only astrology as the subject matter of the course
is not likely to be granted.

\subsection{Location, location, location}

We linger where we have been for decades, slightly above the local midplane of the
galaxy by 23 pc (Joshi 2005) or $19.6 \pm 2.1$ pc (Reed 2006), but also, somewhat more
strangely, above the average vertical locations (in both latitude and galactic coordinates)
of new stars and old clusters found in 2MASS, such that $b < 0$ and $z < 0$ objects exceed
``greater than'' by factors $1.5 - 2.0$ (Mercer et al.~2005), zero being the galactic plane.

And it is our distance from the sun that determines how grim things will become
when our star expands as a red giant, enhancing its wind by $10^5$, 
its total photon radiation
by $10^3$, and its UV flux by a factor 100 (Rybicki 2006). Much the same sorts of things
would be in the offing if we happened to orbit any of about 50 other stars that are quite a lot
like the sun (Holmberg et al.~2006). Melendez et al.~(2006) report on HD 98618 the
second closest to us after 18 Sco, and King et al.~(2005) on HIP 78390, tied with HR 6060
for most sun-like (using a method whereby the sun has a mass = 1.01 M$_\odot$).

The ``late heavy bombardment'' stage would have been even worse if we claimed Vega
as our star (Absil et al.~2006), except that we would not yet have had time to evolve and
so (probably) wouldn't mind very much, along with the deceased astronomers of Sect. 3.
The various sorts of edges of the solar system have moved outwards a good deal in the 50
years since the start of space astronomy, and, with the heliosheath now 75 AU thick
(Czechowski et al.~2006) it seems unlikely that Voyager 1 will get out in our lifetimes.

Observers see (and seers observe) faculae (Sheeley and Warren 2006) and granulation
(Vazquez Ramio et al.~2005) directly on the sun (because they are close to it) but must
infer existence on other stars from micro-variability (Regulo et al.~2005). That the
characteristic time scales for changes are given as $1-4$ minutes for the sun (Vazquez Ramio 
et al.) and $8-10$ minutes for other stars (Regulo et al.) reflects, we suspect, some
difference in definition (perhaps an astrological one), not excessive solar haste.  The
brightness contrast between grains and lanes increases from F to M as convection zones
deepen (Regulo et al.).

To know that the amount of mass ejected in an eruptive prominence ($10^{15}$ g, Gilbert
et al.~2006) is about the same as the mass of a primordial black hole evaporating now or
of a 1 km asteroid will probably not help to protect you if you happen in the wrong
location with respect to any of them.

Even individual astronomers can find themselves in the wrong location. Shirley (2006)
submitted his paper from a well-supported institution, where solar research is not
uncommon, but declares that the work was ``supported by the private resources of the
author.''  The paper concerns an explanation of solar cycles that struck your author
who has seen between 5 and 6 of them as rather odd, though it goes back to G.~H.~Darwin
in 1898. That is, of course, 5-6 cycles, and a much larger number of explanations.

\subsection{Ordinary Things about Ordinary Stars}

But we begin with the oddest, an analysis of line widths of supergiants in the
Milky Way and SMC (Dufton et al.~2006).  The authors report (for both samples) that
the average v sin i increases from 20 to 60 km s$^{-1}$ from B8 to BO I 
in rotational broadening.
There is the expected microturbulence of $10-20$ km s$^{-1}$ (this is nearly always big in
supergiants).  But, in addition, they report $20-60$ km s$^{-1}$ of macroturbulence (also
largest in the earliest types), which they describe as ``large scale supersonic velocity
field of unknown origin and nature.''

This is, we suppose, part and parcel of the on-going absence of a decent theory
(not peculiar to astrophysics) for turbulence, convection, and such.  Most analyses of
stellar spectra are done with mixing length theory, for which the secret word is alpha,
the ratio of the distance a blob travels to local pressure scale height ($1-2$ is typical).
Its fudge aspects are revealed in the different best-fit alphas you get for different
choices of how to normalize composition to the changing one of the sun (Ferraro et 
al.~2006). Full spectrum convection (which allows blobs of different sizes and vertical
durations) must be better in some physically meaningful sense, but not enough better
that you cannot fit most stars with either (Miglio and Montalban 2005).  A 3-d
numerical simulation is still better (Samadi et al.~2006), but has so far been applied
only to the sun, where it improves agreement with measured properties of p modes.

Meanwhile, the cry for ``extra mixing'' (more than expected from the convection that
carries out heat) rises from many throats, including De Laverny et al.~(2006) on carbon
stars in the SMC and IGI, Denissenkov et al.~(2006), on tidal forces in close binaries
as a driver, Palacios et al.~(2006) on population II red giants, for which the
maximum expected from rotational mixing is still not as much as seen in surface
compositions vs. position on the HR diagram, and Chaname et al.~(2005) on underestimates
for slow rotators and overestimates for rapid rotators.

Refreshingly different, Ventura and D'Antona (2005) seem to be calling for less mixing to
account for Mg-Al anticorrelation in surface abundances of globular cluster stars.

Barnard's star, previously famous for rushing by (or, rather, being rushed by, by us)
as a halo star had a large flare like a young dMe star, most unlikely for its supposed
old/pop II status (Paulson et al.~2006), though the authors note that there are no real
measurements of its composition or age.  The flare was caught accidentally in the
McDonald Observatory exoplanet search (Two game wardens ...).

\subsection{Stellar Structure and Evolution Calculations}

Textbooks (including our own) make this sound simple.  Write down the four differential
equations;  put in auxiliary tables for opacities, equation of state, and nuclear energy
generation rates (all as a function of temperature, density, and composition); solve the
equations; and compare your output with data.  Actually it works pretty well, though we
note here some residual discords and new triumphs.

\begin{itemize}
\item[o]{There is still missing opacity, especially since at lower temperatures (Berger 
et al.~2006).}
\item[o]{The two most widely used equations of state differ in some quantum effects,
pre-ionization mechanisms etc. (Trampedach et al.~2006).}
\item[o]{The range of allowed energy generation rates is explored in the 10,000 models of
Bahcall et al.~(2006).}
\item[o]{No one this year disputed the sturdiest of the four equations}
\begin{equation}
	{dM \over dr} = 4 \pi r^2 \rho \ .
\end{equation}
\item[o]{The three standard sets of isochrones widely used for analyzing binary stars and
stellar populations do not entirely agree, and no one can simultaneously fit both
components of some binary systems for which we have high precision data (Armstrong et
al.~2006 on $\theta^2$ Tau; Reiners et al.~200S on an M dwarf pair).}
\item[o]{Thus one should welcome a fourth, independent set. Pietrinferni et al.~(2006)
incorporate better boundary conditions, overshoot, etc. As the authors are from ltaly,
we are going to have to improve our system of nomenclature for the three previous
sets (``Yale'', ``VandenBerg'' and ``Italian'').  Not to be superseded, VandenBerg et 
al.~(2006) have provided a new set of tracks for the commonest sorts of ranges of mass,
Fe/H, $\alpha$/Fe, etc., including convective core overshoot at the empirically-calibrated
level.  And Claret (2005) has added a set of tracks suitable for use in the SMC
(but terrestrial astronomers can use it too).}
\item[o]{Stars initially leave the main sequence very slowly.  Lacy et al.~(2005) discuss
BW Lac, for which isochrones and data come together to show that R is only 1.186 R$_\odot$
for 0.928 M$_\odot$ at 10.8 Gyr.  But then they gallop, so that the Hertzsprung gap is
almost empty.  Not quite though: Boden et al.~(2006) put HD 9939 at 1.072 M$_\odot$,
age = 9.12 Gyr, and T = 5050 K.}
\item[o]{The later phases are harder to calculate.  Dearborn et al.~(2006) are apparently the
first to slog right through the helium flash of a 1 M$_\odot$, solar composition star in
three dimensions.  In our childhood, this was known to be impossible, even in 1-d.
The burning is initially concencrated in a convective shell inside the hydrogen-burning
shell around an inert core.  At the horizontal branch stage, more accurate diffusion
(vs. instantaneous mixing) changes the time scale (Ventura et al.~2005).  And still
later (AGB, second dredge up, carbon ignition), everything depends on everything
(Siess 2006), of which we note only the fact of off-center carbon ignition in the
narrow range $9-11.3$ M$_\odot$ and much less second dredge up for M$>$11 M$_\odot$.
Stars will eventually die all over the place, but first we dredge up (often for at least
the second time) a number of tidbits that come in between, ordered alphabetically in a
language known to none of the authors.}
\end{itemize}

\subsection{The Chemically Peculiar Stars}

Surface chemical peculiarities (any other sort are very hard to observe) can result
either from internal nuclear processes plus mixing or from surface segregation.  This
has been so for more than three decades, with only very rare (and unpersuasive) cases for
``both please'' in a single star.  A very common intrinsic anomaly is a C/O ratio larger than
one, yielding a carbon star (frequently with associated excess of s-process elements like
barium). We caught three papers that indicated three processes that could be responsible,
mixing in the star itself, transfer from a companion, and pollution by a supernova
explosion, probably also of a companion.  (Ryan et al.~2005 = own reactions + AGB
companion transfer; Wanajo et al.~2006a; Allen and Barbuy 2006 = companion as a supernova
to account for barium stars with additional r-process excesses).

Our favorite sort of carbon star has silicates around the outside, and, as in
several previous years, you had your choice between a binary with one of each (Szczerba et
al.~2006 on V778 Cyg, a disk resolved with Merlin, showing that it could not be one star
rapidly evolving or the silicates would have faded in 14 years of observation) or rapid
evolutian of a single star from C/O less than one to greater (Garcia-Hernandez et 
al.~2006a on IRAS 09425-6040).  As many as a third of late type carbon stars exhibit R CrB
like behavior $-$ seemingly random obscuration events ( Whitelock et al.~2006,
Menzies et al.~2006; Feast et al.~2006). Causes are being sought.

About 6.5\% of main sequence A and B stars in the Milky Way (but less than 3\% in
the LMC, Paunzen et al.~2005) have magnetic fields of kiloGice\footnote{
Two goose is geese (Allan Sherman, c. 1960).} or more, and associated
chemical peculiarities. The item we circled in green is that the first magnetic star
found by Babcock (1960) has still the strongest field known, though a second strongest
was found this year.  Again you get a choice, HD 154708 at 24.5 kG (Hubrig 2005, polite
enough to cite Babcock) or HD 137509 at 29 kG (Kochukhov 2006a, and not), vs. $32-34$ kG
for Babcock's star.  HD 137509 probably has the largest, 40 kG, quadrupole field.
These strong fields are invoked in modeling surface chemical peculiarities (Eu
up by $10^5$ and such in $\alpha^2$ CVn), and the papers we flagged this year dealt primarily 
with origins of the fields. The main contenders are fossil fields traceable back to the
interstellar medium 
(Braithwaite and Nordlund 2006; Donati et al.~2006), dynamos at the young stellar object stage
(Wade et al. 2006), or a combination (Kochukhov and Bagnulo 2006). For what it is worth,
$5-10\%$ of Herbig AeBe stars (massive YSOs) already have ordered fields near 1kG (Wade et 
al.~2006a), based, we (and they) admit, on a sample of three.  One of them also already has
composition spots (more, Mr. Kipling, like spots on your skin than like spots in South
Africa).

\subsection{Brown Dwarfs}

Like extra-solar-system planets, these were expected long before they were seen but
now number ``many.''
We indexed 50 papers and will mention a subset that addresses
questions that have been around for some years without actually resolving any of them.

{\sl 6.5.1 Formation.} They form rather like stars, at least in the sense that some have
disks and planets (Jayawardhana and Ivanov 2006 on Oph $112225-240515$, whether J or
B not stated), which, on strict mass grounds, is a BD plus a planet, though with error
bars that probably overlap two of either.  Other relevant items include Jayawardhana and
Ivanov (2006a, BDs with disks in star formation regions), Luhman et al.~(2005 BDs with
disks and planets), and Luhman et al.~(2006 a low mass object called Cha $110913-773444$,
again without indication of Besselian or Julian zero point).

As for why they never grew up to live useful lives as sheep, sorry stars, there
remains some enthusiasm for the idea that they get ejected prematurely 
from regions with accretionable
material (Guieu et al.~2006), but we caught five votes against, with Basu and
Reiners (2006) last in the year.  The generic point is the difficulty of retaining disks
and companions through the ejection process.

Martin and Magazza (2005) is a conference on formation of very low mass stars and
brown dwarfs if you want more.

{\sl 6.5.2 Star-like properties?}  The first brown dwarf with a central hole in its disk
(Muzerollen et al.~2006) could be the result either of photo-evaporation or of planet
formation, both very starlike things to do.  Just a little unstarry is the possible cusp
in the IMF across the MV/BD line in several young clusters (De Wit et al.~2006).
Two more stellar traits - their rotation rates slow as they age (Zapatero Osorio 2005),
and they are sometimes X-ray sources, especially when young and especially when in
binaries (Preibisch et al.~2005, part of a suite of 13 papers on a Chandra survey of the
Orion region; Reid and Walkowicz 2006 on a system most remarkable perhaps for having at least
three names, one declaring it an X-ray source, one a 2MASS source, and one a star of
large proper motion).

And a very un-starlike property.  Brown dwarfs have weather.  That is, dust forms
irregular cloudy structures that change with time.  This interferes with accurate
determination of composition and generally hampers efforts to match models to data
say Heiling and Woitke (2006 on patchiness of dust), Littlefair et al.~(2006 on variability
as weather), and Burrows et al.~(2006a, models at $700-200$ K and the need for more detailed
cloud meteorology).

{\sl 6.5.3.  Binaries, Statistics, and Types.}  The question of brown dwarfs in various
binary contexts is a vexed one.  Fifty percent, same as everybody else, wrote Burgasser
et al.~(2005), with the odd triple or quad (Torres 2006; Simon et al.~2006), and a good
many other binary BD papers that are left on the living room floor.

Now, coming from the other side (the brown dwarf desert), there is undoubtedly a
deficiency of brown dwarfs compared both to normal companion stars and to planets for a
significant range of orbital periods around solar type stars (Grether and Lineweaver 2006)
and white dwarfs (Farihi et al.~2005). But there is no BD desert among single YSOs,
say Oasa et al.~(2006), who have examined 600 embedded sources in S106, though there
are not so many that they will fade into significant dark matter.  Have we understood
this? Only in the sense inspired by a postcard photo of two turtles caught in an
embarassing position: turtles don't do it that way.

We are still looking for a memorable, clean mnemonic for the sequence OBAFGKMLT,
but type L may disappear at low metallicity (Burrows et al.~2006a).  And the next spectral
type coming, when H$_2$O condenses out, you lose the Na and KI doublets, and NH$_3$ and H$_2$
begin to dominate various infrared bands at a temperature of less than 600 K, is to be called Y
(Burgasser et al.~2006a, who also provide standard stars for types T0$-$T8 V).

\subsection{Real Time Stellar Evolution}

We have gradually got over the idea of unchanging heavens, but visible changes in
an observer's lifetime still give us pause.  Supernovae (Sect. 10) and novae and their
cataclysmic relatives (Sect. 8)  can be defined by physical processes, though at the
price of occasionally having to leave open whether a particular variable belongs to the
class.  Here we inventory a very heterogeneous collection of rapid changes and other
oddities whose classes (when they can be classified!) defy this sort of characterization.

Eta Carinae, like the poor, is always with us (anyhow since 1843).  Its long
hypothesized companion has been seen directly in FUSE spectra (Iping et al.~2005),
and the environment shows that star formation continues nearby (Yonekura et al.~2005). You
may, if you wish, regard it as the prototype of the luminous blue variables or 
Hubble-Sandage variables, of which there are about 30 in the Local Group 
(Pasquali et al.~2006)
with typical lifetimes of $10^4$ years. Only astronomers with very long life expectancies
should be allowed to study them, for, while two went off in 2002 and 2003
outside the Local Group (and were accidentally classified as supernovae, Maund et 
al.~2006, though both were faint and slow to expand and change their spectra and had two
prediscovery observations each), M33 Var A (Humphreys et al.~2006) began its present epoch
of excitement in 1950, and an M33 LBV, just now changing its spectral type from O to WNL
(Massey 2006), has been trying to attract our attention for 2000 years (no, it was probably
not a naked eye object in the year zero).

The coveted green circle in real-time stellar evolution goes to Percy et al.~(2006)
and V725 Sgr, whose pulsation period has lengthened from 12 days in 1926 to 90 days
in 2005 as it hurried from being a Cepheid to being a red semi-regular variable\footnote{
And yes, in a preliminary draft this sentence was organized to associate the period
change with Percy, who is in fact a very steady fellow, rather than with the star.}.
The most likely cause was a thermal flash, dragging the star from the AGB in a blue loop
to the Cepheid strip, from which it has now returned.  V725 Sgr therefore joins the
August company of FG Sge (about which we caught no papers this year).  The first study
was by Swope (1937).

V605 Aql (which peaked in 1919, then had a temperature near 5000 K and is the core
of PN Abell 58) was also tagged as a last helium flash by Clayton et al.~(2006). It is
now at a Wolf-Rayet temperature of 95,000 K and has a surface that is about half helium
and half carbon (well mixed, of course) and so is of type WC.  The authors make comparisons
with V 4334 Sgr (peak 1996) also probably a last helium flash.  Each passed through an
R CrB-like phase in composition, but only for a couple of years; and each has continued to
fade away, while R CrB has, on observational grounds alone, been with us far at least
200 years.

We put V4332 Sgr (tagged as Nova Sgr 1994-1) here to maximize potential for confusion.
It has been classed with V838 Mon (Tylenda and Soker, 2006, authors who think both are
stellar mergers), but is not necessarily the same sort of beast, according to Kimeswanner
(2006), who notes that it began to brighten as long ago as $1950-76$.

V 838 itself lives in Sect. 3 (``a little help from our friends''), and we here remind
you that it has been identified as a mass-losing (evolved) red giant (De Guchi et al.~2005),
a young stellar object  of $5-10$ M$_\odot$ (Tylenda et al.~2005), a star merger (Tylenda and
Soker 2006), and a planetophage (Retter et al.~2006), so that to say some other source is
like it may not contribute as much enlightenment as we had hoped.

HD 45166 was assigned the spectral type qWR, where q stands for quasi rather than queer
(Steiner and Oliveira 2005), but we think the star is both.  It is a candidate V Sge star,
of which class it would be the second member.  The presence of some hydrogen in its
atmosphere is inferred from the He II Pickering decrement (think about it; the penny
dropped for us as this was being written, but the paper was read five months earlier).

EK Dra, seemingly a single, active ZAMS star, has been fading for more than 45 years
(Jarvinan et al.~2005).  The fraction of its surface covered by spots varies on a 10.5
year cycle, but there is a flip-flop of active longitudes on half that period.

V1647 Ori, the energy source for McNeil's variable nebula, was green stellar star
of {\sl Ap 04}. This year, it acquired a possible imitator, which powers the ``Braid nebula''
(Movsessian et al.~2006). They are inclined to put both in the FUOr class (young,
sporadic accretion and decrecion from a disk).  The images shown don't look very 
braid-like to the author who most often wears her hair that way, 
but this should not impugn the
interpretation, as we've never really been able to see the Crab either.  FUOr status
was rejected by Kastner et al.~(2006) because the outburst didn't last long enough,
and left open by Aspin et al.~(2006, who found only one previous outburst, in $1966-67$,
on a long series of plates from Harvard and Asiago).  Ojha et al.~(2006) declared it to be
an EXOr, whose prototype is not EX Ori, and Gibb et al.~(2006) said it was not much like
either class. FU Ori itself had its disk resolved by mid-infrared interferometry during
the reference year (Quanz et al.~2006).  The dust consists of amorphous big grains.

The stars in the past seven paragraphs are hereby consigned to the ``John's Other
Wife'' class\footnote{
A Spike Jones item, of which the final line is ``I gotta go away somewhere and figure
this out.''}, where already live the Be stars; both disk and polar stuff contribute
to the emission for Alpha Eri say Kervella and Dominciano de Souza (2006), who are good
friends of the star and call it Achernar.  It cycles at $14-15$ years between ordinary
B type (H alpha in absorption) and Be (Vincicius et al.~2006).  It is the brightest
of the Be stars and so by definition untypical.

Struve (1931) remarked upon the signficance of rapid rotation for Be stars, and
Linnell et al.~(2006) have found V360 Lac at very close to breakup.  The rapid rotation
is due most often to spin-up by mass transfer in close binaries (McSwain and Gies, 2005,
a survey of 48 clusters).  The two members of the Gamma Cas X-ray source class also
have Be type disks (Smith and Balona 2006).  The second member of the class is BZ Cru =
HD 110437, and that large data bases can reveal new phenomena after 80 years should
give the custodians of SDSS pause!

\subsection{Motions in Space}

The sorts of rapid evolution in the previous section are motions in the HR diagram,
often confused with motions in space (``the truckin' star problem'') by students in
Astro 100 (not our sections, of course).  But stars do also move rapidly through space.
Dray et al.~(2005) divide WR and O runaways unequally among the classic mechanisms
(1/3 expelled by clusters, 2/3 liberated by supernova deaths of close companions).  And we
note with pleasure that the originators of both mechanisms, Adriaan Blaauw and Arcadio
Poveda, were both at the Prague IAU.  The three radio  sources now leaving the Orion
Trapezium region (Gomez et al.~2005) started only about 500 years ago, and so count as
rapid in two senses.

Runaway binaries, of which a low-mass, X-ray binary pulsar from the XTE catalog
is an example (Gonzalez-Hernandez 2006), being now 1.85 kpc above the galactic disk,
must belong to the expelled class, since a third hypothetical companion would have been
too far away to produce much of a kick velocity when it exploded.  Martin (2006) notes
that there are no Be stars among the runaways, again some sort of statement about binary
companions.  Whether a subset of the high latitude B stars could have formed in situ
rather than being kicked out of the plane remains open (Mizund et al.~2006).

The high-speed green star is, however, the first of a new class of hypervelocity
stars, moving away from the general direction of the Galactic center at 
$500-700$ km s$^{-1}$.
It was announced late in the previous index year (Brown et al.~2005).  Enhancements this
year include (a) a probable mechanism $-$ binary disruption during close approach to Sgr
A$^*$ (Gualandris et al.~2005), the other star being left typically in a high eccentricity
orbit around the black hole, (Ginsburg and Loeb 2006),  (b) a second group hunting for
hypervelocity stars (Hirsch et al.~2005), (c) an example that may have come (on grounds of
speed, distance and lifetime) from the LMC (Edelmann et al.~2005), though the mechanism
is then not obvious,  (d) recognition of the first example as a slowly pulsating B main
sequence star (Fuentes et al.~2006), and (e) additional members of the class, five in print
(Brown et al.~2006c,d) and many more in preprint, found by the pioneering group.  This
is a sufficient number to conclude that the stars are not all the same age and not, 
therefore, the product of a single star formation burst that badly wanted to expel its
products for some reason.

\subsection{Stellar Rotation and Activity}

These belong together in so far as you agree that most stellar activity is magnetic
in nature; magnetic fields are generally produced by dynamos; and dynamos work best with
rapid rotation, or anyhow with large Rossby number (Rao and Pendharkar 2005).

Our odd couple for the year consists of Arcturus, whose rotation period is two
years (Gray and Brown 2006) and Vega, for which it is about 12.5 hours, less than 10\%
away from break-up, though the star is seen nearly pole on, making line spectral
features misleadingly sharp.  This was an independent discovery from two facilities,
reported in Peterson et al.~(2006) using the NPOI to get color temperature on the disk
and by Aufdenberg et al.~(2006), using CHARA to measure polar vs. equatorial temperature.
The radii reported do not agree within the errors.  The star is unfortunately an MKK
standard for A0 V, $B-V=0$ (unfortunately, because it is half a magnitude brighter than
the other AO V standards and has an anomalously large equivalent radius).  Gray (l988)
first suggested rapid rotation as the cause of various Vegan vagaries.

And the rotating circle goes to HBC 338, a T Tauri star, whose rotation period,
determined from photometry, changed from 5.3 days to 4.6 days between 2000 and 2005
(Herbst et al.~2006).  One's first thought is, of course, the torque exerted by 
$2 \times 10^{22}$
elephants\footnote{
This calculation was done once by the most elephantine author and not checked.  The
assumption is that an elephant could exert a torque equal to its weight times the
available lever arm (stellar radius).  The chief difficulty would seem to be spacing the
elephants at about $10^{10}$ cm$^{-1}$ around the circumference of the star. Well, 
perhaps they are on multiple levels like the oarspersons on a trireme.}
but apparently the explanation is considerable differential rotation plus
a shift (presumably as part of a cycle) from high to low latitude spots.

The idea of spot cycles leaves us at as good a bus stop as any from which to begin
a tour of traditional questions in this area, to most of which the 2006 answers was
``some of them are and some of them aren't'' or ``both please.'' For instance:

Stellar activity cycles?  Apparently none in Alpha Cen (Robrade et al.~2005, or perhaps
X-rays are just not the way to do this), but yes in many other stars, with longer cycle
periods accompanying longer rotation periods (Lorente and Montesinos 2005; 
Froehlich et al.~2006a). 
The latter is also a counter example to the rule that vigorous activity = no
cycles, but the 13.3 yr period they report for 24.35 day rotation is about what you would
expect.

Do stars spin up or down as they approach the main sequence?  Yes (Herbst and Mundt
2005; Irwin et al.~2006).

Is there differential rotation with radius?  Sometimes (Maeder and Meynet 2005).  With
latitude?  Sometimes (Reiners 2006), and sometimes it violates von Zeipel's theorem
(Lovekin et al.~2006).  The Star is Achenar, which we have already met in another context,
and the data date from OAO-2 in 1976, but even then von Zeipel had already been dead for
17 years and unlikely to care.

Is the minimum level of stellar activity set by acoustic or magnetic waves?
Probably (Bercik et al.~2005).

Do optical emission lines (like H$\alpha$), X-rays, and radio flux probe stellar activity
levels and their correlations with star mass, temperature, and rotation rate?  Yes, but
they don't always all give the same answers (Berger 2006).

Can X-rays from stars that are not supposed to have coronae be attributed to late-type
companions or colliding winds?  Sometimes (Stelzer et al.~2006; Van Loo et al.~2005).

Do stars otherwise like Vega (A0 V with infrared from dust) have Vega-like magnetic
fields and X-ray emission?  Yes, if you agree that ``no evidence for either'' is ``same''
(Hubrig et al.~2005; Pease et al.~2006).

Are fully convective stars like the others?  Not entirely. There are dynamos but
no cycles, say Chabrier and Kuker (2006) and Dobler et al.~(2006), which are calculations,
and Donati (2006), which is an observation of Doppler polarimetry of V 374 Peg.

Can stars change their minds, so that instead of ``both please'' we get, first one
then the other?  Emission vs. absorption lines?  (Galazutdinov and Krelowski 2006).
The latitudes (intermediate vs. polar) where most of the flux pops out (Holzworth et
al.~2006)?  The longitude where most of the flux pops out (Mekkaden et al.~2005)?
At least sometimes, these authors  say.

And, finally, peeking between the photospheric and coronal zones just examined, do
chromospheres display a First Ionization Potential effect, such that easily ionized
elements are preferentially lifted up?  Some do, some don't, even for two stars so
emotionally close as 70 Oph A and B (Wood and Linsky 2006).

\subsection{Pulsating Stars}

If, as Howell et al.~(2005) imply, all stars vary, this is going to be a very long
section, though they note that most amplitudes are small, most stars are multi-modal,
and that in 12 days of data, they find very few long periods.  Compare Pojmanski et 
al.~(2005), who are sensitive to longer periods but only somewhat larger amplitudes.
The {\sl General Catalogue of Variable Stars} soldiers on through its 4th, e-only, edition
(Samus et al.~2006), but can never, we suppose, include more than a tiny fraction of
the eligible stars.  A dozen or so of the 53 stars indexed under this heading led to a
shout of egad!

\begin{itemize}
\item[o]{Shapley was right: the population II Cepheids and RR Lyraes in globular clusters 
fall
on the same period-luminosity relation (Matsunaga et al.~2006), along with the W Vir
stars of the LMC, but the RV Tauri stars are brighter and redder, and some are
binaries (De Ruyter et al. 2006).}
\item[o]{That the RR Lyrae stars of globular clusters and the field do not display the same
period-color-luminosity-amplitude relationships has been noted in a  different context
(Sandage 2006) as was their use to trace star streams in the Galactic halo (Vivas and 
Zinn 2006), so you get one pointing out that the Blazhko effect is still not
understood (Chadid and Chapellier 2006).}
\item[o]{We hadn't previously realized that WN8 stars pulsate (well, all right, WRs
in general)
but the discovery (Lefevre et al.~2005) and a reasonable theory (Dorfi et al.~2006)
both crowded into the reference year, leaving no excuses.}
\item[o]{Gamma Doradus stars, whose mere existence was a highlight of {\sl Ap 99}
(Sect.~5.5) indeed have
gas flowing along the surface of the star (Jankov et al.~2006). The mechanism (to be
shared with the Maia stars if they exist, pulsating Be's, and maybe the slowly pulsating
B stars) may be a new class of pulsational instability $-$ retrograde mixed modes, a
hybrid between Rossby waves and Poincare modes (Townsend 2005).}
\item[o]{There has been a gap in the variable HR diagram between the slowly pulsating B stars
and the $\delta$ Scuti stars.  It is slowly filling up (Antonello et al.~2005 on HR 6139).}
\item[o]{We apologize for the need also to keep a bit of space there for the Beta Cephei variables
(a very large fraction of the bright O 9.5 to B 2.5 III-V stars, Telting et al.~2005).
The driving mechanism may require metallicity to be a function of height in the
atmosphere (Handler et al.~2006a), not to mention the roAp stars, for which layering also
matters (Kurtz et al.~2006).  That they thought of using Pr III lines to look for this
strikes us as remarkable.  The record for number of modes is not quite a tie, with 15
for FG Vir (a Delta Scuti star, Zima et al.~2006) and 14 for Nu Eri (a Beta Ceph,
Schnerr et al.~2006).}
\item[o]{Pulsating sdB stars have their choice of p and g modes (respectively EC 14026 and
PC 1716 stars).  A few can do both (Jeffery and Saio 2006).}
\item[o]{Mira her (? - it's hard to tell with stars) self is a binary, which is absolutely a
requirement for resolving the two radio sources (Matthes and Karovska 2006).  The VLA
helped too, and the emission mechanisms are different for the cool and hot components.
For the class as a whole and the related semi-regular variables, the correlations of
periods, luminosities, and other properties are of truly outstanding complexity
(Soszynski et al.~2005) which we approximate very crudely  by noting that there are
five sequences in a near-IR period-luminosity diagram for the LMC, with, at K = +11,
logarithms of the period in days = 1.45, 1.7, 1.85, 2.2, and 2.65 (that is, 28 to
447 days).}
\item[o]{The prize for the most confused star goes to QU Sgr in the globular cluster M71, who
is an SX Phe star, and an Algol with a pulsating donor star (Jeong et al.~2006), and
if he\footnote{
Well it is hard to tell with stars, but Qu sounds to us like a man's name in Chinese.}
should be next door with the binaries, we apologize. The 70 SX Phe stars in
Omega Cen come in two flavors (radial and non-radial modes with large and small amplitudes
respectively) and are simultaneously blue stragglers and rather like Delta Scuti stars
sitting at the end of the Cepheid strip (Olech et al.~2005).}
\item[o]{Each year, a star or two gets promoted from mere oscillation or pulsation to 
astero-seismology, and one or two may also get demoted (like RNe candidates).  Procyon
continues to hover (Aufderheide et al.~2005; Bruntt et al.~2005).  The multitude of modes
is supposed to allow one to reconstruct density (etc.) vs. radius.  Since radial
dependences become more complex for red giants, they ought to be particularly interesting.
Zeta Hya has been examed by Stello et al.~(2006), and there are lots of modes between
60 and 110 $\mu$Hz.  Unfortunately they live only about two days (much less than predicted)
and so cannot be identified well enough to use for interior reconstruction.}
\end{itemize}

Cepheids?  You think Cepheids are pulsating stars and should be mentioned?  Oh, all
right.   The green circle went to Polaris (Vsenko et al.~2005), which is clinging to its
membership in the instability strip on its 3rd or 5th crossing (based on N/(C+O) about
three times solar).  The authors think it may have a binary companion as well.  Polaris
is the second Cepheid to have a CHARA envelop 2-3 times the size of the star (Merand et
al.~2006), which cannot help but be bad for the Baade-Wesselink method of calibrating
the Cepheid distance scale.  As long as we are being gloomy, only one Cepheid in 30
is crossing for the first time (Kovtyukh et al.~2005).  And if you ask whether the masses
of Cepheids determined from their pulsation properties can be made to agree with
evolutionary tracks (Keller and Wood 2006), the answer is they can; and pigs may fly, but
they're unlikely birds.

\subsection{Sad Tails of the Deaths of Stars}

The tails are, of course, the planetary nebulae (though the majority do not have
tail-like morphologies).  They were a disappointment this year, being poor tracers of
dark matter (Bekki and Peng 2006 on NGC 5128, Sambhus et al.~2006 on NGC 4697, Bergond et
al.~2006 on NGC 3379, for which globular clusters are better). In addition, they are (1)
sources of He$^3$, confusing to those attempting to extract a primordial abundance for
cosmological purpose (Balser et al.~2006),  (2) biased distance indicators because the
observed proper  motions are a pattern speed of ionization or shock front, while the
radial velocities are matter motion (always slower than the pattern speed), so distances
will always come out too small by factors of 1.3 to 3, even for spherical expansion
(Schoenberner et al.~2005, who remind us not to bet on expansion ages either),  (3)
unreliable about turning off accretion and nuclear burning on the core, so as to set firm
initial conditions for white dwarf cooling curves (Soker et al.~2006), and (4) able to
seduce respectable astronomers into describing their expansion into existing space as a
``Hubble type outflow'' (Meaburn et al. 2005), which brings us to

\subsection{White Dwarfs}

We green circled two papers.  First, a careful consideration of the 11 WDs in
Praesepe (Dobbie et al.~2006), which is the number you would expect, unlike some other
clusters which are WD-deficient.  The ones seen have cooling times close to 300 Myr, vs. a
cluster age of $624\pm50$ Myr, present masses of $0.72-0.76$ M$_\odot$, and descend from main
sequence masses of $3.3-3.5$ M$_\odot$. By extrapolation, this implies a cut between stars
that form white dwarfs and those that form neutron stars at $6.8-8.4$ M$_\odot$.
The second green circle went to the suggestion (Zhang and Gil 2005) that a white dwarf
with magnetic field near 10 G and a rotation period of 77 minutes could account for
the transient radio source GCRT J1745-3009, which gave forth five 10-minute bursts, 77
min apart last year, and then was seen no more (or before either). In case you have left
your coordinate transformer at the office, 1745-3009 is in the general direction of the
galactic center\footnote{
A coordinate transformer is like a voltage transformer, only with fewer windings,
probably because a 77 minute rotation period is actually a typical one for white dwarfs,
whose v sin i values are mostly 10 km s$^{-1}$ or less (Berger et al.~2005). The magnetic
field on the other hand is at the upper edge of what is seen (Wickramasinge and Ferrario 2005,
a catalog). Individual rotation periods can be found for some of the pulsating WDs from
mode analysis, e.g., 14.5 hr for the ZZ Ceti star G185-32 (Dech and Vauclair 2006).}.
We are, however, saving the word strange for the white dwarfs of
Benvenuto (2006), which have up to 1\% of their mass in the form of central strange quark
matter.

There were also words about many long-standing white dwarf questions. The four
commonest were: more work is needed.   For instance, the mass of Sirius B is 0.978 M$_\odot$
from a spectroscopic $\log g$ and 1.02 from gravitational redshift (Barstow et al.~2005).
Hard to say that $1\pm 0.02$ is not satisfactory agreement, and time perhaps for another
assault on  the astrometric orbit, to make sure it too still agrees.  Procyon B (Gatewood
and Han 2006) admitted only to an astrometric $0.48\pm 0.14$ M$_\odot$, and perhaps something
could be done about its spectroscopic masses.

The luminosity and age of the faintest WDs count  as an independent way of getting
at ages of stellar systems.  Within the SDSS sample of 6000 stars, there is a smooth,
nearby monotonic rise of numbers to M$_{bol}=15.3$, and then an abrupt drop, though whether
all the way to zero is not clear (Harus et al.~2006). This must be relevant to the
nearby population of thin and thick disk stars and the time disk formation began. In
contrast, the faintest WDs in the globular cluster NGC 6397 (Richer et al.~2007) may
(or may not) be telling us something close to the Hubble time.

Is there a clean division in the HR diagram between pulsing and non-pulsing WDs?
Given the range of masses, core compositions, and He and H layer thickness, cleanness
seems unlikely.  Castanehir et al.~(2006) report that all their ZZ Ceti (pulsating DA)
stars fall between 10,850 and 12,270 K.  But Kepler et al.~(2005) point out that purity
is hard to establish (and perhaps not very meaningful) because the temperature excursion
through a pulsation period is typically 500 K, and the integration times of spectrograms
are often longer than the periods.  The pulsating PG 1159 stars (aka GW Vir) are hotter,
less pure, and less well understood, so we are not surprised at 0.05 M$_\odot$ differences
between spectral and seismological mass estimates (Miller Bertolami and Althau 2006).

Puzzles of very long duration pertain to the surface compositions of white dwarfs
vs. temperature and how one type evolves to another, particularly the DA (seemingly pure
H) and DB (seeming pure He, with DC - continuum, not carbon - as their low-temperature
extension). Metals on WD surfaces are far commoner than when we were type DO ourselves,
and in some extreme cases, enough species have been measured to suggest, for instance,
s-process enhancements, left from the DO's own AGB phase, we think (Chayer et al.~2005).
DA metals are generalJy attributed to ISM accretion (Koester and Wilken 2006), but with
three DA's now known to have debris disks (Kilic et al.~2006), late pollution by 
planets to which the stars were once hosts must also enter the inventory.

Oldest and strangest of all is the temperature dependence of the ratio of DB to DA
stars, which drops very nearly to zero between 28 and 48 kK (Einsenstein et al.~2006).
Mixing and settling must somehow alternate, and one might look for insight from how
the phenomenon behaves in clusters of various ages.  Unfortunately, DB/DA drops
nearly to zero in open clusters (Williams et al.~2006) for all temperatures, adding to
the puzzle rather than to the solution.

From time to time there is an assault on the Chandrasekhar mass. Not the number or
the underlying physics, but the appropriateness of the eponym. After all, things are
never named for the guy who discovered them (Bobrowski's law), so it must have been two
other fellows, probably Ralph H. Fowler of Cambridge, Wilhelm Anderson of Tartu, and
Edmund Stoner of Leeds.  This year's discussion comes from Blackman (2006), who notes
that Stoner's stars were indeed rather stone-like, having constant density.  And yes,
electron pressure is also sufficient to resist gravity in domestic stones\footnote{
Our first draft said `domestic scones' and you may well have encountered some with
nearly the rigidity of degenerate matter. - Ironically, the more juvenile author who
caused the mispelling in the draft is member of a biking group who calls themselves \
`scone-age group', referring to their stone-age (mostly in the mid-50s) and their 
preference for scones and coffee rather than the sporty challenge.}, there being
no lower limit to WD masses the way there is for neutron stars.  Chandra's own take on
the story appears on p.~451 of his stellar structure book (Chandrasekhar 1939).
Neutron stars live in Sect. 10.

\subsection{Single Black Holes}

For starters, are there any, and how would they let you know of their existence?
Maeda et al.~(2005) improve an old analysis of the microlens event MACHO-96-BLG-5 and
conclude that the best fit is a 6 M$_\odot$ isolated black hole. To Gupta and Kuman (2005),
however, this could be an ultradense star made of charged Buchdahl-type fluid.  The
density remains as large as $2 \times 10^{14}$ g cm$^{-3}$ right to the surface, 
but the sound speed is
less than $c$ throughout, and the solution joins smoothly onto Reissner-Nordstrom at the
pressure = 0 boundary (Buchdahl 1959, in case you need his fluid for anything).

The conditions required to form a single black hole (minimum main sequence mass,
perhaps conditions on angu1ar momentum) have not been established.  One would like, of
course, to know the turn-off mass of the youngest cluster that has at least one pulsar
in it, but young clusters just don't hang on to their pulsars.  In the binary case, one
can minimally say that the smallest BH progenitor is probably less massive than the
largest neutron star progenitor (Muno et al.~2006).

The other single black hole question that has bled blad blood\footnote{
See {\sl Ap 05} Sect. 12.} through these
pages over the years is the existence of intermediate mass black holes, where intermediate
might mean anything from 30 to 3000 M$_\odot$ and their purpose in life is to be responsible
for the X-ray emission from sources, not at galactic centers, that are brighter than
the Eddington limit for the, say, $5-15$ M$_\odot$ black holes that might reasonably 
come from normal massive binary star evolution (e.g., Celino et al.~2006).

We caught during the year two firm yeses, two firm nos, five probable yeses,
five probable nos, three decline-to-states, plus some more nuanced considerations, for
instance that the properties of the sources brighter than $10^{40}$ erg s$^{-1}$ 
(assuming isotropy)
are different from those of the fainter ones (Liu et al.~2006c).

Assume first that you have cleaned your sample of the (often larger number of)
background QSOs and foreground flare stars (Lopez-Corredoira and Guitierrez 2006), then you
still have to consider various forms of asymmetry and sporadicity in accretion and
radiation before you can really decide whether you have an ultraluminous X-ray source
of the sort that requires an intermediate mass black hole.  We record only the last in the
year each of the gentle yes's (Chakrabarty 2006) and the gentle no's (Lehmer et al.~2006).
This is probably a statistics of small numbers effect, but is there something about
publishing in AJ that inclines authors to gentleness?

The theorists are, of course more than happy to form IMBHs for us (Begelman et 
al.~2006, Freitag et al.~2006, with slightly different mechanisms), starting as early
as $z = 21$ (Kuhlen and Madau 2005), and, paraphrasing a high court judge: ``So, miss, you
are saying that the defendent put his hand on your knee while you were attending the
cinema?'' ``Yes, your lordship.'' ``Well, I see it is the regular time for adjournment,
so we will leave it there over the weekend'' which in our case is Sect. 7.

\section{ASTROBIOLOGY}

In last year's {\sl Ap 05} we introduced a section on astrobiology (\S 7), thereby filling a
gap in our yearly overview of astronomy and astrophysics. I, the junior but
oldest author (CJH), was responsible for that section as I am for the present
one. The tack I took then, and which I continue here, is to discuss but a small
number of topics in some depth. Because I cannot count on them having been
treated in past {\sl Ap XX}s, the style of this section differs from that of the rest
of our review in that many papers that I reference will be from outside the
reference year. Also, the topics are chosen at my pleasure. If I find
something that interests me that can fit into a short section, then I choose it.
Otherwise, you will have to wait until some future year.

\subsection{Reviews,  Books, and Awards}

Volume 98 (June 2006 issue) of {\sl Earth, Moon, and Planets} consists of
nine chapters masquerading as separate papers, thus forming a book of 370 pages.
There are a total of 25 authors and they, collectively, discuss dating methods
and chronometers in astrobiology, the formation and evolution of the Solar
System, the emergence and evolution of terrestrial life, and, finally, a short
summary chapter that concludes with the question of extraterrestrial life.
Two end papers give the curricula vitae of the authors and a very useful
glossary. This is not a ``book'' for  popular audiences but rather should be
useful for the professional.

Since much of astrobiology has to do with biotic evolution, often of life forms
impossible to imagine now, I suggest the classic text of
Mayr (1982)\footnote{Ernst Mayr died on 3 February 2005 at the age of 100, a
mere year after his 25th book was published. As, among other things,
an ornithologist, explorer, and evolutionary biologist, he was a giant. No star
of ours, gold or otherwise, could do him justice.}, the very personal (and
difficult) tome of Gould (2002), and the (perhaps) unique perspective of
Dawkins (2004) as extended reviews of terrestrial evolution.
Related to the last text is the US National Science Foundation's
{\sl Assembly of the Tree of Life} program, which has to do with the
evolutionary development and diversification of species or groups of organisms;
i.e., phylogeny. For a short (and nicely put together) statement of the program
see \texttt{http://www.nsf.gov/bio/pubs/reports/atol.pdf}.

Congratulations are due to my local colleague David Grinspoon for being awarded
the {\sl 2006 Carl Sagan Medal for Excellence in Public Communications in
Planetary Science of the AAS}. (You might wish to visit the website at
\texttt{http://www.aas.org/dps/prizes{\_}sagan.html}.)
Among David's positions is that of Curator of Astrobiology of the Denver CO,
Museum of Nature and Science. Part of this award must have been for his fun book
{\sl Lonely Planets} (Grinspoon 2003). A bit of the flavor of that book may
be sampled on his website \texttt{http://www.funkyscience.net/}. For reviews of
{\sl Lonely Planets} see Drake (2004), Nittler (2004), and Rummel (2004).

\subsection{SETI Lives}

A note to the reader: You will not see any reference here to alien abductions,
UFOs, or visiting aliens building the pyramids. I leave such topics to those
with ``interesting'' mindsets.

It has been over 40 years since Cocconi and Morrison (1959) argued that
Earthlings should be able to pick up radio signals with (then) current
technology if the signals were beamed to us by a distant alien civilization.
They even suggested that the H I 21 cm line might be a promising wavelength
candidate. Shortly thereafter Frank D.~Drake, using
the US National Radio Astronomy Observatory facilities at Greenbank WV,
inaugurated Project Ozma (after a {\sl Wizard of Oz} Princess) to listen to
$\tau$ Ceti and $\epsilon$ Eri at 21 cm (independent of Cocconi and Morrison's
1959 suggestion). Except for one false alarm (due to a military transmission),
he had no luck $-$ as reported in Drake (1961).\footnote{Drake, in two respects,
had the right idea; $\epsilon$ Eri has a planet of Jupiter-like mass
(Butler et al.~2006), and $\tau$ Ceti has a conspicuous disk
(Greaves et al.~2004), although not as massive as that around
$\epsilon$ Eri $-$ which is not surprising because the latter is younger.}

The first real stirrings of SETI, as an organized goal for a group of
professionals, may be dated to 1961 when, at the suggestion of the US National
Academy of Sciences, Drake organized a conference at Green Bank, MD, in 1961.
As he charmingly puts it in Drake (2003) on deciding who to invite, ``I
invited every person in the world we knew of who was interested in working in
this subject $-$ all twelve of them.''  He goes on to discuss the
``Drake Equation,'' of which more later. It wasn't until November 1984 that a
formal organization, the SETI Institute, was incorporated as a non-profit.

The official website of the SETI Institute, \texttt{http://www.seti.org}, lists
subsequent observing programs up through almost the present (but they leave out
the {\sl Allen Telescope Array}; see later).\footnote{The website listed
contains all sorts of information and links, although, for me, it was not that
easy to find what I wanted. On the other hand, you can go to
\texttt{www.seti-store.yahoo.net} to buy such hot items as various kinds of
shirts, hats, pens, mugs, and art glass. None of these items were made by
aliens $-$ as far as I know.}  There were various sources of funding
for these programs, including the US government. However, in October 1993, the
US Congress cancelled funding for the promising {\sl Microwave Observing
Program} (later renamed the {\sl High Resolution Microwave
Survey}), which started observing in October 1992. The reasons for this
cancellation seemed to revolve around the perception of what some called
the ``Little Green Men'' aspects of the program plus the lack of any firm
extraterrestrial signals. I know of some
colleagues who regarded SETI as a drain on ``real science.'' Well, SETI has done
some real science but the perception remains. (See the review of Tarter 2001
for more of the science.) Some others of us have the same
feelings about the extravagantly expensive {\sl International Space Station}.

An interesting 1998 letter to US President Clinton signed by 35 luminaries
asking that federal funding be restored to SETI may be found at
\texttt{http://setileague.org/editor/petition.htm}. Stephen Jay Gould was the
first signatory.

\subsubsection{The ATA}

If you were ranked as the sixth richest person in the world in 2006, as was
Paul G.~Allen (co-founder of Microsoft with Bill Gates in 1975) by
{\sl Forbes Magazine} $-$ and see
\texttt{http://www.forbes.com/billionaires} $-$ -we would look kindly on you if
you gave away some of your cash to worthwhile enterprises.  And so Mr.~Allen
did by donating \$(USD)11.5 million in early 2001 for technology research and
development of what is now known as the {\sl Allen Telescope Array}
(hereafter ATA).\footnote{For a review of Paul Allen's philanthropic activities
see his website \texttt{http://www.paulallen.com}. Oddly enough, he does not
mention ATA there.} In March 2004 he gave an additional \$(USD)13.5 million for
construction although, as we understand it, he requires matching funds $-$ -which
have not been easy to come by.\footnote{Some of the monetary figures you see
here are quoted from a talk Jill Tarter gave in Boulder, CO, USA in January
2007. And for more of the history of
ATA see Tarter (2006).} But enough of this nasty money talk for a bit.

The ATA is sited at the Hat Creek Radio Observatory run by the Radio Astronomy
Laboratory (RAL) of the University of California at Berkeley, CA. It is a
relatively radio-quiet site out in the middle of what appears to be nowhere.
Cattle do roam around, however, and the deer and the antelope play.

The ATA, when finished, will consist of 350 units of 6.1 m offset Gregorian antennas
that, with associated electronics, will cover the band $0.5-11.2$ GHz, all at
once. At 21 cm the field of view will be $2.45^\circ$. What is remarkable is
that, even with the first stage 42 antenna array not quite finished,
it is capable of multiple, and simultaneous, explorations of the radio
sky $-$ including those directly concerned with SETI. Nulling out of
unwanted signals is also possible in many circumstances. (See, e.g., the
reviews of Tarter 2001 and Tarter 2006, for more technical details.) To keep
costs down, much of the hardware consists of off-the-shelf components (such as
simple molded aluminum dishes). Electronic components take advantage of the
equivalent of Moore's law in computer chips and can be (almost) mass-produced:
the processing of signals keeps up apace.

How far is ATA along? Perhaps the best way to tell is to go to the ATA website
\texttt{http://atacam.seti.org} where you can see panoramic shots of the
telescopes and get a status report on what they are up to. I counted reports for
18 dishes, so I suppose that is the count as ATA works toward the first goal of
42. The projected completion year for the entire 350 dish array is 2010 but that
depends on their rate of funding. At an estimated \$(USD)100,000
per dish and back end\footnote{We understand that for \$(USD)50 000 in matching
funds you can have the name of your choice put on a dish.},
the total cost, multiplying two numbers together, should
be \$(USD)35 million, but will probably approach \$(USD)50 million. In the world
of defense budgets or international finance, this is pocket change, but it still
has to be raised. ATA has some help from the US National Science Foundation
(through RAL) and, oddly enough, the United States Navy.

As a parting note on ATA Koenig (2006), in discussing the projected
{\sl Square Kilometer Array} (and that {\sl is} big), has the quote
``The ATA will be of crucial importance to the technology of the SKA.'' We wish
them the best.

Where is SETI headed for the future? I suggest you peruse the SETI publication
{\sl SETI 2020} edited by Ekers et al.~(2002) in which
various authors attempt to forecast where SETI is, or should be, headed.
I won't ask where the funding might be coming from.

\subsubsection{BOINC}

One of the reasons I named this subsection ``SETI Lives'' is that, aside from
its continuing programs, it has captured the imagination of the computing
public. The University of California at Berkeley hosts {\sl BOINC},
meaning the {\sl Berkeley Open Infrastructure for Network Computing} (see
their informative site
at \texttt{http://boinc.berkeley.edu}). {\sl BOINC} is ``a software platform
for distributed computing using volunteered computer resources.'' That is, you
can download {\sl BOINC} software to your home computer (or mainframe) and
join any of a number of worthwhile projects via the internet. Data that you
download from the project(s) are then manipulated by your computer during idle
cycle time and then results are  sent back to the project. (It really sounds
like fun but my old-fashioned phone line is dismally slow.) Among the projects
is {\sl SETI@Home}; i.e., ``SETI at Home.'' It is, in fact, the most popular
project and has, as of 7 February 2007, just under 600,000 participants on over
1,300,000 computers. (Simple division reveals that some people are computer
rich.) This is wonderful. Would that all science could enthuse the public as
well! And, just for the record, the {\sl BOINC} site applauds one volunteer,
Vince Berk, who is contributing ``74 billion [US $10^9$]
floating-point operations per second.'' Bravo.

\subsubsection{The Drake Equation}

I mentioned the Drake Equation earlier. It still lives $-$ all (usually) seven
factors of it. I am not going to go into how useful it really is in practice,
but two authors, one way back in the early 1970s, used Shklovskii and
Sagan (1966, {\sl Intelligent Life in the Universe}) as the text for a
non-science major course in astronomy. Using the Drake Equation as a crutch to
go through practically anything you want to talk about in astronomy was a
great success. My plea is for someone (not me!) to bring that book up-to-date.
That would be a service to those lecturers who find the cookbook textbooks that
try to cover everything (even with pretty pictures and links to
the web) rather boring.

Depending on what numerical factors you put in the Drake Equation, the
probability of there being intelligent life in our Galaxy ranges from virtually
nil to near certainty.  If the true answer leans towards the latter, then
``Where is everybody?,'' as Enrico Fermi is reputed to have said. Hence the
{\sl Fermi Paradox.}  A fairly recent book by Webb (2002) attempts to
examine/explain/resolve (with 50 solutions) the paradox. (And see also \S10.3
of Chyba and Hand 2005.) It seems like a good
try and we suggest you look it over, along with the reviews of the book by
Tarter (2003) and Cirkovic (2005). On a more pessimistic note,
Ward and Brownlee (2000) argue that complex life is uncommon in the universe.
In the meantime, we're still waiting.

Finally, if there are alien intelligences out there, what might they be like?
Simon Conway Morris, Professor of Evolutionary Paleobiology at the University of
Cambridge (U.K.), has some ideas (Conway Morris 2005, or in more expanded form,
Conway Morris 2003). He is regarded by some as a maverick, but one you must
listen to. In his short 2005 paper he first states that ``I think the evidence
we can glean from the evolution of life on Earth compellingly supports the
thesis that a human-like intelligence is evolutionary inevitable [other than
on Earth].'' He then goes on to conclude ``\ldots but I have never said it
should reside in a human-like brain.'' This sounds scarier and scarier.

\subsection{Updates, Mendings, and Miscellaneous}

\subsubsection{Water On Mars?}

I ended the {\sl Ap 05} astrobiology section with a mention of possible evidence for
water on Mars in its distant (and perhaps not so distant) past. The case seems
stronger now with more recent observations by the Rovers and orbiters. Some of
you may wish to look over the article by Bell (2006)\footnote{This is another
{\sl Scientific American} article for the science-aware reader. I include
such references because of the diverse interests of the astrobiological
community.}.

\subsubsection{Archaeal Musings}

About half of the astrobiology section in {\sl Ap 05} was devoted to the discussion of
what are the earliest signs of life on Earth.
Lopez-Garcia et al.~(2006, in the {\sl Earth, Moon,
and Planets} ``book'' referred to earlier) cover much of the same material but
extend it to the emergence of the eukaryotes. We recommend it.

In {\sl Ap 05} I reported on a possible new phylum of the Archaea represented by the
sole hyperthermophilic symbiont/parasitic species
{\sl Nanoarchaeum equitans}\footnote{This is a popular species. Google lists
227,000 hits in English for it. Even with duplications that's a lot.}
(Huber et al.~2002), which cannot be cultivated independently of its
host. I quoted a linear size of 150 nm. It should have been 400 nm. Sorry.
Size aside, Huber et al.~(2002, 2003) and Waters et al.~(2003) 
all agree that {\sl N.~equitans} is the type species for the phylum
Nanoarchaeota, which, along with Crenarchaeota and Euryarchaeota, comprise the
Archaea domain. (We leave out the problematical ``Korachaeota,''  a phylum
postulated on the basis of some ribosomal RNA sequences derived from high
temperature geothermal environments.)

Waters et al.~(2003) succeeded in sequencing the
genome of {\sl N.~equitans} and find it consists of a single circular
chromosome with only some 491,000 base pairs $-$ the ``smallest genome of
a cellular organism sequenced to date [2003].'' They suggest that this
organism diverged very early on, perhaps even before the emergence of the
two other accepted phyla. It isn't primitive, though, because it has a
sophisticated system that ``contains complete versions of the modern
archaeal-genre replication, transcription, and translation systems,''
but it still depends on its host for other functions.

Not everyone agrees that Nanoarchaeota is a true phylum.
Brochier et al.~(2005), using somewhat different strategies than the
above authors, find that {\sl N.~equitans} is probably most closely related
to a member of the Euryarcheaota. They posit that {\sl N.~equitans}, in its
rush to get along with its host, has evolved so quickly that it only appears
to have evolved very early on. Lateral
gene transfer between other prokaryotes may also have confused the issue. (See
Brochier et al.~2005a for a discussion of what is involved.)

Now back to small sizes. Archaea delight in living in tough environments.
Baker et al.~(2006) describe reconstructions of unusual DNA fragments
belonging to ``novel archaeal lineages'' derived from living biofilms found
underground in the Richmond Mine at Iron Mountain, California. What sets my
teeth on edge is that they survive in metal-rich acidic solutions of pH from
$\approx0.5$  to 1.5. (My teeth wouldn't last very long but those little
bugs are tough.) The authors also find, using transmission electron microscopy,
cell-like objects of length $199-299$ nm (mean of 244 nm) and width $129-207$ nm
(mean of 175 nm). If these are, say, cylinders, then the cell volumes
may be less than about 0.006 $(\mu m)^3$, as a Lilliputian population of
Archaea. To be fair, however, the authors point out the possibility that
there may be ``unobserved connections between the objects that appear to
be cells.''

The key word in the last sentence is ``unobserved.'' This leads us to
recommend the papers by Schopf and Kudryavtsev (2005), and Schopf, Tripathi and
Kudryavtsev (2006), who show beautiful three-dimensional pictures of
fossil microorganisms. Taking the two papers together, the authors use
non-intrusive and non-destructive Raman imagery and confocal laser scanning
microscopy of thin slices of precambrian fossils in cherts. The revealed details
(on the micrometer level) are amazing and,
by progressive scanning through the slices, you can get a very good idea of the
overall geometry of the fossils. The techniques are not for amateurs $-$ like me.

Why do I bring up all this business about microfossils? To quote
Brochier et al.~(2005), ``Despite a ubiquitous distribution and a
diversity that may parallel that of the Bacteria [see, e.g., Forterre
et al.~2002], the Archaea still remain the most unexplored of life's
domains.''  If we are still in this primitive state of
understanding the life-forms on our own planet, how confident are we that
life elsewhere can even be recognized as such, as discussed in {\sl Ap 05}?

And, talking about the early Earth, you might look at the short, but lovely,
photo-essay by Lanting (2006) of scenes from the present-day Earth that could
well  have been taken very early on in Earth's history.

\subsubsection{Let's Have a Little Fun}

I have a fine time putting together my minor
contribution to these reviews. For example, I came across the laudatory review
by Frank (2006) of Martin A.~Nowak's book {\sl Evolutionary Dynamics}
(Nowak 2006). That book now sits on my desk along with a thick pad of paper,
a selection of color pens, and mathematical software manuals for my computer.
For those of you with an interest in evolutionary
biology and who can still do your sums, this book looks ideal. How about
predator-prey theory, mutation and adaptation in genome sequences,
game theory as applied to finite populations, or even the evolution of
language? One of the advantages of an Emeritus Professorship  is that you
can pick what you want to do. I look forward to a fun year.

\section{MORE THAN 1 BUT LESS THAN $10^6$}

These are stars grouped from binaries to large globular clusters.  They strain the
concept of ``astrophysical accuracy,'' which is more often expressed as, say, ``more than
30 but less than 100'' (students in one's class for instance) and the standards of
English grammar, which would require ``more than 1 but fewer than $10^6$.''

\subsection{Binary Stars}

Since the sun is not a member of a binary system, and since binary pairs (for reasons
of temperature and orbital stability) are unlikely to be hosts for life-bearing planets, at
most one-third of your authors (assuming quantization of authors) can be interested in
this topic.  Unfortunately, it is the 1/3 that is writing this section.

\subsubsection{Binaries in Stellar Populations}

Many astronomers, once they feel they have understood stars, want to start putting
them together into stellar populations (e.g., Maraston 2005, exploring a range of
metallicity, age, star formation rate vs. $z$, IMF, horizontal branch morphology including
blue metal-rich stars, and AGB tip stars important at $z=2.4-2.9$).  Indeed this final
assembly stage is arguably the last frontier in stellar evolution and nucleosynthesis,
though by no means a completely unexplored frontier.  About like Pike's Peak - lots of
people have been there before, but you can still get yourself killed.  One of the safer
paths traverses the evolutionary phases that dominate light at various times after a
starburst; AGB stars appear at 200 Myr and peak at $500-600$ Myr, while red giants take
over by 900 Myr (Mucciarelli et al.~2006).

Obviously one of the more hazardous snow-bridges is the need to include binaries
in your population. This is an excellent thing to do, agree Zhang and Li (2006), though
not a trivial one, since different methods yield different color indices over a wide
range of ages and metallicities (whether any or all are better than leaving out the
binaries completely, they don't quite say). Dionne and Robert (2006) have confined
themselves to the first 50 Myr after a formation episode and conclude that binary
fraction matters for WR production and energy injection into the ISM, but not for
supernova rates, total luminosity, color, UV line strengths, or mass returned to the
ISM. Ah but just wait for the next 99\% of stellar lives, the turn-on of Type Ia
supernovae and all (they live in Sect.~10 as do their neutron star products).

From the point of view of population synthesis, we need to know the percentage of
stars that form in binaries and the distributions of separations and mass ratios as
a function of primary mass and composition. These are unlikely to be constants of nature.
Indeed Bouy et al.~(2006) report specifically that the fraction of M dwarfs with
companions at $100-150$ AU in the Upper Sco OB associatian is larger than that in the
Pleiades.  This could, of course, be either initial conditions or survival probability
(Wiersma et al.~2006 on the survival issue for closer systems).  Conversely, as it were,
rather than pairs with small mass and wide separation being lost from clusters, Bonnell
and Bate (2005) prefer to start with such systems and, by additional accretion, transform
them into close systems of large total mass and mass ratio generally close to one (as
seen for massive systems).

\subsubsection{Binary Statistics and Triples}

Our binary green circles all more or less address statistical issues pertaining to
initial conditions. First is a truly astonishing deficiency of B dwarf eclipsing systems
in the LMC found by Mazen et al.~(2006) using the OGLE sample.  They report less than
10\% of the Milky Way binary abundance.  This must be either true or the result of a
total1y uncopernican conspiracy (in which most of the systems are face-on and so do not
eclipse) or an almost equally unlikely gross overestimate of the size of the parent
population of B dwarfs in general. We are so sure that MWIN on this item that we are
almost tempted to try it ourselves.  Almost. Meanwhile, if the result is true, then we
are back to one of the verities of the 1960s, that metal-poor populations don't do
binaries. The rarity of CBSs among the hot horizontal branch stars of NGC 6752 (Moni
Bidin et al.~2006) and the out-of-period 47 Tuc data from Albrow (2001) make us feel
young (and binary deficient) again.

Where else are binaries rare?  Among the M dwarfs, says Lada (2006), 75\% of which
are single, contrary to the usual result of 50\% or more binaries among bigger stars.
This is less of a spanner in the evolutionary works than you might suppose, since M
dwarfs aren't  going to do much within a Hubble time anyhow.

Triple systems generally have the third star far enough out that it also makes
little difference to the evolution of the close pair and so is irrelvant in this year's
narrow frame. A collective green circle nevertheless to a trio of papers concluding
that most binaries really are triples, though we learned one in six in our Battenhood
(Batten 1973). Tokovinin et al.~(2006) subjected 165 solar-type spectroscopic binaries
to an adaptive optics search for visual third stars and found $63\pm5\%$. Thirds accompany
96\% of pairs with period less than 3 days and only 34\% of periods greater than 12 days,
leading the authors to conclude that the third star tends to shorten the period (but
couldn't the causality go the other way?). In a small sample of contact binaries,
$59\pm18\%$ have a third star (Pribulla and Rucinski 2006).  And in a third set, also of
contact systems (D'Angelo et al.~2006), third smaller stars are found for at least 
$1/3-2/3$, and may actually be present in all.  There are also more triple Algols than we 
had previously known about (Kabis et al.~2005).

Your brown dwarf triple for the year is Gleise 569B (Simon et al.~2006).  And we
green-starred the triple orbit for K Peg (Muterspaugh et al.~2006a), which used new data
from the PTIO but also dozens of observations from the
1880s to the 1920s, including one each from G.~V.~Schiapparelli and T.~J.~J. See.  The
middle names were Virginio and Jefferson Jackson, reflecting parently aspirations that do
not seem to have been fulfilled.  Lest we forget, the first direct (binary,) measure of
masses, radii, and temperatures for a young brown dwarf pair yielded M$_1=0.054$ M$_\odot$,
R$_1=0.669$ R$_\odot$, T$_1=2650$ K; M$_2=0.034$ M$_\odot$, R$_1=0.511$ R$_\odot$, and 
T$_2=2790$ K
(Stassun et al.~2006, who were less surprised by the smaller mass star being hotter than
we are).

A proper set of initial conditions for population synthesis includes not only
binary fraction vs. primary mass (however uncertain even that may be) but also the 
distributions of separation and mass ratio (and eccentricity if you are really fussy). These
are, of course, both functions of primary mass. Kraus et al.~(2005) note that small
masses have smaller separations but larger mass ratios (nearly all $>0.6$) than bigger
stars, even among the very young ones of the Upper Sco association.  One worries,
obviously, about selection effects.

As for the distribution of mass ratios, your 1/3 author's particular King Charles
head, Pinsonneault and Stanek (2006) conclude that there is a secondary peak at M$_2$/M$_1 
\approx 1$ which makes up 25\% of all binaries right on through to white dwarf and neutron
star pairs (where it is relevant to rates of SN Ia and perhaps GRBs).  Soderhjelm and
Dischler (2005) sneak in backwards by pointing out that you underestimate the number of
eclipsing binaries on the upper main sequence by a factor of more than 10 if you give your
primary a random companion from the initial mass function, implying, thus, that M$_2$/M$_1$
favors values near one.  And no, ``we'' are not cited by either of them. If you like tying
things together (possibly in knots), you could say that this was the answer to the LMC
deficit of eclipsing B dwarfs $-$ they have companions as expected, but distributed like
the general IMF so that only a few secondaries are big enough to cause detectable
eclipses.

The sun has no obvious stellar companion, we said up front, and neither, it seems,
has Epsi1on Eri (Marentgo et al.~2006).  They searched the area with SST and found 460
sources in a 5.7\arcmin\ square field, none likely to be part of a visual binary. 
One would need
radial velocity data and proper motions to be sure.

Incidentally, though we clearly live in a new era of precision astrometry, both the
proper motion people and the radial velocity people want better data.  On the proper
motion side, you need milli-arcsec in order to be able to allow for light travel time
across systems like 61 Cyg and Alpha Cen (Anglada-Escude and Torra 2006).  And better
radial velocity orbits are needed to go with visual binary data from interferometers.
Tomkin and Fekel (2006) are not, by the way, complaining about the km s$^{-1}$ to 
m s$^{-1}$ numbers
that come from radial velocity spectrometers and exoplanet searches, but about the
10 km s$^{-1}$ or worse data in the 25\%  of pre-1950 orbits in the {\sl Eighth Catalogue of
Spectroscopic Binaries} (Batten et al.~1989).  The 9th catalog is, unfortunately, complete
only up to 1989 publications.

\subsubsection{Binary Evolution}

Binaries once formed - and catalogued! - evolve. System periods can change while the
stars sit not far off the main sequence (two papers, advocating need for additional
mechanisms, but from the same group; Lanza et al.~2006 is the second).  Their candidate is
changes in magnetic field structure affecting the quadrupole moment via hydrostatic
equilibrium, and not just transport of angular momentum.  Another of those ``cited but
not thanked'' cases, so we note  here Applegate (1992), of the first mechanism.

Contact, W UMa, binaries earn first prize in the ``well, you can't have formed that
way'' contest.  Extended surveys are beginning to pick up lots of them, for instance 1022
from ROTSE (Gettel et al. 2006), meaning that one star in 300 is a W UMa. All are X-ray
sources, according to Geske (2006), though their sum is a minor contributor to the total
Galactic X-ray flux. W UMa's are only one star in 500 for nearby FGK dwarfs (and less
for the brightest stars) says Rucinski (2006) on the basis of 3374 ASAS W UMa's with periods
less than 0.562 days. He notes that the fraction may be larger in the inner Galactic
disk. We don't object to W UMa incidence varying with stellar ages, metallicities, etc.,
but would just as soon that the northern and southern hemisphere values were the same.

It takes at least 200 Myr and more probably a Gyr for binaries formed in star clusters
to reach the U Ma stage (Hargis et al.~2005). We had always supposed that they came from
short period, detached systems in the same mass range, but there are very few detached
pairs with periods less than 1 day in the ASAS (All Sky Automated Survey) sample of
50,099 variables (Paczynski et al.~2006, who suggest Kozai cycles in triple systems as
a better way of making contact pairs). Somehow we had also thought that W UMa's would all
end up by merging into single rapid rotators (FK Comae stars perhaps), but Qian and Zhu
(2006) say this happens to only a subset, with small mass ratio and high overcontact.
They call the subset AW UMa stars. The meaning is not quite clear.  It is not a 
prototype, for those are BO CVn and SS Com.

The entire enterprise of studies of close binary evolution began as an effort to
understand the Algol paradox (less massive star more highly evolved). Remarkably, then
we caught only one Algol paper this year.  Van Rensbergen et al.~(2006) say you need to
lose quite a lot of mass but very little angular momentum from systems to end up with
the distribution of mass ratios and periods found in the real world.

The common envelop binaries are another set of very old friends (250 Myr for the one
you are just about to meet). WD 0137-349 = BPS CS 29504-0036 (equatorial and galactic
coordinates we suppose) has a period of 0.08 days, and a 0.39 M$_\odot$ white dwarf orbited
by a 0.055 M$_\odot$ brown dwarf (violating the BD around WD desert rule, but never mind), 
which must actually have spent some time inside a red giant, say Maxted et al.~(2006), 
with the
green accolade of a News and Views commentary (Liebert 2006). It is not very bright (though
we think highly of Maxted and Liebert).

The single star section already remarked that binaries provide a particularly fine
opportunity 
for models not to fit observations.  Bagnuolo et al.~(2006) present 12 Per as another
example, with measured masses larger than those expected from other properties of the
stars plus models.  Other examples were Torres et al.~(2006 on V 1061 Cyg and other 
chromospherically active stars): Reiners et al.~(2005, on an M dwarf pair whose stars are
either too massive or too faint for their ages, and not the first case like this);
Jancart et al.~(2005 on HIP 20935 in which only one of two stars of identical mass near
0.98 M$_\odot$ contributes to the spectrum; it's in the Hyades if you want to visit); and
Southworth et al.~(2005, two perfectly nice stars of 1.96 and 1.81 M$_\odot$, T = 7960 and
7970 K, and R = 1.93 and 1.84 R$_\odot$, but fit only by isochrones with more than three
times solar metallicity).

Beta Lyrae is the prototypical binary caught still in its rapid mass transfer phase.
The rate is $10^{-5}$ M$_\odot$ per year say Nazarenko and Glazunova 2006.

Blue stragglers - stars whose position in the HR diagrams of clusters or other 
well-defined 
populations indicate that they are more massive and hence younger than their
neighbors - are, as a rule, binaries, past binaries, or (for all we know) future binaries.
A double handful of papers but you get only the five that say there must be at least five
ways to get there: Warren et al.~(2006a on binary collisions), Tian et al.~(2006 on
mass transfer in primordial CBSs), Ferraro et al.~(2006a, saying not star exchange and not
encounter-enhanced mergers in Omega Cen), Sandquist (2005 in favor of collisions in wide
binaries for the open cluster cases), and the confuson, De Marco et al.~(2005) who
find that a bundle of globular cluster blue stragglers are all in the Hertzsprung gap,
a seeming impossibility given the speed at which stars cross it.

\subsubsection{Cataclysmic Variables}

Indexed papers concerning cataclysmic variables numbered 49.  This is the very last
section being written (wherever it may appear in the final paper), and we are getting
tired, so all you get are about 10 NASA bullets, less well aimed than some other sorts
(though not so badly as the Jezail that simultaneously hit the shoulder and thigh of
Dr. John H. Watson, and no, we don't know what he was doing at the time).

The green circle goes to the discovery that among the (rather few) recurrent novae
each arguably has a fairly stable recurrence time, determined by the two closest-spaced
outbursts caught.  This point of view led Schaefer (2005) to find three missing
for U Sco, increasing considerably the rate at which the WD's mass should grow toward
the Chandrasekhar limit and supernovation.  But Robinson et al.~(2006) struck out on
three other RNe and four suspected RNe sought in the 1913$-$1995 plate files of Maria
Mitchell Observatory.

\begin{itemize}
\item[o]{The first CV whose white dwarf is a ZZ Ceti star (Gaensicke et al.~2006)}
\item[o]{
The first CV in a multiple star system (Vogt 2005).  It is FH Leo = HD 96273 = BD
+07 2411B}
\item[o]{The first example in which an accretion disk instability deposits enough material 
that there was a nuclear explosion almost immediately after, complete with mass ejection.
It happened to Z And in late 2000, when a 2-magnitude flare was followed by brightening
to $10^4$ L$_\odot$. Sokoloski et al.~(2006a) call this a combination nova.}
\item[o]{A couple of very slow novae: BF Cyg, which is finally back to its pre-outburst
luminosity level of 1894 (Leibowitz and Formiggini 2006), and V 723 Cas (Iijima 2005)
which took 18 months from peak to onset of the nebular spectrum phase.}
\item[o]{In contrast, V382 Vel was caught on 22 May 1999, was a supersoft X-ray source 
through December and gone by February 2000 (Ness et al.~2005).}
\item[o]{The Milky Way has more than $2 \times 10^4$ CVs with X-ray luminosity less than
$2 \times 10^{33}$ erg s$^{-1}$ (Ebisawa et al.~2005, from a Chandra survey of the plane).
If this is the
primary reservoir, each must go off once per millenium to keep up our expected nova
rate of 34 per year (Darnley et al.~2006). This seems like awfully hard work for
them and comes perilously close to erasing the distinction between ordinary and
recurrent novae. A recurrence time of $10^{4.5}$ yr would require a pool of $10^6$ systems.}
\item[o]{The first supersoft source with pulsation, at 38.4 minutes for CAL 83 thought to be a
nonradial mode of the white dwarf (Schmidtke and Cowley 2006). It must be noted, however,
that lots of things are supersoft X-ray sources without being massive white dwarfs
en route to a Type Ia supernova explosion. Orio (2006) presents a list of 27 in M31
which includes some SNRs and known classical novae, not to mention background and
foreground objects, which, like the poor, are always with us.}
\item[o]{One does have to go through a supersoft stage to get from a typical CV to a recurrent
nova like U Sco (Sarna et al.~2006).}
\item[o]{A new class of CVs, found exclusively in globular clusters, whose members never went
through a common envelop binary phase.  Authors Shara and Hurley (2006) predict that
their distribution of periods should not have the standard gap at 2-3 hours.}
\item[o]{There are, however, now 11 CVs known in that gap (Schmidtobreick and Tappert 
2006a), but
a shortage of expected dwarf novae that have evolved past the turn-around in orbit
period (Aungwerojwit et al.~2006), a manifestation perhaps of the general phenomenon
that, to understand CV statistics, we need ways of losing angular momentum in addition
to gravitational radiation (Williams et al.~2005a).}
\item[o]{Old novae never die, though the radio from GK Per (1901) is fading (Anupama and 
Kantharia
2005), and the event of 1848 and DI Lac (at age 76) are both now very faint (Engle
and Sion 2005).}
\end{itemize}

And here, with the courtly old gentleman's hand still on the girl's knee, we must
leave you to go do our income taxes.  This is a phenomenon that may not be understood by
non-American readers.  We hear rumors of countries where the government simply tells you
how much they think you owe, like a credit card bill, and (also like a credit card bill)
you dispute it at your peril.  Here, not only does a struggling academic face a marginal
rate near 40\% in California, but we are expected to spend $10-50$ hours per year figuring
out how much to send, with significant penalties if we get it wrong or fail to predict
how much we will owe next year and don't prepay enough.  Last year, the one of your
authors who first saw a pocket calculator over-paid state tax by 83 cents and received a
penalties bill of \$10.71 as a reward.

\subsubsection{Binary Black Holes}

Cygnus X-1 was the very first black hole X-ray binary.  There are now many more
though not the 10,000 that Yungelson et al.~(2006) says we should find.  The total absence
of Type I X-ray bursts (common in neutron star systems) says they have horizons
(Kemillard et al.~2006). The spin parameters stretch fairly uniformly across the
allowed values from 0 to 1 (in suitable dimensionless units), Shafer et al.~(2006)
reporting on two with a/m=0.7 and 0.8, vs. extreme values reported for other systems
in earlier years.

Meanwhile Cygnus X-1 has been displaying a stray, 150 day period for the last 30
years, with constant period and phase.  Lachowicz et al.~(2006) think it is probably
disk precession.

SS 433 hasn't been watched for so long (and its black hole nature is still occasionally
disputed), but it was watched at lots of wavelengths during the index year (Chakrabarti
et al.~2005), at as many of them as possible from sites in India.  Despite the height of
the mountains there, this was a challenge for X-rays, in fact observed with RXTE.  But
we would like to think that at least some of the observations were done when the
satellite was over India!

\subsection{Star Clusters}

It is an artefact of our place and time in the universe that we normally think of
star clusters as consisting of ``open,'' ``globular,'' and ``other,'' where ``open'' = 
in the Galactic disk, some still being formed (and falling apart), relatively few stars, and
irregular morphology; ``globular'' = in the galactic spheroid and halo (+ maybe thick disk)
all quite old, mostly stable, metal poor, $10^{4-6}$ stars, and spheroidal morphology; and
``other'' includes ones around galactic nuclei, massive young clusters being formed in
mergers, and perhaps some transition objects between small galaxies and big clusters
long ago, ``other'' would have been the dominant sort, just as before $z=1-2$ ``irregular''
becomes the commonest sort of galaxy. Each of the three c1asses has a green circle paper,
the first mentioned in its subsection.

\subsubsection{Open Clusters, Moving Groups, and Star Streams}

We were considerably surprised (not having thought of it for ourselves, the source
of most surprises) to learn that many star clusters actually get bluer with age for
$40-80\%$ of their lives (Lamers et al.~2006), because the low mass (red) stars 
preferentially
leave.  Colors are, therefore, very poor age indicators, with 13 Gyr successfully
masquerading as $2-5$ Gyr, (why doesn't Estee Lauder sell this in a tube?) while later in
life, the clusters get red very fast and age can be overestimated. Cluster ages in general
remain a bit of a problem. After living with the Pleiades for many years, astronomers
still can't do better than $70-100$ Myr, from oscillation frequencies of six Delta Scuti
stars (Fox Machado et al.~2006).  Because star formation in a given cloud can extend over
considerable time, stars approaching the main sequence don't just look younger
than stars leaving, they really are (Tan et al.~2006; Subramaniam et al.~2006), though
Lyra et al.~(2006) note that yau can bend your rulers to make all the stars look the same
age. Probably that should say ``bend your clocks,'' but we aren't quite sure how to do that
unless the clock is the old fashioned pendulum sort.

What is a cluster anyhow?  Markov (2006) has struggled to find the 100 M$_\odot$ needed
to bind Alpha Per within 6.5 pc of the center, while Slesnick et al.~(2006), looking at
the Upper Sco OB association, found brown dwarfs popping out all over the place, and
expect 100's of low mass stars by the time they finish the survey.

Perhaps the most important thing to be said about open clusters is that they are
generally not built to last.
Although the senior member of the Galactic disk is $10\pm1$ Gyr old (Berkeley 17, according
to Krusberg and Chaboyer 2006), even M67 is a mere shadow of its former self (Hurley et 
al.~2005), and Orion will unbind soon by blowing its gas away (Huff and Stapler 2006). In a
global sense, you can inventory the clusters in our part of the galactic disk, measure
their ages, get a characteristic dissipation time, and therefore determine the rate at
which stars have formed in these inventoriable clusters. It is considerably less than
the real formation rate (Piskunov et al.~2006;
Gieles et al.~2006a, and several other papers from the same group).
The rational conclusion is that many clusters are no longer recognizable as such by the
time they remove their dust shrouds (Bonatto et al.~2006; Lada and Lada 2003). Notice
that stars live backwards and have shrouds at the beginning rather than the end
of their lives.
The result is that what we see is what is dissipating, not what little survives
(van den Bergh 2006).  Loss goes by means of symmetric tidal tails on both sides of the
cluster (Chumak and Rastorguev 2006).  Some clusters develop noticeable mass segregation
while falling apart (or had it to begin with, Santos et al.~2005 on M11 at 250 Myr),
others do not (Lutkin 2006 on the somewhat younger Pleiades).

The distribution of cluster masses is more or less the same in all galaxies (Lutkin
and Sauvage 2006) at least at the upper end.  This cannot be true at the lower end if we
try to accept simultaneously that M82 has a log-normal N(M) and was born that way (De
Grijs et al.~2006) and that, in other places,  the IMF of stars connects smoothly to
the IMF clusters near 100 M$_\odot$ (Beltran et al.~2006).

\subsubsection{KREDOS}

We have invented this acronym for {\sl Kinematically Recognizable Entities of Dubious
Origins} to signify that disagreements about them sometimes sound almost faith-based
(e.g. Sect. 3.2 on star stream vs. disk warp in the Milky Way). There are, at least,
three sorts, (1) stars (and perhaps gas) moving recognizably through the halo as a result
of disruption or partial stripping of a small galaxy (one of our kindly data-donors notes
that these raise the question of just how small can a galaxy be),  (2) the remnants of
former OB associations still moving together, and (3) assemblages of stars formed over
a considerable time and with a range of compositions that cling together in velocity
space (a topic pioneered by Eggen 1959 and dozens of later papers).

The halo streams have, in the past year or two, reached the status of ``many'',
including these 2006 representatives: Vivas and Zinn (2006, found using  RR Lyrae halo
stars), Abad and Viera (2005, a parallax and proper motion study), Grillmair (2006,
something that looks like a stream from a dwarf galaxy, but with no candidate dwarf to
be seen), and Grillmair and Dionatos (2006, using SDSS data).

Arfyanto and Fuchs (2006) make the important point that it is not always possible
to distinguish tidal debris from non-axisymmetric perturbations of the Milky Way
potential from the Eggen sorts of assemblages, but disintegrating star clusters are
recognizable because they include obviously young stars and live in the thin disk. 
Helmi et al.~(2006) report three new kinematic moving groups in the disk, each with a
range of ages and metallicities, which they describe as debris, though of what is not
perfectly clear.

The Sirius supercluster (also known as the U Ma stream) has kinematical substructure
which its discoverers (Chopina et al.~2006) do not attempt to explain, though they note
that the inventory of members began with Roman (1949).  This, rather than the discovery
that high velocity stars have weak metal lines, was her thesis work.

The stars of Gould's belt march to a different set of Oort constants from other
disk OB stars, but could be just a chance superposition of moving groups (Elias et 
al.~2006). Some of their data go back to Herschel at the Cape (meaning John, who collected
the data in 1834$-$38 and published in 1847; we think that like the younger Fords
and Marriotts, he inherited the business and didn't need to write a thesis).

\subsubsection{Globular Clusters}

In galaxies that have both, the open clusters generally outnumber the globular
ones, but we found 75 globular papers (vs. only 31 open ones) and idiosyncratically
assigned the green circle to the curious factoid that Einstein considered M13 and
concluded that its non-luminous mass did not greatly outweigh its luminous mass. The
paper appeared in a not-very-accessible {\sl 10th Anniversary Festschrift for the 
Kaiser Wilhelm
Gesellschaft} (now {\sl Max Planck}), and so we refer you to Richler (2006), 
who has a number of
other interesting things to say, including the possibility that there could be more than
one formation mechanism for globular clusters.

Indeed this year, let's focus on ``more than one'' as the globular cluster theme, even
when the contenders are nearly as mutually exclusive as A and not-A.  Consider the
proportion of binary stars among cluster members. 
Much smaller for horizontal branch stars in the nucleus of NGC
6732 than in the field say Moni Bidin et al.~(2006), vs. so many along the main sequence
of M80 that you can actually see the parallel one of MS + WD stars in a color-magnitude
diagram (Shara et al.~2005). There are, say Beccari et al.~(2006), both primordial and
collisional binaries to be found, so that the specific frequencies of blue stragglers and
of millisecond pulsars are anticorrelated.

Whatever you may think about the total initial fraction of binaries in globular
clusters, they are a truly excellent place to live if you want to be an X-ray binary,
binary millisecond pulsar, or cataclysmic variable, most of which arise from dynamical
processes in the clusters rather than from isolated close binary evolution (Lombardi et 
al.~2006; Bregman et al.~2006; Xu et al.~2005; Shara and Hurley 2006; Pooley and Hut 2006;
Kim
et al.~2006). The CVs yield few optical outbursts (we remember when it was one per
cubic universe) because of small mass transfer rates (Dobratka et al.~2006). From
the point of view of evolution of the cluster itself, effects of close binaries saturate
if they make up as many as 10\% of the initial stars (Heggie et al.~2006).

The Oosterhof Type I/II dichotomy hasn't really been a clean one for several years,
but we record the intermediate properties of NGC 6441, high metallicity = II, long RR Lyrae
periods = I (Clementini et al.~2005), in part for the pleasure of noting a pre-discovery
by Grosse (1932), to which our attention was called by Sandage (2006). His real purpose
in that paper was to point out the very considerable differences between field and cluster
RR Lyrae stars, a point of view which we indexed under ``Carthago delenda est.''

``The'' second parameter is shorthand far all the causes that might be responsible for
clusters with very similar total metallicity (the first parameter) having horizontal
branches of very different morphology.  This year we recorded two votes for helium
abundance (Mochler and Sweigert 2006, D'Antona et al.~2005), implying dY/dZ=16 or so,
and the puzzle is just where the additional He has come from, not how it will
affect the stars; two votes for something correlated with cluster orbits or location in
the Galaxy (Barbuy et al.~2006; Carretta 2006), for which the issue of how it works seems
to have puzzled the authors almost as much as it does us;  one vote for total cluster
mass (Recio-Blanco et al.~2006) because big ones retain more metals (which sounds like
first parameter); and one vote for time ellapsed since last core collapse in
gravithermal oscillation (Suda and Fujimoto 2006), because the interactions in collapse
phase spin up stellar rotation, yielding extra mixing and a wide range of HB colors.
The chemical compositions we see on the surfaces of globular cluster stars today (well,
anytime this past half-century or more) must be a combination of what was in the gas
from which the cluster formed, stuff produced by massive cluster members when the present
ones were still forming, material transferred from close companions, and up-mixed products
of nuclear reactions in the stars now being studied. The votes:
\begin{itemize}
\item[o]{Pre-RG stuff + current mixing (Smith and Briley 2006)}
\item[o]{Mixing + close binary transfer (Smith and Briley 2005)}
\item[o]{Contributions from stars too massive ($>$8 M$_\odot$) to go through an AGB stage
(Kas'yanova and Shchekinov 2005)}
\item[o]{Two flavors of AGB stars (Yong et al.~2006)}
\item[o]{Some in situ proton capture (Johnson et al.~2005)}
\item[o]{At least initial conditions + reactions in the stars themselves (Letarte et 
al.~2006)}
\end{itemize}
And we left out yet other papers on this topic because they said the same things, or
different things, or things we couldn't quite understand.

Within the Milky Way, the system that most severely raises the question ``what is a
globular cluster and how are they related to other things with stellar masses of
$10^{4-6}$ M$_\odot$'' is Omega Centauri. It has a range of stellar populations like those 
in a
dwarf spheroidal galaxy (Stanford et al.~2006), but on the other hand, it has an M/L ratio
of only about 2.5 like other globular clusters (Van der Ven et al.~2006, using plates that 
date back to the 1930s to get proper motions).  There is also a ``second parameter'' main
sequence with Y = Y$_p$ + 0.12 (Bekki and Norris 2006; Maeder and Maynet 2006). The idea that
Omega Cen is indeed the core of a dwarf, integument stripped away, and is old enough that
the cluster has become a prototype for a class (Ma et al.~2006b). One would like to
know the results of a deep H I search, since the dwarf galaxies quite often have just
about the amount you would expect their stars to have lost (Bouchard et al.~2005), while
some globular clusters have even less than that, 0.3 M$_\odot$ vs. 30 M$_\odot$ 
in the most extreme case (Van Loon et al.~2006).

Now, what about the populations of clusters in galaxies as opposed to stars in the
clusters? Brighter galaxies have brighter clusters (Jones 2006) and more metal rich
ones (Peng et al.~2006a) as well as more clusters total. This would be easier to interpret
though harder to measure if the specific frequency, S, were reported vs. mass rather than
vs. luminosity (Bekki et al.~2006 on why some dwarfs have S = 30). Other papers
emphasized how very much more similar the clusters are than the galaxies that host them
(Beasley et al.~2006; Conselice et al.~2006, which we indexed under ``no downsizing here.'')
A question that bothers your author in academic harness whenever she teaches ``galaxies''
is what happens to S (roughly number of clusters per $10^{8}$ L$_\odot$) when ellipticals 
are made from spiral mergers?  Nothing bad, say Bekki et al.~(2005).

Do young globular clusters exist? No definite answer this year, but a caution that
many candidates (2/3 of those in M31 for instance) are ``asterisms'' - faint associations
with a few bright stars superimposed (Cohen et al.~2005). This is not quite what the Chinese
meant by their equivalent of asterism (more like a compact constellation).  And having
just this week met the non-element asterium, we hope to report further details next
year. At this time, the intermediate age globular clusters of M81 (Ma 2006a) will
be only very slightly older than their current few Gyr.

Total globular populations are frequently bimodal in number vs. color distribution
(Bassino et al.~2006 on NGC 1399 in the Fornax cluster). Harris et al.~(2006) describe the
conventional way of getting double-peaked N(color) with red = metal rich, younger, blue =
metal poor, older. The Milky Way arguably works that way (Bica et al.~2006), with the
red ones coming from an initial monolithic collapse and the blue ones (mostly at larger
galactocentric radii) from mergers with metal-poorer entities. In contrast, Yoon et 
al.~(2006) would like to start with a single age for all clusters and a unimodal distribution
of compositions and introduce the structure in N(color) with non-linear transformations
from metallicity to color.

In Cen A (NGS 5128) the color distribution is a continuum (Gomez et al.~2006) that
extends into the realm of the Faint Fuzzies and the Ultracompact Dwarfs and so leads us
on to the next subsection.

\subsection{Other Sorts of Star Clusters}

The roundest green circle in this territory has already been noted briefly, the idea
that there may well be a continuum of properties which take in
\begin{itemize}
\item[o]{G1obular clusters and dwarf spheroidal galaxies (Leet et al.~2006)}
\item[o]{The faint fuzzy clusters that were new a year or two ago, globular clusters, 
ultra-compact dwarfs, and extended luminous clusters (Gomez et al.~2005)}
\item[o]{UCDs and globular clusters (Bastian et al.~2006; Kissler-Patig et al.~2006)}
\end{itemize}
Defences of the separate classes naturally also appeared (De Rijcke et al.~2006,
Fellhauer and Kroupa 2006), and some specifics, with mechanisms:
\begin{itemize}
\item[o]{UCDs, some = merged young massive star clusters (for instance in Fornax), 
others are
tidal stripping (for instance in Virgo), Mieske et al.~(2006) a ``both please red dot''
paper.}
\item[o]{Diffuse clusters in Virgo E and S0 galaxies, which in the Milky Way would be VERY
difficult to find (Peng et al.~2006)}
\item[o]{Faint fuzzies, on the grounds that they live only in SO0 galaxies 
(Hwang and Lee 2006).}
\item[o]{Things that look like young globular clusters, but are very short lived 
(Gieles et 
al.~2005 on an entity of $10^6$ M$_\odot$ that disrupted in a few Gyr in M51). 
Indeed star burst
clusters are in general short-lived (merriness not recorded by Melioli and Gouvera Dal
Pinto 2006).}
\end{itemize}

Many, perhaps most, galaxies have an accumulation of stars at their centers (a point
to which we return in the black hole/bulge part of Sect. 10), with ages ranging from
$10^7$ to $10^{10}$ yr (Rossa et al.~2006, reporting HST data on 40 galaxies).  
The Milky Way
is obviously not the best or the brightest, but it is the most highly resolved, enabling
study of the kinematics (Paumard et al.~2006). Remarkably, the OB stars within the
central parsec seem to live in two disks, tilted to the main galactic plane and rotating
in opposite directions, each young enough to require star formation in situ.

\section{SILENT, UPON AN PEAK IN DARIEN: EXO AND ENDO PLANETS} 

In {\sl Ap 96}, we threw stars and circles in all the colors of the rainbow at the
announcement of 55 Peg B. Like whiffnium, whaffnium, and whoofnium\footnote{
Substances within which you might discover a new effect, thereby, according to
the late Sam Goudsmit, garnering one or two appearances in Physical Review Letters,
but not three ...} 
the next few were also clearly highlights, but so many have now swum into our ken,
that it has become
difficult for a planet to stand out from the crowd. We mention here a few new extrema,
progress on detection methods, relationships with disks, formation and dynamical evolution
processes, and statistics of the planets and their hosts. None of the items appears to
require major changing of opinions.

\subsection{Exoplanets}

These are not only outside the sun but outside our solar system. Detection of any
outside the Galaxy may have to wait a while.

\subsubsection{Extrema}

The triple Neptune of HD 69830 (Louis et al.~2006; Charbonneau 2006a) has planets
at 0.08, 0.19, and 0.63 AU, the last on the inner edge of its habitable zone. The masses
are 10, 12, and 18 M$_\oplus$, and they must have formed inside the ice line (followed by
migration), so the expected composition is rocks and perhaps gases, but probably not
water, even if it would be stable there now. An infrared excess indicates the presence
of an asteroid belt at $0.3-0.5$ AU, with mass = 25 times ours.

HD 37141 (Vogt et al.~2005) became the fourth triple host as it was announced,
and indeed was not an {\sl ApJ Letter} (wheffnium?). The third planet of Gl 876 (Rivera et 
al.~2005) is, at 7.5 M$_\oplus$ (P = 1.94 days), the least massive found by the 
triumphant radial
velocity technique. It is, however, beaten out for smallest of the year by a 5.5 
M$_\oplus$ microlensing object (Beaulieu et al.~2006). 

Planets found by microlensing generally belong to stars that would otherwise have
been invisible, though the color-dependent centroid of OGLE-2003-BLG235 
= MOA-2003BLG-53
(Bennett et al.~2006) means that both lenser and lensee are contributing photons to the
HST images (and that they are different colors!).  The planet is about three times the
mass of Jupiter, and the name arguably even harder to remember than HD numbers
(especially the part about ``where do the noses - sorry - hyphens\footnote{
''Vere do zhee noses go?'' Was Garbo on kissing in Ninotchka.} go?'').

Another OGLE (2005-BLG169) is their second cool Neptune, leading the authors to declare
(Gould et al.~2006) that the class must be a common one. The planet is about 2.7 AU
from a 0.5 M$_\odot$ star, and the magnification in the event was 800 at the time it was caught.

At least some exoplanets resemble our giants in having significant rocky cores,
for instance HD 149026b, whose radius and density come from a transit (Charbonneau et
al.~2006). Theorists are actually fairly optimistic about the potential for
earth-like planets in earthlike orbits (Li et al.~2005; Raymond et al.~2006), even for
systems with hot Jupiters, like 47 U Ma. Kokubo et al.~(2006) find two terrestrial
planets between 0.5 and 1.5 AU in a typical simulated formation process.

\subsubsection{Search Methods}

Some attempt was made to count these in previous year, reaching up perhaps to 20,
of which three clearly successful (periodic variability of radial velocities, transits,
gravitational lensing).  Additions in 2006 are called N+1, N+2, etc., and subtractions
N$-$1.

N+1 is metals in atmospheres of otherwise hydrogen-surfaced white dwarfs, interpreted
as infall from a debris disk of dying asteroids.  GD 362 is the second (Kilic et al.~2005;
Becklin et al.~2005, and no, neither cites the other) and GD56 the third (Kilic et al.~2006).

N$-$1 is the non-confirmation of H$_2$O masers from exoplanets orbiting some stars,
previously announced, but unpublished. No detections, say Minier and Lineweaver (2006).

N+2 = linear polarization due to scattering in planetary atmospheres by H$_2$O and
silicates; and N+3 = a new sort of Doppler detection, ``dispersed,  fixed-delay
interferometer'' (Ge et al.~2006).

A few discouraging words about classic methods also got printed. To catch a transiting
planet in a star cluster, you need to follow 7400 dwarf stars for more than a month
(Burke et al.~2006). In applying the radial velocity method, you sometimes need to
separate the effects of a planet from stellar oscillations and activity, but deconvolution
of the line bisector won't necessarily work, because different lines have different
bisector shapes (Dall et al.~2005). We remember this one from a long ago effort to
try to determine the distortions of radial velocity numbers due to convection. There
is at least one young star with a periodic V$_r(t)$ due to surface activity (Konig et 
al.~2006). With a period of 4.38 days and an amplitude of 30 m s$^{-1}$, you might have supposed
HN Per to have a very warm Jupiter, but there are 4-5 day periods in other data, implying
activity plus differential rotation. Beta Pic, known to have a (pre/post planetary)
disk shows similar radial velocity variations due to pulsation (Galland et al.~2006).

N+3, spectral signatures of exovegetation, belong to the ``first catch your rabbit''
class (Tinetti et al.~2006), though the idea that there might be chlorophyll analogs
that store up three rather than two photons to make one big h$\nu$ appeals to Wolstencroft
and Raven (2002).

Transits dug out of Hipparcos data (Hebrard and Lecavelier des Etangs 2006) and
direct detection in strong spectral features (Berton et al.~2006) have in common with
exovegetation that it helps a lot if you already know there is something there.

\subsubsection{Statistical Properties, Hosts vs. Planets}

The rarity of brown dwarfs in short period orbits around solar-type stars (the brown
dwarf desert) is seen most clearly in the exoplanet search projects.  By curious parallel
the first radial velocity survey of very young brown dwarfs finds a planet desert around
them for a $<$0.1 AU (Joergens 2006). Binary stars have their fair share of planets
in the restricted range of orbits where they are dynamically stable (Raghavan et al.~2006;
Mugrauer et al.~2006).

In a mid-year snapshot including 172 radial velocity planets within 200 pc of us,
Butler et al.~(2006) report that N(P) peaks at 3 days; N(a) is nearly smooth in log a for
0.03 to 10 AU; and dN/dM$\propto$M$^{-1.1}$. Working from a smaller, earlier sample of 143 
planets
Jiang (2006) finds a continuum in N(M), but three clusters in period-eccentricity space.
Both papers include complete tables of the planets analyzed, in case you want to hunt for
correlations of your own. These samples exclude the transit planets, of which the 10th
appeared at the end of the index year (McCullough et al.~2006). The host is very much like
our sun, and the observing team, with four amateur astronomers, one based at Raccoon Run,
somehow also sounds rather heimlich (meaning homelike, not the maneuver).  But the
meaning of their project name, X, is not explained.

Exoplanet hosts span a wide range of ages, perhaps on average a bit older than spectral
matches with no planets (Saffe et al.~2005). The ``spectral match'' aspect is presumably
meant to eliminate the selection effect against stars with lots of activity (which are, of
course, systematically young). Perhaps it did.

That planet hosts are comparatively metal rich has become a cliche, with residual
disagreement limited to why $-$ is it that metal rich stars find it easier to make planets,
or that planetary leftovers pollute host atmospheres? In a paper sure to offend all true
believers, Ecuvillon et al.~(2006) look for correlations of host composition with 
condensation
tempeartures of individual elements and conclude that some of each possible cause is
needed. Interesting, but arguably consistent with either hypothesis, is a possible
correlation between host metal abundances and the total amount of heavy elements (typically
20$-$100 M$_\oplus$) present in the transit planets whose interior structure can be inferred
from a combination of mass and radius values (Buillot et al.~2006).

An assortment of correlations of planet frequency, mass, and orbit parameters with
various host properties were explained during the year (Lada et al.~2006; Pinotti et 
al.~2005; Boss 2006; Robinson et al.~2006a; Kornet and Wolf 2006), and we skimp on the details
out of the feeling that the opposite correlations could also have been explained. Indeed
in a paper to which we had tentatively awarded the dreaded green Q, Mugrauer et al.~(2005)
appear to find and explain a correlation without making it absolutely clear which
way it goes.

\subsubsection{Formation Mechanisms}

The planetesimal folks and the instability folks have not come to a meeting of the
minds or papers, and so a hybrid mechanism (Cai et al.~2006) is sure to displease all
comers. In contrast, the conclusion that the planet in the triple star system
HD 188753 could only have got there via star exchange seems to have been reached 
independently by two investigations (Portegies Zwart and McMillan 2005 and Pfahl 2005).

The exoplanet most poked and prodded during the year was HD 209458b. The fun
begins with a Spitzer Space Telescope detection of the occultation when the planet goes
behind the star (Seager et al.~2005), the planet being otherwise invisible (Rowe et 
al.~2006). There followed models for redistribution of radiation in it and similar planetary
atmospheres (Barman et al.~2005), a discussion of expected values of radius for a
given amount of heating (Nagasawa and Lin 2005), and a final definitive set of planetary
parameters (Wittenmyer et al.~2005). It is only about 1/4 as dense as Jupiter, and,
given the small albedo of 0.25 (Rowe et al.), we propose a hollow\footnote{
Have we ever told you about the Hollow Dogs of UC Irvine?  They averaged 3 feet in
length and were to be found wandering our halls at a time when Buildings and Grounds
forbade access to non-human animals weighting more than 40 lbs. A limit on weight for
human animals these days might do some good.} structure made of black Lego\copyright\ blocks.

Much was also written during the year about migration, evolution of planet orbits,
resonances, and stability, of which the most disconcerting item was that the two (or
more) systems previously advertized as being in 2:1 mean motion resonances are really
Trojans, with 1:1 mean motion resonances and orbits with the same semi-major axes but
different eccentricities (Tinney et al.~2006; Gozdziewski and Konacki 2006).

Disk thou wert and to disk returnest, thus sayeth the prophet, at least when he (?,
it's hard to tell with disks) is speaking to planets. Of the ``befores'' our favorite is the
early tadpoles of Cressell and Nelson (2006) and of the posts, a bunch that are very much
like our own (Bryden et al.~2006), since they lead inexorably to

\subsection{Endoplanets and the Rest of the Solar System}

These are, of course, outside the sun but inside the solar system.

\subsubsection{Dust and Meteors}

The zodiacal\footnote{
Secretly we like to pronounce this with accents on the 1st and 3rd syllables rather
than the 2nd.  But then you should hear us say Betelgeuse.}
light is our very own debris disk, as demonstrated by the presence
of several young asteroid trails in SST data (Nesvorny et al.~2006) and the dust isn't
all that different from presolar grains and meteorites, at least in some key isotope
ratios (Buseman et al.~2006a).

Large enough grains make meteors when they hit the earth's atmosphere. Properties of
a number of showers were reported during the year, but we note the converse question
(obvious only once someone else asked it) of whether there are any true sporadic meteors,
or are they all part of weak showers (Campbell-Brown and Jones 2006, Jones and Jones 2006).
The most disrupted author thinks the situation is like the gradual transformation of star
clusters to star streams and eventually field stars, so that we have lost forever any
knowledge of which stars formed with our sun. Streams of meteoric dust have no fewer
than three ways to get separated from their parent comets (Vaubaillon et al.~2006). And
pity the poor Quadrantids of January (Koten et al.~2005), who have lost even their parent
constellation (Quadrans Murales, now part of Bootes). The parent body, 2003 EH$_1$, is
dormant too.

Neslusan (2005) has predicted meteor showers for all the terrestrial planets,
coming from close approaches of comet and asteroid orbits.  Everybody gets some, but the
Mercutians never see theirs (no atmosphere to ablate the grains), while the Venerians
had a pair as recently as June and August 2006 (Vaubillon and Christou 2006), made up of
stuff from comet 45P Honda-Mrkos-Pajdusakova ejected between 1943 and 1980.

In amongst all those grains are a few (which we can study because they have been
trapped in meteorites) that preserve excesses of short-lived isotopes. Some of these
must surely have come from a supernova that fed newly-synthesized heavy elements into the
proto-solar cloud (Tachibana et al.~2006), but there were a good many votes during the
year for production of some of the short-livers by particle irradiation by the nearby sun
(Gounelle et al.~2006; Hsu et al.~2006; Wolk et al.~2005, a paper that is part of
a package reporting a Chandra X-ray survey of the Orion region). Fractionation of isotopes
in meteorites is generally attributed to electromagnetic effects, and so should be
proportional to Q/M of the nuclides (and so monotonic in atomic number, A). Fujii
(2006) presents laboratory evidence for a ``nuclear field shift'' that will produce odd-even
rather than monotonic fractionation. Yes, he explains it, crediting the idea to Biegeleisen
(1996). Mass-dependent fractionation was first treated by Urey (1941), a
year that not even we remember very well.

The issue of deciding which meteorites come from which asteroids must be important
because the 2006 papers on the topic appeared in high-profile journals (Mayer et al.~2006
on the LL Chondrite parent body and the 145 Myr old Morokwevy crater in South Africa;
Greenwood et al.~2006 and Clayton 2006 on pallasites, mesosderites, and Vesta).

\subsubsection{Comets and Asteroids}

Creeping upward in size to the asteroids of the main belt, we find one that broke into
6 pieces 450,000 years ago (Nesvorney et al.~2006a). It, or they, are called
Datura\footnote{
Among other things, a poisonous number of the nightshade family, and if the poison
caused personality dissociation, the name would be marvelously appropriate.}
but wait til you meet the fragmented comet a few paragraphs down.

Does the possibility of an asteroid hitting earth keep you awake at night? If so,
you should be comforted by the news (Anonymous 2006b) that there is now ``an IAU Committee
to keep it up to date about asteroids that may pose a serious threat.''  The members'
appointments presumably expire upon impact. Or you could tow the threatening objects
away using purely gravitational forces (Lu and Love 2005). They suggest using the mass of
a spacecraft, though piles of unread journals might be more appropriate.

Venus could have Trojans but doesn't (Scholl et al.~2005). Neptune has a bunch only
recently discovered (Marzari 2006), Jupiter has also acquired more (Yoshida and Nakamora
2005), and its binary, 617 Patroclus, has a newly determined orbit (Marchis et al.~2006).
In case the namers should run out of heros of the Trojan war, we suggest for Neptune
names of seaweeds that follow the tides, and for Venus names of collectors of Marilyn
Monroe memorabilia.

We scanned the {\sl IAU Circulars} for the reference year, hunting out asteroid/comet/KBO
items and have the following to offer: The number of identity switchers (coteroids
and asmets?) has reached the level of ``many'' ({\sl IAUC} 8704 to 8735
and beyond).  The TNOs and
main belt asteroids that acquired satellites or binary companions during the year was at
least half a dozen each. The three named dwarf planets are now 1 = Ceres, 134340 =
Pluto, 136199 = Eris.  We think that the rate of discovery of asteroids compared to the
rate of human population increase is such that eventually everybody could have one, yet another
reason to try to avoid identity theft.

The 177th comet accorded periodic status is 1889 Barnard = 2006 M3 ({\sl IAUC} 8737).
Comets are still being discovered by individual, sometimes amateur, astronomers, as well
as by space missions and consortia, sometimes even several per person, but when we got
to 10 for R.~H.~McNaught ({\sl IAUC} 8721) we stopped counting. And perhaps one should have
stopped counting the bits as comet Schwassmann-Wachmann fragmented, running off the ends
of several alphabets with fragment BS ({\sl IAUC} 8715). We're not making this up, you know,
and miss Anna Russell very much.

Cometary highlights included both the individual and the statistical. Results from
the Deep Impact of comet 9P/Tempel 1 reached the paper literature, recording it as very
dusty (Kuppers et al.~2005; Feldman 2005), with a mix of materials (olivine, fosterite,
pyroxenes, and all) that must have formed at different places in the proto-solar system
and then been mixed together (Lisse et al.~2006; Schulz et al.~2006). Indeed so much
dust was kicked up that it was never possible to image the Deeply Impacted crater and
determine its size with any precision.  Anonymous (2006c) estimated 100$-$250 m. We suspect
that the conclusion that the comet was back to normal only six days afterwards was
reached by folks who weren't there at the time (Schleicher et al.~2006).

On the statistical side Sekanina and Chodos (2005) provide a comprehensive introduction
to groups of comets that have come from fragmentations, some of which are also traced by
meteor showers. The traumas date from before 950 to about 1780.

Hsien and Jewitt (2006) have found the third comet with an orbit indicating its origins
in the main asteroid belt, thereby establishing these as ``a we11 known class of
astrophysical object,'' and indeed they propose that there are three comet reservoirs,
the Oort cloud, the Kuiper belt, and the asteroid zone. Their object has both a comet
and an asteroid name. The smallest nucleus to date,  160 m across, also belongs to a
coteroid/asmet (Jewett 2006) called D/1819Wl (Blanpain) = 2003 WY$_{25}$. Meanwhile a new
member of the Jupiter family, 162P/Siding Spring = P/2004 TU$_{12}$ (Fernandez 2006) has a
hefty 6 km radius, but a V-band albedo of only 0.034.

The second most hyperbolic orbit, e = 1.0106 has been established for C/1853 El
(Secchi) by Branham (2006a).  The author makes the somewhat curious statement that ``to let
a comet named after such an important pioneer remain in a parabolic orbit seems 
unprofessional.''  
He is apparently not disturbed by the most eccentric, C/1980 El (Bowell)
with $e=1.0573$.  Secchi's comet was not identical to C/1664 W1, which would have brought
its $e$ down below l and its semj-major axis up from $-103.03$ AU.

Higuchi et al.~(2006) report that Jupiter did most of the scattering that sent
planetesimals out to the Oort cloud, and Dybczynsk (2006) says that the perturbations that
sent them back in are due mostly to the galactic disk tide.  The next relevant close
stellar approach will be that of Al 170 at 0.2 pc, 1.3 Myr in the future.  Both are
calculations rather than observations.

Most of the long period comets with decent orbits were not observed on their first
visits to the inner solar system (Dybczynski 2006). On the one hand, we think that this
is at least partly due to the potential first observers having been busy knapping flint,
and, on the other hand, it sounds like it ought to be somehow associated with the fact
that 95\% of old comets with periods less than $10^6$ years are missing (Neslusan 2006).
The processes by which comets are captured into periodic orbits seem, however, sufficiently
complex that retrodicting a uniform capture rate over the past million years is surely a
bit risky (e.g., Hahn et al.~2006 on P/2004 Al, which Saturn will hand over to Jupiter
in 2026.

Comet surface chemistry is not yet entirely sorted out.  On the one hand, some
things you might expect to find (ethanol, methyl formate, etc.) are not there, though
methanol is (Remijan et al.~2006).  And, on the other, there are still quite a few
unidentified features, and photoluminescence may be required to model them (Simonia 2005).
The same author has advocated photoluminescence and cathodluminescence of frozen
hydrocarbons to explain features in the spectrum of the planetary nebula NGC 7020 (Simonia
and Mikailov 2006, and elsewhere).

Concerning Kuiper Belt and Trans-Neptunian objects, there is both good news and
bad news. The good news is that 2003 UB$_{313}$, the one that caused all the fuss and bother
about what is a planet is now merely Eris, named for the goddess of discord, and her moon
is Dysnomia, goddess of lawlessness and daughter of Eris.  Those are IAU committee
decisions and so to be trusted as much as any IAU committee decisions. But the LA Times
article reporting the names also declared that Clyde Tombaugh named Charon after his
wife Charlene. Yes, Tombaugh (1906$-$1997) was very much alive when Charon was discovered
in 1978 (by James W. Christy of the US Naval Observatory), but he was not the discoverer,
and his wife was Patricia (nee Edson)\footnote{
This is not something we would necessarily as a rule know, but the semi-final proofs
of the Biographical Encyclopedia of Astronomy (Hockey et al.~2007) currently sit on the
floor of the office of the author who has most books on the floor.}.
You might want to start over and assume that
Christy's wife's name is Charlene and that the LA Times has (at least) one editor like
the one who just processed a chapter of ours for a book on evolution in general.

The bad news is that Eris almost ought not to exist at all. If the numbers of TNOs
implied by occultations of Sco X-1 (Chang et al.~2006; Cooray 2006) are correct, the
little ones should have totally ground each other up by now.  The required number is about
$10^{15}$ with diameters of 10$-$100 m. Optical searches are chasing wildly after the X-ray one,
with Roques et al.~(2006) reporting two events due to 100-m sized objects at or beyond
100 AU. Note the implication that the KBO no longer has a sharp edge at less than
50 AU as advertized a year or two ago, but the full extent and possible structure in
N(a,e) remain to be further determined (Vicente and Alves 2005; Allen et al.~2006, Reid and
Parker 2006).

\subsubsection{Moons}

Titan swept the field this year, because of the return of copious data from the
Cassini mission and Huygens lander. There is an awful lot going on at and above the
surface, for instance:
\begin{itemize}
\item[o]{Impact craters and fluid motion (Elachi et al.~2006).}
\item[o]{Rain and clouds (Hoeso and Sanchez-Lavega 2006).}
\item[o]{Hadley cells and drizzle (Tokano et al.~2006; Griffith 2006).}
\item[o]{Methane lakes, but no seas (Lunine 2006).}
\item[o]{Weather, ethane clouds, and possible snow (Griffith et al.~2006).}
\item[o]{Atmospheric temperature structure with a troposphere and stratosphere rather like
	earth (Flasar 2006).}
\item[o]{A chemically complex atmosphere with 58 molecules, up to C$_2$H$_5$CNH$^+$ 
	(Vuittoni et al.~2006).}
\item[o]{``Titanophysical'' activity revealed by albedo changes (Barnes et al.~2005).}
\item[o]{Preferential longitudes for clouds (Griffith et al.~2005).}
\item[o]{Volatile releases from geysers or cryovolcanoes (Roe et al.~2005).}
\item[o]{An atmosphere dominated by N$_2$ with smog, an earth-like landscape, 
50\% relative
humidity at the surface in CH$_4$, presence of both $^{36}$A (presumably primordial) and
$^{40}$A 
(presumably radiogenic), breezes mild at the surface but up to 360 km/hour at 100$-$150
km above the surface, and surface T = 94 K, P = 1467 hPa (Owens et al.~2005 and the
next seven papers).}
\item[o]{Longitudinal dunes like those in the Namibian desert, consistent with the surface wind
speed of 0.5$-$1 m s$^{-1}$ (Lorenz et al.~2006).  These were actually predicted by Greeley
(1981). The 100$-$300 $\mu$ particles are smaller than Earth sand and of very different
composition.}
\end{itemize}

Why is all this so pleasing? Well, weather forecasting on earth remains at least
as much an art as a science, and it might help to have another very different climate to
practice on.  And only once in our academic 2006 year notebook did we accidentally record
a Triton paper under Titan (it is not in the list above!).

Saturn has bunches of moons, nine more (all retrograde) just this year ({\sl IAUC} 8727),
including 35 with names (Daphnis is the latest, {\sl IAUC} 8730). Cassini (the probe; the
persons are all dead and presumably do not care) also had something to say about Hyperion
(Denk 2005), whose surface is like a sponge (in texture, we presume, not composition),
and also about the surfaces of Phoebe and Iapetus (Brown et al.~2006a) and, especially
Enceladus (Porco et al.~2006, plus 3 preceding and 8 following papers), which has a
complexity of its own in magnetic effects, surface flow structures, and some unidentified
spectral features.

There are, of course, the rings, also with much recent data, but we mention only a
theory paper, Burns et al.~(2006), which suggests (a bit like the Titan weather)
that modelling them should be good practice for study of protoplanetary disks.

As for the rest of the solar system brigade of satellites, let's start at the
outside and work in. Nix and Hydra (of Pluto) were probably produced in the same giant
impact event that made Charon (Ward and Canup 2006; Lissauer 2006).  The category of
impact-born moons therefore now has four members (with Luna) entitling it to ``well known
class'' along with capture and co-formation in protomoonary disks (proto-satellitary, is
worse, isn't it?)

Neptune acquired more of both the common categories (Sheppard et al.~2006), Triton
retaining its status as most massive irregular (meaning large e and i) moon.

Uranus upped its moonventory to 27, plus two new rings fed by Mab ({\sl IAUC} 8648, 8649).

Europa of Jupiter has a highly cratered surface, with up to 95\% of the smaller ones
being secondaries (Bierhas et al.~2005). If this preponderance of secondaries should prove
to be true of Mars as well, it would affect estimates of the ages of various parts of the
surface based on crater numbers.

Two organizations (JPL and ESA-ESOC) currently model and predict the locations of
Phobos and Deimos. Their predictions do not agree, and neither precisely fits the
observations (Oberst et al.~2006).

Only big planets can have big moons. This is not a tautology but a result of
calculations for both the main formation mechanisms (Canup and Wood 2006) and for the impact
mechanism (Wada et al.~2006).

Ours is, of course, the biggest moon around in the relative sense, and ``formation''
papers continue to appear (e.g., Garrick-Bethell et al.~2006), but some of the other items
were perhaps more fun. You can see a full moon at high noon to about $5^\circ$ south of the
Arctic circle, what with its orbit tilt, refraction, and all (McCurdy 2005). Noon is
admittedly not very high there.  The moon can at times (and places) also be circumpolar
and rise a few minutes earlier each day (like the sun) rather than an hour or so later.
Given these evidences of erratic behavior, it is pehaps not so surprising that cultures
for whom the new moon is important have tended to want someone actually to see it, rather
than relying on calculations (Shakat and Syeed 2006; Siddiqi 2006).

Even we have seen a few standard total and partial eclipses of the moon (though
rigorously avoiding solar eclipses, aurorae, and meteors), but there also exist total
penumbral eclipses (McCurdy 2006). We missed the one on 14 March 2006, but there will be
another in 2053, which the vast majority of your current students can reasonably expect
to see.

Moon craters have been impact features all of our lives (and most of its), but Van
Frese (2006) suggests that back-side impacts could have triggered some front side
volcanoes.  The threatened sample of lunar (Apollo) dust was not auctioned off after
all (Anonymous 2006c), but it doesn't matter, since the US is apparently going back and
can get some more, in something to be called Orion (Anonymous 2006d). Some members of the
astronomical community have expressed less than overwhelming enthusiasm about the return
and have been told for our pains that we expected to have an unrealistic degree of
influence on what NASA does (Griffin 2005, 2006). Still, it is hard not to feel some
sympathy for a project that allows McKay (2006) to say that ``the moon is like New Jersey,''
and enables us to present the ``making lemonade award of the year'' to the organizers of
a conference (at STScI in November 2006) called {\sl ``astrophysics enabled by the return
to the moon.''}

\subsubsection{Major Planets, General Confusion, and Colonel Deshafy}

Mercury is almost ready for his close-up from Messenger. Whether Messenger will be
ready we cannot say. Meanwhile, Ksanfomality (2005) has used ground-based speckle images
to examine the side not recorded by Mariner-1 and found a 2000 km basin, implying the same
sort of ``front-back'' or ``top-bottom'' asymmetry that characterizes the other terrestrial
planets and our moon.

Venus has polar clouds (Svedham 2006) seen in infrared absorption and was
recently rediscovered by a very difficult method. ACRIMSAT noticed a 0.1\% solar dimming
during the 2004 June 8 transit of Venus (Schneider et al.~2006a). This is perhaps a
hopeful sign for the possibility of finding terrestrial planets around other stars by
transit observations from space.

Earth remains, marginally, habitable, but Jupiter currently provides only minimal
protection from impacts (Laakso et al.~2006). Other terrestrial items appear in
Sect. 4.

Martian surface features include many that are best interpreted as relics of past
liquid water (Squyres et al.~2006), but also many from impacts and volcanoes (Knauth et
al.~2005; McCollom and Hynek 2005). Continuing copious data returns provided evidence
during the past year that Mars also has (a) polar caps that sublime from bottom to top
rather than top to bottom (Keiffer et al.~2006), (b) aurorae like terrestrial ones (Lundin
et al.~2006), (c) an ionosphere that expands in response to solar flares (Mendillo et 
al.~2006), (d) glaciers (Furget et al.~2006, which is actually a calculation pertaining to
past epochs when the obliquity of the Martian ecliptic was $45^\circ$ or more), and (e) two
sorts of X-ray emission, fluorescence and charge exchange, both to be blamed on the sun
(Dennerl et al.~2006). Marsquakes are not the main exciter of its Chandler and Inner
Core Wobbles (Dehant et al.~2006). And the ``Mars face'' looks rather different in images
taken at different angles and resolutions (Neukum 2006), though the most easily deluded
author still sees the bits of terrain that made up the nostrils, eyebrows, lips, and chin.

Jupiter has a new spot, formed from three smaller ones in 2000, which is now
turning red (Go 2006). He also dissipates tides raised by various satellites at a rate
implying Q=$10^{5-6}$ (Wu 2005). Similar numbers come from (a) extra-solar-system hot
Jupiters and (b) circularization of F-K binaries. One expects the Jupiter-Io system to
act like the Earth-Moon pair and transfer angular momentum from planetary rotation to
moon orbit. This would affect timing of solar eclipses seen from Jupiter if there
were anyone there to see them, but you will recall that there have been no dinosaurs on
Jupiter since the impact of Comet Shoemaker-Levy 9.  Calculating  whether the angular
diameter of Io as seen from Jupiter is large enough to cause solar eclipses is left as an
exercise for the reader.

Saturn rotates. This is not our ``Queen Anne is dead'' award for 2006, but a prelude
to the discovery of a rotation period of 10h 47m for the magnetic field, as measured
by Cassini (Giampieri et al.~2006). The Saturnian field, unlike those of Jupiter and
Earth is not just a tilted dipole (Stevenson 2006, who however says little about what
the actual topology is).

Uranus probably has C$_2$H$_6$ in its atmosphere, and Neptune already did (Hummel et 
al.~2006). Both have axial tilts that probably arose when they passed through orbital
resonances (Tremaine 2006, commenting on Brunini 2006, which is a discussion of the
Jupiter-Saturn case).

Pluto escaped being prototypically P1utonic by a vote of 157 to 158 in Prague. The
exhausted IAU members still present and voting denied any wish for a recount, after
being threatened with deportation to Florida. How, you probably don't wish to know,
did your authors vote on the various Plutonian issues?  Not at all, it turns out, for
the one who was there was skippering the team of students who ran up and down the
aisles counting the raised yellow cards of the voters
(slightly different from raised hackles, but not entirely).

And who is Colonel Deshafy? A somewhat distant relative, retired from the US Air
Force, who has recently taken on a new job that requires project management in difficult
places (arguably not quite so difficult as some of those he visited for the Air Force),
and who therefore seems an appropriate person to associate with the news that asteroid
99942 Apophis will not hit us in either 2029 or 2036 ({\sl IAUC} 8711) according to an orbit
redetermination from Acecibo Doppler and ranging data. Possible glosses include the
thought that it would be a shame to lose Arecibo to either asteroid or funding hits, and
that collisions are presumably even less likely in years other than 2029 and 2036.
These dates are not so very far away if you remember 1984 as ``recent''.

Planet X, or Nemesis, like the little man who wasn't there, wasn't there again today.
Kuz'michev and Tomanov (2006) deduce this from absence of effects on comet orbits (though
false alarms go back to 1949, Schuette 1949). Zakamaska and Tremaine (2005) and Bhalerao and
Vahia (2006) also say no. Among their considerations is the absence of an effect on the
acceleration of the barycenter of the Solar system.

Note that, in light of previous discussions, X in this context is to be pronounced
``nine.''

\section{BIGGISH BANGS} 

There was, of course, only one Big Bang, at least in our universe.  But we feel that
supernovae, gamma ray bursts, flares of active galactic nuclei, and related events and
products ought to count as Biggish Bangs and have grouped them accordingly. The grouping
also helps toward our goal of precisely 13 sections of roughly equal length (and yes,
some animals are more equal than others, according to a Blair who was not prime minister
of England).

\subsection{Supernovae, Their Remnants, and Gamma Ray Bursts}

These have to live together from now on because some events are both SNe and GRBs,
though clearly the general run of core collapse supernovae and of gamma ray bursts
happen in different environments (Fruchter et al.~2006).

\subsubsection{Supernovae}

A number of our prejudices were confirmed during the year, for instance, there are
more being found all the time, our particular annus discordensus taking in 2005ep ({\sl IAUC}
8607) to 2005nb ({\sl IAUC} 8657) and 2006A ({\sl IAUC} 8656) to 2006gz ({\sl IAUC} 8754), 
a total of 403,
or maybe 402, after subtracting 2006U, which turned out to be a $z=0.2$ AGN, or perhaps
403, if you count 1985U found during the reference peciod.  Should you count it? Well,
since the cosmic SN rate probably isn't changing from year to year, but search skills are
improving, perhaps yes. There were, for comparison, only six novae, five Galactic and
one in the LMC.

The progenitors of Type II events are frequently red supergiants. Li et al.~(2006)
report a pre-need image of the SNII-P event 2005cs in M51. It was a $10\pm 3$ M$_\odot$ star
caught as a red supergiant, and, in trio with 2004gd and 2004A (Hendry et al.~2006a),
gives us three with initial masses in the 8$-$10 M$_\odot$ range. The red SGs have strong
winds (Immler et al.~2005 on late-time X-rays from SN 1979C; Schlegel and Petrie 2006 on
1988Z, with mass loss up to $10^{-3}$ M$_\odot$/yr).

The standard calculated core collapse supernovae have cores that collapse just fine,
but they don't explode (Buras et al.~2006), or they don't explode with sufficient vigor
(Moiseenko et al.~2006). But sooner or later, someone was bound to have a better idea.
It is, pause while we put a green circle around the necks of the authors (Burrows et 
al.~2006), acoustic oscillations, which have the advantages over neutrinos that outer gas
absorbs them very efficiently and that they continue to be generated as long as accretion
on to the core continues and drives g-mode oscillations of the core. This has a slight
flavor of one of the recent proposed solutions to the problem of cooling flows in X-ray
clusters.

The progenitor(s) of Type Ia (nuclear fusion) supernovae remain(s) to be firmly
identified. Candidates during the year included (a) a revival of the single-star
hypothesis, at least for SN 2002ic (Imshennik and Dunina-Barkovskaya 2005) and low mass
population III stars (Tsujimoto and Shige\-yama 2006),  (b) binary white dwarfs, our lingering
and not much-loved by others favorite (D'Souza et al.~2006),  (c) RS Oph, a well-established
recurrent nova (Hachisu and Kato 2006;  Sokoloski et al.~2006),  (d) symbiotic stars
(Jahanara et al.~2005), (e) a new class of ``absorbed supersoft CVs'' (Steiner et al.~2006),
(f) an unlikely-sounding candidate, the  AM CVn stars (Anderson et al. 2005), 
based on observations
of the first eclipsing one, SDSS J0926+3624; about 12 of these are now known (more than
the confirmed RNe), but the well-studied ones, though of short period, which is good, have
small total masses, which would normally be thought of as fatally bad, and (g) white
dwarfs of mass either smaller than the Chandrasekhar limit (Stritzinger et al.~2006) or
larger (Howell et al.~2006; Branch 2006, on 2003fg). One or both of (g) may strike you as
unlikely (they did us), but Leonard et al.~(2005) say that it is really only a coincidence
that ignition occurs very close to the Chandrasekhar limit. We recorded one firm vote against
any donor that would have put much hydrogen onto the threatened white dwarf, which rules
out main sequence, subgiant, and red giant companions (Mattila et al.~2006), and two
firm votes for two classes of progenitors such that one class of events goes off quite
soon after a burst of star formation and a second billions of years later (Mannucci et al.~2006;
Greggio 2005).

The ignition process in thermonuclear SNe is traditionally described as deflagration
(flame propagating at less than the speed of sound) or detonation (supersonic propagation).
There was a vote for pure deflagration for 2002cx (Jha et al.~2006), and one for deflagration
turning to detonation (last year's top choice, Badense et al.~2006), one for a smoldering
phase (Stein and Wheeler 2006), and two opinions on the number of points on the white dwarf
surface that ought to ignite at once to get the best bang for the bomb. Kuhlen et 
al.~(2006) said spots bigger than 1 km, and Roepke et al.~(2006) said about 150 spots per
octant. To save you getting out your Qipu, the surface area of a white dwarf allows for
lots of space around 1200 1-km spots.  But the last word on the subject goes to Zingale et
al.~(2005) who have done a three-dimensional calculation, with Rayleigh-Taylor instability,
turbulence, anisotropy, fire polishing and much else, for which we think the right word
may be ``contingency'' in more or less the biological sense that things could go a number
of different ways from very similar initial conditions.  So why are they all so similar?!?

\subsubsection{Some Other Things About Supernovae}

On the eve of its 20th birthday, SN 1997A has hit enough surrounding material for the
X-ray emission to be global rather than spotty (Park et al.~2005b), but it is still also
inspiring models of the progenitor and the explosion process, of which we mention only
Kifonides et al.~(2006) for the pleasure of noting that they cite both Richtmyer and
Meshkov in using the Richtmyer-Meshkov instability to describe some of the mixing and
anisotropies they find.

Cas A is much older, but we are not quite sure by how much because no one saw it,
an aspect of observational constraints on their progenitor model not mentioned by Young
et al.~(2006). The epoch of Flamsteed's star (1680) remains plausible (Fesen et al.~2006).
The event was perhaps a sub-sub-GRB, since the present jet has a kinetic energy near
$10^{50}$ ergs (Laming et al.~2006). You should study it now if you are interested
because the radio has been fading about 0.8\% per year since 1950, when the 38 MHz
observation frequency was called 38 Mc/s (Vinyaikin 2006). And $(1.008)^{324}$ is a
sufficiently large number that Flamsteed could have seen it with quite a modest radio
telescope.

How common are supernovae? One to two per century of the core collapse type in the
Milky Way (Reed 2005), up to $4\pm 2$ per year in Arp 220 (Lonsdale et al.~2006). This
also applies to core collapse events since it is based on watching radio remnants rise and
fall, and no Rolls Royce has ever broken an axel.  Sorry. No SN Ia has ever been caught
as a radio source (Panagia et al.~2006). In between comes M83 with six events since WWI,
four of which are (still) radio sources (Maddox et al.~2006).

What about the Ia rates?  Three new LMC light echos appear all to be Ia's less
than 1000 years old (Rest et al.~2005), which seems like a lot for such a little galaxy.
They also saw the 1987A light echo, further out than it used to be. Who first thought
of looking for SN light echoes?  Zwicky (1940) of course.  Shklovsy (1964) was presumably
independent. The SN 1a rate in general is smaller (except in elliptical galaxies and
other old populations) than the core collapse rate; does not vary a great deal since
redshift 0.8; and can be expressed several ways, for instance that you have to put
1000 M$_\odot$ into stars to get one Ia, or as 0.15 SNU (Neill et al.~2006).
The rate of Type V un-supernovae (aka extreme outbursts of luminous blue variables, which
the star survives) is apparently about one per year among events that get SN names\footnote{
A decision less publicized after the IAU in Prague than the status of Pluto was what
to call supernovae when/if we run off the current system with 2017zz or thereabouts.
Event 703 will be called 2017aaa. How likely this was to be needed remained a bit
uncertain, given that faint fuzzies now get preliminary circular designations and real
SN names only when a little bit is known about them, and that large searches from ground
and space have a somewhat uncertain future.}
Maund et al.~2006 on 2002kg and 2003gm). Indeed 2002kg was probably also V37 in NGC
2403, and its peak absolute magnitude was about $-9$.

Supernovae have many jobs to do.  Accelerating cosmic rays remains part of their
job description (Satyenda 2006; Marcowith et al.~2006).  But two groups of theorists were
prepared to excuse them from primary responsiblity for heating and stirring the interstellar
medium (Haverkorn et al.~2006 in favor of H II regions and Koda et al.~(2006) in favor of
galactic differential rotation). Tasker and Bryan (2006) dissent and hold SNe
responsible for these tasks.

\subsubsection{Supernova Remnants}

You don't get very many this year because the Milky Way inventory is so very
incomplete: the 230 catalogued are 1\% (Koo et al.~2006) to 10\% (Brogan et al.~2006) of
the number expected, with diffuse, faint, and compact ones all likely to be 
under-represented.  
Helfand et al.~(2006) used radio plus infrared data to demonstrate this
unequivocally by finding 49 smaller than 45\arcsec\ in a bit of sky were 7 had been known
before. And he is an X-ray astronomer!

Your Crab Nebula paper of the year (Seward et al.~2006) reports yet another 
non-detection of the missing 2$-$5 M$_\odot$ left when you subtract the mass now present in the
visible nebula + pulsar from the mass needed to trigger a core-collapse supernova. It
is a very deep X-ray search, which does see a dust-scattering halo from the main nebular
source but no other more extended emission.  The authors are kind enough to cite one of
our own non-discoveries of the missing Crab mass.  Whether the intervening number of
failures is 5 or 500 depends on how hard you have to look for it to count $-$ consider
the non-detection of gravitational radiation every time you hold a large spoon up to
your ear and don't hear it ring.

Vela, with pulsar B0833$-$45, is a TeV source (Aharonian et al.~2006a), presumably
via inverse Compton scattering, and it is the first to show a true TeV peak, not just
a continuation of the X-ray spectrum.  The little bit in the corner was this year demoted
to a planetary nebula with a thermal radio spectrum (Reynoso et al.~2006). It is
petitioning for status as a dwarf SNR.

SN1006 and Tycho's remnant are accelerating galactic cosmic ray ions according to
rather indirect arguments (Ellison and Cassam-Chena 2005; Warren et al.~2005). We originally
wrote ``by rather indirect arguments,'' but suspect that it is really by electromagnetic
forces.

When does a supernova become an SNR? When its shock hits dense interstellar
material say Immler and Kuntz (2005), recording this happy event for SN 1970G. Some
other associations among historical SNe, remnants, and pulsars were both advocated and
denied.  There are no compact Chandra sources (central neutron stars) down to $2\times 10^{31}$
erg s$^{-1}$ in six nearby SNRs (Kaplan et al.~2006). PSR B0531+21 and 3C58 are not SN 1181
but much older, according to new radio proper motions (Bietenholz et al.~2006).  The
compact core of RCW 103 (SNR age about 2000 years) has a regular period of 6.67 hours,
which is odd for either an X-ray binary or a magnetar (De Luca et al.~2006). For four
more cases we record only a best-known name and the general comment ``more complex
scenarios'': GX 1+4 (Hinkle et al.~2006), S147 (Gvaramadze 2006), Sgr A East (Park et 
al.~2006), and not even the name for Kothes et al.~(2005).

\subsubsection{Gamma Ray Bursters}

We indexed 84 papers an the GRB page, many associated with hot topics of a year or
two ago, especially associations with Type Ic supernovae, progenitors of those of short
duration, host galaxies, and models for fireballs, radiation mechanisms, jets and such.
We are ``not unaware'' of a high-profile event of a putative 3rd type reported after the
end of the reference year, and it is already in our {\sl Ap 07} notebook. Meanwhile, there
were two earlier new, third classes in 2006.  Cline et al.~(2005) described a group
lasting less than 0.1 s and conceivably attributable to evaporation of primordial black
holes. The signature for this is that spectrum vs. time should chirp toward high
frequencies rather than growling toward low frequencies, but with short enough
duration, you can't tell!  Horvath et al.~(2006) found three clumps of events in the
spectrum-duration plane; their new group is of long duration and intermediate
spectrum.

On the fading issues, mergers of neutron star pairs or NS + black holes continue to
be plausible models for the short duration events in most minds and preprints (Belczynski
et al.~2006, the last of a number of papers during the year), and there are no
associated supernovae (Hjorth et al.~2005).  Most of the short ones are at modest redshift
and not associated with large amounts of recent star formation (Prochaska et al.~2006).

The most distant long duration burst this year, 050904, was at $z=6.295$, not far
short of the QSO and galaxy records (Kawai et al.~2006 and several accompanying papers).
It was bright enough to be seen by the 25 cm TAROT robotic telescope (Boer et al.~2006).

The more distant (long duration) bursts show optical absorption features much like
those in QSO spectra, only more so. Prochter et al.~(2006) find that 14 of 14 GRBs had
Mg II intervening absorption in a redshift interval where only 3 of 14 QSOs would.
They advance five arguments in favor of gas in galaxies that strongly lens the GRBs and
only four against. There are also Lyman alpha forest lines and features due to the
host galaxies (Penprase et al.~2006).

The long duration events do not precisely trace star formation (Le Floc'h et al.~2006),
and there must be at least one other critical factor, for which modest metal
abundance (leading to retention of more envelope and more rapid rotation) is a strong
candidate (Berger et al.~2005).

The GRB-SN Ic connection developed further with 060218 = 2006aj at 135 Mpc (vs.
35 Mpc for 1998 bw, the closest ever, Campana et al.~2006 and accompanying papers).
Events this faint could be quite common.  The spectrum was on the X-ray flasher end of
the distribution of GRB spectra, which may or may not be an orientation effect (Granut
et al.~2005). SWIFT caught its very first X-ray flasher just as the year ran out
(Levan et al.~2006), and we may know more about these in a year or two.

Moving gently in the direction of the path less taken, we find (a) the collision
of two massive stars as a way to make a still more massive one with a rapidly rotating
core - a possible GRB progenitor (Dale and Davies 2006), (b) a couple of named processes
for putting magnetic energy to work making GRBs (Wang et al.~2006c on screw instabilities
plus the Blandford-Znajek process; and Uzdensky and MacFadyen, 2006, making use of a
process described by Lynden-Bell 2004, of which year it was a poorly understood highlight),
and (c) a second phase of the vacuum, degenerate with the one we know, which compresses
baryons into 20 cm dense chunks inside dense stars. These can expand and explode to
make GRBs (Fruggatt and Nielsen 2005).

\subsection{The Pulsars of the Nations}

This was the very first section written, because it has three green circle papers.
First, the biggest one belongs to the fastest
pulsar so far discovered, an eclipsing binary in the globular cluster Terzan 5 (Hessels
et al.~2006) at 1.396 ms. This is 716 Hz, so it is singing something like an F-sharp
(no, not your F\#, the higher one that the fat lady has to sing before we can all go
home). The rotation speed at the equator approaches 1/4 c, and the magnetic field is less
than $10^8$ G (Grindlay 2006). A pulsar that comes too close to break-up rotation (about
1 ms) quickly slows itself down by gravitational radiation, and the previous record,
1.558 ms (between D\# and E) for B1937+21 had stood for so long that many calculators
assumed in the 24 years between that this must be the real physical limit and managed to
match it. Someone has surely by now derived the 1.396 ms of Ter5-ad (though we didn't
catch the paper).  But there is another theorists' task waiting: five of the fastest
pulsars known are in Ter 5, and a sixth in 47 Tuc. The explanation must be some combination
of favorable conditions for formation and a selection effect arising from relatively small
dispersion measures at high galactic latitude.

High F\# is, of course, too fast even for the pulse of a humingbird, but our second
green circle pulsar has P=5.54 s, about right, perhaps for an elephant\footnote{
No, this is not a five-parameter year, and the discovery that elephants can sometimes
recognize themselves in mirrors belongs to next year $-$ first get a really big mirror $-$
so the 2006 elephant item is the sad, rather belated news (Anonymous 2006a) that all the
elephants in Italy were killed during World War II.}. It is
AXP XTE J1810-197, the first anomalous X-ray pulsar to show pulsar-style radio emission
(Camilo et al.~2006). It has pulse substructure down to 0.2 ms, in other words quite
typical in that respect. And we won't deprive you of the fun of doing the arithmatic of
$P$ and $\dot{P}$ ($1.02 \times 10^{-11}$ s/s) to verify that the magnetic fields is at 
least a million times larger than that of the previous item, definitely in the magnetar 
range. The $P/2\dot{P}$ slowing down time is about 9000 years.

Third is a new class of radio transients, which at least look and quack like ducks
(sorry, pulsars), though we are not quite sure about the walking as no proper motions
were reported. The inital report (McLaughlin et al.~2006) called them RRATs
(Rotating RAdio Transients) and described 11 sources, 10 with periods between 0.4 and 7
s (at last, a human pulse!) and three with period derivatives, yielding fields of
2.7, 5.8, and $50 \times 10^{12}$ G. Why were these a notable, indeed a very difficult, 
discovery?
The time between pulses is anything from 4 minutes to three hours, so you have to watch
a long time and be prepared to use a fairly clever method of time series analysis.
These mostly-off pulsars suggest a further class that have turned off completely (and
not just because the periods have lengthened and the fields decayed) and so are even more
difficult to observe.  At least one of the first 11 is also an X-ray source (Reynolds
2006).

Weltevrede et al.~(2006) argue that members of this new class are also exhibiting
a fourth class of pulsed radio emission in addition to normal pulses, giant pulses, and
giant micropulses.  More ordinary giant pulses appear so far in eleven sources, seven
with strong fields at their light cylinders (starting with the Crab) and
four not (Kuzmin and Ershov 2006).

\subsubsection{More Pulsars Singles and Singularities}

Having reached four in two ways, we back up to three, the ``natural'' value of the
pulsar slowing-down-index, $n = P \ddot{P}/\dot{P}^2$ (or $\dot{\nu}=K\nu^n$) 
for magnetic dipole radiation.
Contopoulos and Spitkowsky (2006) predict $n=3$ for most pulsars from a more complex
spin-down process and larger values near the death line.  But in fact all six measured
values are less than three, including 2.65 for the latest and, with $P/2\dot{P}$ less than 884
years, the youngest rotationally powered pulsar (Livingstone et al.~2006).  The observed
numbers can, of course, also be explained (Chen and Li 2006). They favor a gradual
increase in the perpendicular component of the field or torque due to fall-back in a
disk, but present competing mechanisms as well.  Kramer et al.~(2006) and Van den Heuvel
(2006) were unable to report a slowing-down-index for the 0.81 s psr B1931+24, which
is on typically a week or so per month, because of timing noise. But the derivative is
larger when the emission is on than when it is off, implicating a pulsar wind (mechanism
number three) in the slowing.

The discovery of an AXP with a debris disk (Wang et al.~2006a) sounds like it
ought also to be part of this story.  The disk has a lifetime long compared to $P/2\dot{P}$ but
short compared to massive star lives, and so presumably is made of fall-back material.
The authors suspect that this is also the origin of the planets around B1257+12, an idea
originally put forward by Lin et al.~(1991) to explain the (non-existent) planet of
psr 1829-10. Rather direct evidence for fall-back comes from the unique oxygen-neon
atmosphere fit to the X-ray spectrum of 1E 1107.4-5209 (Mori and Halley 2006). It is
isolated, with P = 0.424 s, a surface temperature near $10^6$ K, and is in a supernova
remnant.

We break away from all these threes to report a couple of possible neutron star 
firsts:  crustal oscillations in SGR 1806-20 as the giant flare faded to fit
quasi-periodic oscillations at 92, 626, 18, and 26 Hz (Watts and Strohmayer 2006);
and free precession at 7.1 years for the isolated, radio-quiet J0720.4-3125 (Haberl et 
al.~2006); with a rotation period of 8.39 s and a magnetic field of nearly $10^{14}$ G it is
presumably a Moderately Anomalous X-ray Pulsar.

At least a few things are not pulsars, including the gamma ray sources in Gould's
Belt, after a deep Arecibo search (Champion et al.~2005).  Indeed the association is
at best a two-dimensional one (Popov 2005).  Some are not even neutron stars, a sad
falsification of the neat idea of 2005 that some short  duration GRBs could be the peaks
of giant flares of soft gamma repeaters (Popov and Stern 2006; Nakar 2006), though this
gives a chance to mention that the 27 December 2004 flare of SGR 1806-20 rebrightened from
a coasting to a Sedov-Taylor phase (Taylor et al.~2005 and not the same Taylor, who,
like some of the astronomers of Sect. 3 won't care), when it began to sweep up its
own ejecta, but in 7-20 days, rather than years (like a supernova).  Perhaps not quite
by chance, the paper notebooked just ahead of that one (Park et al.~2005) records day
6200 of SN 1987A, when X-ray brightening showed that the blast wave had reached the
main body of dense circumstellar stuff.  Notoriously, such X-ray observations do not
reveal any compact neutron star (or accreting black hole) at the remnant center.

Other things are not, anyhow, the neutron stars we are used to. Majczyna and
Madej (2005) report a mass of 0.4 to 0.6 M$_\odot$ and R=4.6 to 5.3 km for MB1728-34;
and Guseinov et al.~(2005) suggest masses near 0.5 M$_\odot$ for the full  range of AXPs and
SGRs (but weak fields). Higher masses of 1.2 and 2.1 M$_\odot$ were reported in
higher-prestige journals (Bassa et al.~2006; Ozel 2006; Val Baker et al.~2005; Nice et 
al.~2005). Perhaps the politer (as well as more grammatical) way to say this is that larger
masses were reported in larger journals.

The not-even-neutrons award goes to Cea (2006) for the suggestion that the magnetars
are actually p stars - up and down quarks in beta equilibrium as an ``Abelian
chromomagnetic concentrate.'' He has put forward the idea before, but this seems to be
the first time in the mainstream astronomical literature (assuming that, like all the
rest of us, he cites all of his previous papers).  And if you want to be sure of
becoming a black hole rather than a neutron star, you must begin life with at least 40
M$_\odot$ (Bogdanov et al.~2006, on a magnetar in the cluster Westerlund 3, which still has
main sequence stars heftier than 35 M$_\odot$).

Pulsars en masse (but another trio of papers) display a correlation of the
directions of their space motions with the orientation of their rotation axes (Wang et
al.~2006; Socrates et al.~2005 $-$ no, we don't know what he called Plato, or Pluto $-$
Johnson et al.~2005). In a rare degree of neutronal harmony, the three agree that this
must come from kick velocities, somehow powered by neutrinos over a very short period
of time, and that the mechanism (connected with ongoing radiation) put forward by
Harrison and Tademaru (1975) is neither necessary not sufficient, which is not quite the
same as saying that it doesn't ever happen.  The Crabbiest author has often expressed
puzzlement that the Crab Nebula and its pulsar are both moving in about the same direction
(northwest, or to the upper right in a standard image) and has often felt that the
explanations somehow violated conservation of momentum.  She thinks that Fryer and
Kusenko (2006) are saying that the neutrinos are sufficiently beamed in the other direction
to keep the bean counters happy.

Finally, we caught three papers that look at the entire Milky Way population of
pulsars (Malov and Malov 2006; Faucher-Giguere and Kaspi 2006; Ferrario and Wickramasinghe
2006) and describe them in terms of a birthrate, radio luminosity vs. spin-down
luminosity, progenitor magnetic field, slowing down index, and total number that should
be detectable down to some flux limit like 0.19 mJy. Though the discrepancies are not
enormous, no two of the papers agree about all the numbers.  Taking averages and ranges,
we might come away thinking that recent formation rates of 2-3 per century; radio
luminosity scales as spin-down luminosity to some power between 1/3 and 1/2; total
number = something $\times 10^5$; and so forth and it would seem that the tail of normal pulsar
properties should extend well into the P, dP/dt, and B regime of the magnetars, coming
mostly from higher masses (Ferrario and Wickramasinghe 2006). Thus we should probably not
complain that the observed ranges of P, dP/dt, and B also overlap (Kuiper et al.~2006 on
J1847-0130 at $9.4 \times 10^{13}$ G, but with no detectable X-ray emission). But we wonder how
a given star decides which to do!

\subsubsection{Binary Neutron Stars}

The first neutron star found in a binary (Sco X-1) was discovered at very nearly the
same time as the first single (pulsar) one, nigh on to 40 years ago. It must be so,
since the source inventory now includes transient X-ray binaries with recurrence times up
to 32 years (Galloway et al.~2005; Tomsick et al.~2005). Not surprisingly, they are on
average very faint, since their maximum luminosities are the Eddington value just as
for steadier sources. The binaries appear in about as many papers per year as the
singles down to this day, though the catalogued numbers are fewer, for instance 114
high mass (companion) X-ray binaries in a catalog from Liu et al.~(2006). The actual
data are, of course, e-only, and the paper does not note that one of the authors died
some years ago. Single pulsars number about $10^3$.

The primary catagories are binary pulsars (divided into young and recycled millisecond
and also by whether the companion is an ordinary star, a white dwarf, a neutron star, or
even another pulsar) and X-ray binaries (divided into those with high and low mass
companions, with Be stars an optional subset of the former). Each of the classes has
some aspect imperfectly understood, not to mention the difficulties systems find in
navigating among the classes, and we note a subset here.  All classes except those with
short-lived companions are greatly, over-represented in globular clusters (c.f. Irwin
2005).

{\sl Getting there.} X-ray binaries with low mass donors (LMXRBs) will surely end up as
recycled pulsars when enough angular momentum has arrived and mass transfer stops
(Bogdanov et al.~2005). Chen et al.~(2006) present a nice scenario for this, but do not
address the ancient problem that you don't get enough this way. And the amount of accretion
against which the neutron star can defend its strong magnetic field seems to range from
almost nothing up to half a solar mass (Zhang and Kojima 2006).

{\sl Proliferating classes.} By finding examples two and three, Masetti et al. (2006) 
have transformed 
the LMXRBs with M-giant donors into ``a well known class of astrophysical object.''
The same this year could be claimed for (1) the highly absorbed HMXRBs (Beckmann et 
al.~2005, actua11y the seventh example), (2) an overlapping bunch with supergiant donors
and often rotation periods of 140-1300 s (Walter et al.~2006), and (3) a class of HMXRB
transients with red supergiant (rather than the commoner Be) donors (Sguera et al.~2006)
of wnich there are now about 10, four with optical IDs.

{\sl At most one.} In contrast, XRBs with Wolf-Rayet (massive helium star) donors are
expected to be very rare (Dray et al.~2005), perhaps only one per cubic Milky Way.
(Lommen et al.~2005). Ours is Cyg X-3, and we should also have one WR + BH system with
detectable wind accretion.  It has not (yet) been recognized.  For a brief moment there was
also only one variable galactic TeV source attributable to a neutron star binary, but
very shortly after Aharonian et al.~(2005) reported B1259-63 (which orbits a Be star),
Albert et al.~(2006) chimed in with LSI +61 303. There is variation at the orbit period
(Mirabel 2006). The first is a HESS discovery (which the owners like to have written
as H.E.S.S., though {\sl Astronomy and Astrophysics} doesn't always cooperate), and the second 
came from MAGIC (M.A.G.I.C. if you wish, though the day we saw it, there were more sand
grains than periods flying around).

{\sl Bursting LMXRBs.} For most of these, the fuel is predominantly helium, though
Chenevez et al.~(2006) and Weinberg et al.~(2006) make clear that some hydrogen is also
required. The time development of the bursts has generally been explained by a photosphere
that expands and falls back (Wolff et al.~2005, and implied also by Nakajima et al.~2006),
but Bhattacharya and Strohmayer (2006) advocate flames spreading over the surface instead.
And then there are superoutbursts, supposed to be powered by explosive carbon burning,
though two papers (Cooper et al.~2006 and Cumming et al.~2006) told us this works only if
the initial surface carbon abundance is a good deal larger than solar, to catalyze initial
hydrogen burning in the accreted fuel. Doing their best to help, Nelemans et al.~(2006)
report the first system, XB 1916-05, in which the material being transfered from the
donor star is helium and nitrogen rich. They say that others are enhanced in C and O.

{\sl Testing General Relativity.} Well, if it fails (e.g., in the report from Gravity Probe
B expected in April 2007) you won't need us to tell you about it.  But the year saw a
couple of modest binary NS successes.  First, the emission of gravitational radiation
by source X-2 in the globular cluster M15 should be enough to drive the required mass
transfer of $10^{-10}$ M$_\odot$/yr (Dieball et al.~2005). Second, a recently discovered
double neutron star
pulsar, Y2127+11C (also in M15) now sets limits to post-Newtonian parameters about as
tight as those that come from ``the'' binary pulsar, B1913+16 (Jacoby et al.~2006).  And there
are comparable numbers for B1534+12 (Stairs et al.~2002). Sadly, the M15 pulsar limits
cannot be made tighter, because the limiting uncertainty in variable Doppler shifts 
are due to acceleration of the system in the cluster.

{\sl Surprise.} The binary millisecond pulsar population shows a gap in the distribution
of orbit periods at 20-60 days (D'Antona et al.~2006), which the authors suggest is a
magnetic field effect that might be related to the 1-2 hour gap in N(P) for the
cataclysmic variables.  There was a day when we thought we understood that one, but it
must have been a Tuesday.

\subsection{Black Blogs: QSOs, AGNs, and All}

The list of standard questions upon which to comment each year now has about 13
entries (unlucky for both writer and reader), so we hit first the items about which there
seems to be a reasonable degree of agreement, for the range from rather puny LINERS
and Seyferts on up to powerful quasars, some perhaps with associated star bursts.

\subsubsection{More Certain Answers}

Do they contain black holes?  Yes, and sometimes you can measure the mass (1-3 $\times$
$10^9$ M$_\odot$
from reverberation mapping for the Seyfert NGC 4151, Lyuty 2005), though Schild
et al.~(2006) vote for a black hole alternative, the horizonless Magnetospheric Eternally
Collapsing Object. They use MECO as an abbreviation, and some of our resistence may well
be due to the images of meconiums conjured up (You do have a dictionary, don't you?)

Maybe two black holes?  Yes, say Rodriguez et al.~(2006) with an example separated
by only 7.3 pc (VLBA resolution) for which optical velocities yield a total mass near
$1.5 \times 10^8$ M$_\odot$  and a period of $1.5 \times 10^5$ years. The most widely 
mentioned case is OJ287 (Wu et al.~2006a; Valtonen et al.~2006) at 12 years.

Maybe even three black holes? Hoffman and Loeb (2006) present some indirect
evidence, and there is a much more highly Pressed candidate out of period.
How do b1ack holes grow?  By eating gas (Fathi et al.~2006 on NGC 1997), stars
(Ivanov and Chernyzkova 2006, a calculation of how massive the BH can get before the process
turns off, and some optical candidates in 2007), and each other (Holt et al.~2006, on
ensuing rapid accretion), one to three per galaxy, say Merritt et al.~(2006).

Does the Milky Way have one?  Yes, and it has a horizon.  Otherwise the accretion
required to power the submm flux would heat the surface until we could see near IR emission
(Broderick and Narayan 2006).  The green black circle this year, however, belongs to the
discovery of correlated infrared and X-ray flares from the vicinity of Sgr A$^*$.  Data
first:  The 2.1 and 3.8 $\mu$m fluxes brightened from 1 to 10 mJy in 9 minutes (Ghez et 
al.~2005, using adaptive optics on Keck II), and the 2-10 keV X-rays recorded by XMM
brightened a factor 40 to $9 \times 10^{34}$ erg s$^{-1}$ in an hour (Belanger et al.~2005).
Additional
data, concerning longer wavelengths and  additional possible correlations appear in
Eckart et al.~(2006), Krabbe et al.~(2006), and Yusef-Zaden et al.~(2006). Do the
theories outnumber the data points?  Not quite, but you have the choice of (1) Rossby
wave instabilities on a compact hot disk (Tagger and Melia 2006), (2) stochastic
electron acceleration (Liu et al.~2006a),  (3) energy input from a nearby supernova about
the time Huygens recognized the rings of Saturn (Fryer et al.~2006), and (4) whichever
ones we managed to miss.  Gas sources and accretion rates covered a wide range over the
year, but everybody agreed that Sgr A$^*$ had some of each, so we note only the 
non-detection of gravitational bending of light at 3.5 mm (Shen et al.~2005) and the
prediction that something of the sort ought to turn up at slightly shorter wavelengths
(Reynolds et al.~2005, who are, however, being overly optimistic in including LISA in
a list of ``current'' experiments).

M31, having a bigger bulge than Mr. Bigger's Baby ought also to have a larger
central black hole, but it just isn't doing its job. We caught one mildly colored paper
(vs. 18 for Sgr A$^*$) reporting that a previously-advertized candidate for the expected
central X-ray source is not actually central (Garcia et al.~2005), but that the authors
have found a possible alternative at a Chandra luminosity of $10^{36}$ erg s$^{-1}$. And just
what job is it that M31$^*$ isn't doing?  Keeping astronomers, referees, editors, and
all employed.

Do QSOs have host galaxies?  Yes, and of a dozen papers on properties and correlations
we note only that the host halo masses haven't changed much with redshift (Myers et 
al.~2006), but long ago they were the rare big halos found in rich clusters (Kajisawa et 
al.~2006), while now they avoid high density environments (Coldwell and Lambas 2006, a paper
we originally indexed under ``3C lives,'' though they are talking about optically selected
SDSS AGNs).

Do QSOs (etc.) evolve?  A firm and nearly unanimous yes, if you are asking whether
there were more in the past, the comoving density peaking around $z=2.5$ (Richards et 
al.~2006; Brown et al.~2006b). The details were described as luminosity-dependent density
evolution (La Franca et al.~2005). You get a slightly less firm and unanimous no, if you
are asking whether the typical properties have changed much with redshift (Shemmer 2005,
De Vries et al.~2006, both X-ray results), and one herring hanging on the wall in which
Barger and Cowie (2005) say that there were fewer X-ray sources with $L=10^{43}-10^{44.5}$ 
erg s$^{-1}$ at $z=2-3$ than there are now. History requires that this be described as
anti-evolution (but not, we trust, as intelligent design).

Do all big black bugs have stellar bulges/spheroids around them? Yes (or all but
$2-3\%$, Libeskind et al.~2006), but the contrary would be difficult to demonstrate, except as an
upper limit to gravitational lensing by invisible lenses, and the proportionality  of
BH to bulge mass has become enough of a cliche that most papers focussed on why or on
which came first.  On the why side, we mention only the last paper of the year (Escala et
al.~2006, gas transport processes).  And on the which came first issue, you get one chicken 
paper and one egg, more or less the first of the year, since the issue surely will not go away
(Tamura et al.~2006, BH first; Kawakatu et al.~2006, stars first).

Ah, but do all stellar bulges/spheroids have black holes inside them?  No! And a
cluster of green circles to Cote et al.~(2006) for the determination that the central
entity switches from a black hole to a nuclear star cluster at galaxy luminosity fainter
than M$_B=-20.3$ (see also Ferrarese et al.~2006),

Do most active galaxies vary?  Of course they do, with 3C 454.3 having set a new
cosmic brightness record at M$_B=-31.4$ during the year (Villata et al.~2006). Reasons for
variability include star collisions (Faure et al.~2005) and shocks in jets (Brinkmann et
al.~2005). Total brightness and rate of change sometimes reach the point where one has
to invoke a good deal of beaming and relativistic motion (or coherent radiation processes)
to escape the inverse Compton limit on brightness temperature at $10^{12}$ K (Kovalev et 
al.~2005). Some very subtle variability is dubbed nanolensing by Paraficz et al.~(2005).
They attribute it to $10^{3-4}$ M$_\odot$ lumps in the accretion disk. Still more subtle is the
apparent constancy over $10^6$ yr of the UV luminosity of a $z=2.8$ QSO (Adelberger et 
al.~2006), if a Lyman-alpha-emitting blob 380 kpc away is fluorescing from the central UV
flux. More or less the opposite has been claimed for Sgr A$^*$ (Lu et al.~2006).

Turning off is a sort of variability, and it has been clear since the 1960s that,
if every galaxy has a blig black hole, they must be ``resting''\footnote{
Resting is a technical term in theatrical circles, meaning ``unemployed, but still
hoping.''} most of the time. One
quantitative estimate during the index year comes from Wang et al.~(2006f). They examined
nearly 11,000 QSOs in the SDSS data base and find that bright AGNs are on $10-100\%$ of the
time at $z\approx 2$, but only $0.1-1\%$ of the time now.

And of course they are blue in standard color systems, aren't they?  Or at least
they used to be, so a circle of somewhat uncertain hue to Dobrzycki et al.~(2005) who
say, ``QSOs are typically redder than other variability-selected objects'' from the OGLE
survey.

\subsubsection{Less Certain Answers}

Now, keeping in mind that those were the more or less settled issues, let us turn
to the unsettled ones, most of which are also very old, and for which you will typically
be told about only one paper from each viewpoint, or one that says ``both please.''

Is there a strong correlation between nuclear activity and vigorous star formation?
In a magnitude-limited sample, two kinds of light sources are probably brighter than one,
so yes for many ULIRGs (Bekke et al.~2006a), and if you wish to look at the faint end,
the Milky Way is pretty feeble in both.  But the plurality of votes this year said that
the two can happen separately and in varying proportions (Beelen 2006; Kim et al.~2006a,
Lipari and Terlevich 2006, who conclude that outflow is also important so that orientation
functions as a second parameter).

This brings us, of course, to ``unification'' $-$ the on-going mild dispute over just
how important orientation is compared to black hole mass, accretion rate, black hole spin,
host galaxy properties, disk structure, magnetic fields, and whatever else you want.
Having indexed five papers that describe phenomena for which orientation is paramount and
five for which it is not, we cheerfully call attention to Ogle et al.~(2006) who say
some of them are and some of them aren't, in the sense that 45\% of narrow-lined radio
galaxies are misaligned quasars and the other 55\% are not.

Equipartition is the question of which sources and which parts of sources have energy
more or less equally divided between relativistic electrons and magnetic field (which
also comes very close to the conditions for minimum total energy). A coveted purple
patch to Erlund et al.~(2006), not so much for their conclusion that, where inverse
Compton on the CMB is seen at high photon energies, many of the sources could be in
equipartition, but for providing the relevant equations in particularly useful form,
suitable for reviewers of very little brain.  Since there are contexts where practically
all the energy is kinetic and coming out in jets or beams (Taylor et al.~2005), the
initial question probably needs to be rephrased anyhow.

Radio loud vs. radio quiet.  Once upon a time there were QSRSs (quasi-stellar
radio sources) and radio galaxies, plus the much commoner QSOs (quasi-stellar objects)
and Seyfert galaxies, whose ratio L$_r$/L$_{opt}$ was much smaller than for the first classes.These days there are also radio-intermediate sources and radio-bright Seyferts, but we
would still like to know why!  There are correlations which are not somehow very
explanatory $-$ radioluminosity scales with both optical and X-ray luminosity (Shen et 
al.~2006) but with very large scatter, and the radio/optical proportionality can be seen back
to $z=5$ (Cirasuolo et al.~2006).  Bigger black holes are noiser (Liu et al.~2006b).
Near-maximal BH spin seems to be important (Wang et al.~2006g).  And last year's
highlights included correlation of radio emission with core (loud) vs cusp (quiet) central
structure in the host galaxies.  Capetti and Balmverde (2005) present a firmer version.
But there would not be 25 more papers under this heading if the answer were in!

The importance of jet strength (vs. orientation) appears in Gregg et al.~(2006), which
presents the 8th known broad-line absorber quasar with Fanaroff-Riley II structure and
in Buchanan et al.~(2006) and Wang et al.~(2006h) who conclude that the radio intermediate
QSOs suffer from jet weakness. Komossa et al.~(2006) give us numbers 5 to 11 of the
radio-loud narrow line Seyfert 1 galaxies, and conclude that they are not a separate
class but part of a smooth distribution of L$_r$/L$_{opt}$. This, we guess, could reflect 
either jet strength or jet orientation.

Also once upon a time, proper active galactic nuclei had spectra with broad emission
lines, or the lines were obscured and the sources got called Type 2 (Seyferts, quasars, or
whatever), and once in a great while a specific source was allowed to change types
(Pronik and Metik 2005 on Seyfert NGC 3227 which lost its broad emission lines between
1967 and 1992. Recent refinements (or confusors, if you think of AGNs as a multiple
choice exam) include:
\begin{itemize}
\item[o]{Two possible sources of obscuration - a dusty host or an accretion torus, the latter
obviously most relevant to ``unification'' (Martinez-Sansigre et al.~2006).}
\item[o]{The existence of ``natural born 2's'' with no obscuration (Wolter et al.~2005).}
\item[o]{What we think may be mostly mis-applied terminology, including Blazars with strong
emission lines and radio quasars hosted by FR II galaxies (Land et al.~2006).}
\item[o]{An excess of obscured over non-obscured sources at $z=0.5-1.5$ (Trenter et al.~2006);
similar remarks appear in various attempts to understand the X-ray background
(Sect.~12).}
\item[o]{The discovery that BAL QSOs actually have larger column depths opposing their photons
than do mere ordinary Type 2 AGNs (Punsly 2006, apparently the first comparison of
this sort).  The gas is not purely equatorial in either class of source.}
\end{itemize}

This naturally leads to our last major query: Do AGN photons get absorbed by
stuff ?  The answer is a sort of yes, but ...

\subsubsection{An Absorbing Topic}

Our last ``some of them are and some of them aren't'' AGN question of the year is:
Do QSO absorption features have anything to do with QSOs?  Not very much, most of the
2006 astronomical voices said, though they are excellent tracers of large scale structure
(Putman et al.~2006) and how it correlates with that traced by galaxies (Ryan-Weber
2006) and of the slowly evolving composition of low density gas clouds various places in
and out of galaxies (Lehner et al.~2006) and how it differs from that of galaxies at the
same redshift (Vladilo and Peroux 2005).  Richeter et al.~(2006) draw analogies between
QSO absorbing clouds around other galaxies and the high velocity clouds of the Milky
Way.

There are also absorption clouds associated with the QSO host galaxies and the clusters
in which they live (Russell et al.~2006)  and an enormous lore of the Broad Absorption
Lines, which come from stuff being blown out in real time (North et al.~2006; Sulentic et
al.~2006).

But the three green circles belong to three items that triggered ``Oh! I should have
known that would happen and written it up myself'' reactions. (Yeah, sure).  First:
four QSOs with Mg II absorption from gas associated with other QSOs along the line of
sight to us (Bowen et al.~2006).  There is gas in the host, in companion galaxies, and in
tidal debris, echoing the customary association of AGN feeding with galaxy interactions.

Second is ''real time quasar evolution.''   Well, not exactly, but a new Na I
absorption feature at $z=1.173$ appeared in APM 08279+5255 in 2.6 years between observations
(Kondo et al.~2006). This requires, for instance, a cloud 100-200 AU moving across at
50-260 km s$^{-1}$.  There are other, stable Na I and Mg II lines as well.

Third is the transverse proximity effect.  Eh?  Well, the normal proximity effect
shows up as a deficiency of absorption features at redshifts just a bit less than that
of the QSO and is due to extra ionization by UV from the QSO itself.  Worseck and
Wisotzky (2006) have found that the ionization level of the Lyman alpha and He II
forests probed by a background QSO is affected by other QSOs along the line of sight.
Now isn't that more fun than most of the dogs you know?

\section{OAO} 

This traditional abbreviation for the status of a potential significant other can
mean ``one and only'' or ``one among others,'' and so is a fit and proper heading for our
section that incorporates firsts, extrema, large numbers, and devices, effects, and
algorithms of which at most one exists, including a few for which one may already be
too many (though see Sect.~12 for other examples of this phenomenon).

\subsection{Countdown}

This year we have a green circle number, 6220, and so will count upward from it to
a googleplex and then back down to two.

\refe{6220 eclipsing binaries in the largest catalog to date (Malkov et al.~2006), special not
because N is truly enormous but because it was achieved by the old-fashioned method of
culling the literature from many observatories rather than as a single sweep from a
Major Mission.}

\refe{7400 stars in a cluster you need to monitor for one month to catch one planetary
transit (Burke et al.~2006).}

\refe{10,000 solar models computed to explore the range of input physics that can fit data on
neutrino production and solar oscillations (Bahcall et al.~2006). Results include
best age = 4.57 Gyr, M = $3.8418 \times 10^{33}$ g, fractions of luminosity from pp I chain
= 88\%, pp II = 10\%, pp III = 1\% (and, we suppose, CNO = 1\%), plus a very interesting
discussion of how the paper was written, the models assembled, and so forth.}

\refe{10,864 groups including at least four galaxies in the SDSS-3 data release (Merchan et
al.~2005).}

\refe{11,509 variable stars from a portion of the ASAS survey (Pojmanski et al.~2005). The
whole available sky from declination $-90^\circ$ to $+28^\circ$ should include about 
50,000 variables.}

\refe{23,781 variable stars in M31 from a pixel lensing project (Fliri et al.~2006). Their
37 RV Tauri stars are a lot for that experiment, and there is evidence for extinction
associated with the arms.}

\refe{38,000 dollars in the salary of an average US postdoc (Sigma Xi 2006). We got \$6500
in 1969 and took it to an institution where it made us the second best paid person
on the staff (after the professor and director) according to the Keen Amateur Dentist,
who in those days was a mere postdoc himself.  Correction.  He was never mere.}

\refe{39,088 SDSS ga1axies of which 40\% have GALEX and 2MASS (UV and IR) counterparts, but less
than 1\% have ROSAT and GB8 (X-ray and radio) ones (Obric et al.~2006).}

\refe{39,320 SDSS early-type galaxies to $z=0.15$ (Bernardi et al.~2006a), mostly in passive
evolution after relatively recent cessation of star formation.}

\refe{46,420 SDSS QSOs, of which 3366 (7\%) are to be found in ROSAT X-ray and/or FIRST radio
samples, a much larger non-optical yield than for normal galaxies (Shen et al.~2006).}

\refe{114,218 new double star measures made in 1889-2005 of 47,00 systems from 140 catalogs
added to the {\sl Washington Double Star Catalog} (Wycoff et al.~2006).}

\refe{351,507 SDSS lenses (Mandelbaum 2006), with the percentage of baryons put into stars a
declining function of halo mass, so that halos larger than $10^{13}$ M$_\odot$ all house 
ellipticals and ones smaller than $10^{12}$ M$_\odot$ all spirals.}

\refe{354,822 dwarf stars from the Tycho 2 catalog intended as candidates foc planetary
searches (Ammons et al.~2006).}

\refe{371,781 stars in M31 (and 146,622 in M33) with photometry good enough to investigate
stellar populations and history of star formation (Massey et al.~2006).}

\refe{720,199 characters were chiseled for this review, and we are very happy that we have
electronic wordprocessing these days to ease editing, proofreading, and alphabetizing
of references\footnote{Ironically or not, the klutzy author at the handicapped-unfriendly
institution still bangs out text on a typewriter, causing the journal editor to struggle 
with optical character recognition (OCR) software that misreads at least 10\% of the 
characters, resulting in text corrections that take a time equivalent of 4 weeks 
sun-tanning at the Waikikki beach. The reader who offers the most efficient OCR software 
before the next review will be awarded a fine bottle of Bordeaux rouge. - MA.
A journal willing to take paper submissions will be awarded page charges! - VT}.} 

\refe{2,670,974 stars in a proper motion catalogue for the Bordeaux zone (declination 
=$+11^\circ$ to $+18^\circ$ of the Carte du Ciel (Ducourant et al.~2006), and we can hear 
Bigourdan and
all saying ``I knew it would be useful some day'' not only across the Atlantic (American
observatories did not participate) but across a wider barrier.}

\refe{200,000,000 citizens of Europe who have English, French, or German as their native
languages, and 300,000,000 who don't (Mantere 2006). The item is actually about the
European Patent Office, which probably does not accept applications in Ruthenian.}

\refe{$5 \times 10^8$ transitions in a line list for H$_2$O (Barber et al.~2006), 
and if you will come over
here where the chancellor of the exchequer can't hear us, we will reveal that it is
the third commonest molecule in the universe after H$_2$ and CO. No listing was provided
for the favorite molecule of our youth, methyl ethyl wickacol.}

\refe{133,941,114,847,215, the coordinates of a ZZ Ceti star ({\sl A\&A} {\bf 365}, 969).  
If there are
fewer than $10^{12}$ stars in the Milky Way, then every one can have 100 numbers of its
very own.}

\refe{$10^{500}$ advertized as being the number of possible universes in a particular 
rendering of the multiverse concept ({\sl Nature} {\bf 439}, 10), which is described 
as being ``more than the number of atoms in the observable universe.'' 
A hell of a lot more, is all we have to
say on that, although you might want to respond to ``the cosmological constant, which
describes how fast the universe expands.''  But stay away from the quantum computationist
who recently announced in a UC Irvine colloquium that $2^{200}$ is more than the number 
of atoms in the universe.}

\medskip
\par\noindent And, working our way back down again:
\refe{4784 BAL QSOs (Trump et al.~2006) which may have broader emission lines than non-BAL
QSOs, and we honestly don't know whether this argues for or against orientation being
important in visibility of broad absorption lines.}

\refe{4131 candidate nearby stars (Lepine 2005), of which 63 are within 15 pc, 
539 within 25 pc,
and 18\% of stars closer than 25 pc are still unknown (presumably on some assumption of
uniform distribution in space and velocity).}

\refe{3169 SDSS spirals, of which l5\% are bulgeless (Kautsch et al.~2006).}

\refe{2728 SDSS ellipticals, whose colors are functions of age, metallicity, mass, velocity
dispersion, and [$\alpha$/Fe] (Chang et al.~2006a).}

\refe{2080 nearby ($z<0.4$) SDSS QSOs (Serber et al.~2006) which tend to live in L$^*$ galaxies
whose overdensities extend to about 100 kpc.}

\refe{1616 X-ray sources in the Orion region (Getman 2006, and the next 12 papers). Of these
Chandra ultradeep sources, 1382 are members, 159 are AGNs, and 16 are foreground stars.
This leaves 39, enabling us to offer them to Jack Benny as a birthday present, one for
each year of his age.}

\refe{1022 W UMa binaries cataloged by ROTSE, leading to the conclusion that one main sequence
star in 300 is a W UMa (Gettel et al.~2006).  Numbers in past years have ranged from
1/5000 to 1/50, and there is probably real variation among populations. All the W UMa
stars are X-ray sources at the ROSAT/XMM level (Geske 2006), but their sum is a small
fraction of the total Galactic luminosity.}

\refe{785 RR Lyrae stars (Wils et al.~2006) found by ROTSE-1, of which 188 are new, 34 show
a Blazhho effect, and 7\% are double moded. Their search partially overlaps the
ASAS zone and misses some stars, especially RRc's.}

\refe{666 legs on the {\sl millipede Illacme plenipes} (Marek and Bond 2006). 
She was the first of her
species seen since the 1928 discovery, and males have only 318-402 legs, which is a
good thing, since guys hate buying shoes.}

\refe{654 or 650 open clusters in the solar neighborhood (Piskunov et al.~2006; 
Bonatto et al.~2006). This is a perfectly fascinacing pair of papers, implying a total 
cluster population of about $1-4 \times 10^5$, two groups by life expectancy of $10^8$ 
and $10^9$ years, and,
most curiously, that less than 10\% of disk stars around now appear to have came through
this route, meaning, it seems, not rampant formation of isolated stars but rather a
very large fraction of proto-clusters that disintegrate before they ever become
disembedded from the clouds where they formed so as to get counted (Lada and Lada 2003).
Lamers et al.~(2005) also report a larger star formation rate in embedded clusters than
in visible ones, though less extremely so, on the basis of 520 clusters.}

\refe{504 is the star number af HD 98618 in the Allen Telescope priority list for SETI
(Melendez et al.~2006). It is the 2nd closest solar twin after 18 Sco, whose number
in the list (Turnbull and Tarter 2003) we really meant to look up for you.}

\refe{496 WD + M V SDSS spectroscopic binaries with reasonable data on star temperatures and
spectral types (Silvestri et al.~2006).}

\refe{476 dwarf elliptical galaxies in Virgo (Lister et al.~2006). 5-10\% have clear disks or
bars or arm structure, and these are distributed in the cluster like spiral and
irregular galaxies rather than like E/SO's and ``pure'' dE's.  They get called dEdi.}

\refe{450 MACHO events in the direction of the galactic bulge (Popowski et al.~2006), 
implying an optical depth near $2 \times 10^{-6}$.}

\refe{398 DA white dwarfs, the largest sample with 3-d kinematics.  Some indeed belong to the
thick disk and halo, but they are a negligible contribution to the dark matter supply
(Pauli et al.~2006).}

\refe{235 protoclusters with masses of 102-320 M$_\odot$ in a star formation region, 
leading to the
very interesting conclusion that the IMF is continuous across the conceptual divide
between large stars and small clusters (Beltran et al.~2006).}

\refe{230 supernova remnants catalogued in the Milky Way (Koo et al.~2006). It seems like a
nice, hefty number except that, note the authors, we expect more like 20-30,000 from a
formation rate of a few per century.  They add one new one 16 kpc from the galactic
center and $3 \times 10^5$ yr old (both under-sampled parts of parameter space), so only
19,770 papers to go.}

\refe{209 narrow fluorescent H$_2$ lines in the spectrum of one T Tauri star, pumped by 
Ly $\alpha$ (Herczeg et al.~2006).}

\medskip
\par\noindent
No fewer than 11 papers record numbers between 199 and 101.  You cannot have them all, and
we are ruthless in noting only:

\refe{175 the age of Harriet the tortoise at the time of her death in an Australian zoo
(Anonymous 2005d).}

\refe{170 the estimated IQ of John Quincy Adams (Simonton 2006), at the top of the presidential
pack. Harding sets the low for the 20th century, below the 125 of G.~W.~Bush. Notice
that in a random sample of humans, these wou1d all be respectably high numbers, which
says something about something.}

\refe{153 the number of abstracts at a meeting of the Astronomical Society, of India 
({\sl BASI} {\bf 37}, 337-415), covering topics from comets to cosmology. A large January 
AAS runs to about
10 times that, which strikes us as roughly proportional to the sizes of the two
communities.}

\refe{122 systems of galaxies in the Shapley supercluster, of which 60 are new 
(Ragone et al.~2006). Their masses are in the range $10^{13}-10^{15}$ h$^{-1}$ M$_\odot$.}

\refe{101 years of radial velocity data used to establish the orbit periods of the triple
star Epsilon Per (Libich et al.~2006).}

\refe{100 distance of Voyager 1 from the Sun in AU, as the year ends.}

\refe{100 Bayesian evidence must be a pure number, because you can take its logarithm 
(Bridges et al.~2006), sort of like chi-squared, we guess, only bigger is better.}

\medskip\par\noindent
And there were no fewer than 78 items smaller than 100 (but bigger than 1), 
though not uniformly distributed, with 10 threes and 17 twos, so once again a subset.

\refe{72 well-established solar observatories considered for site selection of the ATST.  
In the end they looked hard at six (Socas-Navarro et al.~2006).}

\refe{66 supercentenarians (folks 110 or older) alive on a particular day. All but two were
women, and we seem to have lost the reference, evidence perhaps that we are not
destined to join them.}

\refe{63 highly ionized high velocity clouds (Fox et al.~2006). They are also fairly high off
the Galactic plane, but still associated with the Milky Way, according to FUSE data.}

\refe{53+ the ionization state of Xe, heaviest ion with calculations of its dielectronic
recombination (Badnell 2006).}

\refe{47 stars with Babcock-type strong magnetic fields (Ryabchikova et al.~2006, and, of
course, Babcock 1960).}

\refe{43 American experts on AIDS who were eventually allowed to go to the biennial congress 
in Toronto (Burklow 2006). In the lead up, the number had varied between 77 and 25.
Canada, though fairly inexpensive to get to, counts as a foreign country (for which,
we believe, they are generally grateful).}

\refe{40 R CrB variables in the Milky Way, with five new ones presented by Zankewski et 
al.~(2005). They predict another 250 should be lurking, absorbed, in the galactic bulge.}

\refe{30 luminous blue variables in the whole Local Group (Pasquali et al.~2006).}

\refe{22 sources of ultrahigh energy cosmic rays near the galactic center (Aharonian et 
al.~2006). Most, they say, are supernova remnants and pulsar wind nebulae.}

\refe{19 millisecond pulsars in the globular cluster 47 Tuc (Bogdanov et al.~2006a).}

\refe{16 microquasars in the Milky Way (meaning, more or less, thase with relativistic jets) 
if you count both low and high mass donors and both neutron star and black hole recipients
(Dermer and Bottcher 2006, who were  modelling the UHE gamma emission from one).
Personally we count only the sunny hours.}

\refe{14 is the largest number of red supergiants in any one star cluster (Figer et al.~2006).}

\refe{13 double-double radio galaxies (Saikia et al.~2006). A typical one was turned off about
20 Myr between radio-active episodes.}

\refe{12 American universities in the world's top 20 as counted by some bean (Anonymous 2006f),
plus one each in China, Japan, Australia (Melbourne), and France, and four in the UK.}

\refe{11 dwarf novae with periods in ``the gap'' (Schmidtobreik and Tappert 2006). Perhaps it
is that long-needed gap in the literature.}

\refe{10 authors on a paper whose thank you's begin, ''I would like to acknowledge...''
({\sl ApJ} {\bf 648}, 1246). We are thinking of establishing a Walt Whitman award for this sort
of thing.}

\refe{9th transiting hot Jupiter (Bouchy et al.~2005). It was HD 189733b, with a density of
0.75 (that is, less than Jupiter, but not so much so as some of the others), and there
were more before the year was out.}

\refe{8th broad absorption line quasar that is also a Fanaroff-Riley Type II radio source
(Gregg et al.~2006).  The radio luminosity and absorption line strength are anti-correlated 
in the class, and the authors conclude that this is not an orientation effect
but jets and lobes battling to emerge from a cocoon of BAL stuff.}

\refe{7 white dwarf + red subdwarf M binaries (Monteiro et al.~2006). We chose this from
among about 7 7's partly for the contrast with number 496 above.}

\refe{6th unambiguous determination of the relative angle of inclination in a triple system
(Muterspaugh et al.~2006). It is $24^\circ$ for V819 Her, and coplanarity is mildly favored
for the ensemble.}

\refe{5 supernova light echos; 2003gd (Sugerman 2005) having been added to 87A, 91T, 93J, and
98bu.}

\refe{4 periods to be seen in earth nutation (Londrak et al.~2005). They are Chandler wobble
(435 days), retrograde free core nutation (430 days), prograde free core nutation (1020
days), and inner core wobble (2400 days).  And a 4th very complete set of stellar
evolution tracks and isochrones (Pietrinferni et al.~2006). It is the first to include
the effects of non-solar $\alpha$/Fe in the boundary conditions and color transformations
(from the theoretical plane to the observed, as they should be) as well as in nuclear
reactions, opacities, and equations of state.}

\medskip
\par\noindent
We found roughly $3^2$ 3's, but limit you to 3! of them.

\refe{The third Type IIP supernova with a probable red supergiant as its ``pre need'' 
counterpart (Hendry et al.~2006). The mass was about $9^{+3}_{-2}$ M$_\odot$, much like 
the others.}

\refe{The third clear case of GRB = SN, 060218 = 2006ej (Cobb et al.~2006). It was at 
$z=0.053$, and three is too many for these faint things just to be off axis. The authors
propose a separate class of faint GRBs, which are commoner than the famous sort, and
very bad if you want to use them as standard candles.}

\refe{Numbers 3 (and 2) each of LMXRB with the donor an M supergiant (Masetti et al.~2006);
they have only coordinates, not names.  And Cepheids with envelopes considerably larger
than their photospheres (Merand et al.~2006).  They have only names, not coordinates,
Polaris and $\delta$ Cep itself. The first was 1 Cas.}

\refe{A third GRB with an X-ray scattering halo (Vaughan et al.~2006). It was 050724, and
the dust is 139 pc from us.  Another appeared before the year was out, 050713A, with
the dust at 364 pc say Tiengo and Mereghetti (2006) who describe another 12 events with
no such halo.}

\refe{These are three bonds between guanine and cytosine in DNA (Wain-Hobson 2006). 
This is not
exactly news. The hot flash is that Watson and Crick showed only two in 1953,
and it was Pauling and Corey who correctly said 3, 3 years later. It took us a moment
to figure out why that notebook line also says Miller-Urey, Death of Stalin, and
Casino Royale.}

There are also many 2's, and we regretfully leave out the number of articles about Brad
Schaefer in the 20 January Issue of {\sl Science} and the chap who described himself as
chancellor of two fine universities in order to focus on second examples and such that
may help to define new astronomical classes. Please read (or say as appropriate) ``the
second'' before each of the following items:

\refe{Confirmed fossil cluster based on X-ray data (Cypriano et al.~2006).}

\refe{X-ray emitting pulsar with a neutron star companion (Kargaltsev et al.~2006).}

\refe{Set of groups of galaxies on track to became a cluster (Brough et al.~2006, who examined
6dF data; this one has a total mass of $7 \times 10^{13}$ M$_\odot$).}

\refe{Exoplanet with very high eccentricity orbit (Jones et al.~2006). HD 6601 with 
$e=0.92\pm0.03$ overlaps the first at $e=0.93$.}

\refe{Member of the class of stars of which Gamma Cas is the prototype. The new one is BZ
Cru = HD 110432 (Smith and Balona 2006).  They are characterized by X-rays, rapid rotation,
Be disks, and chaotic variability in hours with cycles near 100 days.}

\refe{Interstellar source of acetone [(CH$_3$)$_2$CO] emission (Friedel et al.~2005) 
in Orion KL. The first was Sgr B2.}

\refe{Measured radius for a star with mass less than 0.1 M$_\odot$. It is OGLE Tr-1136 
at 0.085 M$_\odot$ and R=0.113 R$_\odot$ (implying a mean density of 56 g cm$^{-3}$, 
Pont et al.~(2006).}

\refe{Galactic LBV still in its birth cluster. There are 24 stars earlier than B3 
(Pasquali et al.~2006), and it is WRA 751.}

\refe{Supershell that has blown out on both sides of the galactic plane (McClure-Griffith et
al.~(2006). Both are Parkes H I sources with 7-digit telephone numbers\footnote{
Who first used this phrase for coordinate-type source names?  Well, your author who
has changed phone numbers most often says she first heard it from Bohdan Paczynski,
and he always said he first heard it from her. We mention this only because the page
is being typed less than four hours after we heard of his death.}}.

\refe{Largest Ap star magnetic field, 24.5 kG for HD 154708 (Hubrig 2005). Curiously, HD
215441 (Babcock 1960) still holds the record.}

\refe{Star with multiple linear combinations of modes (O'Toole et al.~2006). It is an sdB star
from the Palomar-Green survey.}

\refe{Accreting neutron star with three cyclotron lines (Pottschmidt et al.~2005). These come
at 27, 51, and 74 keV.}

\subsection{First}

A good many similar items appear in other sections by subject matter, but here are
some strays, roughly from large scale down to small.  Say ``the first'' as usual before each.

\medskip
\refe{Detection of rotation measure in a supercluster of galaxies (Xu et al.~2006). If the
electrons are presumed to be those from the WHIM plus leakage from radio galaxies, then
the (partially coherent) magnetic field is B$\le 0.3$ $\mu$G (l/500 kpc).}

\refe{E or S0 giant galaxies with the core helium-burning (HD, red clump) stars resolved
(Kejkuba et al.~2005).  It is NGC 5128 observed with HST.}

\refe{X-ray emission from a Type Ia supernova (Immler et al.~2006). It was the underluminous
2005be and gives the impression that a red supergiant of $\lapprox 10$ M$_\odot$ blew off
an outer
layer about $10^4$ yr before accretion (from it) pushed a white dwarf companion over the
ignition limit.}

\refe{Formation of an intermediate-mass black hole miniquasar ought to have happened at 
$z=21$ (Kuhlen and Madau 2005).}

\refe{Molecular line survey outside the Milky Way (Martin et al.~2006). Remarkably this was
NGC 253, not the Large Magellanic Cloud, and no organics more complex than CH$_3$CN and
CH$_3$CCH were found.}

\refe{Extragalactic detection of H$_3^+$ outside the Milky Way and also not in the LMC 
(Geballe et al.~2006). The host is IRAS 08572+3915.}

\refe{LMC star cluster with an age (near 9 Gyr) between that of the very old globular clusters
and the recent star formation epoch (Mackey et al.~2006).}

\refe{DY Per star in the SMC (and the 5th ever), Kraemer (2005).}

\refe{Extragalactic W UMa (Kaluzny et al. 2006) returning us to the LMC.}

\refe{Dust in a dwarf spheroidal, in one of the four planetary nebulae in IGI (Zijlstra et 
al.~2006).}

\refe{Dust emission from a high velocity cloud, seen with a combination of the SST and the 
GBT 21 cm survey (Miville-Dechenes et al.~2005). The authors, in best astrophysical tradition,
use this sample of one to conclude that most infall gas is cold (but see the FUSE HVCs
in countdown at N = 63).}

\refe{Cataclysmic variable in a multiple star system, FH Leo (Vogt et al.~2006).}

\refe{X-ray binary in a supernova remnant (Williams et al.~2005). There is, of course,
SS433 = W50, but the nebula in that case has quite possibly been blown by the XRB rather
than being left from the formation event.  The present system also has several names,
of which the most mysterious are DDB1-15 and r3-63.  It is in M31.}

\refe{Eclipsing high mass black hole X-ray binary, M33 X-7 (Pietsch et al.~2006).}

\refe{YSO accretion disk with two components, probed with two different molecules, rotating
in opposite directions (Remijan and Hollis 2006). It is IRAS 16293-2422, and the
authors offer neither surprise nor explanation.}

\refe{Brown dwarf with a central hole in its accretion dish (Muzerollen 2006).}

\refe{The 1.3rd L type subdwarf (Reiners and Bastri 2006). Don't ask us; it's their paper:
we're just trying to report it.}

\refe{University professor of chemistry, 1609, Marburg, Johann Hartmann.}

\refe{Successful resolution of optical polarization with long-baseline interferometry
(Rousselet-Perraut et al.~2006).}

\refe{1=value of $q_0$ adopted briefly in {\sl ApJ} {\bf 548}, 81 (it gets better later).}

\refe{1/2: When the universe was 1/2 its present age, cluster ellipticals were half that age
({\sl ApJ} {\bf 644}, 30). And the mean elliptical is now half the age of the current universe,
a coincidence they say.}

\refe{1/3: ``A third of all papers are never cited'' ({\sl Nature} {\bf 442}, 344), 
stated without source,
in the context of a study by Harry Collins of replication (or non-replication) of
scientific experiments. We bring to this specialized knowledge of two forms.  Even
casual consideration of the
astrophysical literature reveals that the percentage of papers totally uncited after
three years is more like 3\% than 33\% (in a sample that includes journals of both high
and low prestige).  In addition, there is a discussion of the impossibility of
perfect replication, in which Collins quotes a physicist: ``It's very difficult to
create a carbon copy ... if what is critical is the way he glues on his transducers ...
the technician always...''  This deals with Weber bar detectors for gravitational
radiation; and he always glued on his own transducers (though indeed some cements
were better than others).}

\medskip
We have three candidates for the Lincoln's Doctor's Dog's Favorite Jewish Recipes
award: (a) ``the first time in history that a new superluminal component has been
detected at the predicted time and angle'' (Pyatunina et al.~2006), (b) the first brown
dwarf science from laser guide star adaptive optics (Liu and Leggett 2005), and
(c) the first detection of large scale magnetic fields in an Sa galaxy in the radio
range (Krause et al.~2006). It is NGC 4594, the Sombrero.

\subsection{Extrema}

Initially, the two notebook pages for extrema (84 and 85) were called human and
inhuman.  Brief consideration led to the conclusion that astrophysical and 
non-astrophysical was perhaps a better division. In any case, the human side includes:

\begin{itemize}
\item[o]{The largest single research facility in the world, NIH in Bethesda  
({\sl Nature} {\bf 441}, 1).}

\item[o]{The commonest disease, diagnosis, according to Kraus (2006).  He is arguing against
routine treatment of ADHD and high cholesterol (with which your most attention-deficited
author agrees, but she claims that Viagra and hormone replacement, to which
he also objects, fall in the class with eye glasses and hearing aids; well maybe he
disapproves of those as well).}

\item[o]{The largest state? We were happy to hear that the Tex-Mex conference reported in
{\sl Rev. Mex. A\&A Conf. Ser. No. 23} took in UC Santa Cruz, NASA Ames, Vanderbilt U.,
and Univ. of Oklahoma.\footnote{And it must be the Tex half that has expanded to take
in Santa Cruz and Ames, because Mex deaccessions Alta California more than 1.5
centuries ago.}}

\item[o]{The oldest new world writing (Del Carmen Rodriguez Martinez et al.~2006). It is attributed to
the Olmec culture about 2900 years ago (an uncalibrated C-14 date).  There are 62
signs in the sample, including 28 distinct ones. Some are clearly pictographs - an
ear of corn, dart, fish, and insect, etc., but unclear whether ideograms, a syllabary,
alphabet, or something else.}

\item[o]{Youngest person to discover a supernova, probably Jennifer Tigner at 18 (SN 2005de,
Cerevolo 2005). It was number 101 for the team, and she is now a University of
Victoria student in physics and astronomy.}

\item[o]{The most names taken in vain in a cosmological model award goes to Ghafarnjad (2006)
for Brans-Dicke (scalar-tensor), Klein-Gordon (wave), deBroglie-Bohm (particle),
Minkowski (background), Hamilton-Jacobi (equations).}

\item[o]{The most stable optical clock (Bergquist et al.~2006). It is a single atom of mercury.
They didn't say which one (and if you attempt to tell us about identical particles you
will discover that some of them can get more annoyed than others).}

\item[o]{The previously most stable optical clock drifted at a comparable rate  to that of the
215 second pulse period of the white dwarf G117-B15A (Kepler et al.~2005a), so that one
had to subtract off its slippage in s/s to recover the WD slowing of 
$5.57 \pm 0.82 \times 10^{-15}$ s/s. And yes, we also understand that you can cancel the 
seconds up and down
stairs to get a pure number, but we don't much like c=G=1 relativity papers either.}

\item[o]{The largest earthworm, {\sl Megascolides australis}, reaches 11 feet in length and 
can squirt liquid to a distance of 18-24\arcsec ({\sl Nature} {\bf 441}, 167, 
reprinting an item from the 12 May 1956
issue).  We did not enquire just what the liquid was, and if you find out, we'll thank
you not to tell us.}
\end{itemize}

\subsection{Astronomical Extrema}

Starting big, we find, most unsurprisingly, that the largest measured redshifts get
a Little Bigger\footnote{ 
No. The Bigger family have gone to a spa in North Carolina.}
each year. In 2006 came $z=6.96$ (Iye et al.~2006) for a galaxy
imaged with Subaru. It is about 700 Myr old (or rather it was when the photons left),
has a luminosity of $10^{43}$ erg s$^{-1}$, and a star formation rate near 10 M$_\odot$
yr$^{-1}$ (Bouwens and
Illingworth 2006). The number of detectable galaxies seems to drop off very steeply
above $z=6$ (McMahon 2006). The Subaru program is described by Shioya et al.~(2005).

The record for a radio galaxy is only $z=5.2$ (Overzier et al.~2006). The source is
within an overdense region of Ly $\alpha$ emitters, and so will perhaps also do for the
most distance protocluster. No X-ray emission was reported.

In contrast, the largest and smallest metal (meaning oxygen in this context)
abundances associated with galaxies have held remarkably steady in recent years. In
customary units\footnote{
Whose meaning is about as transparent as for the customary units of heat conductivity
used by American builders - BTU-inch/(ft$^2$ - $^0$C-hour) which, as a radio announcer 
recently
explained in connection with the new speed record set by the French TGV train, is about
160 drachma.}
the minimum hovers around 12+log(O/H)= 7.12-7.17 (Izotov et 
al.~2006 reporting on SBS 0335-152W, in competition with I Zw 18) and the maximum for spirals
at 8.75, close to the galactic value (Pilyugin et al.~2006). Be warned, however, that
the number was 9.4 in earlier analyses, and we have spotted solar numbers including 8.56
and 8.65 since the most recent oxygen drop.

The brightest QSO had M$_B=-31.4$ during a recent, year-long outburst (Villata et 
al.~2006). The fastest superluminal radio sources have v/c only about 25 (Piner et al.~2006),
perhaps a natural speed limit, since it applies to three sources.  A case reported with
40c a few years ago involved an overestimated redshift.

The largest galactic velocity dispersions reach 400 km s$^{-1}$, though there are also
accidental superpositions in that range.  They are not anomalous in the fundamental
plane, but simply very massive and compact (Bernardi et al.~2006), and have also very
massive black holes, the authors conclude. The closest pair of black holes in an AGN is
separated by only 7.3 pc and has been resolved by the VLBA (Rodriguez et al.~2006). In
combination with optical line velocities, this implies a total mass near $1.5 \times 10^8$
M$_\odot$ very much as expected.

Moving on inside galaxies, we find the strongest spectral line in the universe comes
from [C II] at 158 $\mu$m (Rodriguez-Fernandez et al.~2006).  The particular galaxies they
are looking at turn out to have very little obscured star formation. The 158 $\mu$m line is,
of course, redshifted like everybody else, to 900 $\mu$m in a $10^{13}$ L$_\odot$ QSO 
examined by Iono
et al.~(2006). It is a less important coolant, only 0.04\% of the far IR luminosity,
at large redshift.

Which are the biggest star clusters? Gieles et al.~(2006), examining the distributions
of cluster masses and luminosities in a number of galaxies, conclude that there is
a natural physical maximum at $2 \times 10^6$ M$_\odot$, but that very few galaxies have 
enough clusters
to include even one of those.  Theirs are in M51 and in NGC 4028/39 and should be globular 
clusters 10 Gyr from now.  By way of gumming up the works, Ma et al.~(2006) present a
12.4 Gyr old cluster in M31 which, if bound (as its age would seem to require) has a
total mass near $3 \times 10^7$ M$_\odot$ and a normal IMF. The universe, we are forced
to conclude, isn't what it used to be. They predict a velocity dispersion of 72 km s$^{-1}$, 
which, we trust, will have been measured long before you read this.

Among individual stars, the largest masses are 140-160 M$_\odot$ (again, not just
statistics, but running off a natural limit, Koen 2006), and the smallest with a
circumstellar disk is a brown dwarf not much heftier than Jupiter, but alone in the
world in Chameleon (Lugman et al.~2005).

The earliest eclipsing binary is V1182 Aql (Mayer et al.~2005). This does not mean
that it can be expected to eclipse by 7 AM at the latest, but that it consists of two
O 0.5 stars, the primary being less massive than you would expect for its luminosity
and temperature, while the secondary is on the ZAMS.

The brightest stars reach $10^6$ L$_\odot$ (Heap et al.~2006), and the faintest ... well, 
there
they are in an extreme population II sample from the nearby globular cluster NGC 6397
(Richer et al.~2006).  There is both a main sequence trickling off toward the hydrogen
burning limit at 0.083 $M_\odot$ and white dwarfs extending a magnitude or two fainter.

So how much is that in drachma? Annoyingly, the article gives everything in HST
apparent magnitudes, indicating nowhere what they think the distance to the cluster or
the bolometric corrections might be. Well, less than $10^{-4}$ L$_\odot$ anyhow. Richer et 
al.~(2006) believe they have seen also the end of the WD branch, consisting of stars with 
cooling times equal to the age of the cluster, though there are other possibilities. The WD
sequence crooks back to the blue in infrared colors, as expected.

What the authors call the darkest bright star is simply a Chandra lower limit to
the X-ray luminosity of Vega at $2 \times 10^{25}$ erg s$^{-1}$ (Pease et al.~2006), 
which is smaller than expected in some models.

The brightest subdwarf B star rejoices in the name BALLOON090100001 (Telting and
{\O}stensen 2006). And the coolest (hence probably faintest) subdwarf M is LEHPM2-59
(Burgasser and Kirkpatrick 2006) at 3000$\pm$200K. Brown dwarfs, of course, pass through
much lower temperatures on their way to invisibility, for instance 2MASS0939-2448
(Burgasser et al.~2006) at something like 700 K and log L$_{bol}=-5.4$.

Our main conclusion is that star names just aren't what they used to be, and we
wish these could be called Aleph-1 Supellex Cubiculii and Aleph-2 Cochlea piscatoris,
following to their logica1 extensions the numbering system of Flamsteed and the
constellations of Moore (2005)\footnote{
No, these names are not quite right. Readers skilled in the use of globes will have
noticed that we would have had to go home and get our Latin textbook to report the
correct genitive forms of the constellation names, except perhaps Vaccae.}.

At the beginning of stellar life, we find the smallest bipolar molecular outflow
(500 AU) driven by a very faint young stellar object (Bourke et al.~2005) and at the
end, the oldest pulsar/SNR combination (Kothes et al.~2005), whose 15 minutes of fame
came $10^7$ years ago, when all the observers were busy grooming their thesis advisors
(la plus \c{c}a change ... ). It is in the general direction of Vacca Volitans, Flamsteed
number uncertain.

\subsection{Familiar Physics, Expected Effects, and Wonderful Widgets}

The underlying idea here is, we think, that there is no such thing as useless
knowledge.  Thus, if you have previously met supernova remnants, Venn diagrams, and the
Small Magellanic Cloud, you cannot be led too far astray by Venn diagrams of SNRs in
the SMC (Filipovich et al.~2006). On the other hand, when Abad and Viera (2005) describe
star streams identified by using Herschel's method, without citing Herschel, you might be
led to suppose the method means extreme exploitation of sibling labor.

\subsubsection{Physical Principles}

We found four candidates for the ``if there is anyone for whom this isn't true,
please don't tell me about it'' certificate\footnote{
This is generally awarded in introductory astronomy courses in connection with the
idea that most of us have spent nearly all our lives on earth.}.
First, inverse Compton and synchrotron
radiation in astrophysical sources violate Lorentz invariance by less than 6 parts in
$10^{20}$ (Altschul 2006). Second, Fuzfa and Alimi (2006) are proposing a form of dark energy
that violates both the weak and strong equivalence principles, but currently passes the
standard GR tests. Newton's almost constant, G, will vary with time. E=mc$^2$ (or,
rather, of course, as always $\Delta$E=$\Delta$mc$^2$) to within 5 parts in $10^7$ for 
(n,$\gamma$)
reactions on Si$^{29}$ and S$^{30}$ (Rainville et al.~2005). And, fourth, Kirchoff's laws 
for circuits can be violated in quantum devices (Gabelli et al.~2006).

There are reversible processes in the real world, if you are careful and don't try
to go too far (Pine et al.~2005). It was nearly the end of the 19th century before the
concept of temperature in astronomy was well enough established that people stopped
publishing outrageous numbers for the sun (Hughes 2005).  The most recent reported is
5775.9 K.  The last digit should decrease sometime in the next million years.

The four forces seem to be in pretty good condition. General relativity passed,
as usual, all tests thrown at it (Jacoby et al.~2006), but a few cosmological alternatives
appear in Sect.~12. Electromagnetism saw a new, more precise value of the fine structure
constant (137.035999710)$^{-1}$ (Gabrielse et al.~2006). The corresponding calculation
required 891 Feynman diagrams. Alpha first deviated from 1/137 in the mind of Dirac (1928)
and in the laboratory of Kusch and Foley (1948). The most deviant author started
kindergarten that year and did not read the paper until later. We caught three papers
during the year on changes in alpha, all upper limits, but Levshakov et al.~(2006) say that
it could oscillate with time and vary with position as well.

The ratio of proton to electron mass, 1836.15267261 is also known to a good many
decimal places, and is reported to have changed by $\Delta \mu/\mu=2.6\pm0.6 \times 10^{-5}$
since $z=2.8\pm0.2$
(Reinhold et al.~2006).  This could be related to extra dimensions (Barrow
2006), and if anyone is in the market for these, we would be glad to dispose of a bit
of extra breadth.  The interface between electromagnetism and quantum mechanics (and
between physics and astronomy) brought us a new and better calculation of the Paschen-Bach
effect for molecules (Berdyugina et al.~2005) and an atomic analog of the Hanbury-Brown
and Twiss intensity interferometer (Schellekens et al.~2005).

The strong interaction remained strong enough to sustain (a) a gradual laboratory
creep up toward the superheavy island of stability at $z=114$, N=184 (Herzberg et 
al.~2006), which is probably not reached under even the most extreme stellar conditions, (b)
a better understanding of laboratory quark-gluon plasmas (Asakawa et al.~2006), in which
``nearly perfect liquidity'' arises during expansion, likely, they say to be relevant to
the early universe, (c) at least four assaults on baryogenesis, none of which seems to
have left the world with even one more baryon than it had before, so we mention only the
(seemingly) most complex, in which inflation with CP-component gives rise to elliptically
polarized gravity waves, which produce lepton asymmetry, which is then responsible for
the baryon asymmetry (Alexander et al.~2006), and (d) calculacions of big bang 
nucleosynthesis in which variable $\alpha$, $\Lambda$, and G can be fine-tuned to recover 
the standard
results from constant constants (Landau et al.~2006). Five-parameter elephant joke
goes here.

The neutrino (which one?), exemplar of the weak interaction, turned 50 during the
year, and there was at least one celebratory conference. It is, however, not 100\% certain
that the neutrino was there, since the two presentations we recorded dealt with 
high-precision lunar laser ranging as a test of general relativity (Dwali 2006) and Q balls
as candidates for dark matter and dark energy (Roszkowki 2006).

On the edge between gravitation and electromagnetism, you can find the wormholes of
Kardashev et al.~(2006), which are kept open by an electromagnetic field.  From outside
they look like macroscopic magnetic monopoles, and because there is neither a horizon nor
Hawking radiation, ones of less than $10^{15}$ g can still exist. They are a good way to 
reach other universes if we all arose out of chaotic inflation. Few ideas this year brought
forward the thought ``Don't go there!'' more strongly.

\subsubsection{Processes}

Raman association forms H$_2$ when a photon scatters off two nearby, unbound atoms,
leaving them bound. At $z \approx 10^3$ the process can convert $10^{-4}$ of the hydrogen to 
molecules, (Dalgarno and Van der Loo 2006), but it is gone by $z=100$.

Fractals describe the distribution of star formation (De La Fuentes Marco and De La
Fuentes Marcos 2005, not a new result), but not the large scale distribution of galaxies
(Joyce et al. 2005, also not a new result, though fairly new to that team).  Come to
think of it, there must be some length scale beyond which star formation is no longer
fractal either, perhaps the diameter of a galaxy.

Though computing grows ever more powerful, Anonymous (2006g) opines that LSST and
LHC will be the installations that challenge current capability for massive data processing.
It is probably significant that L stands for Large in both cases (Large Synoptic Survey
Telescope and Large Hadron Collider). In case it is not obvious, it is the survey and
the collider that are large, not the synopses or the hadrons.  But the green computing
circles goes around De Val-Borro et al.~(2006), who have compared a number of SPH codes
that are designed to model protoplanetarv disks and planet formation. Typically there is
agreement at the 5-10\% level (one or two planets out of 20?), and they provide some salutory
advice for others attempting similar projects in other fields.

``Lucky imaging'' stacks the best of many 10-second frames to improve angular resolution
(Law et al.~2006).  Unlike adaptive optics, it works at visible wavelengths.  ``Contour
binning'' as an alternative to square pixels (Sanders 2006a) makes for much prettier
pictures, almost as nice as real, silver halide photographs.

Confusion limits have become much more stringent over the years.  Ilyasov (2006) says
that a careful radio astronomer wants 200 beams per source, vs. the 25 we grew up with
in the days of 3C.  The Eddington effect (Eddington 1913) is not the same as Malmquist
bias, but, like having too few beams per source, it can also lead your counts astray,
as indeed can using the wrong correction method where your counts are incomplete (Geijo
et al.~2006). But our favorite, Oh dear!, of the year is the inevitable result
that, if you do a survey and then later on do a more sensitive one, most of the variable
objects will have been brighter the first time around.  Panessa et al.~(2005) surely
cannot be the first discovery of this.  They were considering ROSAT sources recoveced by
XMM and Chandra.

\subsubsection{Catalogs}

We wonder whether Henry et al.~(2006) reporting 442 sources in an ``undistinguished
spot of moderate galactic latitude and extinction'' is the last ROSAT catalog ever.
At least 30 other catalogues appeared (or, sometimes, only e-ppeared) during the year.  A
few were virtual (Gomez de Castro and Wamsteker 2006 in the UV; son of 2MASS, 
Cabrera-Lavers et al.~2006; on beyond SDSS, Hikage et al.~2005). Others with remarkably many or
few entries appear in the Countdown section 11.1. And here is a subset that seem to
invite further attention to either the contents or the methods:
\begin{itemize}
\item[o]{Unidentified radio sources; yes there are still some (Ciliegi et al.~2005).}
\item[o]{Unidentified gamma ray sources, of which there are lots (Dean et al.~2005).}
\item[o]{Recurrent X-Ray transient sources (Sguera et al.~2005) which seem to be rather a dog's
dinner, but include at least one new class at low galactic latitude and durations
less than three hours.}
\item[o]{Radio transients, which are rather rare (Gal-yam et al.~2006).}
\item[o]{Optical transients, dearest to our heart but rather discouraging: Shamir and Nemiroff
(2006, one satellite glint); Kulkarni and Rau (2006), a ``dense foreground fog'' of
active M dwarfs\footnote{
Otherwise summarized as:
\par Two dwarf novae, seven flare stars, and a bird, 
\par That flew across our dome slit on the 3rd.}.}
\item[o]{A data set called the {\sl Washington Fundamental Catalog}, although no catalog was ever
released under that name ({\sl AJ} {\bf 132}, 50).}
\item[o]{The 12th QSO catalog (Veron and Veron 2006), with 85,221 QSOs and an assortment of
Blazars, Seyferts, and all. The first in 1971 had 202 sources.}
\item[o]{The SST extragalactic catalog (Fadda et al.~2006) with 17,000 sources.}
\item[o]{Variables old and new from POSS plus SDSS (Sesar et al.~2006), of which 20\% are QSOs
vs. only 0.6\% of all point sources (most of the rest are called stars).}
\item[o]{A radial velocity catalog for HIPPARCOS stars culled from many sources (Gontcharov 2006).}
\item[o]{And absolutely oodles of recognizable kinds of binaries in the ASAS data base (Paczynski
et al.~2006) plus a good many non-recognizable ones.}
\end{itemize}

\subsubsection{Sites and Observatories}

Old friend Dome C received the most attention - at least five papers ending with
Geissler and Masciardri (2006), plus a nod toward Dome F (Swain and Gallee 2006), which
gets better if you can suspend your observatory from a sky hook at least 20 meters above
ground. The high Atlas mountains of Morrocco were a new entrant in the ``why come here?''
stakes (Bienkhaldoun et al.~2005). In briefest possible summary, their 1.05\arcsec is Mauna
Kea's 0.75\arcsec .

Only 24 observatories appeared in the 2006 notebooks and index, out of nearly 400
that contributed to the astronomical literature of a typical recent year  (Trimble
and Ceja, work in progress), and these include a good many that don't yet actually exist,
of which the world supply must be very nearly infinite. Mt. Stromlo is recovering slowly
(Sackett 2006). The Heterogeneous Telescope Network is up and running (Naylor et 
al.~2006) with, we hope, less difficulty than the 666 leg millipede (Sect.~11.1). There may
come a time, around 2050, when contrails plus additional clouds from global warming
render most ground-based optical observatories useless (Gilmore 2006). Lead times for
observing proposals grow ever longer, but even so, you probably won't have to apply for
2050 time before Christmas.  Another year ended with no word from Gravity Probe B
(Marsden et al.~2005 on its guide star, IM Peg), though we are not precisely sure what
probes say when left on their own after dark.

The National Astronomical Observatory of Japan is a long-established, productive
institution, whose director is a person very much after our own mind (Miyama 2006), and
you will have to go there to find out what he said.

Competition for the site of the national underground lab has been reopened (Stanley
2006), thereby, we suspect, postponing the need to spend any money now, but increasing
the eventual cost.  If this reminds you of the life cycle of some other projects, it
cannot be helped.  Griffin (2005) is a report on the status of JWST somewhat confused
by the writer's use of elegant variations rather than precise repetition of the exact
names of each of the five advisory committees involved.

More hopeful sound PanSTARRS (Kaiser et al.~2006), APEX (Guesten et al.~2005) and
GTC, the Grand Telescope Canarias (Hidalgo-Gomez et al.~2005).

\subsubsection{Widgets}

Some 37 papers dealt with almost as many devices.  Three green circles attached
to papers for which we said wow or why or who?  (1) A radio-band automated photometric
telescope in Japan (Kuniyoshi et al.~2006).  It has actually found a few transients, which
result in funny looking fringes.  (2) Integral field spectrscopy as a method of exoplanet
detection (Berton et al.~2006).  No, the method in general is not new, but it did seem to
blossom in the literature this year.  (3) And what Badacke-Damijni and Roselot (2006)
describe as a modern astrolabe.  As companions, we refer you to an inventory of 
pretelescopic optical instruments (Egler 2006).  The astrolabe, of course, but also the
nocturnal, armillary spheres, cross staff, quadrant, dioptra, and (on p. 193) that
funny-looking star-burst on a stick that old astronomers (we mean 15th century or something,
not ourselves) are sometimes shown holding.  The oldest Indian instrumentation seems to have
been stone and brick circular platforms,  calendar stellae, and cave paintings (Rao 2005).

What else was there?  Telescopes are, we suppose, at the heart of astronomy. Much
to our surprise, Alfred Russel Wallace thought about telescope building, including the
possibility of producing very flat glass by floating it on mercury (Smith 2006). Borra
et al.~(2006) have designed, but not yet built, a new sort of liquid mirror device, in
which a ferro fluid (covered by a reflecting layer of nanoparticles) is kept in shape
by loops of current-carrying coils. Terrestrial ones are limited to 15-44 meters (It's
hard work fighting gravity), but larger mirrors could be supported on the moon.  For a
50-m mirror, each mm of thickness requires a tank truck of fluid.  As for how to decide
how good the surface is, you may or may not be comforted to hear ({\sl PASP}  {\bf 118}, 1165) 
that ``An experienced optician can detect low-order aberratians by looking at the defocussed
image of a point source.''  Who is this experienced optician, and where was he when we
needed him?

We caught no new sorts of detectors this year, but only a sad lament of how hard it is
to do 1\% photometry with CCDs (Stubbs and Tonry 2006).

It has come, of late, to seem almost immoral to have a telescope without adaptive
optics or interferometry, especially a large one, though the first science for laser guide
star AO on an 8-10 m class mirror came only in late 2004 (Wizinowitch et al.~2006).
Artificial guide stars have, so far, normally been sodium lasers that illuminate a layer
at 90 km (and tend to worry pilots, the airforce, and users of nearby telescopes).
Lloyd-Hart et al.~(2005) describe a system with five laser beams that Rayleigh scatter at
24 km, and are time-gated to that altitude.  They didn't say whether this is better or
worse than sodium from the interference point of view.

Optical and infrared interferometry tramp forward, e.g., Perrin et al.~(2006 from
Keck) and Poncelet et al.~(2006, from the VLT).  The latter yielded ``the first direct
observation of distribution of dust around an AGN central engine with the first N-band
and VLBI observatory of an extragalactic source.'' It was NGC 1068, and another clear
candidate for the LDDFJR prize. Ofir and Ribak (2006) note that 30 years have passed
since R. Hanbury Brown tackled optical intensity interferometry from Narrabri, reaching
V = 2.5.  They say it is time to try again and think that with larger collecting area,
modern detectors, and such, one could get to m=14.4.  Do not sneer.  When we first
heard about this sort of thing, even the sun seemed a bit faint.

Does closure phase exist for nulling interferometry with three or more telescopes
(one of the schemes in the air for exoplanet imaging)?  Danchi et al.~(2006) provide
reassurance that it does. If this enables you to sleep better at night, you may have
to switch to some waveband other than optical astronomy that is done by day.
Plucking one from each end of the spectrum reveals (a) that RHESSI has seen gamma ray
polarization  of 10-20\%  for a couple of solar flares (Boggs et al.~2006), when a
gamma ray scattered in one detector is stopped in another (the same method invoked in
the GRB polarization false alarm a couple of years ago) because the direction of
scattering depends on orientation of the dominant E vector, and (b) that some radio folk
have carried out a wide-field polarization survey by undoing the interferometry feature
of the Australia Compact Telescope Array (McConnell et al.~2006).

A wow from outside the strict bounds of astronomy was the development of an ``invisibility
cloak'' by which radio waves could be bent around an obstacle and reassembled as
if there had been nothing there (Leonhardt 2006; Pendry et al.~2006). The paper is
somewhat difficult going, but it would be dishonest to pretend that the main difficulty was
the use of boldface B to mean both the magnetic induction field and the Poynting vector.
A rudimentary device was built not long after the end of the model year.

The materials gang has produced stuff with negative index of refraction (Dolling et
al.~2006; Behring et al.~2006), all the better not to see you with, as it were.
Though even the group velocity of the light is negative, apparently you cannot employ
the material to bring back your message the previous night, or ``investors'' would be
lined up lab-bench deep to enter stock prices in Morse code. Oh. Maybe the only problem
is that Morse code was voted out of existence a couple of years ago?

Widgets that everyone needs included an Ancient Egyptian mousetrap (Effland 2006)
for which, we suppose, the world will beat a path to your pyramid. And among the widgets
that no one needs, at least in science and engineering, is Powerpoint (Tufte 2006).  You
are undoubtedly certain that email is essential. Curiously, it has not much changed the
way people prioritize and respond to incoming messages (Oliveira and Barabasi 2005, after
examining the Darwin and Einstein archives).

Two devices we hope you don't think are essential: (a) pipe organs, the building of
which is restricted in Europe because they exceed limits on lead content for products
that work on electricity (Anonymous 2006h) and (b) accurate watches, since ones that
are designed to correct themselves from the GPS can drift several seconds between updates
(Huziak 2005).

On a solemn note, a technique to monitor nuclear reactors for illegal Plutonium
production saw its first test at San Onofrio (Bernstein and Bowden 2006). This happens
to be our friendly, neighborhood reactor. Friends of the late Anna Russell 
and Phil Morrison\footnote{
The Morrison connection is his having noted long ago that significant evidence for
plate tectonics had come from seismometers meant to watch for underground tests.  And
Anna Russell, of course, described two lectures on (a) how to make patchwork quilts
from old skirts and (b) how to make skirts from old patchwork quilts.  Grandmother
Farmer used to make braided rugs from old skirts and petticoats from old dish towels,
and we sincerely hope that you are well enough paid that you don't have to do either.}
will be pleased to hear that earthquake data could be used to monitor bomb tests
(Chaudhry 2006).

And for our colleagues who will soon be tooling up for another Decadal Review, a
piece of good news that synthetic aperture radar (which descends from synthetic aperture
radio astronomy - Martin Ryle and such) is useful for long-term monitoring of
post-earthquake relaxation, which can last 50-90 years (Gourmelen and Amelung 2005).

\section{COSMOLOGY} 

The universe comes at the end this year, enabling us to say we have saved the biggest
for last. Is it also the best?  Well, make up your own mind over the next 25 pages or
thereabouts, or skip to the end for our nuanced vote.

\subsection{A Child's Garden of Cosmological Models}

We freely admit to having stolen this title from Peebles (1971), who arguably
borrowed it from Robert Louis Stevenson. But Peebles meant the Lemaitre, 
Robertson-Walker sort, 
and we mean The Other Sort (conventional cosmology appearing in Sect.~12.2).
Twenty-four discrete and indiscrete names appear in the notebooks, a few covering more
than one idea, and a few ideas represented by more than one name. Some appeared in
{\sl Physical Review Letters} or {\sl Astrophysical Journal Letters} 
(two journals that, as Michael
Turner noted a number of years ago, each publish some papers that the other wouldn't
touch with a 10-foot referee) and are in or close to the main stream. Others are, again,
The Other Sort, and appeared in journals with lower impact factors.

\medskip\par\noindent
A Bianchi IX viscous fluid model in which lambda decreases with time (Pradhan et al.~2005).

\medskip\par\noindent
A seemingly nameless cosmology where the lensing probability does not depend on the total
matter density but is in rough agreement with observations (Abdel-Rahman and Hashim 2005).

\medskip\par\noindent
Some non-cosmological, intrinsic redshifts in normal spirals, so that late types are
redshifted relative to early types in clusters (Russell 2005).

\medskip\par\noindent
A Kaluza-Klein universe with an equation of state that varies with time so that as 5d
gives way to 4d, entropy is increased to the current value (Bhui et al.~2005).

\medskip\par\noindent
A scale expanding cosmology, in which all four dimensions expand, as do bound objects, and
some of the redshift comes from tired light. It explains the Pioneer effect and violates
Newton's first law (Masreliez 2005).

\medskip\par\noindent
A viscous, Kasner-type universe, with teleparallel gravity (which has appeared in 
{\sl Physical Review}) and total value of all conserved quantitites = 0 (Salt 2005).

\medskip\par\noindent
A form of extended or higher-dimension gravity that requires $J \propto M^2$ 
and predicts that
spins will decay as the universe expands, testable from the Tully-Fisher relation at
large redshift, says Wesson (2005a).

\medskip\par\noindent
A locally rotationally symmetric Bianchi I universe, filled with a bulk, viscous
cosmological fluid, in which G and c increase with time, and $\Lambda$ is negative and
decreasing (Belinchon 2005).

\medskip\par\noindent
Projective Unified Field Theory (PUFT, Schmutzer 2005).

\medskip\par\noindent
Chirality that remains from primordial symmetry breaking and determines the rotation of
spiral galaxies, the handedness of amino acids, and the difference between neutrinos and
anti neutrinos (Copozziello and Lattanzi 2006).

\medskip\par\noindent
An assortment of examples of quantized redshifts: Godlowski et al.~(2006) on the Local
Group; Bell and McDiarmid (2006) concerning SDSS QSOs (46,400 of them, so poor statistics
is probably not an issue).  The latter is contradicted by Tang and Zhang (2005), who
include Bell in their acknowledgements, making them good guys in our simple classification
scheme.

\medskip\par\noindent
The blueshift of the year, $z = -0.489$ for five optical and X-ray features in the spectrum
of the neutron star 1E1207.4-5209 (Basu 2006). And blueshifts with an (almost) 
conventional explanation (Basu 2006a), described as slingshot ejection of a 3rd black hole
(carrying its gas along for accretion) by a close black hole pair at the center of a galaxy
merger product. Somehow, however, both $z = -0.38$ and $z = -0.62$ result.

\medskip\par\noindent
A new version of the multiverse, called superstring landscape, in which we live in the
corner of parameter space where we can (Linde 2006 for, Richter 2006 against).

\medskip\par\noindent
The possibility of forming astronomical objects from fragmented macroscopic superstrings
as seeds (Brosche and Tassie 2006).  And here we depart from roughly chronological
order to tuck in a few other sorts of strings.

\medskip\par\noindent
A tangle of superconducting strings as the source of 511 keV emission from the bulge of the
Milky Way (Ferrer and Vachaspati 2005).

\medskip\par\noindent
You should probably not try to breathe a string gas (Battefeld and Watson 2006), and we
mention it largely to justify paying for a subscription to that journal, whose astrophysics
coverage nearly disappeared about the time one of us was terminated as one of its editors.

\medskip\par\noindent
With unusual specificity, Plaga (2005) surrounds the X-ray cluster Abell 194 with an
Einstein-de Sitter vacuole, invoking Einstein and Strauss (1946), and we are not sure
whether to count that as among the RMP astrophysics papers of the year or not.

\medskip\par\noindent
A solution to the cosmological moduli problem (Yokoyama 2006) and the suggestions
that it is even harder to solve than you thought (Endo 2006). We, inevitably, didn't know
we had this problem and indexed it under ``aluminum siding'', which solves another problem
we didn't know we had, according to a very persistent telemarketer.

\medskip\par\noindent
The top-down cosmology of Hawking and Hartle (2006) awarded the oak cluster, 2nd class of a
News and Views in {\sl Nature} (Bojowald 2006).

\medskip\par\noindent
Conformal gravity declared a failure for the cluster Abell 2029 (Horne 2006), because the
dispersion implies a total mass of $1.4 \times 10^{12}$ M$_\odot$ within that theory, 
but about $10^{13}$ M$_\odot$ of hot gas is required to emit the X-rays seen. 
Chances are, the primary advocate of the theory (Mannheim 2006) would not agree.

\medskip\par\noindent
A probability of less than $10^{-12}$/yr that the new generation of accelerators will produce
something that triggers our collapse into a lower vacuum state (Tegmark and Bostrom 2005).
The estimate comes from very high energy cosmic rays not having triggered the collapse
in the past.

\medskip\par\noindent
Gravity with a Yukawa potential that has a scale length close to the Hubble radius,
yielding accelerated expansion (Signore 2005).

\medskip\par\noindent
Metric skew tensor gravity, described by the authors as better than MOND (Brownstein and
Moffat 2006, the second of two papers in the year), because it fits both X-ray clusters
and galactic rotation curves without any dark matter, and reverts to Newton and Kepler
at distances in excess of 350 kpc to fit SDSS clustering data. We guess that ``better
than GR'' should be taken as given.

\medskip\par\noindent
MOND itself, on which the last word during the year was no, because dynamical friction
would have long ago dragged the globular clusters of dSph galaxies into their cores
(Sanchez-Salcedo et al.~2006).  An additional discouraging word came from Pointecouteau and
Silk (2005), who say that MOND will do for X-ray clusters only if there is a sterile
neutrino of mass larger than 1.74 eV, which is bad for the CMB.  Some supporters naturally
still support at various levels.  Sanders (2005) has modified MOND to deal with WMAP
and all by making it into a tensor-scalar-vector theory with some dark matter in the
form of very soft bosons (which don't cluster) and a prediction of the Pioneer effect.
McGaugh (2005) would prefer dark matter, if there is some, to interact repulsively
with baryons.

\medskip\par\noindent
MOND is not the same as Bekenstein's theory according to Famaey and Binney (2005)
who are not, we think, enamoured of either.  But a coveted green circle to Zhao (2005)
for noticing that the Roche geometry is different in MOND, with the lobes more squished.
This would affect only systems like satellite galaxies and globular clusters around large
galaxies, not close binary stars.

\medskip\par\noindent
And on to Bekenstein. An historical objection to MOND was that you couldn't really
predict much of anything. He (Bekenstein 2004) moved the program forward with a specific
scalar addition to gravity, enabling some predictions. The current situation seems to be
that,  with scalar, tensor, and vector  parts one might fit everything (Skordis et 
al.~2006) or at least lensing, supernova data, and CMB fluctuations, but not all with the same
values of the parameters (Zhao 2006). If the difference from GR is enough to deal with
rotation curves, there must be deviations from 1/r$^2$ in the outer solar system (Sanders
2006). And data with 10-30 cm precision would  distinguish dark matter (within relativity)
from either MOND or a Yukawa potentia1 (Sereno and Jetzer 2006).

\medskip\par\noindent
The tests for extra dimension theories proposed by Wesson (2005) sound rather similar.
There is also a possible test to separate dark energy from non-relativistic gravity from
a scalar field from VWLS (Bertschinger 2006), but the least scalar\footnote{
Least scalar meaning least like the cat who walked by himself and all places were alike
to him. Real cats are also typically much attached to places.}
author cannot claim to
understand it. The names of Brans and Dicke are still occasionally invoked in discussion
of these issues, but we don't think they would have recognized or embraced the universes
of Kim (2005) or Makamura (2006).

\medskip\par\noindent
A number of respectable brane universe alternatives appeared in the 2006 literature
associated with the names of Dvali, Randall and Sundrum, and Cardassian, on which we give
the only word to Fay (2006), who points out that, whatever such universes were like during
the first $10^{-32}$ or even $10^3$ seconds, they must now look very much like $\Lambda$ CDM.
But the cuteness award goes to baby branes (Flachi and Tanann 2005), which arise from branes
bending around small black holes (which escape into extra dimensions with brane 
reconnection). 
And we are nobly refraining from telling you about Mr. Bigger's baby (who is a little
bigger, and resurfaced in the literature after 90 years of burial with Capt'n Billy's
Whizbang.) Not, admittedly, the astrophysical literature.

\medskip\par\noindent
Loop quantum cosmology (Ashtekar et al.~2006) has a Big Bounce rather than a Big Bang. Our
notes say that its scalar field acts like a clock, or possibly a dock.

\medskip\par\noindent
``Creation lives'', though not the creation of Hoyle or Bondi and Gold, in Komiya et 
al.~(2006). The rate is about the same.  Perhaps there aren't really a lot of
alternatives to a critical density per Hubble time.  The world is not yet prepared for the
story that prompted a thoughtful colleague to wonder whether prolonged contemplation of
a steady state universe might predispose people to think that money also came from a
C-field.

\medskip\par\noindent
Another local anomaly is the local static space-time of Chernin et al.~(2006). It takes
in the Local Group, accounting for smallness of nearby velocity dispersions and lack of
infall into the LG and other nearby groups of ga1axies. That space-time could, they say, be
imbedded in a larger Friedmann universe, and be tested with lots of good distances and
velocities for nearby dwarf galaxies.

\medskip\par\noindent
There are perhaps no completely quantum cosmologies (or cosmologists) at present, but
Pfenniger and Mccione (2006) say that quantum entanglement of neutrinos
with rest masses less than 1 eV leads to higher neutrino pressure in the universe than
when the effect is neglected (by 68\% at present). Be careful, therefore, not to
over-inflate your bicycle tires with neutrino gas, lest they come to resemble the
scones of footnote$^{28}$.

\subsection{The Conventional Universe}

You don't need to be told that there was a giant press release based on the results
of three years of observation with the Wilkinson Microwave Anisotropy Probe (Page 2006;
Bennett 2006; Spergel 2006). The most notable changes from the first year were modest
reduction in all of (a) the normalization of density contrast, $\sigma_g$, below one, (b)
the spectral index of the initial perturbation spectrum, $n$ down to 0.95 or so, and (c)
the optical depth to electron scattering of the post-recombination CMB, meaning that
reionization doesn't have to start until $z = 12-15$ or so (vs.~17). All have been
incorporated into numerous papers, quite a few of them published, without the actual WMAP3 
package appearing in the journal to which it was apparently submitted.

The concordance/consensus model is, on balance, safe (Bennett 2006a, and the next four
papers), or as Carroll phrased it (2006), ``we are stuck with the Universe we have.''
On various occasions, it is the better part of wisdom to remember this in connection with
the students who come to our classes, the public that reads (or does not read) our
outreach efforts, and so forth.

Is there anybody against this standard model universe? Yes, of course, and in
accordance with our desire to be fair to the 2+2=5 party, we note Lieu et al.~(2006), who
found, by cross-correlating the WMAP1 data with X-rays that there is no evidence for
Sunyaev-Zeldovich upscattering of microwave background photons passing through clusters.
If you would like some reason to be a bit frightened by this, we warn you that the same
group appears to have been right about being able to rule out certain kinds of space-time
foam with interferometer observations of distant sources (Christiansen et al.~2006;
Pearlman 2006).

At least two papers denied evidence for a cosmological constant arising from observations
of supernova brightnesses as a function of redshift. Schild and Dekker (2006)
prefer reduced transparency of intergalactic space due to Lyman $\alpha$ clouds nucleated on
dark matter primordial planetesimals.  Balazs et al.~(2006) prefer internal extinction in
the host galaxies, and we dare not ask how they make their dust. But an earlier method
of lambda-disparagement has collapsed before the siege engine\footnote{
This at an intermediate stage came out as a search engine, and indeed
we are planning to develop one to seek out and destroy both medieval castles
and medieval cosmologies (i.e., all but one of those mentioned in this section).} 
of additional thought. Horest
et al.~(2005) point out that large scale lensing comes out OK after all with the correct
choice of N(M) of clusters vs. redshift.

Meanwhile, traditional ways of measuring and modeling the universe soldier (sailor,
Marine, and airforce) on. Only 8 values of the Hubble constant appeared, ranging from
50 (Van de Steene et al.~2006) to 77 km s$^{-1}$ Mpc$^{-1}$ (Bonamenta et al.~2006), 
the latter from the just disparaged Sunyaev-Zeldovich effect.

If you would like to know why $\Lambda$ is small, Steinhardt and Turok (2006) and Vilenkhin
(2006) provided a cyclic and an anthropic argument respectively for why most galaxies should
form in regions and times where/when $\Lambda$ is comparable with the matter density. 
Lambda is
non-negligible from the length scale of clusters upward (Balaguear-Antolinez and Nowakowski
2005). No special desire to disagree with the main conclusion, but once they had
invoked Newton-Hooke space-time they could at least have provided appropriate citations.
Is that ``our'' Newton and ``our'' Hooke?

The normalization constant, $\sigma_g$ was the most-papered parameter of the year (12, 
some in
combination with the power spectcum slope, $n$, and total matter density). The idea is the
rms density fluctionation on a scale of 8 h$^{-1}$ Mpc, and the scale was originally chosen to
yield a value near one.  The first and last published during the year happen more or less to
agree with each other and most of the rest on $\Omega_m = 0.24$ and $\sigma_g = 0.85$ 
(Tkinder et al.~2005; Stanek et al.~2006), so that is all you get.

As for $n$, considerations of large scale structure (especially from the Sloan
Digital data bases) hit below 1.0, independent (more or less!) of WMAP3 at, e.g. 
$0.95\pm 0.04$
(Viel and Haehnelt 2006). It helps, of course, that the theorists have given us
a new, lower target to aim at, 0.95 to 0.98 says Boyle (2006).

Bias is the extent to which light doesn't trace mass, and, once you have a large enough
data base, you would undoubtedly join Pollo et al.~(2006) in concluding that it varies with
redshift and with the luminosities of the galaxies you look at, not always in quite the
ways you would have guessed (Yahata et al.~2005).

Among the cosmological concepts, ink was spread most widely over reionization (15
papers) and ``where are the baryons?'', if the total number is set by considerations of
Big Bang nucleosynthesis (16 papers).  Reionization means what happens between 
(re)combination at $z$ near 1000 when essentially all the gas went neutral and now, when 
intergalactic space is quite transparent (ionized) most places. It is probed by electron
scattering of the microwave background, of which there is rather more than we had come
to the party expecting.  This means that somebody needs to make quite a lot of UV photons
fairly early (Pop III stars in proto-galaxies at first; AGNs later is the majority view)
and get them out away from their sources. A review by Barkana (2006) presents no 
show-stoppers, so we won't either.

Where are the baryons that are not obviously in stars and gas in galaxies?  Well,
some each in damped Lyman alpha systems (Rao et al.~2006)\footnote{
It never ceases to astound us that the damping is authentic, quantum mechanical
broadening of the lines, a case where astronomy on large scales meets quantum mechanics
on small scales without anybody getting angry.}, warm/hot Lyman alpha
clouds (Richeter et al.~2006), in gas that has been blown back out of ellipticals (86\%
of what they started with say Hoekstra et al.~2006), and, in a sort of symmetry, rather
more in the coronae of spirals than you might have thought (Fukugita and Peebles 2006).
But you are encouraged to keep hunting in your favorite corner, under the rugs, etc.

Finally we note a few cases where the cosmological effects are just what you would
expect, and a couple where they are not. There is aberration of the CMB, 20\arcsec same as
for everybody else (Burles and Rappaport 2006); and it was hotter in the past, 
$7.2 \pm 0.8$ K
at $z = 1.776$ (Cui et al.~2005). On the other hand, the dependence of angular
diameter distance on redshift nearly eliminates 21 cm absorption at $z > 1-2$ (Curran and
Webb 2006). And the best-buy age of the universe (13.7 Gyr or so) crowds awkwardly
against a l6 Gyr estimate for the age of the globular cluster M92 (Rue las-Mayorua
and Sanchez 2005). But the final green dot goes to Mukherjee et al.~(2006) for pointing
out that, at least in principle, choosing the best model to describe an assortment of
data, where the models can have different numbers of physical quantities in them, is not
the same as parameter fitting.

\subsection{Distance Indicators and Gravitational Lensing}

Supernovae and lots of other bright things provide luminosity distances, though we
have picked a bit of bad news, that by $z = 1.5$ gravitational lensing degrades the number
of useful supernovae by a factor near three (Holz and Linder 2005). Hellaby (2006) has
an angular diameter distance of the following sort: ``The maximum in the diameter  
distance is the only point on the past null light cone that corresponds to a 
model-independent mass.''
The angle for a fixed physical diameter has a minimum at the same 
redshift for all relativistic models, and it can be used for various cosmic, or at least
cosmological purposes. The author is the first to want to use this for anything, but the
concept was known to Hoyle at the time of the 1960 Varenna Summer School, and the
correct equation for an Einstein-de Sitter universe appears in McCrea (1934).

It is still just barely possible to publish a discussion of the Cepheid distance
scale that disagrees with the Hubble Space Telescope Key Project version. Paturel and
Terrikorpi (2006) and Saha et al.~(2006) both tend to decrease H.

M31 now has a distance from ana1ysis of a single eclipsing binary (Ribas et al.~2005).
It is $772\pm 44$ kpc, and you should not adjust your Hubble meter until a few more are in
the binary basket.

Of cosmic significance only if you had doubted the conventional wisdom is a QSO in the
field of the Magellanic Bridge.  Its spectrum shows absorption features at the wavelengths
expected for the Bridge gas (Smoker et al.~2005). This also applies to direct evidence
that things have moved apart from each other with time, deduced from statistics of the
Ly $\alpha$ forest between $z = 2$ and 4.5 (Rauch et al.~2005).

The first item on gravitational lensing is bad news $-$ even at redshifts less than
1, one-third to one-half of the structures analyzed have interlopers belonging neither to
the intended target nor to the intended lens (Momchesa et al.~2006 on strong lensing and
Clow et al.~2006a on weak lensing).  Conversely, as it were, lensing by large scale
structure can mimic clustering of QSOs (Scranton et al.~2005). They call this cosmic
magnification. And one Mikado\footnote{
``The most dreadful thing has just happened. It seems that you are the son of the Mikado.
Yes, but that happened some time ago.''}
item. Ivanova and Khovanskaya (2005), in a discussion
of how lensing can interfere with cosmological tests refer to a calculation by
Zeldovich (1964).  This is unambiguously before the completion date of the thesis of
Gunn (1967), which we had always supposed and sometimes said was the first treatment of
the topic.

\subsection{Dark Matter}

We recorded about 23 candidates this year and present them in the order (a)
familiar and ``generally recognized as distinguished'' (a possible criterion for election
to membership in the Cosmos Club), (b) familiar but unlikely (often in the form of
upper limits), and (c) unfamiliar.  The edges are somewhat fuzzy.

The classic WIMP, if its mass is near 60 GeV c$^{-2}$, has a cross-section for interaction
with silicon or germanium less than $1.6 - 30 \times 10^{-43}$ cm$^{2}$ 
(Profumo et al.~2006; Fayet
et al.~2006), though a firm lower limit on the mass is considerably smaller, at 10 MeV or
so. If small compact galaxies are all $3 \times 10^{7}$ M$_\odot$, then the particles whose 
gatherings
host them should have a mass three times that of the proton (3 GeV) according to Gilmore
(2005). If the gamma ray excess seen by EGRET at $0.1-10$ GeV is a product of WIMP
annihilation in the halo and disk of the Milky Way, then $50-100$ GeV is the likely WIMP
mass (De Boer et al.~2005a,b).

Axions have perhaps been seen in the lab say Zavettini et al.~(2006) and Lamoreaux
(2006). The manifestation is rotation of the plane of polarization of 1.064 $\mu$ laser
light in a 5 Tesla magnetic field. The rotation is $4 \times 10^{-12}$ radians. If this 
is due to
axion production, then a very small mass is fine, but the interaction strength is very
much larger than expected. The experiment sounds eminently repeatable.

SuperWIMPs were the official UCI candidate in {\sl Ap 03}, because they can remove excess
Li$^7$ left from the early universe (Sect.~5). This year they are back as a solution to
the problem of excess small scale structure in models of galaxy and cluster formation
(Cembranos and Orlem 2005).

Candidates that are not the whole banana, or even the peel, include:
\begin{itemize}
\item[o]{White dwarfs (Pauli et al.~2006).}
\item[o]{Strings, but data in favor of the existence of one as a gravitational lens 
(Sazhin et al.~2006).}
\item[o]{Primordial black holes. Titarchuk and Chardonnet (2006) suggest some around 
Sgr A$^*$
as a source of electron-positron pairs, but would be happy to make the whole Galactic
halo out of PBHs.}
\item[o]{Intermediate mass black holes of $10^3-10^5$ M$_\odot$ are less than 1\% of the 
mass of the Milky 
Way found in baryons, from the absence of X-rays when they cross the disk (Mapelli et
al.~2006).}
\item[o]{MAssive Compact Halo Objects (MACHOs), which could be as much as a quarter 
of the Galactic halo,
and, if so, have masses of 0.3-0.8 M$_\odot$ (Holopainen et al.~2006). For what it is
worth, the mean mass from microlensing in the host of QSO SDSS 0924+0219 could be
anything in the range 0.015 to 1.0 M$_\odot$.}
\item[o]{Limits on the MACHO components of both the Milky Way and M31 come from 14 events
(Ingrosso et al.~2006; De Jong et al.~2006), though the actual light curves are a bit
difficult to fit with either self lensing in Andromeda or lenses in the MW.}
\item[o]{Gravitational radiation at frequencies to which LIGO is sensitive 
(Abbott et al.~2005).}
\item[o]{Sterile neutrinos of a few keV, favored for their contributions to 
pulsar velocities
and to early reionization (Biermann and Kusenko 2006; Goerdt et al.~2006).  Now we are
virtually certain that we caught during the year two papers setting contradictory upper
and lower limits to the masses of such neutrinos from other considerations. The upper
limit was 14 keV (Mapelli and Ferrara 2005), and the lower limit was larger, but darned
if we can find the paper, though somewhere in the 163 page index, it has a green circle.}
\end{itemize}

\medskip
We are approaching the surface between ``OK if you must'' and ``not in my journal'' with
Ahn and Shapiro (2006) on self-interacting dark matter with a very large cross-section
of 200 cm$^2$ g$^{-1}$, and on beyond to
\begin{itemize}
\item[o]{More than weakly interacting DM (remember SuperWIMPs are less than weakly) put 
forward
by Chuzhoy and Nusser (2006). The former is said to be good for structure formation,
the latter for reheating of cooling flows when protons scatter off the DM particles.
This was meant to say the former and latter of low and high cross-section DM, not the
former and latter of Chuzhoy and Nusser.}
\item[o]{Annihilating dark matter as the source of the radio halo around the Coma cluster
(Colafrancesco et al.~2006).}
\end{itemize}

\medskip
And, as we roll off the edge,
\begin{itemize}
\item[o]{Boson fermion stars of masses from $10^{18}$ g up to whole galaxies (Henriques 
and Mendes 2005).}
\item[o]{Baryons compressed in 20 cm chunks of a different, second phase of the vacuum, 
degenerate with ours (Fruggatt and Nielsen 2005).}
\end{itemize}

\medskip
Press releases and green circles showered on the X-ray double cluster IE 0657-558
this year, when Clowe et al.~(2006) announced that the full range of observations could be
fit only with some non-dissipational form of dark matter. The two groups of galaxies look
very much as if they had passed through each other, leaving most of the gas in the middle.
And a weak lensing map says that most of the mass is where the galaxies are. A rebuttal
(Zhao 2006) says that some of the non-GR theories of gravity also work, including TeVeS
(relativistic MOND) and MOG.

Whatever the dark stuff is, there are a few galaxies made only of it according to
Karachentsev et al.~(2006a).  The evidence is a few otherwise isolated galaxies that appear
to have been gravitational disturbed by a neighbor. But the luminous outnumber the dark
by at least 20:1. We suppose that the dark galaxy is probably also disturbed and distorted,
but aren't quite sure how to check this.

\subsection{Dark Energy}

The need for a cosmologica1 constant, quintessence, dark energy, or something else
that is currently accelerating the expansion of the universe and adding to the matter to
yield total flatness appears in essentially all the concordance data sets.  Candidates
come and go. In 2006, k essence was out (Bonvin and Durrer 2006), and various scalar fields
were in. Some of our favorites:

One that interacts with neutrinos so that the neutrino mass will vary with time
(Brookfield et al.~2006).

The sort with P/$\rho$ different from $-1$ (in c = G = 1 units) and time variable, so that
(a) dark matter could win in the long run and deceleration set it (Carvalho et al.~2006) and
(b) energy is not exactly conserved in spherical collapse to galaxies, because the dark
energy doesn't virialize with the dark matter (Wang 2006), (c) source counts as a function
of redshift are affected (Nunes et al.~2006), and (d) the dark energy can cluster and
affect structure formation (Perciva1 2005).

The line between some of the more imaginative dark energy candidates and alternative
theories of gravity strikes us as a bit fuzzy.  Mena (2006) says that a generalized
modified gravity along the lines suggested by Carroll et al.~(2004) can replace dark energy,
but you will still need dark matter.  You must not let this drive you back to the
previous section, or we will never get home for the weekend.

\subsection{The Backgrounds}

Yes, of course there is the 3-degree, isotropic, microwave, cosmic, relict, thermal
background radiation.  But there are also backgrounds in every photon band that can be
explored, not to mention in neutrinos, gravitational radiation (we suppose), and cosmic
rays (next subsection). The two green circles live at long wavelength radio and short
wavelength gamma addresses. But let's dispose immediately of the upper limits to high
energy neutrinos (Barwick et al.~2006). This was a test flight of the downward-looking
ANITA balloon project.  Why look down?  Because they want to catch photons emitted when
neutrinos hit Antarctic ice. Thus the project will have to be completed while there still
is some Antarctic ice! And the primordial gravitational radiation background contributes
less than $10^{-5}$ of closure density or we would already know about it from distortions 
to the
CMB (Smith et al.~2006), seeing which is part of the goal of current and future missions.
These also illustrate the fundamental distinction between backgrounds that are the sums
of sources (high energy neutrinos from AGNs, GRBs, and such) and ones that are truly
diffuse, like the CMB and the primordial gravitational radiation.

Our low frequency circle falls at 178 MHz and (the circled aspect) is old enough
that it may well have originally been reported as pertaining the 178 Mc/s (Bridle 1967),
with special thanks to Dwek and Barker (2005) for pointing out that this 40 year old
number is still the best available, $32 \pm 8$ K. And this is perhaps the least worst place
to mention that the most cycled author\footnote{
Not quite the most lunar cycles, perhaps, but the most east-west cycles, since she has
moved cross-country 68 times since receiving her PhD in 1968, and had never moved at
all before that. Is there a cycling club for this, and do they eat scones?} 
has provided an overview of the discovery and interpretation of all the sorts of
backgrounds she could think of (Trimble 2006). 

Half a dozen or so AGNs with redshifts up to $\approx 0.2$ have been seen at energies
approaching and exceeding 1 TeV, and one supposes that there would be a background due to
the sum of distant sources if they could been seen. But they cannot. Their photons meet
intergalactic infrared (mostly) photons and make e$^\pm$ pairs. This is called the GZK 
(Greisen
1966; Zatzepin and Kuzmin 1966) effect. Indeed there is some difficulty in understanding
how even the UHE gammas we do see get here if there is any more intergalactic IR than
that due to known galaxies (Aharonian et al.~2006c, on H.E.S.S. sources at $z = 0.165$
and 0.185, Massaro et al.~2006).

We have several reasons for concluding that all is probably well. First the sources
at larger redshifts have apparently softer spectra in proportion, as you would expect
(Schroedter et al.~2005) and only out at 3-6 TeV do you see intrinsic turn-downs in the
lower-z spectra due to the Klein-Nishina cross-section cutting in (Massaro et al.~2006).
Second, some of the IR previously credited to the intergalactic background is probably
zodiacal light improperly subtracted (Stecher et al.~2006). And, third, one actually
sees some attenuation of 20-50 TeV photons within the Milky Way due to the (arguably
better known) IR background here (Zhang et al.~2006) and even in microquasars, some of
which are also TeV sources (Bednarek 2006).

A bit down in photon energy, GeV photons have to fight ultraviolet and X-rays in
both black hole and neutron star X-ray binaries (Dubas 2006). The GeV background is
virtually always said to be the sum of (mostly) AGNs plus a bit of emission from normal
galaxies (Pavlidov and Fields 2005; Bhattacharya and Sreekuman 2005; Stawarz et al.~2006).
 A modest
fly in that ointment is the conclusion (Narumoto and Totai 2006) that only 1/4-l/2 of the
unresolved background can be unresolved AGNs. This in turn has, however, a good deal of
uncertainty arising from the half of EGRET sources that are still unidentified, so that
extrapolation to larger redshift is necessarily done using counts of X-ray or radio active
galaxies. GLAST still won't resolve everything, Narumoto and Totai conclude, though
Giommi et al.~(2006) indicate that just the Blazars could provide 100\% of what we see at
0.5-500 MeV.

In the INTEGRAL regime of 100-150 keV, only 3\% of the background is resolved down to
a flux level of 1 mCrab (Bazzano et al.~2005), and indeed it seems that there are not
enormous numbers of sources - 213 in a sample presented by Bassani et al.~(2006).

X-ray has historically meant 10 keV down to 2 or 0.2 keV. Any overview must be
prefaced by a reminder that there is nearly a factor of two disagreement about the
normalization, from 1.57 to $2.35 \times 10^{-11}$ erg cm$^{-2}$ deg$^{-2}$ 
(Revntsev et al.~2005). But a
relatively safe summary, once you have summed Chandra images and the deepest surveys, seems
to be that above 8 keV less than half the background is resolved; down around 2 keV all or
most of it is, and at 1 keV you may even have too much coming from AGNs if you want to let
many of the baryons be in a Warm-Hot Intergalatic Medium.  Those words are an imperfect
summary of the following conclusions, not all concordant:

\begin{itemize}
\item[o]{The 2-10 keV range needs an additional population of sources (normal or star burst
galaxies?) or truly diffuse emission (Hickox and Markevitch 2006).}
\item[o]{Stacked images of Chandra and XMM fields containing norma1 galaxies add up to less 
than 40\% of the missing 6-8 keV flux but most or all of the 0.5-6 keV flux.}
\item[o]{The WHIM ought to contribute $1.6 \times 10^{-12}$ erg s$^{-1}$ cm$^{-2}$ deg$^{-2}$
at 0.5-2 keV but the upper
limit allowed by observations is only $1.2 \times 10^{-12}$ erg s$^{-1}$ cm$^{-2}$ deg$^{-2}$
(Roncarelli et al.~2006)}
\item[o]{Above 8 keV is largely unresolved, though AGNs are still a good bet (Peterson et 
al.~2006)}
\item[o]{A new class found by stacking XMM images of SDSS galaxies, which has anyhow the 
right, hard, spectrum (Georgakakis et al.~2006).}
\item[o]{And you must remember to include both type 1 and type 2 sources 
(Wilkes et al.~2005)}
\end{itemize}

Ultraviolet can mean photons able to ionize hydrogen and even helium, in which case
from now back to $z$ at least 3, they really all come from AGNs, according to two clever
papers whose authors have managed to reconstruct the ionizing spectrum by looking at a
bunch of transitions from different ions in QSO absorption systems (Reimers et al.~2006,
who find a he1ium Gunn-Peterson trough, and Agafonava et al.~2005). Takeuchi et al.~(2005)
report a local UV energy density of $2.7 \times 10^7$ L$_\odot$ Mpc$^{-3}$, vs. $10^8$
for infrared, based on
galaxies observed with GALEX, IRAS, and the Spitzer Space Telescope. Both energy densities
were larger at $z = 1$ than now, and the IR/UV ratio then was 15:1 vs. 4:1 now. The 
implication is that 50-85\% of star formation is hidden over this redshift range.

Ultraviolet can also mean softer stuff from the Lyman limit to 1750 \ang . In that case
then it all comes from the Milky Way (Edelstein et al.~2006, followed by eight more letters
on results from the same mission, the first (south) Korean satellite, STSAT-1, launched
on 2003 September 27, though by whom they did not say).

There must be visible light between the galaxies that can be estimated by adding up
all the galaxies in the 6dF or SDSS survey (no, of course they don't quite agree).
Heath Jones et al.~(2006) say $2 \times 10^8$ L$_\odot$ Mpc$^{-3}$ in visible light, 
and up to $9 \times 10^8$ in
infrared, tiresomely more than the numbers we just reported from Agafonova et al.~(2005).

As a result, pride of place and the green circle go to the energy density in starlight
within the Milky Way. Just add up all the stars, though Zagury (2006) wickedly did not
acknowledge the authors they say they are refuting (Gordon et al.~1998). The result of the
addition is a radiation density of equivalent black body temperature 4.212 K (Pecker and
Narlikar 2006). The number is notable in two ways $-$ it is not much larger than the one found
by Eddington (1926, 3.18 K).  And, second, it is within striking distance of the current
CMB temperature, which, you will not be surprised to hear, these authors think might be
significant.

Some more biggish infrared numbers, pushing the total toward $10^9$ L$_\odot$ Mpc$^{-3}$
(e.g.
Loaring et al.~2005) left us eager for a definitive Spitzer result, given that it is said to
resolve about 3/4 of the mid and far IR (Dole et al.~2006). May we have the envelope,
please ... 24 nW m$^{-2}$ sr$^{-1}$. Well, we complained about the units used for such 
things in {\sl Ap 05},
so it is up to you to figure out whether this is more or less than 160 drachma. Another
10 or so infrared background papers are left on the floor of notebook page 73, except that
we circle back to the UHE gamma ray issue to note the SST conclusion of Kashlinsky et 
al.~(2005), that the IR includes some photons from population III stars. The commentary by
Ellis (2005) allows ``conceivably,'' given the problems of subtracting zodiacal light and
Galactic cirrus.

Submillimeter these days generally still means SCUBA, and the good word is that the
modest number of sources seen, if extrapolated below the confusion limit, takes care of
nearly all the background.  (Knudsen et al.~2006). The sources are clustered (Scott et 
al.~2006) and include a mix of extremely red galaxies, obscured AGNs, and other (Knudsen 
et al.~2005; Lutz et al.~2005).

And sinking back down to millimeter photons, we find that very little of the background
has been resolved (Maloney et al.~2005) but that Blazars would seem to add 5-10 K to the
temperature of the CMB.

\subsection{Cosmic Rays}

Suppose with the large majority (different from the ``great majority,'' who are dead)
of our colleagues that cosmic rays with energies small enough to be confined by the
Galactic magnetic field (less than about $10^{15}$ eV) are indeed Galactic and 
accelerated here
by some combination of supernovae and shocks associated with their expanding remnants. The
details of how they propagate through the Milky Way and eventually leave, on time scales
of millions of years, remain in the MWIM category (Shibata et al.~2006; Ptuskin et 
al.~2006) because the current models do not entirely agree with observations of the 
ratios of
secondary to primary nuclides vs. energy. LiBeB/CNO is the best known, but in general,
odd/even ratios like F/Ne are enhanced because of spallatian en route.  Aublin and
Parizot (2006) have put forward what they call a holistic model for acceleration of all
cosmic rays, in which a single sort of source always contributes particles with 
$N(E) \propto E^{-2.3}$.
The low energy ones are our own, and the higher energies are the sum of leakage
from all galaxies. This would seem to predict constant composition across the entire
energy range, which is manifestly not true, leading to the descriptor ''odd'' in our notes.

Most of the ink and electrons expended during the year concerned the ultrahigh energy
cosmic rays, $\approx 10^{19\pm1}$ eV, their total numbers, arrival directions, and origins, 
driven by
worry that they will find it hard work travelling far through the cosmic microwave 
background
radiation. The problem was known to Zeldovich (1965), which discouraged him from
considering a hot big bang at a critical moment in cosmic, or at least cosmological 
history.

An important step forward this year was the first report of data from a new detector
called Auger (for Pierre, Anonymous 2005j). The geographically most challenged author
apologizes for having moved it, in a publication in Another Journal, from Argentina to
Chile (for which no arrival direction is relevant).  The highlight here is that, in
preliminary fashion, Auger reports a flux on the low side of the range found by previous
experiments, considerably alleviating the transport problem. Meanwhile, back in Utah
(which we have so far not succeeded in moving to Nevada, though we suspect both places
might benefit)\footnote{
It was Will Rogers who described the dust bowl migrations of the 1930s United States
as having raised the IQ levels both in the places they left (the Ozark mountains) and
the place they settled (California). Collective not individual IQs are meant, which is
presumably also the case in the claim of Lake Woebegone that all their children are above
average.},
Japan, the Crimean peninsula, and so forth, there were several reports
of correlated arrival directions, but not the same directions from any two detectors
(Uryson 2005; Abbasi et al.~2006; Farrar et al.~2006).

Galactic cosmic rays have at least two jobs - causing mutations and heating dark, dense
interstellar clouds.

\subsection{Very Large Scale Scructure and Streaming}

Investigations can begin with visible galaxies, X-ray sources, radio sources (which
in the days of 3C weren't clustered at all), or QSO absorption clouds, but all of these
require the universe  to be reasonably transparent, and so fail until the era of 
reionization 
($z = 6-17$ or thereabouts, depending on just what you want to look for). To trace
the development of structure from recombination at $z = 1000$ (which we see in the CMB) down
to the oldest galaxy found so far (redshift just a smidge less than 7, Iye et al.~2006), the
only strategy so far devised is to look for very weak absorption or emission at wavelengths
longer than about 140 cm, arising from redshifted neutral hydrogen slightly out of equilibrium
with the CMB (Hirata et al.~2006; Furlanetto 2006; Furlanetto et al.~2006a).

The corporate memory of astronomy is good enough to trace the idea to Field (1959,
and a year earlier in a non-astronomical journal Field 1958), but it is owing entirely to
his gentlemanly insistance on citing a very short AAS meeting abstract (Wouthuysen 1952)
that current writers speak of the Wouthuysen-Field effect and that we (in our usual
plonking way) were able to find Wouthuysen's real paper (Wouthuysen 1952a). In it he
promises a fuller discussion soon, but seems never to have published this. It was far
aField from most of the rest of his work, about which you can learn by prowling through
{\sl Physics Abstracts} from the late 1940s to late 1950s and reading the papers (a web site,
generously excavated for us by Jonathan Pritchard, has rather little).
We wasted a whole afternoon this way; so green circles to all who are working on this very
difficult topic (which is one of the drivers for LOFAR etc.), for George Brooks Field,
Siegfried A. Wouthuysen, and the librarians of UCI who, very reluctantly, consent to keep
things like {\sl Physics Abstracts} and journals more than half a century old on our shelves.

As for the rest, a summary of the good news appears in Bland-Hawthorn and Peebles (2006).
It is that the most advanced simulations (e.g., Springel et al.~2005) look very much like
data from the Sloan Digital Sky Survey out to $c z = 25,000$ km s$^{-1}$. The mass points 
in the calculation ($10^{10}$) currently outnumber the galaxies by a sizable ratio, but a big model
galaxy has thousand of model point masses in it. A computational result that will be
more difficult to check is that, of the initial point masses, 99\% end up in sheets, 72\%
in filaments, and 46\% in halos, which are generally triaxial (Shen et al.~2006a), though an
out-of-period stab at tracing the three-dimensiona1 distribution of dark matter via weak
gravitational lensing is clearly a start.

We begin up close and personal (on the grounds that the Local Group is as much entitled
to personhood as are corporations like General Motors, whom we have known ever since he was
a Boy Scout), the local flow is remarkably cold (that is, has only very small deviations
from smooth Hubble expansion) out to 10 Mpc (Karachentsev et al.~2006), which Teerikorpi et
al.~(2006) attribute to the effects of dark energy. (We won't vote on that one.). The
limit from SNe Ia is a good deal less restrictive ($\sigma_v$ less than 486 km s$^{-1}$, 
Wang et al.~2006d)
but remarkable that it can be done at all!  At least as distressing as the local
chill is the apparent misalignment between the directions of peculiar velocities and the
structures that supposedly cause or arise from them (Whiting 2006).

In contrast, by the scale of the Virgo supercluster, peculiar velocities (called
``infall'' in this context) reach beyond 1000 km s$^{-1}$ and imply that the cluster must have
M/L larger than that of our nearby spiral-dominated field (Mohayaez and Tully 2005).

The next step out, 100-300 Mpc, takes us to the Shapley Concentration and the Great
Attractors, both of  which continued to exist and indeed grew in number of members during
the year.  They, to a considerable extent, account for our nearly 600 km s$^{-1}$ motion 
relative
to the rest frame of the background radiation, though there was some disagreement about
how far out you have to look to see all the relevant stuff. The units will be h$^{-1}$ Mpc,
so that the Great Attractor is out at 80-100 h$^{-1}$ Mpc, more or less crossing the 
Galactic
Plane, with the larger Shapley Concentration further out in roughly the same direction
(a difficult one in which to look in visible light!), and so, we are told,

\begin{itemize}
\item[o]{The dipole converges (that is, there is enough mass to account for our peculiar 
motion) by 60 h$^{-1}$ Mpc (Erdogdu et al.~2006).}
\item[o]{Well, for the most part by 100 h$^{-1}$ Mpc (Romano-Cruz et al.~2005).}
\item[o]{The Great Attractor dominant structure is filamentary and extends beyond 
150 h$^{-1}$ Mpc (Radburn-Smith et al.~2006).}
\item[o]{Nagayama et al.~(2006) report the second biggest cluster in the GA after Abell 3627,
lying in the zone of avoidance and found with X-ray and infrared techniques.}
\item[o]{The Shapley Concentration includes 44 recognizable clusters and as much more material
in intercluster galaxies, enough to affect Local Group motion; the morphology
is roughly flat (Proust et al.~2006).}
\item[o]{Ragone et al.~(2006) up the ante to 122 systems of galaxies of $10^{13}-10^{15}$
h$^{-1}$ M$_\odot$ each.}
\item[o]{Kudrya et al.~(2006) conclude that 60\% of our peculiar motion is due to the SC.}
\item[o]{Kocevski and Ebeling (2006), after considering 810 X-ray clusters, roughly, concur,
attributing 44\% of our dipole to the GA, and 56\% to more distant material at 130-180
h$^{-1}$ Mpc, about half of which is in the SC.}
\end{itemize}

There were a dozen papers on the development of VLSSS with redshift.  Perhaps the
right thing to say is that, like the time the alarm clock sounds, it gets a little earlier
every year.  Examples include Zheng et al.~(2006) on a $z = 5.8$ QSO with associated
galaxies; Overzier et al.~(2006a) on the most distant radio galaxy with $z = 5.2$ and a
cluster of Lyman alpha emitters around it; and Kashikawa et al.~(2006) and on $z=4-5$
pre-clusters that already host multiple bright Lyman break galaxies.

Evolution with redshift of the clustering would seem to be spottier than models
predict (or perhaps we have not yet found quite the right probes), Hildebrandt et 
al.~(2005) say there was almost none in comoving cluster scale length between $z = 4$ and 3,
while from $z = 1$ down to the present, weak lensing shows the expected change in matter
power spectrum (Bacon et al.~2005).

The topology you see with baryonic tracers depends on the level of density contrast
explored relative to the cosmic average.  Bright galaxies live in meatballs, with faint
galaxies in a more bubble-void like structure (Part et al.~2005c), and the less dense
clouds of the Lyman alpha forest in pancakes and voids (Demianski et al.~2006), like
those originally hind cast by Zeldovich (1970). The void shapes are more complex than
spheres or ellipsoids (Shandarin et al.~2006).

Alignments of neighboring entities is another marker for large scale structure.
Satellites, for instance, are typically elongated toward hosts (Agustsson and Brainerd
2006).  Red galaxies point at other red galaxies (Donoso et al.~2006, well, we never told
you galaxies were color blind).  Satellites prefer prograde orbits both observationally
(Azzaro et al.~2006) and theoretically (Warnick and Knebe 2006).  We really don't know
what to make of alignment of quasar (radio) polarizations on gigaparsec scales.  The
authors Hutsemekers et al.~2005) say it could be the rotation of the universe, but they
favor dichroism and birefringence due to photon pseudoscalar oscillations in a magnetic
field.  The po1e of their alignment is roughly the CMB dipole, which, you heard a few
paragraphs ago is mostly a local effect of lumps of matter, leading us to suspect
some tiresome local source of error in these observations.

On the very largest scales, distribution of the SDSS galaxies shows the same baryon
acoustic peak in its correlation function at 100 h$^{-1}$ Mpc as is found in the relict
radiation (Eisenstein et al.~2005), enabling us to say that we live in the best of all
possible worlds, at least on the largest scales.

\section{MISTEAKS WERE MADE} 

And, truth be told, some of them were made by us, and the section, as usual,
begins with those.  It continues with items from the astronomical and related literature
intended to amuse, instruct, and enlighten (if only on what to avoid), and ends with
quotes from some favorite colleagues.

\subsection{We Done It} 

Normally, this would have two subsections, full-fledged errors and differences
of opinion.  As there was only one of the latter this year, it is included with the
definite goofs, in section order.

Sect. 3.2.1: Is Pluto a planet?  We wrote, of course, before the IAU decision
in August, 2006. But an expert theorist, who has been involved in decíding just which
bodies truly dominate their regions of the solar system, suggests that, ``the best
analogy is one of immigration policy. We admitted Pluto into the family of planets
because we thought it was the right thing to do, and now we've discovered that he has
hordes of unwashed cousins right outside our borders who want to come in as well.

Sect. 5: The section heading is a take-off on Henny Youngman's line, ``Take my
wife, please.''  Not, you may say, a kindly sentîment to perpetuate, but a colleague
who remembers the era suggests that the wife in question may have been nagging at her
spouse about some domestic issue just before he was supposed to go on stage and be on
his mental toes for 20 minutes or so.

Sect. 5.3 on dark energy and its equation of state provoked dissappointment in an
uncited (but very highly regarded) colleague.  He drew our attention to three papers
none of which was, or could have been, in the data base: an astro-ph carrying a 2006
date, a paper in Physical Review D (not one of the 21 fully read, or 12 partly-read
journals), and a letter to Monthly Notices of the Royal Astronomical Society after
they eliminated Letters from their paper journal.

Sect. 5.4 on distance scales: The cluster responsible for the large value of the
Hubble constant was  Abell 611 not Abell 64.   Apologies to both Max and George.

Sect. 9.  We cited a 1959 JETP paper by Parenago and are told it should have been
a 1950 {\sl A. Zh.} paper ({\bf 27}, 150). 
The information came from one of the world's true mavens
of astronomical literature archiving and retrieval and so is very probably correct, but
we are not quite sure that JETP is in the relevant archive.

Sect. 9.5.1: ``There were general agreement about ...''  Oh, there were, was there?
Caught by us while looking for the Paranago paper.

Sect. 9.9.5: A nucleosynthetic expert notes that $^{45}$Sc is the only stable isotope
of this element, though it is, nevertheless, rare.

Sect. 10.2: Not only shou1d the rare process we accidentally called double beta
decay have been double proton decay, but the authors whose work we quoted as having
some bearing on the possible gradual change of the fine structure constant tell us
that it applies
only to two specific redshifts, 1.15 and 1.84 (not 1.88), and should not to be further
interpreted at this time. The other portion of their message leaves
us confused, as they say ``We never regarded Keck results as evidence for varying alpha.
Quite the opposite ... we stressed that the systematic shift is present in the Keck data
with a probability of 95\%.

Sect.~12.1 on getting Type Ia supernovae from white dwarfs in binary systems.
Efforts to determine who first proposed in print that mass transfer from a close companion
was the key process have failed, several of the pioneers having, charmingly, credited
each other. The winning candidate will have to precede both Whelan and Iben and
Hansen and Wheeler. A few more speculative comments from others dealt with (a) who
might have discovered general relativity if not Einstein (his own view was Langevin
for special relativity), and (b) who all had
a linear velocity-distance relation before Hubble (a third-hand review
appears in {\sl PASP} {\bf 108}, 1073.

Section 13.6. An astronomer with the gift of tongues points out that the nonce
word ``idoneous'' surely descends from the Latin ``idoneus'' (meaning suitable or fit)
via the Spanish Idoneo, who we always thought was a character in an opera by Mozart.
And the preferred spelling of the sort of mind we are frequently accused of being out of is
Fershlugginer, probably derived from an early German word (the modern one is
Verschlechtener) via Yiddish verschleuniger.  

\subsection{They Done It}

In previous years, we have classified these on the basis of whether words, numbers,
concepts, or logic had been twisted. Many examples this year seem to involve more than
one, and they are arranged, as it were, soup to nuts of a dog's dinner.  Relatively few
include an author's name, because it isn't always clear whether author, editor, or
typesetter had the last word. Item one is an exception.

``The scientific community expects to have too large a role in prescribing what
work NASA should do'' (Griffin 2006).

``ISS is like an old suitcase whose handle is missing $-$ it is totally useless,
but you just can't bear to part from it'' ({\sl Nature} {\bf 437}, 1214).

``Confirming the ongoing demand for observatory reports and justifying the council's
decision to discontinue the publication of observatory reports.'' ({\sl Bulletin of the
American Astronomical Society} {\bf 38}, 291).

``The heat of the sun is due to gravitation, not radium'' Lord Kelvin in {\sl Nature}
on 20 September 1906  (reproduced {\sl Nature} {\bf 443}, 279). Neither of the above, you 
might opine, so

``neither ... LTE nor non LTE model fits'' ({\sl A\&A} {\bf 455}, 315).

``LSST (the Large Synoptic Survey Telescope) will be able to peer back to the
beginning of time - 15 billion light years'' ({\sl Science} {\bf 310}, 777). Well, 
Einstein said space and time were not separable, and so

``speeding between the galaxies, the pulsar is moving at 110 km/hour'' ({\sl Astronomy
and Geophysics} {\bf 46}, 5.5) is presumably right in some coordinate system or reference
frame. {\sl Observatory Magazine}'s Here \& There column also picked up this one.

``oxygen abundance in the {\sl Sloan Digital Sky Survey}'' ({\sl A\&A} {\bf 453}, 487, 
title). It
is, anyhow, enough to keep a very large number of authors alive.

``the well-known Pskowski Phillips relation''  ({\sl MNRAS} {\bf 369}, 1949, abstract).
Sufficiently well known, it seems, that Pskowski is not cited.

``FK Comae Berenices, king of spin''  
({\sl ApJ} {\bf 644}, 464).  Surely at least this shou1d
have been queen of spin.

``APS member honored for intelligence'' {\sl American Physical Society News} {\sl 15},
No. {bf 4}, p. 5.

CaN is a molecule (\apj {\bf 634}, L201).  CaN DO is presumably two molecules.

``... normalized color of the dust inside the coma in the north-south direction is
measured to be $\approx 20-30\%/100$ nm''  ({\sl A\&A} {\bf 445}, 1151, abstract).

``Russia builds reliable spacecraft rockets''  ({\sl Nature}, {\bf 437}, 789, editor). 
Unfortunately not for the Solar Sail experiment.

A summary of the JENAM 2005 session on asteroseismoloxygy appears in {\sl EAS Newsletter}
No. {\bf 30}, p. 2.

``The decision to remove the section about the research left a well-balanced
paper,'' according to the President of the Canadian Medica1 Association, concerning a
censored editorial  ({\sl Nature} {\bf 440}, 10).

``I should like ... I acknowledge ...'' Ä {\sl ApJ} {\bf 648}, 1246, but the paper has 
10 authors.

``The peculiarity feature at 6200 \ang '' ({\sl A\&A} {\bf 454}, 171).

{\sl Physical Review Letters} {\bf 96}, 051102 recommends the use of ``established 
methods of seismic forecasting'' to predict violent space weather, ``on the grounds 
that quakes and solar flares share a similar powerlaw correlation.'' Our earthquake
insurance company will be very interested.

``Some bursts appear dark because their afterglow is faint''  ({\sl ApJ} {\bf 636}, 361, 
abstract).

sBzKs = massive star forming galaxies; pBzKs are passive  ({\sl ApJ} {\bf 638}, 72).

``we use a proper combination of ... observations'' ({\sl AJ} {\bf 131}, 2551, abstract)
clearly goes with ``data that spanned all sensible wavelengths'' ({\sl ApJ} {\bf 646}, 
642, introduction).

``EB thanks the Israeli Army for hospitality during the last month of this project''
({\sl MNRAS} {\bf 368}, 1716, acknowledgements). Other displaced authors include the 
one of {\sl ApJ} {\bf 640}, 801 who lists two current addresses 3700 miles apart  
(and presumably has lots of frequent flier miles); the chap who said ``I am now a 
chancellor of two fine universities''  ({\sl Nature} {\bf 441}, 691), the object of 
``an emeritus scientist with lab privileges is unusual'' at NIST ({\sl Science}
{\bf 312}, 1307), and the author of an {\sl MNRAS} preprint which was ``accepted 
2006 March 1; Received 2006 January 31; in original form 6 March 2006''.
She is apparently lost in time.

A few people who might have appeared in some other section include, Wilhelm
C.W.O.F.F. Wien, 1868-1928, who must have spent some appreciable fraction of those
60 years just learning his own name.  The dean retiring under a considerable cloud, of
whom another administrator said, ``Augusto has deserved all the rest and relaxation he
will now have in retirement'' (from the {\sl Los Angeles Times}). And Rupert Sheldrake,
whom {\sl Nature} had accidentially placed at Trinity College, Cambridge, ``I think he is a
former fellow of Claire, which should have undiluted credit'' (Rees 2006).

``Following the death of the plant or animal, the C$^{14}$ in its tissues will decay
to C$^{12}$'' ({\sl American Scientist} {\bf 94}, No.~1, p.~63), an example of double 
neutron decay?

``Does it really make a material difference whether references are arranged by
authors (alphabetically) or numerically?''  ({\sl Nature} {\bf 437}, 1232). Oh, only to the
grammatical structure of about every third sentence.

``An African fertilizer summit'' ({\sl Science} {\bf 312}, 31), and the Directed Energy
Directorate ({\sl AJ} {\bf 130}, 2262, footnote); the ``confusing scenario'' ({\sl A\&A}
{\bf 447}, 946) and
``the backlitting of QSOs'' ({\sl A\&A} {\bf 450}, 971), might have been phrased more 
elegantly.

But the following paper titles, we think, genuinely do not convey as much information
as they might about the subject matter:  ``The dark that didn't bark'' ({\sl MNRAS}
{\bf 368}, 1833); ``What did the horse swallow?'' ({\sl MNRAS} {\bf 369}, 120); 
``One ring to rule them all'' ({\sl A\&A} {\bf 455}, 953);
and ``The evidence of absence'' ({\sl MNRAS} {\bf 366}, 467).
That three of these four come from the same journal is perhaps just small number 
statistics.

A few source names could have been clarified: ``NGC 1344 is also known as NGC 1340''
({\sl ApJ} {\bf 635}, 290). IGR J16320-4751 = AX J1631.9-4752 could be precession 
({\sl MNRAS} {\bf 366},
274) but is perhaps more likely wavelength-dependent stubborness. And Alpha Persei =
33 Per = HD 20902 = HR1017 = SAO 038787 = HIP 15863 ({\sl PASP} {\bf 118}, 636) 
is really Markab, but you don't hear anybody call it that nearly as often as
Beta Per gets called Algol.
Conversely, as it were, in Figure 2 after page 573 of {\sl A\&A} {\bf 449}, 
the horizontal axis
numbers are 0, $5.0 \times 10^4$, $1.0 \times 10^5$ etc. and the label is ``star'', 
but they do not seem to be HD numbers.

And some other numbers that seem to have suffered in the translation: ``galaxies
in the field of redshift range $0.5 \lapprox z \lapprox 1.5$ ultra steep radio sources 
selected
from ...`` (appeared in {\sl MNRAS} {\bf 368}, but we have lost the page number); 
``four redshift bins over 0.1 Mpc $< z < 3.0$ Mpc'' ({\sl ApJ} {\bf 645}, 68, abstract); 
``It is becoming routine
to see patients over 227 kg in weight''  ({\sl Nature} {\bf 443}, xiii, Schwarz 2006) 
is clearly
political correctness for 500 lb run amok; 227 kg is anyhow a mass not a weight;
and ``ESA is the only space agency to have science operating around four planets: Venus,
the Moon, Mars, and Saturn''  ({\sl Nature} {\bf 440}, 975 and apparently returning to 
the Ptolemaic universe presented in {\sl Ap 05}, Sect.~5).

These seem to be qualitative rather than quantitative: ``a full set of partial
results can be accessed from ...'' ({\sl A\&A} {\bf 447}, 389, abstract); 
``strong MHD turbulence may
be of very small amplitude'' ({\sl ApJ} {\bf 638}, 811, footnote); a major merger 
is defined as a
merger between two galaxies with mass ratio 1:3 or lower ({\sl ApJ} {\bf 638}, 686, 
footnote).
And take a deep breath, because there are two parties involved before either you or
us on this one, an author and the reviewer of the book, the author trying to explain
acidification of seawater as atmospheric CO$_2$ level rises ``... sea water turning 
acidic,
its pH increasing by half to one unit'', and the reviewer noting ``this kind of error is
rather basic'' ({\sl Nature} {\bf 440}, 28).

The Making Lemonade award goes to STScI for its November 2006 workshop on
{\sl ``Astrophysics enabled by the return to the moon''}.

The sun rose this morning award goes to the press release from the National
Solar Observatory reporting that Solar cycle 24 officially began in June, 2006,
based on observations by their Synoptic Optical Long-term Investigations of the Sun
facility (SOLIS for short, we guess).

And our top ten of the year:

``Three of the systems have very bright primaries as a result of their high
temperatures and large radii'' ({\sl Acta Astron.} {\bf 56}, 127, abstract). Well, it was 
that or sigma or pi that had to be larger. Oh, or the four. We forgot the four.

``Once the diversity  of the microbial world is catalogued, it will make astronomy look
like a pitiful science'' ({\sl Nature} {\bf 438}, 384). But then consider the miserable 
state of cosmology, which has only one object to study.

``...rules out intrinsic X-ray weakness causing a lower detection rate of sources in the
X-ray surveys'' ({\sl A\&A} {\bf 446}, 87, abstract).

From a press release on the difficulties of observing Venus,``You can observe it for
about two hours at most. Then the sun rises and blinds the telescope, or Venus sets,
depending on the time of year.''  The bit indicating that the synodic period of Venus
is 365.34 days was, admittedly, outside the embedded quotation attributed to a planetary
scientist at New Mexico State University.

``Competition for a dwindling number of birds and bees is hindering pollination''
({\sl Nature} {\bf 439}, 380), not to mention sex education.               .

From an appeal for donations from a Well Known Educational Organization (where
your befuddled author flunked napkin rings) ``To qualify for Gift Aid, what you pay in
income tax or capital gains tax must equal the amount we will claim in the tax year.''
Equal or exceed was probably intended.

``A systematic approach to star formation and minority representation'' {\sl Science}
{\bf 312}, 1300 reporting a Cottrell Scholar Award.

From the last minute instructions for Prague IAU: ``As far as I know, their server
does not support direct use of ssh, so come prepared having putty in your notebook.''
We had always associated putty with window panes, but perhaps it could also have been
used to re-attach Pluto to the solar system.

``We demonstrate we have colored green the names that we have assumed are surnames.
If any of these are wrong, please let us know so that we can amend the tagging that
the log linear relation does not provide an adequate description'' ({\sl MNRAS} 
{\bf 365}, 1082), and yes the sentiments are separable.

And last, the quote you have been waiting for, a closing thought from the
astronomer who has provided our final words now for a number of years, ``I already think
on this for my small field of astrophysics and come to a conclusion that hier is
observed some calm! None novae at whole year, R CrB itself is very quite! All are
in future!''

\acknowledgements
As always, some brave colleagues have shared their thoughts on highlights of the
year and generally provided a nice deragement of epithets. We are duly and alphabetically
grateful to Eric Agol, Ivan Andronov, Alan Batten, Sidney van den Bergh, Max
Bonamente, Howard Bond, Jean Brodie, Kem Cook, Steven R. Cranmer, Francesca D'Antona,
Tatyana Dorokhova, Faustian Acquaintance, Luigi Foschini, Robert Gehrz, Howard Greyber,
Ethan Hansen,
Peter Harmanec, George Herbig, Bambang Hidayat , Robert S. Hill, John Huchra, Susana
Iglesias-Groth, Sebastian Jester, Fred M. Johnson, Stefan Jordan, Victoria Kaspi, Keen
Amateur Dentist, Vladislaw Kobychev, Michael Kurtz, Sergej Levshakov, Richard Lieu,
Oleg Malkov, Oliver K. Manuel, Stephen Maran, Medical Musician, Robert Nemiroff, Igor
Novikov, Thanu Padmanabhan, Richard Panek, Jonathan Pritchard, Mercedes Rich\-ards,
Nancy Roman, Alexander Rosenbush, Joshua Roth, Irakli Simonia, George Wallerstein, Spencer
Weart, J.~Craig Wheeler, Patricia Whitelock, Ralph Wijers, Mutlu Yildiz, and Barbarina
Zwicky. MJA and CJH acknowledge the NASA Astrophysics Data System
(ADS) and thank the numerous colleagues who provided preprints.
The work was partially supported by NASA contracts from the
TRACE, RHESSI, STEREO, and LWS TRT (Living With a
Star $-$ Targeted Research \& Technology) Programs.


\end{article}
\end{document}